\documentclass{ws-ijmpe-mod} 

\usepackage{graphicx}
\usepackage{epstopdf}  

\renewcommand\Re{\operatorname{\mathfrak{Re}}}
\renewcommand\Im{\operatorname{\mathfrak{Im}}}

\begin{document}

\markboth{S.\ Leupold, U.\ Mosel, V.\ Metag}{Hadrons in strongly interacting matter}

\catchline{}{}{}{}{}

\title{HADRONS IN STRONGLY INTERACTING MATTER
}

\author{\footnotesize STEFAN LEUPOLD$^a$, VOLKER METAG$^b$ AND ULRICH MOSEL$^a$\\stefan.leupold@theo.physik.uni-giessen.de}

\address{a: Institut fuer Theoretische Physik \\ b: II. Physikalisches Institut\\ Universitaet Giessen, Giessen, D-35392, Germany}

\maketitle


\begin{abstract}
We review the current status of theories and experiments aiming at an understanding and a determination of the properties of light vector and scalar mesons  inside strongly interacting hadronic matter. Starting from a discussion of the relevant symmetries of QCD and their connection with the hadronic description through QCD sum rules we then discuss hadronic models used to calculate the in-medium self-energies of hadrons and their spectral functions.
The difficulties to link these calculated properties to actual observables are emphasized. Finally, we review in detail all the running experiments searching for in-medium changes of vector and scalar mesons, both with relativistic heavy-ion reactions as well as with elementary reactions on (cold) nuclei.
Inconsistencies among experimental results are discussed.
While almost all experiments observe a considerable broadening of vector mesons
inside the nuclear medium, no evidence for mass changes is observed in the majority
of the experiments.
\end{abstract}
\tableofcontents

\section{Introduction}

It is nowadays commonly accepted that hadrons, the strongly interacting particles,
are composite objects made out of quarks and gluons. In principle, the dynamics
of the latter is governed
by Quantum Chromodynamics (QCD). In practice, the non-perturbative sector of QCD
which covers the formation of hadrons out of quarks and gluons is still not
fully understood, in particular on the quantitative level.

In vacuum, a single hadron
has specific properties like its mass and life time. It is rather
suggestive that the properties of a composite object are modified once it
is placed in a piece of matter (e.g., a nucleus) or in a heat bath (e.g., a
fireball emerging in a relativistic heavy-ion collision).
At least if the density of the system
is so high that the size of the composite object is comparable to the average
distance between the constituents of the system, then the intrinsic structure
of the composite object starts to play a role. Obviously one probes in that
way more and more the fundamental degrees of freedom, i.e.\ the quarks and gluons.

This interplay between hadrons on the one hand and quarks and gluons on the
other hand has identified the in-medium changes of hadrons as one of the key aspects
in understanding the non-perturbative sector of QCD. Consequently, the necessary
studies are both fascinating and demanding. Two issues become important here:
First, to make qualitative and quantitative predictions concerning the change of
properties of a given hadron, and second, to figure out how to measure these
changes in a situation where the detectors are, of course, outside of the dense
strongly interacting system. For the latter issue the finite life time of the
considered hadron becomes important. Only if the decay happens inside of the strongly
interacting matter one can hope to see an in-medium modification of the properties
of the hadron. In addition, it is most advantageous to produce in this decay
final state particles which leave the strongly interacting system without further interactions. In that way one can make sure that the information about the
in-medium modifications reaches the detectors without distortion. Otherwise
one needs a good understanding of the final-state interactions
(see the corresponding discussion in Sect.\ \ref{sec:hadspec-obs} below).

These considerations highlight the role of dileptons as good candidates for
final-state particles. Emerging from an electromagnetic process they can directly
couple to hadrons which have the quantum numbers of a photon. The hadrons which satisfy
this condition are the vector mesons which will be the main focus of the present review.
The fact that vector mesons provide interesting candidates concerning the observation
of in-medium modifications has already been stressed in the early eighties
of the last century.\cite{Pisarski:1981mq}
The field has been further stimulated in the early nineties by spectacular
predictions for drastic in-medium modifications --- most notably dropping masses,
as obtained from a scaling conjecture\cite{Brown:1991kk} and from QCD sum rule
considerations\cite{Hatsuda:1991ez}. It should be stressed, however, that
in particular for light vector mesons earlier calculations based on a quark model
did not yield significant mass shifts.\cite{Bernard:1988db}

Besides possible mass changes also modifications of the life times of nucleons and
pions in a thermal bath have been addressed
from a hadronic point of view in Ref.\ \refcite{Leutwyler:1990uq}
using the low-density expansion and chiral perturbation theory. The results indicate
significant widths but little mass modifications.
Subsequently, the scope has been widened by studying also other hadrons and by
considering more generally the spectral shape
of a given hadron.
Such questions can be tackled more easily from a hadronic point of view since
a change of the life time is due to modifications of the hadronic decay channels
(Pauli blocking, Fermi motion, \ldots) and the appearance of additional processes,
in particular ``collisional broadening'' due to hadron-hadron scattering
events.\cite{Bugg:1974cz}
In quark models a reliable description of such decay and scattering processes is
notoriously difficult. Nowadays it is common practice to use hadronic many-body
calculations with different degrees of sophistication to calculate hadronic
in-medium spectra. Nonetheless, the connection to quark-based approaches like
quark models or in particular QCD sum rules remains important to bring in the
aspects of the fundamental degrees of freedom hidden inside the hadronic composite objects.

A very recent review\cite{Hayano:2008vn}, similar in spirit to the present one, covered a very broad range of in-medium effects with an emphasis on a discussion of the various experiments. Here --- in contrast --- we concentrate on the light vector and scalar mesons. Other recent reviews include Refs.\ \refcite{Cassing:1999es,Rapp:1999ej,Rapp:2009yu} which all deal with particular aspects of this topic, but are mostly limited to the discussion of in-medium effects in heavy-ion collisions while we discuss experiments both with heavy-ion and with elementary probes, mainly photons.

It is the purpose of the present work to review the current status of the physics
of in-medium modifications of hadrons both from the theoretical and the experimental side.
We will cover the range from fundamental QCD-based considerations via hadronic models
and the identification of observables finally to the actual observations.
We, therefore, start in Sect.\ \ref{sec:leupold} with a discussion of the relevant symmetries of Quantum Chromodynamics (QCD) and then link them to observable properties of hadrons via the QCD sum rule method. One main result of this discussion is that QCD sum rules can constrain hadronic in-medium properties, but they cannot unambiguously predict them.

The following Sect.\ \ref{sec:mosel}, therefore, outlines the methods used to calculate hadronic in-medium properties using ``conventional'' hadronic interactions. We point out that the nucleon-resonance excitations play a dominant role in determining the in-medium properties of vector mesons through their coupling to the meson-nucleon final states. This is the same physics as that determining the Delta-hole model that was so successful in explaining the properties of pions and Delta excitations in nuclei (for a review see Ref.\ \refcite{Oset:1981ih}).

A relatively short Sect.\ \ref{sec:mosel2} then deals with the connection between the calculated spectral functions and actual observables. Since the in-medium properties of a meson have to be inferred from measurements of its decay products, the latter should reach the detector with as little distortion by the medium as possible. From this point of view ideal probes are  dileptons ($e^+ e^-$ or $\mu^+ \mu^-$) which, however, suffer from the small electromagnetic coupling constants. On the other hand, hadronic final states can get distorted by the surrounding nuclear medium; in this case a realistic, state-of-the-art treatment of the final-state interactions is mandatory.

In Sect.\ \ref{sec:exp} we finally discuss at some length the various experiments searching for in-medium properties of scalar and vector mesons. Here we discuss in detail experimental methods, limitations and successes of both heavy-ion collisions and nuclear reactions with elementary probes.
Consistencies and discrepancies between different experimental results are evaluated.
The importance of detector acceptances for the comparison of experimental data and for the sensitivity to theoretical predictions is emphasized.
We find that the results from ultra-relativistic heavy-ion collisions are compatible with those from photon-induced reactions: both classes of experiments show a significant broadening of vector mesons in the medium, but --- so far --- no mass shift has been established.
We provide an outlook on expected results from presently running and planned future experiments.

Finally our conclusions are presented in Sect.\ \ref{sec:conclusion}.

\section{Symmetries of QCD, Hadron Properties, and QCD Sum Rules}
\label{sec:leupold}

In the present section we review our current understanding of the symmetry
pattern of QCD and the expected changes in a strongly interacting medium,
the connection to the in-medium properties of hadrons,
and the bridge between hadrons and QCD condensates based on QCD sum rules.

\subsection{Symmetry pattern of QCD}
\label{sec:sympat}

Quantum chromodynamics (QCD), the by now accepted theory of the
strong interaction,\cite{Peskin:1995ev}
has two faces: At high enough energies one can observe quarks and gluons as active
degrees of freedom,
e.g.\ in deep inelastic scattering and in jet production.
The quarks and gluons also constitute the elementary building blocks which enter the
QCD Lagrangian. On the other hand, it is not possible to isolate single quarks and
gluons. Instead, at low energies one observes hadrons
which are understood as composite objects formed by the elementary building blocks.
The quarks and gluons are {\it confined} within hadrons.

\subsubsection{Center symmetry}
\label{sec:cent}

Confinement is related to the approximate center symmetry of
QCD.\cite{'tHooft:1977hy,'tHooft:1979uj}
A non-abelian
gauge group governs the theoretical description of the interaction between
quarks and gluons. If one restricts
the Lagrangian to static quarks, the QCD action possesses a symmetry with respect to the
center of this non-abelian gauge group. For a system at finite temperature $T$ one can
introduce the Polyakov loop\cite{Polyakov:1978vu}
\begin{equation}
\label{eos-leupold-eq:Polyakov}
L(\vec x) = \mbox{Tr} \ {\cal T} \exp
\left[ i \int\limits_0^{1/T} dt \, A_0(t,\vec x) \right]
\end{equation}
where the ``time'' integral concerns a Euclidean
time,\footnote{One makes use here of the formal similarity between a time evolution
operator $\exp(iHt)$ and a finite-temperature density
operator $\exp(-H/T)$.\protect\cite{Matsubara:1955ws}}
${\cal T}$ denotes time ordering of the exponential, $A_0$ is the gluon field
and Tr denotes the trace over the
color indices. In general, the Polyakov loop is not invariant with respect to the center
symmetry. Under a center transformation an additional phase is obtained,
\begin{equation}
  \label{eq:poltrafo}
  L \to e^{2\pi i n/3} \, L \qquad \mbox{with} \quad n \in \{1,2,3\}  \,.
\end{equation}

One can show that the thermal expectation value of the Polyakov loop
is related to the free energy $F$ of a single static quark,\cite{Polyakov:1978vu}
\begin{equation}
\label{eos-leupold-eq:trafoPol}
  \langle L \rangle = \exp(- F/T)  \,.
\end{equation}
This aspect connects the center symmetry to confinement: If the expectation value of $L$
vanishes, the phase in (\ref{eq:poltrafo}) does not matter. Both the QCD action and the
system are center symmetric. In this case the free energy of a single quark is
infinite. In other words, one needs an infinite amount of energy to isolate a static
quark from the rest of the system, i.e.\ quarks are confined.
In particular such a statement concerns the
system with vanishing temperature, the vacuum:
If the center symmetry is also realized for the ground state,
the vacuum, and not only for the QCD action, then the energy of a single isolated
quark is infinitely large. This would yield
an explanation for the confinement mechanism. In reality, however,
i.e.\ for non-static quarks, things are more complicated. The center symmetry is
explicitly broken by the finite, i.e.\ non-infinite, quark masses. It is not so obvious
how good the static-quark approximation actually is. Phenomenologically, one observes
the following: If one tries
to tear away a single quark from the rest of a hadron, it does not become more
and more expensive in energy. Instead, the hadron splits into two or several hadrons
as soon as there is enough energy for the additional hadron(s) to be generated. In that
way confinement is dynamically realized. Nonetheless, the center symmetry provides
qualitative insight and is an
important aspect for theoretical descriptions within lattice QCD.\cite{Karsch:2001cy}
For further details we refer to the literature.\cite{Holland:2000uj}

\subsubsection{Chiral symmetry}
\label{sec:chir}

In the following we discuss in some depth a second approximate symmetry
which significantly influences
the low-energy face of QCD. For massless quarks, QCD exhibits
a chiral symmetry:\cite{Scherer:2002tk,MoselBuch}
The interaction is the same for left- and right-handed quarks
and for different flavors. In reality, the masses of up and down quarks are very
small as compared to typical hadronic scales.\cite{Amsler:2008zzb}
Also the mass of the strange quark
is moderately small. In the following, we mainly restrict our discussion to the two
lightest flavors.\footnote{When we come to our main subject of in-medium modifications
we often concentrate on cold nuclear matter or nuclei as the strongly interacting medium.
Here a flavor symmetry including
strange quarks is anyway heavily broken explicitly by the medium which contains nucleons
but no strange baryons.}
For the hadronic world it is more convenient to understand
the combined left- and right-handed flavor symmetries as an invariance with respect
to vector and axial-vector flavor transformations.\footnote{We do not discuss here
the axial non flavor-changing $U_A(1)$ transformations. They constitute a symmetry
of the classical QCD Lagrangian, but not of the quantum field theory emerging from
it. For a discussion of this $U_A(1)$ anomaly we
refer to Ref.\ \protect\refcite{Peskin:1995ev}.}
For two flavors the vector
symmetry is just the isospin symmetry which leads to the appearance of isospin
multiplets in the hadronic spectrum. In contrast, there are no (approximately)
degenerate single-hadron states which correspond to the axial-vector symmetry.
This symmetry would
demand the existence of ``chiral multiplets''
where the members of one multiplet possess different parity. However,
there are, e.g., no known hadrons with negative parity and the mass of the nucleon
or with positive parity and the mass of the pion or $\rho$ meson.\cite{Amsler:2008zzb}
One explains this
finding by the conjecture that chiral symmetry is spontaneously broken --- on top of
the explicit breaking by the finite quark masses. In other words, the vacuum of QCD
is in the symmetric Wigner-Weyl phase concerning the vector flavor symmetry and in the
symmetry-broken Nambu-Goldstone phase concerning the axial-vector flavor symmetry.
Formally:
\begin{equation}
  Q_V^a \vert 0 \rangle = 0  \,, \qquad Q_A^a \vert 0 \rangle \neq 0 \,,
  \label{eq:breaking-nonbr}
\end{equation}
with the QCD vacuum $\vert 0 \rangle$, the isospin charge
\begin{equation}
  Q_V^a := \int \! d^3x \, j_V^{0\,a}(x)  \,,
  \label{eq:def-iso-charge}
\end{equation}
the axial isospin charge
\begin{equation}
  Q_A^a := \int \! d^3x \, j_A^{0\,a}(x)  \,,
  \label{eq:def-ax-iso-charge}
\end{equation}
and the corresponding currents
\begin{equation}
  j_V^{\mu\,a} := \bar q \, \gamma^\mu \, \tau^a \, q \,, \qquad
  j_A^{\mu\,a} := \bar q \, \gamma^\mu \, \gamma_5 \, \tau^a \, q \,,
  \label{eq:def-sym-currents}
\end{equation}
defined in terms of the quark fields $q = (u,d)$ and the
isospin\footnote{A generalization to more than two flavors is
straightforward.} matrices $\tau^a$.

A spontaneously broken symmetry implies the existence of massless modes, the
Nambu-Goldstone
bosons. However, chiral symmetry is not only
spontaneously, but also
explicitly broken. The light-quark masses are non-vanishing, but small compared
to hadronic scales.\cite{Amsler:2008zzb}
As a consequence the Nambu-Goldstone bosons
need not be exactly massless, but should be light compared to all other
excitations. Indeed, the pions satisfy this requirement\cite{Amsler:2008zzb}
and are consequently interpreted as the Nambu-Goldstone bosons of the
spontaneously broken chiral symmetry of QCD.

\subsubsection{Order parameters of chiral symmetry breaking}
\label{sec:orderchiSB}

A phase with a broken symmetry can be characterized by an order parameter.
Typically one can find several order parameters which might differ in their
usefulness in practice. Due to the two faces of QCD one can introduce order
parameters at the quark-gluon and at the hadronic level. A much celebrated
order parameter at the fundamental level is
the (two-)quark condensate\cite{GellMann:1968rz}
\begin{equation}
  \label{eq:def-2quark}
  \frac12 \, \langle 0 \vert \bar q q \vert 0 \rangle =
  \frac12 \, \left(\langle 0 \vert \bar u u \vert 0 \rangle
    + \langle 0 \vert \bar d d \vert 0 \rangle \right) \approx
  \langle 0 \vert \bar u u \vert 0 \rangle \approx
  \langle 0 \vert \bar d d \vert 0 \rangle
  \neq 0 \,.
\end{equation}
Below we introduce also other condensates which characterize the QCD vacuum.
Only some of them are related to the chiral symmetry of QCD. It is easy to show
that the operator $\bar q q$ which appears in (\ref{eq:def-2quark}) is not invariant
with respect to chiral --- to be specific --- axial-vector flavor transformations.
Thus, the quark condensate can only be non-vanishing, if the chiral symmetry is
spontaneously broken.

On the hadronic level an appropriate order parameter is the pion-decay constant
$f_\pi \approx 92$ MeV.
More generally, this decay constant is defined for a pseudoscalar state (ps)
with momentum $k$
and isospin characterized by the index $b$ in the following way:
\begin{equation}
  \label{eq:defpiondec}
  \left\langle 0 \vert j_A^{\mu \, a} \vert
    {\rm ps}^b(k) \right\rangle = i\, k^\mu \, f_{\rm ps} \, \delta^{ab}  \,,
\end{equation}
with the axial-vector current defined in (\ref{eq:def-sym-currents}) above.
In QCD one finds the partial conservation of the axial-vector current (PCAC):
\begin{equation}
  \label{eq:cons-ax-vec}
  \partial_\mu  \, j_A^{\mu \, a} =
  i \, \bar q \, \{ {\cal M} , \tau^a \} \, \gamma_5 \, q \,.
\end{equation}
Here the curly brackets denote the anticommutator and the quark-mass matrix is given by
\begin{equation}
  \label{eq:qmmat}
  {\cal M} = \left(
  \begin{array}{cc}
    m_u & 0 \\ 0 & m_d
  \end{array}\right)  \,.
\end{equation}
Using (\ref{eq:defpiondec}) and (\ref{eq:cons-ax-vec}) it is easy to see that the
combination $M_{\rm ps}^2 \, f_{\rm ps}$ is proportional to the quark masses, i.e.\
vanishes in the chiral limit:
\begin{equation}
  \label{eq:vancomb}
  M_{\rm ps}^2 \, f_{\rm ps} = 0 \qquad \mbox{for} \quad {\cal M} \to 0  \,.
\end{equation}
This relation holds irrespective whether the chiral symmetry is broken or not.
However, which of the two quantities $M_{\rm ps}$ or $f_{\rm ps}$ actually vanishes
depends on the phase.
If there is no massless pseudoscalar
state in the spectrum, the decay constants of all pseudoscalar states vanish ---
this is the Wigner-Weyl phase, i.e.\ the chirally restored phase. On the other
hand, in the chirally broken Nambu-Goldstone phase, there must be a massless pseudoscalar
state (in the chiral limit), the pion. Here, the corresponding decay constant,
$f_\pi$, does not vanish. Consequently, the pion-decay constant is an order parameter
of chiral symmetry breaking.

As a consequence of PCAC, the two order parameters identified so far are actually
related by the
Gell-Mann--Oakes--Renner relation:\cite{GellMann:1968rz}
\begin{equation}
  \label{eq:GOR}
  f_\pi^2 \, M_\pi^2 \approx - \bar m_q \, \langle 0 \vert \bar q q \vert 0 \rangle
  \approx - 2 \, \bar m_q \, \langle 0 \vert \bar u u \vert 0 \rangle
  \approx - 2 \, \bar m_q \, \langle 0 \vert \bar d d \vert 0 \rangle
  \,,
\end{equation}
which holds up to corrections of higher powers in the quark masses.
Here $\bar m_q$ denotes an averaged quark mass.

\subsubsection{In-medium changes of the symmetry pattern}
\label{sec:changespatt}

For a symmetry which is spontaneously broken in the vacuum, one typically finds
that it becomes restored at some finite temperature. More generally, we expect
changes in the symmetry pattern of QCD, if the vacuum is replaced by strongly
interacting matter. Corresponding to the exact conservation laws for energy and
baryon number we characterize such a medium in the following discussion by
temperature and baryo-chemical potential. For large temperatures the typical momenta
and therefore also the typical momentum exchanges in reactions are large.
Consequently, quarks and gluons should become relevant degrees of freedom. Indeed,
lattice-QCD calculations suggest that for temperatures larger than about $200\,$MeV
a quark-gluon plasma (QGP) is formed.\cite{Karsch:2001cy,Aoki:2006we}
We recall that in the vacuum the center
symmetry is
realized (confinement), while the chiral symmetry is spontaneously broken
(non-degenerate chiral multiplets, ``massless'' pions). At high temperatures, i.e.\ in
the QGP, this picture is reversed:\cite{Karsch:2001cy}
The center symmetry becomes spontaneously broken
leading to deconfinement. Chiral symmetry is restored. Correlation functions
(cf.\ Sect.\ \ref{sec:cur} below) related by chiral transformations become
degenerate. Due to deconfinement they might not necessarily contain quasi-particle
structures any more (cf.\ Ref.\ \refcite{Leupold:2008ne} and Sect.\ \ref{sec:cur} below).
The situation is summarized in Table \ref{tab:sym}.
\begin{table}
  \centering
  \begin{tabular}{|l|c|c|}
    \hline
    & vacuum & high temperature  \\
    \hline \hline
    {\bf chiral symmetry} & broken & restored  \\
    \hline
    $\begin{array}{l}
      \mbox{order parameters:} \\ \mbox{ two-quark condensate,} \\
      \mbox{ pion-decay constant}
    \end{array}$
    & non-vanishing & vanishing  \\
    \hline
    consequences &
    $\begin{array}{l}
      \mbox{existence of Nambu-} \\ \mbox{ Goldstone bosons,} \\ \mbox{non-degenerate spectra} \\
      \mbox{ of chiral partners}
    \end{array}$
    &
    $\begin{array}{l}
      \mbox{no Nambu-Goldstone} \\ \mbox{ bosons,} \\ \mbox{degenerate spectra} \\
      \mbox{ of chiral partners}
    \end{array}$ \\
    \hline \hline
    {\bf center symmetry} & unbroken & broken \\
    \hline
    $\begin{array}{l}
      \mbox{order parameter:} \\ \mbox{ Polyakov loop}
    \end{array}$
    & vanishing & non-vanishing \\
    \hline
    consequence & confinement & deconfinement  \\
    \hline
  \end{tabular}
  \caption{Symmetry pattern of QCD.}
  \label{tab:sym}
\end{table}

So far, lattice-QCD calculations are restricted to small baryo-chemical potentials.
Therefore, at low temperatures, but large baryo-chemical potentials it is more
speculative
which properties the system has and how the symmetry pattern changes.
Qualitatively, one can also expect here that with increasing density the momentum
transfers become larger, involving quarks as active degrees of freedom. The role
of real gluons is suppressed at low temperatures. They do play a role, however,
as exchange particles. The quarks form a Fermi sea. Their interaction
is largely suppressed by Pauli-blocking effects. Consequently, the most interesting
processes happen close to the Fermi surface. Here, an attractive interaction
leads to an instability which drives the system in a (color)
superconducting phase.\cite{Rapp:1997zu,Alford:1997zt}
For further considerations, one has to
become more quantitative. At least for asymptotically high chemical potentials
one can reliably use QCD perturbation theory to determine the interaction between
quarks near the Fermi surface. The dominant attractive interaction leads to a special
kind of color superconductivity, the
so-called color-flavor locking (CFL) phase involving up, down and strange quarks on equal
footing.\cite{Alford:1998mk} Here, chiral symmetry is also spontaneously
broken, albeit quantitatively the breaking is smaller than in
vacuum.
Interestingly, the excitations in that particular phase resemble the QCD vacuum
excitations, i.e.\ the hadrons. Therefore, it has been speculated that no real phase
transition is required, if there is a direct transition from the low-density hadronic
to the high-density CFL state.\cite{Schafer:1998ef} However, there are also
alternative scenarios, since it is not clear down to which densitities the perturbative
QCD calculations can be trusted. There might be other color superconducting phases
in between the hadronic and the CFL matter. In particular, in a two-flavor color
superconducting phase, chiral symmetry would be restored. Recently, the existence of a
quarkyonic phase has been suggested.\cite{McLerran:2007qj} There, the excitations
are of hadron type, i.e.\ quarks are confined, but chiral symmetry is restored.
In any case, it is suggested that chiral symmetry breaking becomes lifted or at least
significantly weakened in a dense strongly interacting medium,
no matter whether high (energy) density
is achieved by large temperatures or large chemical potentials or both.

\subsubsection{In-medium changes of the order parameters}
\label{sec:changeord}

Already at low densities one finds a drop of the order parameters of chiral symmetry
breaking. If the density is low enough, one can neglect the interactions
between the medium constituents. The order parameter can be regarded as a probe
which interacts independently with the states which form the medium.
We discuss two special cases for which one can make exact, model independent
statements:
(i) A baryon-free low-temperature system described by a gas of pions and
(ii) cold nuclear
matter described by a Fermi sphere of nucleons.
For the first system it is assumed that
the temperature is so low that no other particles are significantly
populated\footnote{Note that the pions are the lightest degrees of freedom due to their
Goldstone-boson character.} and that the pions do not interact. Actually the latter
assumption can be dropped since the interaction between the pions can be systematically
included using chiral perturbation theory.\cite{Gerber:1988tt}
For the second system we assume that the temperature is
exactly zero and that the baryo-chemical potential, i.e.\ the Fermi energy, is lower
than the mass of the first excited baryon. Thus, the system contains only nucleons.
Their interaction can be neglected for not too large densities.

We discuss the
change of the two-quark condensate and of the pion-decay constant: The in-medium
two-quark condensate is just given
by $\langle \Omega \vert \bar q q \vert \Omega \rangle$ with the state
$\vert \Omega \rangle$ characterizing the in-medium system. The in-medium
generalization of the pion-decay constant defined for the vacuum case
in (\ref{eq:defpiondec}) requires some care, since the medium
introduces an additional Lorentz vector. Following Ref.\ \refcite{Meissner:2001gz}
we use for the definition the zeroth
component of the axial-vector current, since it is directly related to the charge
(\ref{eq:def-ax-iso-charge}) which in turn governs the chiral transformations:
\begin{equation}
  \label{eq:deffpimed}
\langle \Omega \vert j_A^{0 \,a} \vert \pi^b(k) \rangle =
i \, k^0 \, \delta^{ab} \, f_\pi^{\rm med} \,.
\end{equation}

For a pion gas described by chiral perturbation theory the change of the two-quark
condensate has been evaluated
in Ref.\ \refcite{Gerber:1988tt} up to three loops.\footnote{``One-loop'' corresponds to
non-interacting pions.} The result in lowest order in the pion density (i.e.~neglecting
interactions) is
\begin{eqnarray}
\frac{\langle \Omega \vert \bar q q \vert \Omega \rangle_{{\rm pionic\; med.}}}%
{\langle 0 \vert \bar q q  \vert 0 \rangle}
& = &
1 - \frac{ \rho_\pi}{f_\pi^2}  \,,
  \label{eq:fintemp2q}
\end{eqnarray}
with the scalar pion density
\begin{equation}
  \label{eq:piondens}
\rho_\pi = 3 \int \! \frac{d^3 k}{(2 \pi)^3 \, 2 E_k} \, \frac{1}{e^{E_k/T} - 1} \;
\stackrel{M_\pi \to \, 0}{\longrightarrow} \; \frac{1}{8} \, T^2 \,,
\end{equation}
the temperature $T$ and the pion energy $E_k = \sqrt{\vec k^2 + M_\pi^2}$.

The leading-order change of the pion-decay constant with temperature
is given by:\cite{Gasser:1986vb,Gasser:1987zq}
\begin{equation}
  \label{eq:fpitempdep}
  f_\pi^{{\rm pionic\; med.}} = f_\pi \,
  \left(1-\frac{2 \rho_\pi}{3 f_\pi^2} \right) \,.
\end{equation}
Naively, one might conclude from the Gell-Mann--Oakes--Renner relation (\ref{eq:GOR})
that the square of the pion-decay constant could scale with the two-quark condensate.
As can be seen from (\ref{eq:fintemp2q},\ref{eq:fpitempdep}) this is not the case.
$f_\pi^2$ drops faster by a factor $4/3$ for low densities. At the same time the in-medium
pion mass increases such that at the temperatures considered in
Ref.\ \refcite{Gasser:1986vb} the
Gell-Mann--Oakes--Renner relation (\ref{eq:GOR}) still holds.

Cold nuclear matter is characterized by its density
\begin{equation}
  \label{eq:nucldens}
\rho_N = 4 \int \! \frac{d^3 k}{(2 \pi)^3} \, \Theta(k_F - \vert \vec k \vert)
\end{equation}
with the Fermi momentum $k_F = \sqrt{\mu^2-m_N^2}$ and baryo-chemical potential $\mu$.
For the in-medium change of the two-quark condensate
one obtains\cite{Drukarev:1991fs}
\begin{equation}
\label{eq:dropfindens}
\frac{\langle \Omega \vert \bar q q \vert \Omega \rangle_{{\rm nucl.\; med.}}}%
{\langle 0 \vert \bar q q \vert 0 \rangle}
 =
1- \frac{\rho_s \, m_N \, \sigma_N}{ f_\pi^2 \, M_\pi^2}
\approx 1 - \frac{1}{3} \frac{\rho_N}{\rho_0}
\end{equation}
with $\rho_0$ denoting nuclear saturation density and with the scalar nucleon density
\begin{equation}
  \label{eq:scalnucl}
\rho_s = 4 \int \frac{d^3 k}{(2\pi)^3} \,
\frac{1}{\sqrt{\vec k^2+m_N^2}} \, \Theta(k_F - \vert \vec k \vert)
\approx \frac{\rho_N}{m_N}  \,,
\end{equation}
where the last relation holds for small enough densities. The nucleon sigma term,
\begin{equation}
  \label{eq:defnuclsig}
\sigma_N =   \bar m_q \, \frac{d m_N}{d \bar m_q} \approx 45\,{\rm MeV} \,,
\end{equation}
which appears in (\ref{eq:dropfindens}) can be obtained within the framework of
chiral perturbation theory from low-energy $\pi N$ scattering.\cite{Gasser:1990ce}

\begin{figure}[ht]
  \begin{center}
    \includegraphics[keepaspectratio,width=0.35\textwidth]{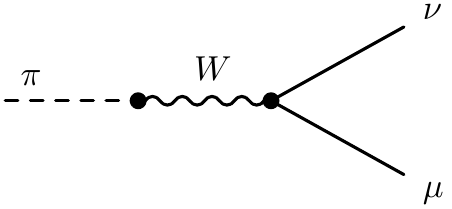}  \\[2em]
    \includegraphics[keepaspectratio,width=0.7\textwidth]{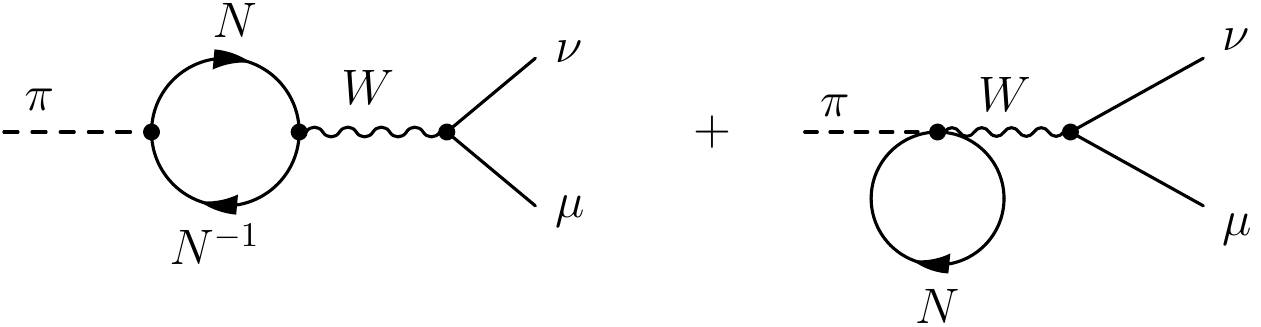}
  \end{center}
  \vspace*{8pt}
  \caption{Pion decay into muon and neutrino in vacuum (top) and in a nucleonic medium
    (bottom). See main text for details.}
  \label{fig:piondec-vac-med}
\end{figure}

For cold nuclear matter the leading-order change of the
pion-decay constant is related model independently to
some low-energy constants of pion-nucleon scattering. Graphically this is depicted
in Fig.\ \ref{fig:piondec-vac-med}. We stress that
the additional processes depicted in the lower part of Fig.\ \ref{fig:piondec-vac-med}
are just the scattering of a pion on a nucleon from the medium into a weak current
and a nucleon, i.e.\ standard hadronic effects which can be determined model independently
within chiral perturbation theory.
The change of the pion-decay constant has been calculated,
e.g., in Ref.\ \refcite{Meissner:2001gz} with the numerical result:
\begin{equation}
  \label{eq:fpidensdep}
f_\pi^{{\rm nucl.\; med.}} =
f_\pi \, \left(1-\frac{\rho_N}{\rho_0} (0.26 \pm 0.04) \right) \,.
\end{equation}
Again, we observe that the square of the pion-decay constant drops faster than
the quark condensate (cf.\ also Ref.\ \refcite{Kwon:2008vq}).

We conclude that already in a hadronic medium the order parameters of chiral
symmetry breaking start to drop. It is important to stress that at least this onset
of chiral restoration is caused by standard hadronic interactions.
It is the main purpose of Fig.\ \ref{fig:piondec-vac-med} to illustrate that point.
The corresponding statement holds also for the finite-temperature
effect in (\ref{eq:fpitempdep}). Hence chiral restoration does not happen on top
of ordinary hadronic effects. At least at low densities it is the presence of the
hadronic medium constituents which is responsible for the onset of changes of the
symmetry pattern of QCD.

\subsection{Hadronic and current-current correlation functions}
\label{sec:cur}

Before we discuss further consequences of symmetry changes on hadronic properties
we introduce a proper framework which is suitable to characterize hadrons and to make
contact with QCD. The central objects are
correlation functions. Corresponding to the two faces of QCD we introduce
hadronic and current-current correlation functions. The currents which enter the latter
are formed from quark (and, if necessary, gluon) fields.

\subsubsection{Hadronic correlation functions}
\label{sec:hadcur}

We start with the description of a single hadron in the hadronic language.
We concentrate on mesons in
the following, but the extension to baryons is straightforward.
It is convenient to introduce
a field $h(x)$ with the quantum numbers of the hadron under consideration.
The properties of the meson can be obtained from the retarded hadronic correlation
function or propagator $D_{\rm ret}$.\footnote{In vacuum one might use the
time-ordered correlator as
well. In a medium the appropriate object to characterize {\em one} state is the
retarded correlator. The time-ordered object contains information about the state
{\em and} about its population.\protect\cite{Danielewicz:1982kk}}
Schematically, i.e.\ suppressing all other quantum numbers besides
the space-time information, one has
\begin{equation}
  \label{eq:hadrcor}
  D_{\rm ret}(q) = i \, \int \! d^4x \; e^{iqx} \, \Theta(x_0) \,
  \langle \Omega \vert
    [h(x),h(0)]
  \vert \Omega \rangle  \,.
\end{equation}
We recall that $\Omega$ characterizes the in-medium system. The retarded correlator
is analytic in the upper half of the complex energy plane. It satisfies a dispersion
relation
\begin{equation}
  \label{eq:disp-hadr}
  D_{\rm ret}(q) = \int\limits_{-\infty}^{+\infty}  dk_0 \;
  \frac{{\cal A}(k_0,\vec q)}{q_0 - k_0 + i\epsilon}  \,,
\end{equation}
where we have introduced the spectral function
\begin{equation}     \label{Aspect}
{\cal A} = -\frac{1}{\pi} \, {\rm Im}D_{\rm ret}~.
\end{equation}
It is essentially given by the Fourier transform of the commutator $[h(x),h(0)]$.

The
important message from (\ref{eq:disp-hadr}) is that all information is contained in
the spectral function. A single hadron appears as a quasi-particle peak in the spectral
function. From the peak position and the peak width one can obtain mass and life time
of the hadron. This extraction is fairly unambiguous as long as the life time is long,
i.e.\ the width is small (as compared to the mass). For large widths it is
insufficient to characterize the hadron by mass and life time. The whole spectral
shape becomes important. A large width implies that there are other (hadronic) states
which couple strongly to the considered hadron. These other states can also interact
without exciting this hadron described by $h$. This causes a
physical background which interferes on the amplitude level with the resonant amplitude.
Actually the hadron field $h$ can be redefined such that part of the background is
included. In other words, background can be shifted back and forth and the
choice of
the field $h(x)$ to characterize the hadron of interest is not unique
(see, e.g., Ref.\ \refcite{Ecker:1989yg} concerning vector mesons).
Therefore, strictly speaking, the spectral function changes if a different field is
chosen (see also, e.g., Ref.\ \refcite{Fearing:1999fw}).
Again, this
is of no concern as long as the hadron is a quasi-particle state, i.e.\ if the width
is sufficiently small. From a theoretical point of view there is no problem with the
mentioned ambiguities. The spectral function provides a tool to calculate observable
quantities. The latter, of course, are independent of the chosen field.

In a medium the spectral function can become more complicated. In many
hadronic models one typically finds significant increases of the widths of hadrons.
In addition, several models predict additional structures in the spectral functions
caused by many-body effects like resonance-hole
excitations.\cite{Friman:1997tc,Rapp:1997fs,Lutz:2001mi,Post:2003hu,Riek:2004kx,Muehlich:2006nn,vanHees:2007th}

\subsubsection{Current-current correlation functions}
\label{subsec:curcurcor}

We now turn to the description of a hadron and its properties in the
quark-gluon language. Here, one introduces a current with the quantum numbers
of the hadron of interest.\cite{Shuryak:1993kg}
As an example we discuss the $\rho^0$ meson. A
generalization to other hadrons is straightforward. A quark
current with the quantum numbers of the $\rho^0$ is given
by\cite{Shifman:1978bx,Shifman:1978by}
\begin{equation}
  \label{eq:what-defjrho}
j^\mu = \frac{1}{2} \left( \bar u \gamma^\mu u - \bar d \gamma^\mu d \right) \,.
\end{equation}
We note in passing that this is one of the symmetry currents
appearing in (\ref{eq:def-sym-currents}).
In complete analogy to the formalism described in (\ref{eq:hadrcor},\ref{eq:disp-hadr})
one introduces the (retarded) current-current correlation function
\begin{equation}
  \label{eq:curcur}
  \Pi^{\mu\nu}_{\rm ret}(q) = i \, \int\! d^4x \; e^{iqx} \, \Theta(x_0) \,
  \langle \Omega \vert
   [j^\mu(x), j^\nu(0)]
  \vert \Omega \rangle  \,.
\end{equation}
A dispersion relation similar to (\ref{eq:disp-hadr}) can be formulated. In other
words, all information is contained in ${\rm Im}\Pi_{\rm ret}$ which is
called ``spectral distribution''.\footnote{We do not use the phrase ``spectral function''
in the present section, to avoid mixing up hadronic and current correlators.
Note that in a free-field
theory the spectral function satisfies a normalization condition whereas the spectral
distribution does not.}
All hadronic objects which have the quantum numbers
specified by the current enter this spectral distribution. Typically there are
several hadrons with the same
quantum numbers. For example, there are excited mesons with the quantum numbers
of the $\rho^0$ at about $1450\,$MeV and $1700\,$MeV.\cite{Amsler:2008zzb} A given current
couples to several hadrons (and of course also to hadronic many-body states with the
appropriate quantum numbers). Hence, one can only obtain information about a single
hadron from the corresponding current-current correlator, i.e.\ the spectral distribution,
if it is sufficiently far away from the other excitations with the same quantum
numbers. In a strongly interacting medium many-body excitations become more important.
The same effects which change the spectral function of a single hadron, of course,
also influence the corresponding spectral distribution. Peaks which appear in the
vacuum spectral distribution typically become broader. They also can get shifted.
Additional structures might show up.

For special currents the spectral distribution of the
current-current correlator is directly accessible by experiment:
The currents
\begin{equation}
  \label{eq:el-cur}
  j^\mu_{\rm el} = \frac23 \, \bar u \gamma^\mu u - \frac13 \, \bar d \gamma^\mu d
  + \ldots =
  \underbrace{\frac12 (\bar u \gamma^\mu u - \bar d \gamma^\mu d)}_{\mbox{isospin 1}} +
  \underbrace{\frac16 (\bar u \gamma^\mu u + \bar d \gamma^\mu d)+  \ldots}_{\mbox{isospin 0}}
\end{equation}
and
$j^{\mu\,a}_{\rm weak} = j^{\mu\,a}_V-j^{\mu\,a}_A$
are involved in the electromagnetic or weak interaction, respectively. Here the dots
denote the contributions of other quark flavors and the currents defined in
(\ref{eq:def-sym-currents}) come into play again. Electron-positron annihilation
and $\tau$-decay processes are used to obtain the corresponding spectral distributions
in the vacuum. The dilepton production from a strongly interacting medium is
a frequently used tool to study potential in-medium changes of hadrons (better to say,
of the quark currents).\cite{Rapp:2009yu}

Also in lattice QCD the central object to study hadronic properties are the
current-current correlators. There one does not study the Fourier transform of
the current-current correlator like in (\ref{eq:curcur}), but directly the
object in coordinate space. One is restricted to space-like distances, since time-like
distances would lead to path integrals where the integrands are not positive definite.
Such integrals cannot be computed with Monte-Carlo methods. Consequently the
full spectral distribution is not directly accessible by lattice-QCD calculations
(see also Ref.\ \refcite{Asakawa:2000tr,Hatsuda:2005nw} for the use of the
maximum-entropy method).
Typically one is restricted to the lowest (or lowest few) excitations in the respective
channel.\cite{Durr:2008zz}
Another theoretical approach which uses the current-current correlation functions
is the QCD sum rule method which is introduced in Sect.\ \ref{sec:inmed} below.

\subsubsection{Connections to chiral symmetry}
\label{sec:conchircur}

Concerning chiral symmetry breaking and its restoration, the current-current correlators
play an interesting role. By a chiral transformation a given current can mix with or
fully transform into a current with different quantum numbers and in particular
opposite parity. This ``chiral partner'' current would contain the same spectral
distribution as the current one started out with, if chiral symmetry was not
spontaneously broken.\footnote{Of course, this
statement holds up to effects from explicit symmetry breaking.} In turn, an observed
difference between such two spectral distributions signals chiral symmetry breaking.
The additional observation of a reduction of this difference in a medium
would signal chiral symmetry restoration. The best explored pair of currents in that
respect --- for the vacuum --- are the charged vector and axial-vector currents given in
(\ref{eq:def-sym-currents}): Studying the decay $\tau^+ \to \nu_\tau +\,$hadrons,
one probes the $V-A$ current combination
$\bar d \, \gamma^\mu \, u - \bar d \, \gamma^\mu \, \gamma_5 \, u$.
Due to $G$ parity the vector part of this combination couples to an even number
of pions whereas the axial-vector part couples to an odd number. Consequently,
one has direct experimental access separately on $\bar d \, \gamma^\mu \, u$ and
$\bar d \, \gamma^\mu \, \gamma_5 \, u$.
But a chiral transformation can turn the current
$\bar d \, \gamma^\mu \, u$
into the current
$\bar d \, \gamma^\mu \, \gamma_5 \, u$.
As can be seen from the $\tau$-decay data the spectral distributions contained
in $\bar d \, \gamma^\mu \, u$ and $\bar d \, \gamma^\mu \, \gamma_5 \, u$ are very
different\cite{Schael:2005am} which is a clear sign of chiral symmetry breaking.
Unfortunately,
$\tau$ decays are not so easily accessible for in-medium situations. But from a
theoretical point of view it is an interesting task to study the in-medium behavior of
the difference between the spectral distributions of
$\bar d \, \gamma^\mu \, u$ and $\bar d \, \gamma^\mu \, \gamma_5 \, u$
or more general of the currents
given in (\ref{eq:def-sym-currents}). At least the neutral vector current enters
the electromagnetic current (\ref{eq:el-cur}) ---
together with the isospin-0 part --- and is therefore
also experimentally accessible for in-medium situations from dilepton
production. In this review we will frequently come back to dilepton production below.

A final remark is in order, concerning the spectral distributions of
chiral partner currents
and the interplay between currents and single hadrons: Note that not
only a given quark current couples to different hadrons,
but also a given hadron couples to different currents. This leads to ambiguities
concerning the chiral partner of a given hadron. For example, besides the
quark current (\ref{eq:what-defjrho}) which has the quantum numbers of the $\rho^0$ there
is also the current
$\tilde j^\mu = \partial_\nu (\bar q \, \sigma^{\mu\nu} \, \tau^3 q)$ with the same
quantum numbers. A chiral
transformation mixes (\ref{eq:what-defjrho}), e.g., with
$\bar d \, \gamma^\mu \, \gamma_5 \, u$ which has the quantum numbers of the (positively
charged) $a_1$ meson. On the other hand, $\tilde j^\mu$ is mixed with
$\partial_\nu (\bar d \, \sigma^{\mu\nu} \, \gamma_5 \, u)$ which has the quantum
numbers of the (positively charged) $b_1$ meson.
This ambiguity exists for a given hadron.\cite{Caldi:1975tx}
For a given quark current there is,
of course, no ambiguity. The latter are the ones which are explored, e.g., in vacuum in
$\tau$ decays and in electron-positron annihilation.

\subsubsection{In-medium changes of correlation functions}
\label{sec:changecor}

We have discussed in Sect.\ \ref{sec:sympat} that a change in the symmetry
pattern of QCD is expected, if the vacuum is replaced by dense strongly interacting
matter. Presumably this change does not happen suddenly at the transition point,
but shows precursor effects. Hence, we expect that hadronic properties change already
in a hadronic surrounding. On the other hand, not every in-medium change might
be related to symmetries. More generally, due to the two-face appearance of QCD,
one can have two different points of view on in-medium modifications of hadrons,
a quark-gluon and a hadronic one: From a purely hadronic
point of view, the change of properties of a hadronic probe in strongly interacting
matter is caused by the interactions of the probe with the hadrons which form this medium.
On the other hand, in
the QCD language a hadron is an excitation of the QCD vacuum. In turn, the vacuum
is characterized by condensates. As can be seen, e.g.,
from (\ref{eq:fintemp2q},\ref{eq:dropfindens}), the condensates change in a
strongly interacting medium. One might say that the underlying vacuum is modified and
consequently the properties of the hadronic excitations of the changed vacuum
are also modified.

It is clearly an exciting task to predict and experimentally
confirm hadronic in-medium changes caused by the modified vacuum structure.
However, one may wonder whether the two different points of view formulated above
are not just different languages for the same physics. At least qualitatively
standard hadronic effects are in line with the expectation from chiral restoration
and deconfinement: As we have discussed in Sect.\ \ref{sec:changeord},
already the presence
of the hadronic medium constituents --- even if they do not interact with each other ---
leads to a drop of the chiral order parameters. Also an onset of deconfinement can
be observed in the correlation functions with hadronic quantum numbers: As we have
already discussed, in vacuum these correlation functions contain quasi-particle
structures, i.e.\ the hadrons with the corresponding quantum numbers. In contrast,
in a deconfined system, the correlation functions contain a rather structureless
quark-antiquark or three-quark continuum. Hence one expects that structures are washed
out.\cite{Dominguez:1992dw} Indeed, hadronic many-body
calculations predict increasing widths for hadronic states and also experimentally
this seems to be confirmed (see below). However, to really figure out up to which point
a hadronic description makes sense, more quantitative calculations are necessary.
In particular, as already pointed out, QCD predicts that the correlators of currents
which are connected
by chiral tranformations (``chiral partners'') become degenerate in a chirally
restored system. Hence, one
task is to figure out to which extent purely
hadronic descriptions show an onset of degeneracy for the spectra of chiral partner
currents.\cite{Leupold:2008ne}

Alternatives to standard hadronic many-body models are provided by approaches which involve
quarks in one or the other way, e.g., quark models or approaches involving QCD condensates.
They seem to offer a more microscopic view on the symmetry changes.
One should be aware, however, that it is not so easy to translate the results
of such quark-type models to hadronic observables.
In particular, there are models which predict dropping masses for
some or all hadrons (except for Nambu-Goldstone bosons), e.g., the famous Brown-Rho
scaling conjecture.\cite{Brown:1991kk} On the other hand, there are
quark models which include the phenomenon of chiral restoration and
do not predict a change of the vector-meson masses (for an early application
of a generalized Nambu--Jona-Lasinio model\cite{Nambu:1961tp}
see Ref.\ \refcite{Bernard:1988db}).
If the hadronic spectra get
much broader, it is not so clear in the first place what a dropping mass actually means
(cf., e.g., Ref.\ \refcite{Kwon:2008vq}).
But not only the quark language has its limitations:
At least close to the transition to a new state of matter
(QGP or quarkyonic matter or color superconducting matter) the hadronic setup
might become rather uneconomical, if not inappropriate, since the relevant degrees
of freedom on the other side of the transition are missing and since hadronic
$n$-body scatterings with $n$ large might become more and more important.
As a conservative approach one should at least try
to figure out how far one comes with an interpretation of the
experimental results based on a standard hadronic description.
Obviously, much more work on the
theory side is needed to reconcile quark and hadronic models and to base them deeper
on QCD. One connection between hadronic properties and QCD condensates is provided by
the QCD sum rule method to which we turn next.

\subsection{QCD sum rules and in-medium changes of hadron properties}
\label{sec:inmed}

We have seen so far that hadronic properties and their potential in-medium changes
are encoded in correlation functions. However, neither the hadronic nor the
current-current correlators are easily calculable in the non-perturbative
regime of QCD one is interested in. On the other hand, the non-perturbative features
of QCD are also characterized by the non-vanishing condensates.
In a strongly interacting system also the QCD condensates change their values.
We will see
in the following that at least to some extent these in-medium modifications of
condensates can be evaluated. In addition, the current-current correlation functions
can be connected to the QCD condensates by the QCD sum rule method. This imposes
constraints on possible in-medium changes of hadronic properties. We will find that these
constraints are actually not strong enough that unique conclusions can be drawn
for a single hadronic property, like the mass or life time of a given hadron.
However, non-trivial correlations can be obtained. We will also comment on the
interrelations between hadron properties, condensates and chiral symmetry.

\subsubsection{Generic structure of QCD sum rules}
\label{sec:genQCDSR}

Originally QCD sum rules were
introduced for the vacuum\cite{Shifman:1978bx,Shifman:1978by}
but later on generalized to in-medium
situations.\cite{Bochkarev:1985ex} In particular the work of
Hatsuda and Lee\cite{Hatsuda:1991ez} which points towards drastic in-medium changes
of $\rho$ and $\omega$ mesons has triggered a lot of activities both on the experimental
and on the theory side.

To show how QCD sum rules work we consider again the $\rho^0$ as an example. As a strongly
interacting medium we take cold nuclear matter characterized by a baryo-chemical
potential $\mu$. All connected quantities have already been introduced in the
paragraph of Eq.\ (\ref{eq:nucldens}). We concentrate
on a meson at rest with respect to the nuclear medium. This is most appropriate
since in a realistic situation where the medium has only a finite spatial extension
(e.g.\ a nucleus), one would be interested in a probe which stays in the medium as long
as possible. Note that we use the case of a $\rho^0$ at rest in cold nuclear matter
as an illustrative and important example. However, the qualitative conclusions which
we draw apply also to other hadrons and to other in-medium situations. In the following
we will frequently also comment on the case of finite temperature.

The sum rule method starts with the current-current correlation function defined in
(\ref{eq:curcur}). Instead of the Lorentz tensor $\Pi^{\mu\nu}$ it is more convenient
to study the scalar and dimensionless quantity $R = -\Pi_\mu^\mu/(3\,q^2)$. We note
in passing that for a probe which was not at rest, more care would be needed on that
stage. In vacuum the imaginary part of $R$ is directly linked to the ratio of
cross sections for $e^+ e^- \to \,$hadrons and $e^+e^- \to \mu^+ \mu^-$,
respectively.\cite{Shifman:1978bx,Shifman:1978by}

There is a regime where the quarks and gluons become the
relevant entities, namely at large energies/momenta. There, the correlator $R$ can be
evaluated using QCD perturbation theory. The purely perturbative result can be
improved by the inclusion of expectation values of local operators, the condensates.
The technical tool here is the operator product expansion (OPE) which works in the regime
of large space-like momenta.\cite{Wilson:1969zs} On the other hand, the whole information
expressed by the current-current correlation function (\ref{eq:curcur}) or equivalently
by $R$ is
contained in its spectral distribution, i.e.\ the values of its imaginary part along
the real energy axis.\footnote{Note that the space-like momenta where
the OPE works can be achieved by imaginary values for the energy. This is appropriate
since we assume that the studied probe is at rest with respect to the medium,
i.e.\ that its three-momentum vanishes.} The spectral distribution in turn contains all
the hadrons which couple to the considered current. Therefore two expressions for
the same quantity can be matched for large space-like momenta: one obtained from the
OPE and one from a dispersion integral. Schematically (ignoring, e.g., possible
subtractions) one finds
\begin{eqnarray}
  \int\limits_0^\infty \frac{ds}{\pi} \;
  \frac{{\rm Im}R^{{\rm HAD}}(s)}{s+Q^2}
  & = &
  R^{{\rm OPE}}(Q^2)
  \label{eq:disp-OPE}
\end{eqnarray}
where we have made explicit by the superscripts
where the respective information is supposed
to come from. This dispersion relation corresponds to the one
given in (\ref{eq:disp-hadr}).
In addition, we have introduced $Q^2 = -q^2 \gg 0$ for large space-like momenta, and
we have rearranged the energy integration using the fact that the considered meson
is its own antiparticle.

On the left-hand side of (\ref{eq:disp-OPE}) the information about all hadrons with
the quantum numbers of the considered current enters. The integral covers arbitrarily
large
energies $\sqrt{s}$. On the other hand, one is particularly interested in the
low-energy regime, where distinct hadronic resonances have been identified.
Technically one can enhance the importance of the low-energy part by a Borel
transformation.\cite{Shifman:1978bx,Shifman:1978by}
While the high-energy part is only power suppressed by $(s+Q^2)^{-1}$ in
(\ref{eq:disp-OPE}), the Borel transformation achieves an exponential suppression.
For a $\rho$ meson in cold nuclear matter the QCD (Borel) sum rule
reads\cite{Hatsuda:1991ez,Asakawa:1993pq,Leupold:1997dg}
\begin{eqnarray}
&&  \frac{1}{\pi M^2} \int\limits^\infty_0 \!\! ds \;
  {\rm Im} R^{\rm HAD} (s) \; e^{-s/M^2}  =
  \frac{1}{8\pi^2}\left(1+\frac{\alpha_s}{\pi} \right)
  \nonumber \\ &&  \hspace*{5em} {}
  + \frac{1}{M^4} \left(
    \bar m_q \, \langle \Omega \vert \bar u u \vert \Omega \rangle
    + \frac{1}{24} \, \langle \Omega \vert \frac{\alpha_s }{\pi} G^2 \vert \Omega \rangle
    + \frac{1}{4} \, m_N \, a_2 \, \rho_N
  \right)
  \nonumber \\ && \hspace*{5em} {}
  + \frac{1}{M^6} \left(
    -\frac{56 }{81} \, \pi\alpha_s \langle \Omega \vert {\cal O}^V_4 \vert \Omega \rangle
    -\frac{5}{24} \, m_N^3 \, a_4 \, \rho_N
  \right)
  \nonumber \\ && \hspace*{5em} {}
  + {\cal O}(1/M^8)  \,.
\label{eq:botr}
\end{eqnarray}
Here $M$ is the Borel mass which takes the role of the large space-like momentum
variable $\sqrt{Q^2}$ after Borel transformation. On the left-hand side of
(\ref{eq:botr}) one sees the already
announced exponential suppression of large-$s$ contributions. One can expect
that the integral is most sensitive to the low-energy part of the spectral distribution.
On the OPE side, the right-hand side of (\ref{eq:botr}), an
expansion in inverse powers of the Borel mass $M$ appears. It originates from the OPE
which is an expansion in $1/Q^2$. The leading term is the result from QCD perturbation
theory. $\alpha_s$ is the running coupling of QCD.
Simply for dimensional reasons the contributions from QCD condensates are
suppressed by powers of the Borel mass $M$.
In the following, we discuss the terms appearing on the right-hand side
of (\ref{eq:botr}) one by one, focusing on three aspects:
First, of course, we define their meaning, second we discuss their numerical
importance and third we comment on their possible connection to chiral
symmetry breaking and restoration. Actual values for the terms appearing in
(\ref{eq:botr}) are provided in Table \ref{tab:condval}.
\begin{table}
  \centering
  \begin{tabular}{|c|c|c|c|c|c|}
    \hline
    $\bar m_q$ & $\langle 0 \vert \bar u u \vert 0 \rangle$ &
    $\langle 0 \vert \frac{\alpha_s }{\pi} G^2 \vert 0 \rangle$  &
    $\alpha_s$ & $a_2$ & $a_4$ \\
    \hline
    6 MeV & $-(240 \, \mbox{MeV})^3$ & $(330 \, \mbox{MeV})^4$ &
    0.36 & 0.9 & 0.12 \\
    \hline
  \end{tabular}
  \caption{Actual values for the ingredients of the operator product expansion.
  Numbers taken from Refs.\ \protect\refcite{Leupold:1997dg,Leupold:2003zb}.}
  \label{tab:condval}
\end{table}

\subsubsection{QCD condensates of dimension 4}
\label{sec:cond4}

The first term which multiplies $1/M^4$ in (\ref{eq:botr})
is the already introduced two-quark condensate. Its density dependence
in leading order of the nucleon density $\rho_N$ is given in (\ref{eq:dropfindens}).
The non-vanishing of the two-quark condensate signals chiral symmetry breaking. However,
its numerical influence on the sum rule is small in vacuum and also
at finite nuclear densities.

One of the terms numerically important in vacuum is
the gluon condensate
$\langle \Omega \vert \frac{\alpha_s }{\pi} G^2 \vert \Omega \rangle$.
Together with the four-quark condensates, discussed below, the gluon
condensate determines the vacuum properties of the $\rho$
meson.\cite{Shifman:1978bx,Shifman:1978by}
The medium dependence of the gluon condensate for finite baryon densities
but also for finite temperatures can be obtained from the
trace anomaly.\cite{Hatsuda:1991ez,Hatsuda:1992bv}
It turns out that this density dependence is very weak. Hence while the gluon condensate
is important for the vacuum properties, it is not responsible for strong in-medium
changes
--- at least not for low densities. The gluon condensate is invariant with respect
to chiral transformations. Thus, there is no direct connection between chiral symmetry
breaking and the non-vanishing of the gluon condensate.

\begin{figure}[th]
  \begin{center}
    \includegraphics[keepaspectratio,width=0.49\textwidth]{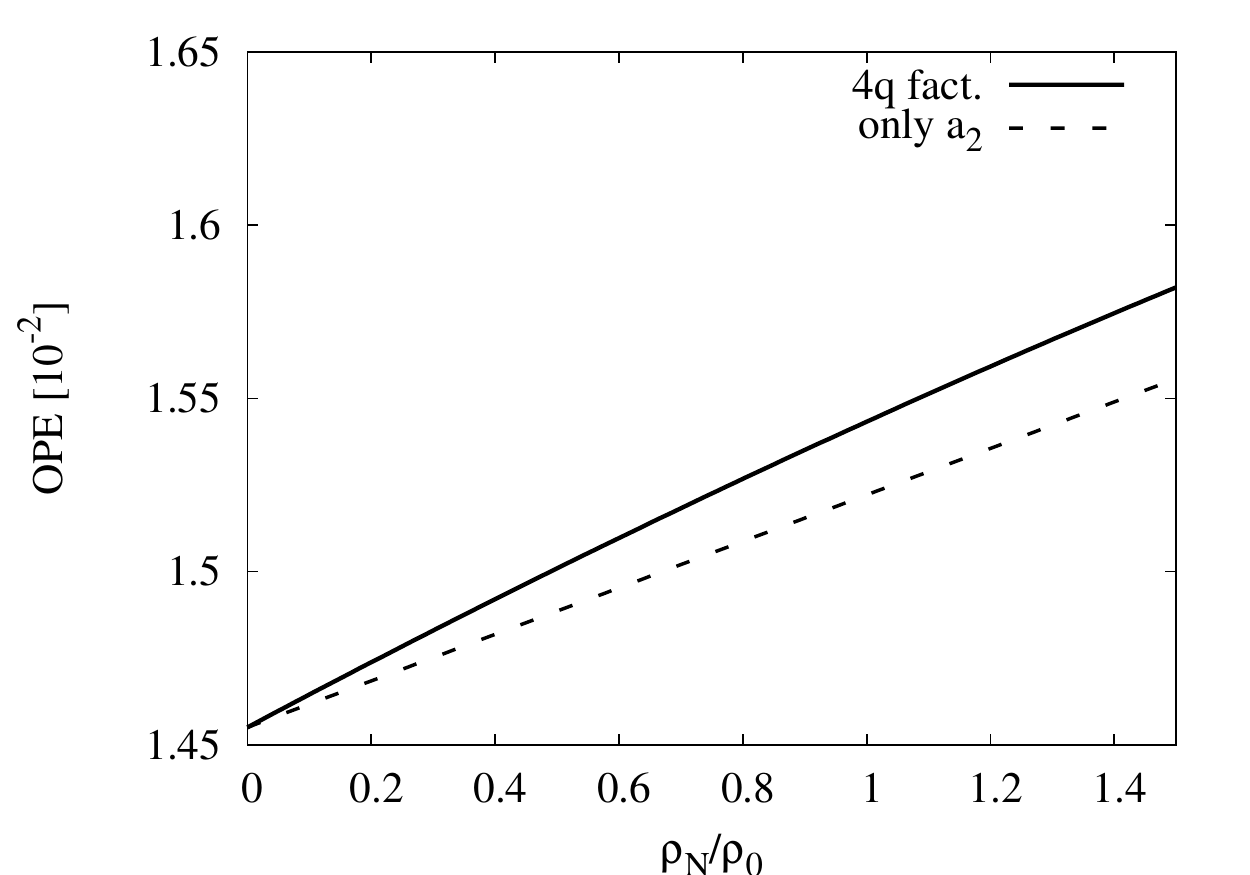}
    \includegraphics[keepaspectratio,width=0.49\textwidth]{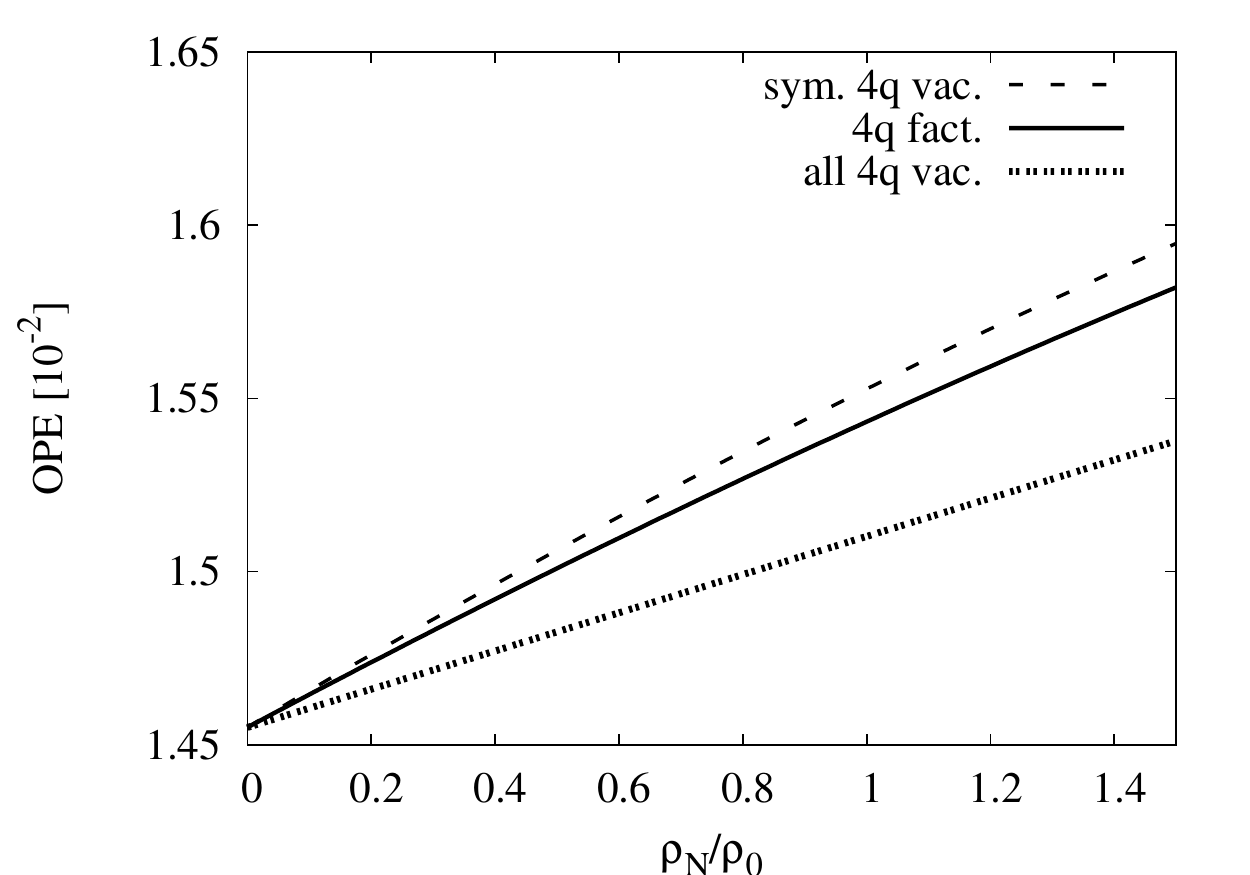}  
    \includegraphics[keepaspectratio,width=0.49\textwidth]{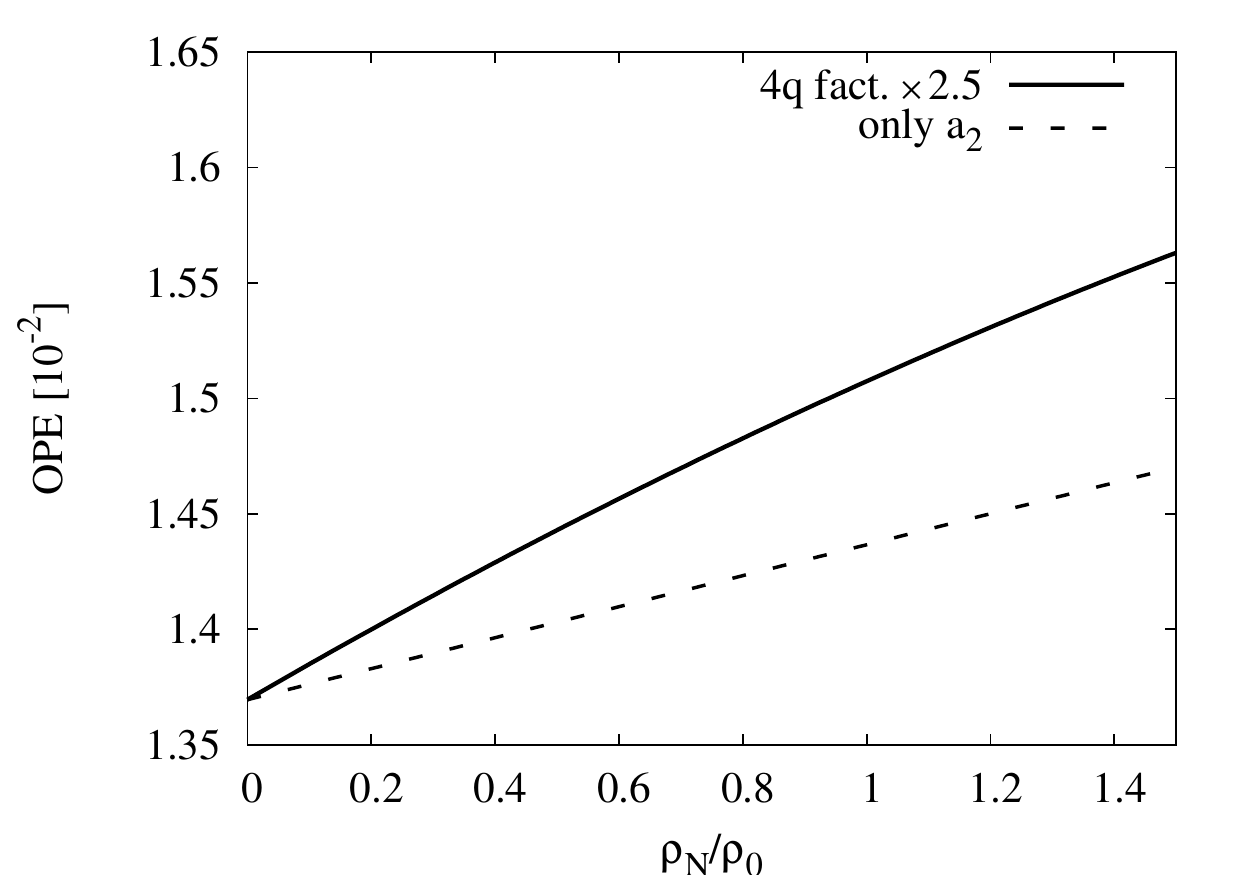}
    \includegraphics[keepaspectratio,width=0.49\textwidth]{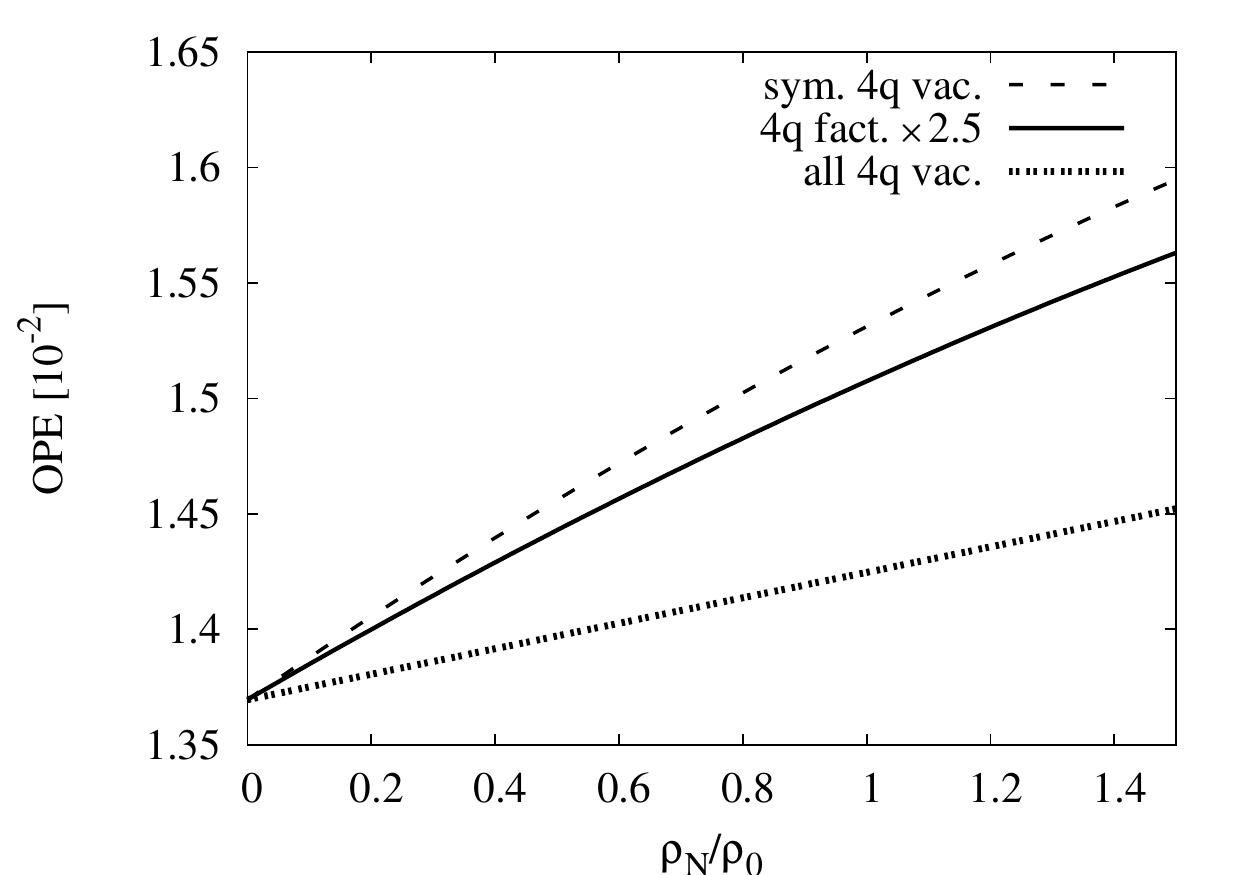}
  \end{center}
  \vspace*{8pt}
  \caption{The right-hand side of (\ref{eq:botr}) as a function of the nuclear
    density for a generic value of the Borel mass, $M = 0.8\,$GeV, and
    for different assumptions concerning
    the four-quark condensates.
    {\em Top left:} Factorization (\ref{eq:what-four-two1}) is assumed for all
    four-quark condensates. The upper line shows the full in-medium
    result.
    The lower line is obtained using vacuum values for all OPE contributions except
    for the $a_2$ term. {\em Top right:} For all lines vacuum factorization is assumed
    for all four-quark condensates. The highest (lowest) line is obtained using
    the respective vacuum
    value also for the chirally symmetric (all) in-medium four-quark condensates.
    For the middle line factorization is assumed for all
    four-quark condensates. From top to bottom these are the cases (iii), (i), and (ii)
    discussed in the main text. {\em Bottom:} Same as top panels, but instead of
    factorization an additional factor of 2.5 is multiplied to obtain four-quark
    condensates from the square of the two-quark condensate.}
  \label{fig:OPE}
\end{figure}

The last term which multiplies
$1/M^4$ only shows up in the presence of a medium.\cite{Hatsuda:1991ez,Hatsuda:1992bv}
It originates from the non-scalar twist-2 condensate
$\langle \Omega \vert \, \bar q \, (\gamma_0 D_0 - \frac14 \,
\gamma_\mu D^\mu) \,q \, \vert \Omega \rangle$. This condensate can be evaluated
for the two in-medium situations discussed in Sect.\ \ref{sec:changeord},
i.e.\ a pion gas (finite temperature) or a nucleon Fermi sphere
(cold nuclear matter). In (\ref{eq:botr}) we had specified the in-medium situation
to the latter case.
The value for $a_2$ can be obtained from the parton distributions determined by
deep-inelastic scattering reactions. Hence, there is not much uncertainty about the
size of this term. It turns out that the numerical contribution of this term is large.
Since the corresponding condensate is chirally invariant, it points towards an in-medium
modification which is not intimately connected to chiral symmetry. A numerical
evaluation is presented in Fig.\ \ref{fig:OPE}, left panels.
The lines start out on the left-hand side from the vacuum value.
If one adds {\em only} the $a_2$
term one obtains the lines with the
label ``only $a_2$''. Obviously, the influence of the $a_2$ term is quite large.
Further details are given below after the four-quark
condensates have been introduced.

\subsubsection{QCD condensates of higher dimensions}
\label{sec:cond6}

The first term accompanied by $1/M^6$ contains a special combination of four-quark
condensates:
\begin{eqnarray}
\langle \Omega \vert {\cal O}^V_4 \vert \Omega \rangle &= &
\frac{81}{224} \left\langle \Omega \vert
(\bar u \gamma_\mu \gamma_5 \lambda^c u - \bar d \gamma_\mu \gamma_5 \lambda^c d)^2
\vert \Omega \right\rangle
\nonumber  \\ && {}
+ \frac{9}{112} \langle \Omega \vert
(\bar u \gamma_\mu \lambda^c u + \bar d \gamma_\mu \lambda^c d)
\sum\limits_{\psi = u, d, s} \bar\psi \gamma^\mu \lambda^c \psi \,
\vert \Omega \rangle  \,,
  \label{eq:fourqdef}
\end{eqnarray}
with color matrices $\lambda^c$. The size of four-quark condensates,
even in vacuum, is still under debate. Originally it was
assumed that in the vacuum, $\Omega = 0$, the four-quark condensate can be approximately
factorized:\cite{Shifman:1978bx,Shifman:1978by}
\begin{equation}
  \label{eq:what-four-two1}
  \langle \Omega  \vert {\cal O}^V_4 \vert \Omega \rangle \stackrel{?}{\approx}
\langle \Omega \vert \,\bar u u \, \vert \Omega \rangle^2  \,.
\end{equation}
Indeed, in the limit of a large number of colors this factorization can be
justified both for vacuum and for finite nuclear densities (but not for finite
temperatures).\cite{Leupold:2005eq}
If one chooses the size of the four-quark condensates such that their order of magnitude
is given by (\ref{eq:what-four-two1}), one obtains a decent description of the vacuum
properties of the $\rho$ meson. A factor of two change in this size does not influence
the $\rho$-meson properties in a drastic way. Thus the vacuum sum rule is not extremely
sensitive to the detailed values of the four-quark condensates, provided the order of
magnitude is correct. The change of the four-quark condensate with density
and the $a_2$ term discussed above have the most influence on the changes of the
OPE side of (\ref{eq:botr}). We will come back to that point below.

Concerning chiral symmetry, relation (\ref{eq:what-four-two1}) --- if true for finite
densities --- seems to suggest that
four-quark condensates drop together with the two-quark condensate and, in particular,
that the four-quark condensates vanish for a chirally restored system. On the
other hand, the second term on the right-hand side of (\ref{eq:fourqdef}) is chirally
symmetric. Hence, there is no strict reason why this quantity should vanish at
chiral restoration. Things are different for the first term on the right-hand side
of (\ref{eq:fourqdef}). This term changes under chiral transformations. Thus, there
is an influence of chiral symmetry breaking and maybe restoration on the OPE side and
therefore on the current-current correlator induced by specific four-quark
condensates. However, the quantitative evaluation is not straightforward. To explore
the uncertainties one might study the following extreme points of view:
(i) The four-quark condensates change at finite density
according to (\ref{eq:what-four-two1},\ref{eq:dropfindens}).
(ii) None of the four-quark condensates
given in (\ref{eq:fourqdef}) changes in the medium. (iii) The chirally invariant
second part of the right hand side of (\ref{eq:fourqdef}) does not change in the
medium while the first
term changes according to the factorization assumption. These possibilities are
explored in Fig.\ \ref{fig:OPE}. In addition, we have studied the consequences of larger
values for the four-quark condensates as compared to the factorization result.
For the other condensates we take typical values from the
literature,\cite{Leupold:1997dg} see also Table \ref{tab:condval}.
Obviously, the variation due to the uncertainties in the four-quark condensates
is significant. Nonetheless, in all explored cases there is a sizeable in-medium change
of the OPE side. It is interesting to observe that case (iii) discussed above always
leads to the largest deviation from the vacuum (respective top line in the right
panels of Fig.\ \ref{fig:OPE}). On the other hand, this case rests on a rather
conservative assumption: It is plausible that at least the four-quark condensate
which is not chirally symmetric changes similar to the two-quark condensate.

For a finite-temperature system instead of a cold nuclear system one can use PCAC
relations to reduce all in-medium expectation values of four-quark operators
to vacuum condensates. This leads to the much celebrated Dey-Eletsky-Ioffe
mixing formula.\cite{Dey:1990ba} It rests on the assumption that the momenta
of the pions, which form the finite-temperature system, can be neglected.
For the OPE side this is a reasonable assumption as momenta of the order of the
temperature are soft compared to the hard OPE expansion parameter ($Q^2$ or
$M^2$). We note in passing, however, that the mixing formula is sometimes also
used for the spectral side where for typical values of the temperature on the order
of 100 MeV it yields misleading results like the appearance of the $a_1$-meson peak
in the vector channel. A kinematically proper treatment of the non-negligible
pion momenta is required here as performed, e.g.,
in Refs.\ \refcite{Steele:1996su,Urban:2001uv}. The result is a broadening of
the spectral distribution and in particular the appearance of strength at low
invariant masses connected to the Dalitz decay $a_1 \to \pi + \,$dilepton.

We turn to the second term which multiplies $1/M^6$ in (\ref{eq:botr}).
It also stems from a
non-scalar twist-2 condensate. In contrast to the $a_2$ term discussed above the
contribution of the $a_4$ term is numerically small. We note in passing that we
have not displayed non-scalar twist-4 contributions. They have been estimated to be
small.\cite{Hatsuda:1995dy,Leupold:1998bt}

In (\ref{eq:botr}) we have neglected all condensates which come with $1/M^8$ or
higher inverse powers of the Borel mass. The sum rule is formulated for large
space-like momenta which translates to large Borel masses. Therefore, it should be
justified to neglect terms which are accompanied by higher powers in $1/M$.
Nonetheless, the use of QCD sum rules in particular at finite
nuclear densities is still a matter of
debate.\cite{Eletsky:1996jg,Hatsuda:1997gj,Eletsky:1997rz}

\subsubsection{Consequences for the spectral distribution}
\label{sec:consspec}

To summarize, the numerically most important contributions to the
OPE side of (\ref{eq:botr}) come from the
gluon condensate, the four-quark condensates, and the (purely density dependent) $a_2$
term. Concerning especially the density dependence, the four-quark condensates and
the $a_2$ term have the dominant influence. This is shown in Fig.\ \ref{fig:OPE}.
The $a_2$ term stems from a chirally symmetric operator, i.e.\ does not have a
direct relation to chiral restoration. In contrast, the numerically
dominant part of the four-quark condensate breaks chiral symmetry.
The largest
uncertainties come from the four-quark condensates, both concerning their vacuum values
and their in-medium modifications. However, since also the $a_2$ term plays an important
role for the in-medium changes, it would be misleading to conclude that the uncertainties
in the four-quark condensates are so large that the sum rules are meaningless.
Clearly, Fig.\ \ref{fig:OPE} contradicts such a pessimistic point of view.
One sees that
independently of the detailed values for the in-medium change of the four-quark
condensates, the OPE side changes significantly at finite density. Therefore, the
hadronic side of the sum rule (\ref{eq:botr}) also has to change to account for the
changes on the OPE side. Presumably, part of these changes is intimately
connected to chiral
restoration, while another part is not, corresponding to one four-quark condensate
and to the $a_2$ term, respectively.

\begin{figure}[ht]
  \begin{center}
    \includegraphics[keepaspectratio,width=0.7\textwidth]{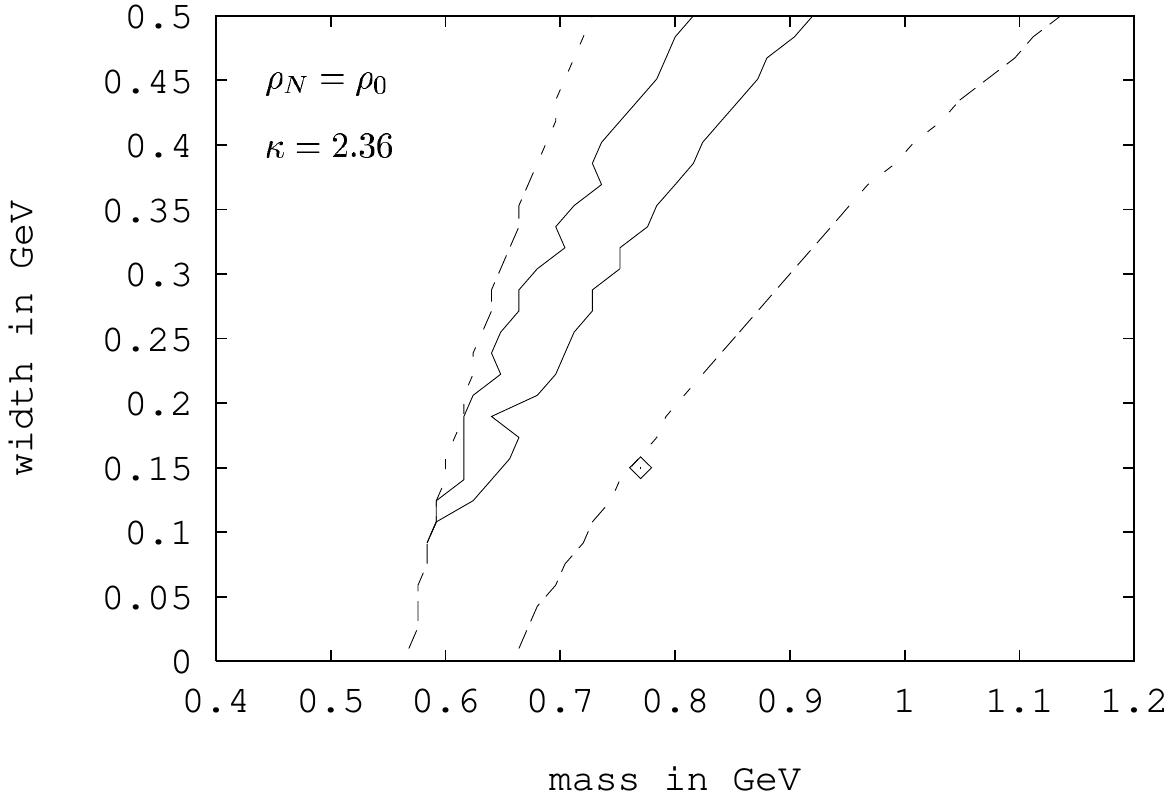}
  \end{center}
  \vspace*{8pt}
  \caption{Correlation between mass and width of a $\rho$ meson placed
    at rest in cold nuclear matter ($T=0$, $\rho_N = \rho_0$)
    as obtained from a QCD sum rule analysis. All mass-width pairs inside the corridor
    given by the full lines are allowed. It is assumed that the low-energy
    spectral distribution essentially consists of a single peak with {\it a priori}
    arbitrary mass and width. For orientation the vacuum values for the $\rho$-meson
    mass and width are given by the diamond.
    Figure taken from Ref.\ \protect\refcite{Leupold:1997dg}.
    See this reference for further
    details.}
  \label{fig:qsr-med-width.eps}
\end{figure}

We have obtained a firm, yet qualitative result from the QCD sum rule approach:
The low-energy spectrum has to change in a medium. It is, however, not so clear which
change is demanded by the sum rule. At this stage specific qualitative assumptions
about the in-medium shape of the spectral distribution have to enter. {\em If} one
assumes that the $\rho$ meson remains a quasi-particle, i.e.\ that its width can be
ignored for the sum rule analysis, then the changing OPE side demands for a dropping
$\rho$-meson mass.\cite{Hatsuda:1991ez}
However, this is not the most general in-medium scenario
one can think of for a spectral distribution. Indeed, one can explore the more
general case where one allows for an in-medium $\rho$ meson which
appears as a peak with {\it a priori} arbitrary mass and arbitrary width in the
low-energy
spectral distribution of the corresponding current-current correlation function.
With such an input, it turns out that the sum rule is not predictive enough to fix
mass and width
of the in-medium $\rho$ meson.\cite{Leupold:1997dg} Instead, a correlation between
in-medium mass and width is obtained. Of course, the details of this correlation
depend to some extent on the assumptions for the vacuum and in-medium four-quark
condensates. An example is shown in Fig.\ \ref{fig:qsr-med-width.eps}.
Qualitatively the mass can drop or the width can grow or both. More generally, the
in-medium sum rule demands for the appearance of additional strength at lower
invariant masses. One possibility to achieve that is a dropping mass. But also a broader
distribution even without a mass drop is compatible with this requirement.
The point here is that the spectral distribution does not enter the sum rule
with equal weight for all invariant masses. Instead, the exponential suppression
of the high-energy part in the spectral integral of the sum rule (\ref{eq:botr})
puts a larger weight on the low-energy part of a broad distribution.
Finally we note that also other assumptions besides mass and width changes,
like the appearance of additional
peaks, are explored in the literature.\cite{Leupold:2001hj,Leupold:2004gh}
In that respect, the corresponding case of the $\omega$ is also very interesting:
Also here the OPE side demands for a shift of strength to lower invariant masses.
However, there is already a large strength (at space-like momenta) caused by
nulceon--nucleon-hole modes (also called Landau damping contribution in the
literature).\cite{Hatsuda:1991ez,Steinmueller:2006id}
At low densities this contribution to the current-current correlation function
can be calculated model independently. It might already overcompensate the demand
from the
OPE side.\cite{Klingl:1997kf,DuttMazumder:2000ys,Thomas:2005dc,Steinmueller:2006id}

So far, we have put most of our attention to the $\rho$ meson. One reason was that
the sum rule method is better explained with a typical example at hand. In addition,
the $\rho$ meson is an interesting probe concerning in-medium modifications. We
briefly comment on other hadrons: For hadrons made out of light quarks one
typically finds that the OPE side changes sizeably in a medium. Hence,
qualitatively there must be a significant change of hadronic properties. Quantitatively,
the OPE has to be worked out for each hadron separately. In addition, as already pointed
out, the analysis depends on the specific assumptions for the qualitative
shape of the spectral distribution. In any case, the sum rule method can be used as
a consistency check for a given hadronic model --- provided the hadronic model yields
a current-current and not only a hadronic correlation function.

Concerning chiral symmetry breaking and restoration we have seen for the $\rho$ meson
that part of the
numerically important in-medium changes of the OPE side comes from chirally invariant
terms, while another part stems from symmetry breaking terms. Such a
finding depends, at least quantitatively, on the considered hadron. For example,
for the $\omega$ meson the whole four-quark condensate structure is chirally
symmetric. This is easy to understand since the $\omega$ meson is a chiral
singlet.\footnote{This statement concerns the chiral symmetry for two flavors
discussed here. Concerning three flavors the $\omega$ is not a chiral singlet.}
Consequently, all condensates
appearing in the OPE are either chirally symmetric or break the symmetry explicitly,
i.e.\ contain a quark mass. The latter terms are then numerically small.
Hence, from the sum rules one cannot deduce an intimate
relation between chiral restoration and in-medium changes of the $\omega$ meson.
This does not imply that such a relation does not exist. The phenomenon of chiral
restoration might also lead to in-medium changes of chirally symmetric condensates.
To pin down such more complicated interrelations requires a much better microscopic
understanding of non-perturbative QCD.

There are other hadrons where a connection to chiral symmetry breaking is more
direct. For example, for the nucleon,
but also, e.g., for open-charm $D$ mesons,\cite{Hilger:2008jg}
the two-quark condensate appears more directly, i.e.\ is not multiplied by a
light-quark mass (which made the contribution rather small for the $\rho$-meson case).
It should be stressed, however, that also here there are other numerically important
contributions emerging from chirally symmetric
operators.\cite{Birse:1996qp,Thomas:2007gx}
We repeat our firm, but qualitative statement: Since the OPE side sizeably
changes in a medium --- in part caused by chiral symmetry breaking
operators\footnote{Except for chiral singlets like the $\omega$.} --- there must be 
significant changes of the hadronic properties of the considered probe when it is embedded in the medium. To pin down these
changes for various probes/hadrons is an important task of in-medium hadron physics.
Since part of the in-medium changes of the OPE side are caused by chirally invariant
operators, an observed change of hadronic properties does not necessarily imply
that a footprint of chiral restoration has been detected.

\section{Hadronic Models}
\label{sec:mosel}
In the preceding section we have discussed the ``duality'' between quark-based models and classical hadronic models. In particular, we have stressed that
QCD sum rules, that work in a quark-gluon world, give valuable constraints on hadronic spectral functions, but cannot predict their detailed shape.\cite{Leupold:1997dg,Leupold:1998bt} It is thus necessary to model these spectral functions, based on our present understanding of meson-baryon interactions.

The theoretical challenge is then to calculate the self-energy $\Pi$ of a meson embedded in a strongly interacting medium.
Early calculations for the in-medium self-energy of the $\rho$ meson\cite{Asakawa:1992ht,Herrmann:1992kn,Rapp:1995zy,Rapp:1997ei} started by dressing the decay pions within the framework of the $\Delta$-hole model\cite{Oset:1981ih}. This would obviously change the in-medium width, i.e.\ the imaginary part of the self-energy. The change of the real part could then be obtained from dispersion relations. Later it was realized that the direct $\rho N$ coupling could also contribute to the imaginary part of the self-energy and that this contribution was more important than the change of the decay width. Subsequent calculations of the vector-meson--nucleon scattering amplitude combined a chiral model with vector-meson dominance;\cite{Klingl:1997kf,Klingl:1997tm,Klingl:1996ps} these calculations suffer from the neglect of any nucleon excitations beyond the $\Delta(1232)$ and the use of the heavy-baryon approximation. The latter was later shown to be grossly unreliable.\cite{Eichstaedt:2007zp} In addition, there were no experimental constraints on the essential coupling constants contained in the effective Lagrangian.
Resonance excitations were first considered in Ref.\ \refcite{Friman:1997tc} and then worked out in detail in Refs.\ \refcite{Rapp:1997fs,Peters:1997va,Post:2003hu,Post:2000qi,Post:2001am}. These latter papers generated major effects from coupling the vector meson to nucleon-resonance--nucleon-hole loops. For the latter one can draw on a long experience with calculations of the in-medium pion self-energy that are based on the so-called $\Delta$-hole model in which the free pion propagator in the nuclear medium is dressed with $\Delta$-resonance--nucleon-hole excitations.\cite{Oset:1981ih} In this case the determining quantities are, first, the coupling constant for the $\pi N \Delta$ vertex and, second, the strength of the short-range correlations; the latter are crucial for obtaining the correct self-energies. The same type of model can be used for the vector mesons.

The starting point for these
studies is the so-called $t\, \rho$ approximation in which the self-energy $\Pi$ of a meson in a medium is determined by the interaction of the hadron with all the surrounding particles. While at temperature $T=0$ these are only baryons, at higher temperatures mesons, mainly pions, and also antibaryons will fill the space around the hadron in question. The interaction of the hadron with all these particles, baryons and mesons, determines its self-energy.

For the sake of simplicity in the following we write down the relevant expressions only for cold nuclear matter; the self-energy as a function of the energy of the hadron, $\omega$, and its momentum $\mathbf{q}$ is then given by\cite{Post:2003hu}
\begin{equation} \label{trho1}
\Pi_{\rm med}(\omega, \mathbf{q}) = \int_{\rm F} \frac{d^3p_N}{(2 \pi)^3 2 E_N}\, t_{\rm tot}(q,p_N) ~.
\end{equation}
Here $t_{\rm tot}(q,p_N)$ is the
meson-nucleon forward scattering amplitude dependent on the hadron's four-momentum $q$ and the nucleon momentum $p_N$ and the integration extends over the Fermi sea of occupied nucleon states. At sufficiently low densities ---
and correspondingly low Fermi momenta --- this expression for the self-energy can be simplified by pulling the scattering amplitude out of the integral if the scattering amplitude varies only smoothly with momentum.
This gives
\begin{equation}  \label{trho}
\Pi_{\rm med}(\omega, \mathbf{q})= \frac{1}{8 m_N} \rho \, t_{\rm tot}(q,p_N) ~.
\end{equation}
where $\rho$ is the baryon density and the nucleon's momentum is simply given by $p_N = (m_N,\mathbf{0})$. A priori it is not clear up to which
densities this so-called \emph{low-density} approximation is valid; this depends primarily on the strength of the nucleon-meson coupling.
Eq.~(\ref{trho}) allows for an easy interpretation: the self-energy is equal to the meson-nucleon forward scattering amplitude, multiplied with the
probability to find a nucleon for an interaction.

For vector mesons there are two distinct polarization directions; the transverse and longitudinal self-energies can be obtained from the self-energy $\Pi$ by projecting on to the transverse and longitudinal degrees of freedom.\cite{Post:2003hu} For simplicity we have not written any spin or isospin factor in (\ref{trho1}).

The expressions just given describe the self-energies of vector mesons in the nuclear medium. If these vector mesons couple also to nucleon resonances then the properties of the latter are obviously also changed. This is so because 1.\ the nucleon resonance can itself collide with a nucleon and 2.\ the decay legs (in this case vector-meson--nucleon) get modified. The problem to determine the self-energy of the vector meson in a medium is thus intimately coupled to the problem of calculating the self-energy of the nucleon resonances and their decay products in the medium.\cite{Post:2003hu}

\begin{figure}[ht]
  \begin{center}
    \includegraphics[keepaspectratio,width=0.5\textwidth]{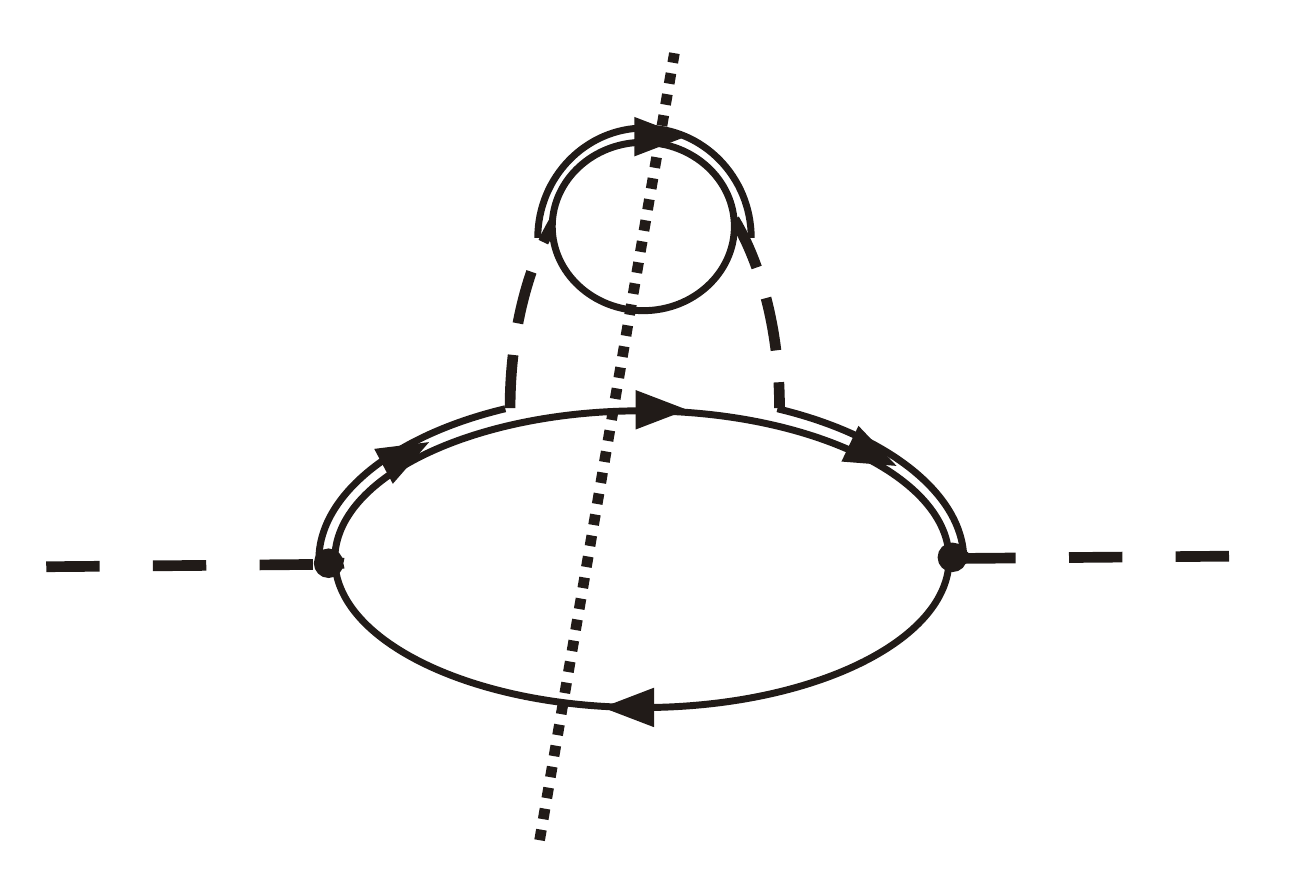}
  \end{center}
  \vspace*{8pt}
\caption{In-medium self-energy of a meson. The single lines represent nucleon propagators, the double lines nucleon resonance excitations. The internal dashed lines denote the propagators of either the vector meson or other mesons, such as the pion. The straight, dotted line shows the cut leading to the imaginary part of the meson's self-energy.}
\label{N*h}
\end{figure}

Diagrammatically, the resonance-hole excitation contribution to the meson self-energy can be represented as shown in Fig.~\ref{N*h}.
This figure contains an important feature: the upper half shows that the resonance propagator itself is dressed due to interactions with the medium. Thus, the determination of the meson self-energy and the resonance self-energy are coupled. In addition, the internal meson lines in Fig.~\ref{N*h} can either represent a meson of the same type as the incoming meson or other mesons as well that couple to the resonance. For example, for the $\rho$ meson the resonance propagator could be that of the N(1520) $D_{13}$ excitation and the internal dashed lines could be propagators either of the $\rho$ or the $\pi$ meson.

The dotted line in Fig.~\ref{N*h} illustrates the cut of the diagram that defines the imaginary part of the self-energy. The latter is given by
\begin{equation}
\Im \Pi (\omega,\mathbf{q})  = I_\Pi \left(\frac{f}{m_V}\right)^2 \int \frac{d^3 p}{(2\pi)^3} \frac{\Theta\left(p_F - |\mathbf{p}|\right)}{2 E_N(\mathbf{p})} \, \Im \frac{\Omega}{k_0^2 - E_R^2(\mathbf{k}) - \Sigma_{\rm med}(k)} ~.
\end{equation}
Here $I_\Pi$ is an isospin factor. The four-momentum of the resonance is denoted by $k = p + q$, where $p$ is the four-momentum of a nucleon and $q$ that of the meson. The factor $1/(2 E_N)$ comes from the hole part of the relativistic nucleon propagator. The numerator $\Omega$ contains the vertex factors as well as the traces over the internal polarization degrees of freedom. Details can be found in Ref.\ \refcite{Post:2003hu}.

The real part of the meson self-energy is then obtained by using a dispersion relation which guarantees that the spectral function of the meson remains normalized
\begin{equation}
\Re \Pi(q_0,\mathbf{q}) = \mathcal{P} \int_0^\infty \frac{d\omega^2}{\pi} \frac{\Im \Pi(\omega,\mathbf{q})}{\omega^2 - q_0^2} ~.
\end{equation}

The effects of the self-energies are usually discussed in terms of \emph{spectral functions} which are defined as the imaginary part of the retarded propagator (cf. Eq.\ (\ref{Aspect}))
\begin{equation}
\mathcal{A}(q_0,\mathbf{q}) = - \frac{1}{\pi}
\frac{\Im \Pi(q_0,\mathbf{q})}{\left(q^2 - m_V^2 - \Re \Pi(q_0,\mathbf{q})\right)^2 + \left(\Im \Pi(q_0,\mathbf{q})\right)^2}~
\end{equation}
with the self-energy denoted by
\begin{equation}
\Pi(q_0,\mathbf{q}) = \Pi_{\rm vac}(q^2) + \Pi_{\rm med}(q_0,\mathbf{q}) ~.
\end{equation}
While the spectral function in vacuum depends only on the invariant
mass $q^2$ of the hadron, in a medium it does depend separately on its
energy and three-momentum.

In the low-density approximation the imaginary part of the forward scattering amplitude $t_{\rm tot}$ in (\ref{trho}) can be related to the total cross sections by means of the optical theorem and the self-energy becomes
\begin{equation}
\Im \Pi_{\rm med}(q_0,\mathbf{q}) = - \rho \sigma_{\rm VN}\, |\mathbf{q}| = - q_0 \Gamma ~,
\end{equation}
and thus the width $\Gamma$ assumes the classical value for a collisional
width\cite{Bugg:1974cz}
\begin{equation}    \label{Gammacoll}
\Gamma = \rho \sigma_{\rm VN} \, v ~,
\end{equation}
where $v$ is the relative velocity of nucleon and meson and $\sigma_{\rm VN}$ is the meson-nucleon cross section. Thus in this low-density case the in-medium cross section is directly related to the imaginary part of the in-medium self-energy.

One of the determining factors in all these calculations is the strength of the resonance-nucleon-meson vertex. In principle, this quantity can be derived from nucleon resonance studies such as $\gamma N \to N \rho$. Several nucleon resonances, such as the N(1520) $D_{13}$, the N(1650) $S_{11}$ and the N(1700) $D_{13}$ are known\cite{Amsler:2008zzb} to have sizeable decay branches into $\rho N$ which can be used to extract the relevant coupling constants.

The situation is more difficult for the $\omega$ and the $\phi$ mesons; the listings
of the Particle Data Group (PDG)\cite{Amsler:2008zzb} do not contain any nucleon resonances that decay into nucleon plus these vector mesons so that the coupling constant cannot be easily obtained. We will discuss this further in Sects. \ref{subsec:omega} and \ref{subsec:phi}.

Besides the resonance-hole excitations there are other contributions to the in-medium self-energy. An obvious candidate are nucleon-hole
excitations which are driven by the nucleon-meson cross section going
to channels that do not (dominantly) involve resonance excitations. In
addition, the dressing of any decay products of the vectors mesons
affects directly their width and thus also --- through dispersion
relations --- the mass. However, even if the in-medium width $\Gamma$ is large this alone does not necessarily imply also a large shift of the pole mass; the latter is sensitive only to the mass dependence of the width and not to its absolute size as such.

Another, less certain contribution may also come
from  in-medium higher-order interactions, i.e.\ from interactions of
the meson with two (and more) nucleons. In this case the low-density
approximation ((\ref{trho}), (\ref{Gammacoll})) no longer holds and, in particular, the meson-nucleon cross section cannot be obtained from the self-energy.
Taking this argument around it is also clear that --- in the presence of sizeable many-particle interactions --- a small meson-nucleon cross
section cannot be used to argue that the meson-nucleus interaction is
small as well. There is no general rule up to which density the low-density approximation is valid. Post et al.\ have shown\cite{Post:2003hu} that e.g.\ for the eta meson the scaling of the self-energy with density is quite good, but it fails already at rather low density for the $\rho$ meson.

\subsection{$\rho$ meson}
\subsubsection{$\rho$ meson in cold matter} \label{subsec:rho}

\begin{figure}[ht]
  \begin{center}
    \includegraphics[keepaspectratio,width=0.92\textwidth]{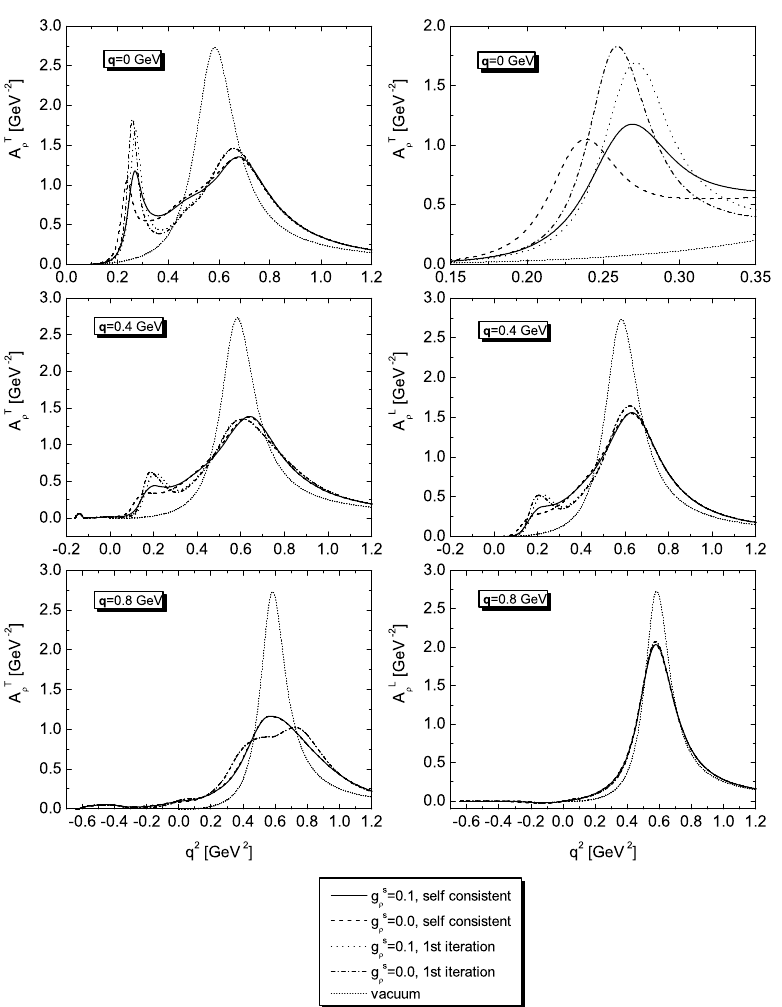}
  \end{center}
  \vspace*{8pt}
\caption{Spectral function of the $\rho$ meson at normal nuclear matter density. Shown are the transverse and the longitudinal spectral functions, $A_\rho^{\rm T}$ and $A_\rho^{\rm L}$, which are degenerate at $\mathbf{q} = 0$. Also shown are the effects of iterating the spectral functions and of varying the short-range correlation parameter $g^s_\rho$. The uppermost picture on the right shows a zoom on the low-mass peak in the spectral function. The inset explains the various curves, obtained for different values of the short-range correlation strength $g_\rho^s$ as well as in the low-density approximation (``1st iteration'') and in the full calculation (``self consistent'') (from Ref.\ \protect\refcite{Post:2003hu}).}
\label{rho_spectral}
\end{figure}

Fig.\ \ref{rho_spectral} gives the results of a calculation\cite{Post:2003hu} in the
resonance--nucleon-hole model for
the $\rho$ meson. Both the transverse and the longitudinal spectral
functions are given there for various values of the three-momentum
$\mathbf{q}$. It is seen that the spectral function can have a quite
complicated structure that cannot be parametrized in terms of a
Breit-Wigner shape. Instead, the spectral function for a $\rho$ meson
at rest in nuclear matter exhibits a distinct lower peak, at a mass of
about 500 MeV whereas the main, dominant peak appears at about 800 MeV.
This complicated structure is a direct consequence of the coupling of
the $\rho$ meson to resonance-hole excitations with the same quantum
numbers leading to level-repulsion. In this particular case, the lower-mass peak is generated by
the coupling to the N(1520) $D_{13}$ resonance, so that the
resonance-hole excitation has quantum numbers $J^\pi = 1^-$, i.e.\ the
same as the $\rho$ meson at rest. This structure moves to lower $q^2$ and becomes weaker with
increasing three-momentum of the $\rho$ meson. This is due to the
special coupling of the meson to nucleon resonances. The latter also
explains that at high momenta the transverse $\rho$ is considerably
broadened compared to the spectral function of a free $\rho$ meson
while the longitudinal $\rho$ resembles at high momenta the
free meson, with hardly any broadening; for a detailed discussion and
explanation of these effects see Ref.\ \refcite{Post:2003hu}.
The result obtained by Cabrera et al.\cite{Cabrera:2000dx} is qualitatively very similar to the $\rho$ spectral function shown here; in particular, it also exhibits the lower peak due to the resonance coupling (see also Lutz et al.\cite{Lutz:2001mi}). The results of the calculations in Refs.\ \refcite{Friman:1997tc},\refcite{Rapp:1997fs} are similar, but they did not contain the essential $D_{13}$N(1520) nucleon resonance.

The lower peak in the $\rho$ spectral function leads to a
downward shift of the first moment of the spectral function, but it is
also evident from this figure that this downward shift appears
only at low meson momenta. It should also be recalled that while the QCD sum rules require additional strength of the $\rho$ meson's spectral functions at lower masses (see Fig.\ \ref{fig:qsr-med-width.eps}) there is nothing in them that forces a downward shift of peak
masses or first moments. This then limits any attempts\cite{Kwon:2008vq} to justify the Brown-Rho scaling\cite{Brown:1991kk}.

The actual strength of this low-mass peak depends sensitively on the $NN^* \rho$ coupling strength. The PDG value of a partial decay width of the N(1520) resonance into $\rho N$ of 15-25\% is remarkably strong considering that the decay can proceed only through the tails of the $\rho$ spectral function since for the peak mass of the $\rho$ the resonance lies about 200 MeV below threshold. This large decay width then translates directly into a very large coupling constant. It is, therefore, important to realize that the $\rho N$ decay channel of this resonance has never been directly observed, but has been indirectly inferred from an analysis of $\pi N$ scattering.\cite{Manley:1992yb} While the DAPHNE experiment\cite{Zabrodin:1999sq} and the TAPS  experiment\cite{Langgartner:2001sg}, both at MAMI, seem to indicate a $\rho$ decay branch of the N(1520) resonance there is so far no reliable quantitative determination of this coupling strength available. For a detailed understanding of the in-medium $\rho$ spectral function a precise determination of the $N(1520) \to N \rho$ decay width is necessary and of high importance.

\begin{figure}[ht]
  \begin{center}
    \includegraphics[keepaspectratio,width=0.5\textwidth]{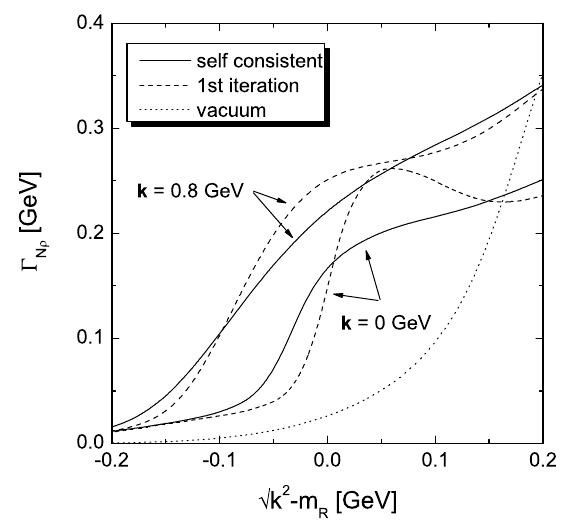}
  \end{center}
  \vspace*{8pt}
\caption{Decay width of the N(1520) $D_{13}$ resonance to the $N\rho$ channel as a function of the invariant mass of the resonance.The thin dotted curve gives the width for the free resonance. The dashed lines show a first-order result without selfconsistency for the two momenta indicated, the solid curves give the results of a selfconsistent
calculation (from Ref.\ \protect\refcite{Post:2003hu}).}
\label{N1520_spectral}
\end{figure}
As discussed above any in-medium change of the vector-meson's spectral
function will also lead to a change of the resonance spectral function.
For the $\rho$ meson which couples ``subthreshold'' to the N(1520) this
is particularly evident: any shift of strength down to lower masses
will increase the phase space for this decay and thus broaden the
resonance. The result of a coupled-channel calculation\cite{Post:2003hu} is shown in
Fig.~\ref{N1520_spectral}.
The figure illustrates nicely the effect just discussed. At the peak
mass the decay width into $N\rho$ increases dramatically compared to
the free case (dotted curve). This increase in the partial decay width
translates into an increase of the full width of the resonance. While
this increase of the resonance width cannot be directly seen in exclusive
particle-production experiments (see Ref.\ \refcite{Lehr:2001ju} and
Sect.\ \ref{subsec:obs-broad} below) it may play a major
role in the experimentally observed disappearance of resonances in the
second and third resonance region in the total photoabsorption cross
sections.\cite{Bianchi:1995vb}

In summary, the $\rho$ meson becomes significantly broader in a medium, at low momenta ($\le 1$ GeV) picking up mainly strength at lower masses due to the coupling to the nucleon resonances. The first moment of the mass distribution is thus shifted downward at low $\rho$ momenta, but for momenta larger than about 1 GeV only a nearly symmetric broadening survives with little shift of the pole mass; this is in line with early predictions in Ref.\ \refcite{Bernard:1988db}.

\subsubsection{$\rho$ meson at finite temperatures}

In heavy-ion reactions higher densities than in cold nuclei can be reached. However, such a compression is always connected with an increase of temperature; at very high, ultrarelativistic, energies a very hot fireball with nearly zero net baryon density is created. At this point a transition from hot hadronic matter to the quark-gluon plasma is expected, in which chiral symmetry is restored. It is, therefore, of considerable interest to also study the temperature dependence of the $\rho$ meson's spectral function.

\begin{figure}[ht]
  \begin{center}
    \includegraphics[keepaspectratio,width=0.65\textwidth,angle=-90]{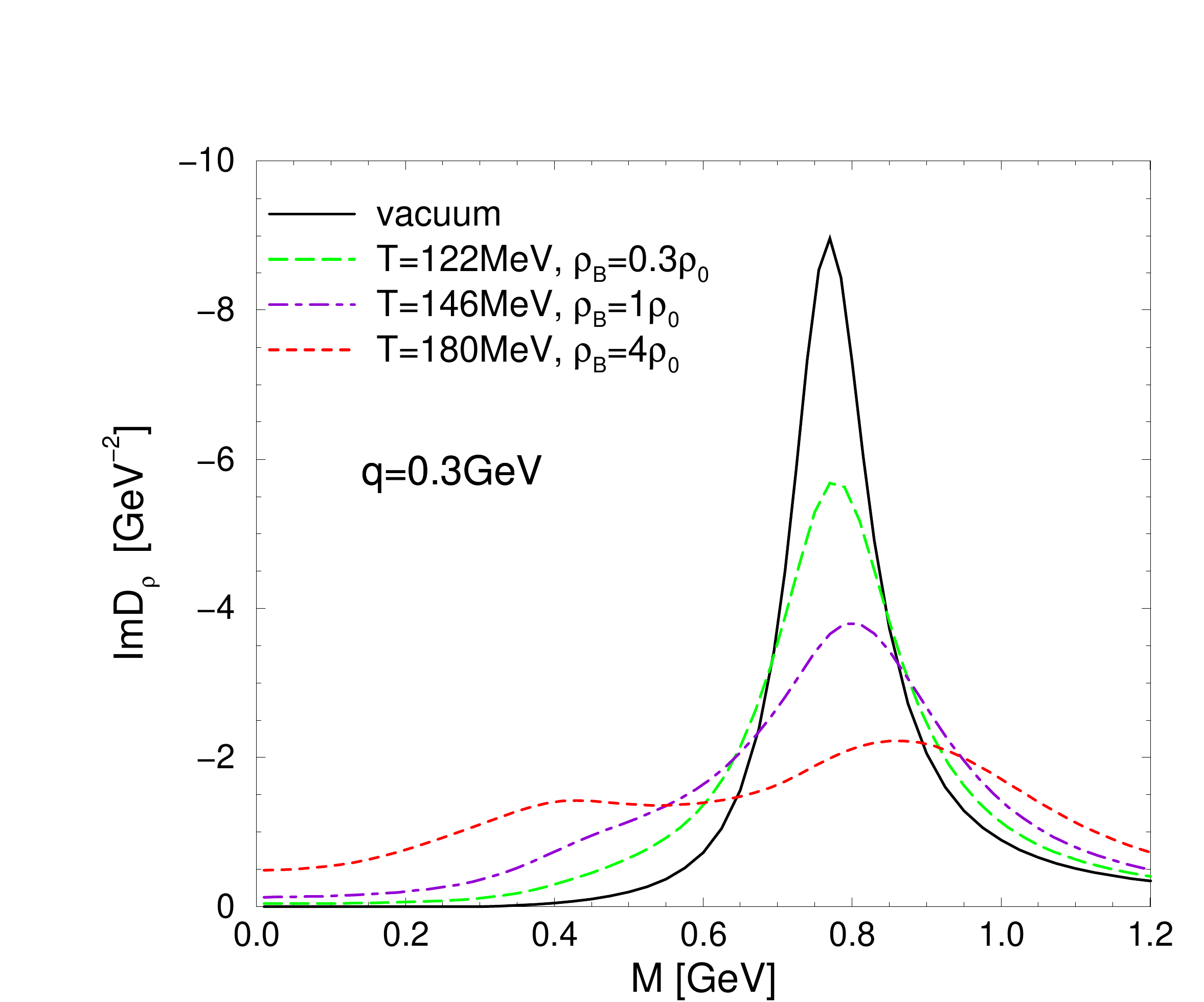}
  \end{center}
  \vspace*{8pt}
\caption{Polarization-averaged $\rho$-meson spectral function at fixed three-momentum $q = 0.3$ GeV in hot and dense hadronic matter at the temperatures and densities indicated in the figure (from Ref.\ \protect\refcite{Rapp:1999ej}).}
\label{RWrho}
\end{figure}
In a hot hadronic environment $\rho$ mesons pick up in-medium changes of their spectral functions from three different interactions: first, the external decay legs, the pions, get modified in a hot hadronic environment, second there is a contribution from direct $\rho$ scattering off thermal pions and third, there is $\rho N$ scattering, where the nucleons are thermally distributed in momentum space. Such a model has been worked out by Rapp and Wambach.\cite{Rapp:1999us,Rapp:1999ej} Fig.\ \ref{RWrho}, taken from their work, shows that the spectral function becomes very broad
when the density and temperature approach the expected transition into the quark-gluon plasma phase. Overall the $T$-dependence broadens the spectral function; raising the temperature thus has an effect similar to that seen when raising the $\rho$-meson momentum in cold matter (cf.\ Fig.\ \ref{rho_spectral}).

\subsection{$\omega$ meson}
\label{subsec:omega}

For the $\omega$ meson the theoretical situation is more complicated
than for the $\rho$. This is due to the fact that the elementary $\omega$
production cross section close to threshold is still a problem of
ongoing research and that, correspondingly, there is much less known
about the coupling of $\omega$ mesons to nucleons. Indeed, the PDG does
not list any nucleon resonance with an $N \omega$ decay branch. Thus a simple application of a resonance-hole model is not possible in this case.

\begin{figure}[ht]
  \begin{center}
    \includegraphics[keepaspectratio,width=0.65\textwidth]{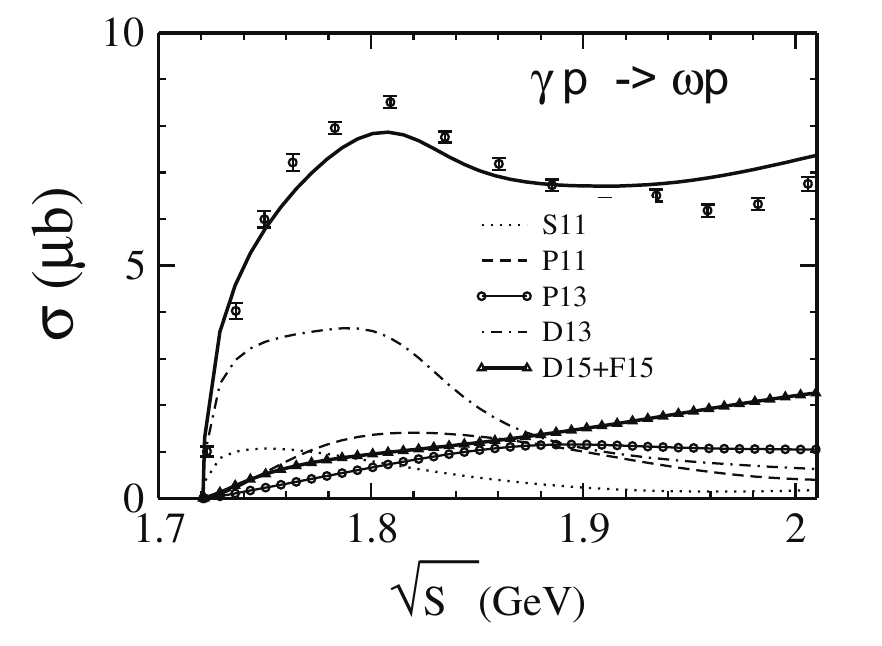}
  \end{center}
  \vspace*{8pt}
\caption{Total and partial wave contributions to the
$\gamma + p \to p + \omega$ cross section (from Ref.\ \protect\refcite{Shklyar:2004ba}).}
\label{K-omega}
\end{figure}

The uncertainties in extracting the resonance contributions from measured cross sections can be illustrated for the $\omega$ meson where recent $\gamma + N \to N + \omega$ data have been analyzed in the framework of a coupled channel $K$-matrix calculation, see Fig.\ \ref{K-omega}.
The figure shows that various resonances, plus $t$- and $u$-channel background terms, all contribute coherently to the observed production cross section. There is thus no obvious resonance that sticks out through its strength.\cite{Shklyar:2004ba} The largest contributions at threshold come from the $D_{13}$ and the $S_{11}$ partial waves.

At present, calculations of the $\omega$ in-medium self-energy therefore still
have to start from theoretical analyses of $\pi N \to \omega N$ and
$\gamma N \to \omega N$ cross sections. In particular the latter have
much smaller experimental errors.\cite{Barth:2003kv,:2008gs} In the model of Refs.\
\refcite{Lutz:2001mi,Lutz:1999jn,Wolf:2004hr}, fitted to photo- and pion-production
experiments invoking vector-meson dominance from the start, resonances are generated dynamically. A rather
strong coupling of the N(1520) $D_{13}$ and the N(1535) $S_{11}$ to the
$\omega N$ channel is found. Thus, here the situation is similar to that
for the $\rho$ meson, and indeed a second peak appears in the in-medium
spectral function at lower masses. This calculation was restricted to the $\omega$ meson at rest in the
nuclear medium, while actual experiments lead to rather large momenta
of the $\omega$ meson produced. The authors of Ref.\ \refcite{Muehlich:2006nn}
have overcome this restriction by using a unitary $K$-matrix approach to
the analysis of $\pi N$ and $\gamma N$ data using the same Lagrangian
and the same parameters for both types of reactions.\cite{Penner:2002ma,Penner:2002md} The $K$-matrix
method is a unitary coupled-channel approach and gives cross sections for the reaction under study. Resonance parameters and
coupling constants are obtained by a fit to measured cross sections.
Thus all the couplings are constrained by data.

\begin{figure}[ht]
  \begin{center}
    \includegraphics[keepaspectratio,width=0.65\textwidth]{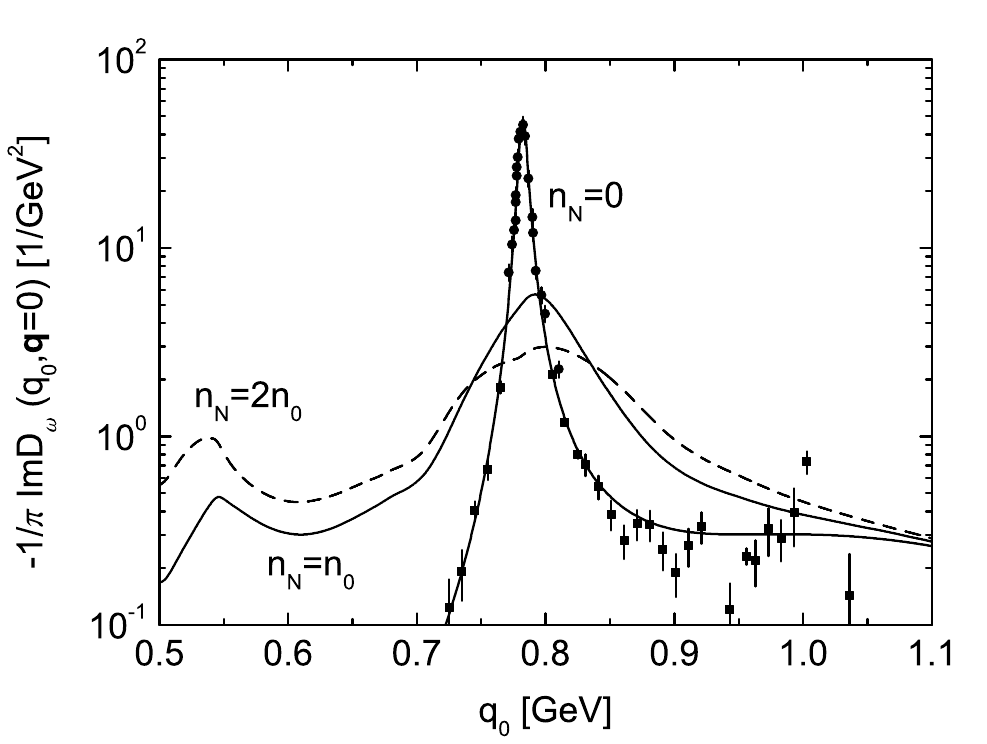}
  \end{center}
  \vspace*{8pt}
\caption{The $\omega$ spectral function for an $\omega$ meson at rest. The solid curve through the data describes the spectral function of a free $\omega$ meson. The solid and dashed curves show the spectral function for density $\rho_0 = 0.16$ fm$^{-3}$ and $2 \rho_0$, respectively (from Ref.\ \protect\refcite{Muehlich:2006nn}).}
\label{ominmed}
\end{figure}

The results of these studies show --- as for the $\rho$ meson --- again a strong influence of couplings to nucleon resonances; $t$-channel graphs
alone give a much too small contribution to the cross section at backward angles.\cite{MuhlichDiss} Contrary to the $\rho$ meson case, however, there is much more interference
between resonance and $t$-channel amplitudes and contributions from
several overlapping nucleon resonances. Fig.~\ref{ominmed} shows the
result for the $\omega$ spectral function, both in vacuum and at
densities of $\rho_0$ and $2\rho_0$.
The in-medium spectral function of the $\omega$ meson is again
double-humped, but the lower-mass hump is much less developed than for
the $\rho$ meson, reflecting the lower coupling strength to the
resonances around 1500 MeV; qualitatively the result is similar to that obtained by the authors of Refs.\ \refcite{Lutz:2001mi,Lutz:1999jn,Wolf:2004hr} in a related study. The lower mass peak is due to the
excitation of the N(1535) $S_{11}$ and the N(1520) $D_{13}$ resonances. These resonances couple in
a relative $s$-wave to the $N\omega$ channel and dominate the spectrum
at low $\omega$ momenta. The in-medium spectral function shows only a
very small upwards shift of the main peak, but the spectral function at
the free mass is considerably broadened to roughly 60 MeV at saturation density $\rho_0$ and vanishing $\omega$ momentum.
It has also been shown in sum rule analyses that a downward shift of the $\omega$ mass requires rather extreme assumptions about the in-medium changes of the four-quark condensate.\cite{Zschocke:2002mp,Steinmueller:2006id}
The near constancy of the peak mass shown in
Fig.\ \ref{ominmed} indicates that such a strong in-medium change of the four-quark
condensate is not required to be in line with hadronic interactions.

With increasing $\omega$ momentum the width grows due to collisions
with the nucleons of the surrounding medium. Again, it is the
transverse degree of freedom that gets significantly
collision-broadened, the longitudinal width stays nearly constant
 with $\omega$ momentum.\cite{Muehlich:2006nn}
While these calculations are
done only in the low-density approximation we expect that taking into
account higher-order meson-baryon interactions will not change the
picture qualitatively. Selfconsistency at higher densities tends to
smoothen structures in the spectral functions; we thus expect even
slightly larger broadening.

Among the more recent calculations there is now general agreement that the $\omega$ mass is only very weakly changed in a medium.\cite{Muehlich:2006nn,Lutz:2001mi,Lutz:1999jn,Wolf:2004hr,Riek:2004kx} Earlier calculations that gave a sizeable downward shift of the mass\cite{Klingl:1997kf,Klingl:1997tm,Klingl:1996ps} are either unreliable in the methods used\cite{Eichstaedt:2007zp} or neglect the essential dispersive corrections\cite{Saito:1998wd}. As a consequence of the more recent hadronic calculations also the sign of the $\omega N$ scattering length is not expected to change much from its free value which comes out to be negative in these calculations. This indicates that contrary to other earlier calculations a bound $\omega$-nucleus state does not exist. The newer calculations agree in their prediction of a width of about 60 MeV at saturation density $\rho_0$ for an $\omega$ meson at rest. This width can directly be accessed through nuclear transparency measurements (see Sect.\ \ref{omega_el} below).

\subsection{$\phi$ meson}
\label{subsec:phi}
For the $\phi$ meson the situation is different from that for the
lighter vector mesons. Because of the pure strange quark composition $\bar{s} s$ of the $\phi$ there are no nucleon resonances that couple to $\phi N$ in the $s$-channel. Also the $\phi N$ cross section is theoretically expected to be
quite small (7-13 mb). As a consequence the main change of the
spectral function is due to the dressing of the kaons in the $K
\bar{K}$ decay channel of the $\phi$. This implies that the main in-medium change here is in the width since the vacuum width of the $\phi$ is very small; its mass shift should be negligible. Indeed, one of the latest calculations gives a mass
shift of $-6$ MeV and a broadening of about 28 MeV at saturation
density\cite{Oset:2000eg,Cabrera:2002hc} for a $\phi$ meson at rest. The calculation indeed gets its effects mainly from
a dressing of the kaon loops in the $\phi$ self-energy.

In summary, the mass of the $\phi$ meson should hardly change in a medium. Instead the main effect is that of a collisional broadening. The latter affects directly the imaginary part of the self-energy and can thus be extracted from nuclear transparency measurements.

\subsection{Scalar-meson ($\sigma$) production on nuclei}
\label{subsec:sigma-prod}

The restoration of chiral symmetry is connected with an equality of the spectral properties of the $\pi$ and $\sigma$ mesons which are chiral partners. Hatsuda et al. and Bernard et al.\ have pointed out that chiral symmetry restoration in hot and dense matter should then lead to a lowering of the $\sigma$ spectral function.\cite{Hatsuda:1985eb,Bernard:1987im,Bernard:1987sx}. As a further consequence the --- in vacuum dominant --- $2\pi$ decay channel of the $\sigma$ meson should become suppressed because of phase-space limitations. This in turn leads to an increase of the life time of the $\sigma$ meson. This conjecture that a partial restoration of chiral symmetry in ``normal'' nuclear matter should cause a softening and narrowing of the $\sigma$ meson has indeed been worked out by Hatsuda et al.\cite{Hatsuda:1999kd} and in Refs.\ \refcite{Chiang:1997di,Rapp:1998fx} and has triggered experiments to measure the $2\pi$ invariant mass distribution near threshold in pion and photon induced reactions on nuclei\cite{Bonutti:1999zz,Starostin:2000cb,Messchendorp:2002au}. Even though the particle character of the $\sigma$ in vacuum is questionable, it might develop a much narrower peak at finite baryon density due to phase-space suppression for the $\sigma \to \pi\pi$ decay and, in particular, a shift of strength towards lower masses.

 This suggestion has led to further theoretical studies of the in-medium $2\pi$ propagator. Models with chiral constraints and realistic pion interactions in nuclei indeed show effects on the $\pi$-$\pi$ interaction and thus on the propagator with a resulting shift of the invariant mass distribution to lower masses.\cite{Chanfray:2004vb,Kaskulov:2005kr,Cabrera:2005wz} The main effect here is caused by the coupling of the pion to $\Delta$--nucleon-hole excitations.

\section{Hadronic Spectral Functions and Observables}
\label{sec:mosel2}
In the preceding section we have discussed the theoretical approaches to calculate in-medium self-energies of vector and scalar mesons. We now discuss the problem how to relate these self-energies and in-medium spectral functions to actual observables.

\subsection{Non-equilibrium effects}
All the calculations of self-energies that we discussed in the preceding section start from the basic assumption of complete equilibrium: a hadron is embedded into nuclear matter at fixed, constant density $\rho$, without any surface effects,  and temperature $T$. This yields the self-energy $\Pi(\rho,T)$ as a function of density and temperature and this self-energy is then --- in a local-density approximation --- used to describe the hadron also in finite nuclei at higher excitation. This procedure is not without problems in the case of heavy-ion reactions where the dynamics of the collision involves also non-equilibrium phases and where the properties of the equilibrium phase change with time. This question has so far hardly been tackled (see, however, Ref.\ \refcite{Schenke:2005ry}).
 On the other hand, this may be a reasonable approximation to hadron production reactions on cold, static nuclear targets where the whole reactions proceed close to equilibrium.

\subsection{Nuclear transparency}
\label{sec:nucl-transp}
The imaginary part of the in-medium self-energy $\Im \Pi_{\rm med}$ of a vector meson can be determined from nuclear transparency measurements where the total hadron yield produced in a reaction of an elementary projectile ($\gamma, \pi, p$) on a nuclear target is compared with that on a free nucleon. The nucleus then acts both as a target and as an attenuator. For the specific case of photoproduction, the total vector-meson production cross section on a nuclear target then reads in a simple Glauber approximation\cite{Muhlich:2005kf}
\begin{equation}     \label{att}
\sigma_{\gamma A} = \int d\Omega \int d^3r\, \rho(\vec{r}) \frac{d\sigma_{\gamma N}}{d\Omega}\, {\rm exp}\left(\frac{1}{|\vec{q}\,|} \int_0^{\delta r} dl \Im \Pi_{\rm med}(q,\rho(\vec{r}\,')) \right) P(\vec{r} + \delta \vec{r})
\end{equation}
with
\begin{equation}
\vec{r}\,' = \vec{r} + l \frac{\vec{r}}{r} \,, \quad \delta \vec{r} = v \frac{\gamma}{\Gamma_{\rm vac}} \frac{\vec{q}}{|\vec{q}\,|}  \,,
\end{equation}
and the (local) nucleon density $\rho(\vec{r})$; here $\Gamma_{\rm vac}$ is the free decay width of the hadron in its restframe.
Finally, $P(\vec{r})$ is the probability for the final-state hadrons to be absorbed. The imaginary part of the in-medium self-energy that determines the attenuation in (\ref{att}) is connected to the collision width of the vector meson by
\begin{equation}
\Gamma_{\rm coll} = -  \frac{1}{\omega} \Im \Pi_{\rm med} \approx  \rho \sigma v \,,
\label{Gammacoll-reit}
\end{equation}
where the last (classical) expression follows only in the low-density approximation. In all these equations $\vec{q}$ is the three-momentum of the vector meson and $\omega$ its energy. If one now assumes that the low-density limit holds, i.e.\ that the collisional width is determined by two-body collisions of the vector meson with a nucleon, then $\Gamma$ is connected to the effective in-medium two-body cross section by (\ref{Gammacoll-reit}). However, it is worthwhile to remember that already for the pion only one half of the absorption is due to two-body collisions whereas the other half involves three-body interactions.\cite{Oset:1986yi}

Eq.\ (\ref{att}) contains only absorption effects and no sidefeeding (regeneration) of the channel under study which in principle might affect the measured transparency ratios.\cite{Sibirtsev:2006yk} However, detailed comparisons with full transport calculations that contain such effects show that the Glauber approximation works very well for the cases studied here.\cite{Muhlich:2005kf,MuhlichDiss} Regeneration of meson channels through secondary interactions plays a role only for weakly absorbed mesons,
such as e.g.\ the $K^+$ (Ref.\ \refcite{Effenberger:1999jc}) or low-energy pions (cf.\ also the remark on uncharged pions in Sect.\ \ref{sec:hadspec-obs} and the corresponding discussion on the
results for 2$\pi$ production in Sect.\ \ref{subsec:exp-sigma}).

\subsection{Influence of branching ratios}
Experimental determinations of the full in-medium spectral function that involves imaginary \emph{and} real parts of the self-energy always rely on a reconstruction of the spectral function from the measured four-momenta of two decay products (e.g., $\rho \to 2\pi$; $\omega \to \pi^0 \gamma$; $\phi \to K^+K^-$; $\rho,\omega,\phi \to e^+e^-$). In a direct reaction of a microscopic probe, such as a photon, with a nucleon $N$ at a center-of-mass energy
$\sqrt{s}$ the production cross section for a vector meson $V$ with invariant mass $\mu$ is given by\cite{MuhlichDiss,Gallmeister:2007cm}
\begin{equation}
\frac{d\sigma_{\gamma N \to VN}}{d\mu} = 2 \mu \frac{1}{16\pi s|\mathbf{k}_{cm}|} |\mathcal{M}_{\gamma N \to VN}|^2 \mathcal{A}(\mu) \, |\mathbf{q}_{\rm cm}|
\end{equation}
Here $\mathcal{M}$ is the transition matrix element, $\mathbf{k}$ and $\mathbf{q}$ are the momenta of the incoming photon and the outgoing vector meson, respectively, in the cm system and $\mathcal{A}(\mu)$ is the spectral function of the vector meson\footnote{In the nuclear medium Lorentz invariance is not manifest and thus the spectral function depends both on $\mu$ and on the vector meson's three-momentum $\mathbf{q}$.}. Assuming a decay of the vector meson into two final particles $p_1$ and $p_2$ the cross section for the production of the final state, again with invariant mass $\mu$, is given by:
\begin{equation}    \label{dsigmadmufinal}
\frac{d\sigma_{\gamma N \to N(p_1,p_2)}}{d\mu} = \frac{d\sigma_{\gamma N \to VN}}{d\mu} \times \frac{\Gamma_{V \to  p_1 + p_2}}{\Gamma_{\rm tot}}(\mu) \times P_1 P_2 ~.
\end{equation}
Here $\Gamma_{\rm tot}$ is the total width of the meson $V$, obtained as a sum of the vacuum decay width, $\Gamma_{\rm vac}$, and an in-medium contribution:
\begin{equation}
\Gamma_{\rm tot} = \Gamma_{\rm vac} + \Gamma_{\rm med} ~.
\end{equation}
 The ratio $\Gamma_{V \to  p_1 + p_2}/\Gamma_{\rm tot}$ represents the branching ratio into the final state $p_1,p_2$ and $P_i$ gives the probability that the particle $p_i$ survives absorption or rescattering in the final state (we neglect here possible channel couplings). In the nuclear medium $\Gamma_{\rm tot}$ increases; it is essential that this increase is contained both in the spectral function $\mathcal{A}$ and the branching ratio.

Eq.\ (\ref{dsigmadmufinal}) shows that the invariant-mass distribution reconstructed from the four-vectors of the final particles always contains effects not only from the spectral function, but also from the branching ratio and from the final-state interactions (fsi) which also depend on $\mu$. For the case of dilepton final states the latter two are known:\cite{MuhlichDiss} the decay width goes like $1/\mu^3$  and the branching ratio like $1/(\mu^3 \Gamma_{\rm tot})$; the dilepton fsi can be neglected. The strongly $\mu$-dependent branching ratio shifts the observed mass distribution significantly to lower masses. For hadronic ($\pi\pi$, $KK$) or semi-hadronic ($\pi^0\gamma$) final states the mass dependence of the decay branching ratio is often not very well known and has to be modeled. While there are theoretical studies available\cite{MuhlichDiss} experimental determinations of hadronic spectral functions for the $\omega$ meson have so far not taken this into account. This is particularly critical if new particle thresholds open in the mass region of interest. This increases the total width and leads to a strong fall-off of the branching ratio with increasing mass.
For example, the branching ratio for the decay $\omega \to \pi^0 \gamma$ is strongly influenced by the opening of the $\rho \pi$ channel just in the $\omega$ mass region.\cite{MuhlichDiss} In this special case a further complication arises from the fact that the $\rho$ meson gets broadened in a medium so that as a consequence the total decay width of the $\omega$ meson may change in the nuclear medium.

In summary, any extraction of the spectral function that relies on the determination of the invariant mass distribution $\mathcal{P}(\mu)$ from the four-momenta of the final particles
\begin{equation}
\mu = \sqrt{(p_1 + p_2)^2}
\end{equation}
requires that the partial decay width is divided out
\begin{equation}
\mathcal{A}(\mu) = \mathcal{P}(\mu) \, \frac{\Gamma_{\rm tot}}{\Gamma_{V \to  p_1 + p_2}} ~.
\end{equation}
This complication is independent of any final-state interactions that the decay products may experience.

\subsection{Observability of collisional broadening}
\label{subsec:obs-broad}

Even if theory predicts a significant broadening of hadronic spectral functions in the nuclear medium it is not clear if this broadening can be directly observed.\footnote{The in-medium width which leads to this broadening can be obtained from transparency measurements as discussed in Sect. \protect\ref{sec:nucl-transp}.}
First of all, the in-medium hadrons do not all experience one given density.
Instead, because of the nuclear density profile, the relevant densities range from saturation ($\rho_0$) down to zero in the nuclear surface. Most nucleons are embedded in densities of about 1/2 to 2/3 of normal nuclear density $\rho_0$. Thus it is immediately clear that hadrons in nuclei will not exhibit properties corresponding to saturation density, but instead to a lower one. Even more important is, however, another not so straightforward effect that suppresses contributions from higher densities. We briefly outline this effect here; more details can be found in Refs.\ \refcite{Lehr:2001ju} and \refcite{MuhlichDiss}.

The semi-inclusive cross section for the production of final states, e.g., via a vector-meson resonance
in a photon-nucleus reaction, is obtained by
integrating Eq.\ (\ref{dsigmadmufinal}) over all nucleons. The result involves a factor
\begin{equation}
\label{inmedXsection}
\mathcal{A}(\mu)\, \frac{\Gamma_{V \to \rm final\;state}}{\Gamma_{\rm tot}}
 = \frac{\mu \, \Gamma_{\rm tot}}{(\mu^2 - m_V^2)^2 + \mu^2 \Gamma^2_{\rm tot}}\, \frac{\Gamma_{V \to \rm final\;state}}{\Gamma_{\rm tot}} ~
\end{equation}
with\begin{equation}
\Gamma_{\rm tot} = \Gamma_{\rm vac} + \Gamma_{\rm med} ~.
\end{equation}
The in-medium width $\Gamma_{\rm med}$ depends on density; in the low-density approximation it is linearly proportional to the density $\rho$ (cf.\ Eq.\ (\ref{Gammacoll})),
\begin{equation}
\label{Gammamed}
\Gamma_{\rm med}(\rho(r)) = \Gamma_{\rm med}(\rho_0) \, \frac{\rho(r)}{\rho_0} ~.
\end{equation}

\begin{figure}[ht]
\begin{center}
    \includegraphics[keepaspectratio,width=0.6\textwidth]{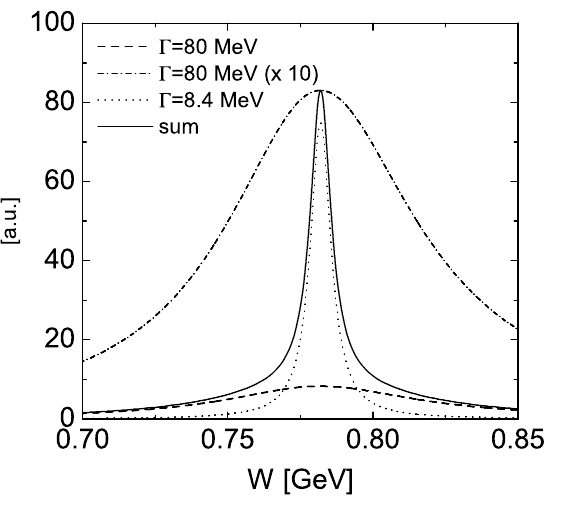}
  \end{center}
  \vspace*{8pt}
\caption{Relativistic Breit-Wigner distributions with equal integrated strength. Their widths are 8.4 and 80 MeV, as given in the figure. The summed spectrum has a fitted Breit-Wigner width of about 12 MeV (from Ref.\ \protect\cite{MuhlichDiss}).}
\label{om-inmedwidth}
\end{figure}

If $\Gamma_{\rm med} \gg \Gamma_{\rm vac}$, as it is the case at least for $\omega$ and $\phi$ mesons, the density dependence of the in-medium width drives the sensitivity of a meson-production experiment towards the surface. Contributions from higher densities are suppressed by order $1/\rho^2$; they are significantly broader and lower in their maximum. Inspecting Eq.\ (\ref{inmedXsection}) one sees that at the peak position, $\mu = m_V$, one suppression factor $1/\rho$ comes from the spectral function and one from the branching ratio. The effect can be seen in Fig.\ \ref{om-inmedwidth} where simply two Breit-Wigner spectral functions for the $\omega$ meson with widths differing by about a factor of 10 and equal integrated strengths have been superimposed. Since the significantly broader distribution is suppressed $\sim 1/(\Gamma_{\rm vac} + \Gamma_{\rm med})$ it changes the summed distribution only in the outer tails. In an experiment this change is difficult to separate from a background contribution.
This comparison contains only the effects of the density dependence of the spectral function. The effects of the additional density dependence of the branching ratio will only enhance the observed behavior, contributing an additional suppression factor $1/\rho$. In the example shown only the width was increased, but it is clear that the same suppression will also take place if -- in addition -- there is a shift in the peak mass. The final-state interactions, not taken into account in this argument, will actually lead to an even further suppression of signals from higher densities, if the decay channel involves strongly interacting particles.

\begin{figure}[ht]
\begin{center}
    \includegraphics[keepaspectratio,width=0.7\textwidth]{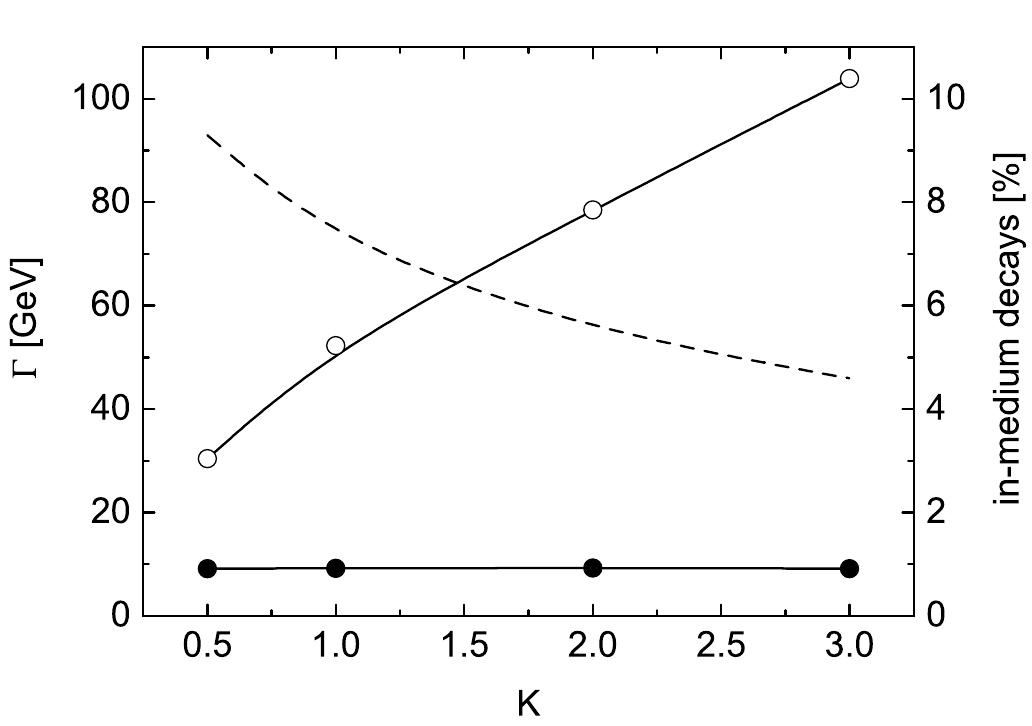}
  \end{center}
  \vspace*{8pt}
\caption{Fitted Breit-Wigner (BW) width of the $\pi^0 \gamma$ spectrum from transport simulations for the reaction $\gamma + \,^{40}$Ca $ \to \omega + X$ at 1.5 GeV photon energy. The full symbols show the BW width of the full spectrum, the open symbols that from events with in-medium decays ($\rho > 0.1 \rho_0$) only. The dashed line (right axis) shows the percentage of in-medium decays; here no cut on low momentum $\omega$ mesons has been applied (from Ref.\ \protect\refcite{MuhlichDiss}).}
\label{om-inmedwidth1}
\end{figure}

Fig.\ \ref{om-inmedwidth1} shows the results of a full simulation for the reaction
$\gamma + \,^{40}$Ca at $E_\gamma = 1.5$ GeV.
Plotted is the width of the $\pi^0 \gamma$ spectrum as a function of a $K$ factor that multiplies the in-medium width in Eq.\ (\ref{Gammamed}) to account for a possible increase of the in-medium $\omega N$ cross section. While the total observed width stays nearly constant when $K$ is increased by a factor of 3, the width of the in-medium decay events (solid line) indeed increases with $K$, but its relative contribution (dashed line) decreases at the same time.

The suppression of higher-density contributions is less pronounced if total cross sections are considered instead of semi-inclusive or exclusive ones. In this case $\Gamma_{V \to \rm final \; state}$ in (\ref{inmedXsection}) has to be replaced by the total width $\Gamma_{\rm tot}$, effectively removing one power of $1/\rho$  so that there is less suppression for higher densities. If in addition the in-medium width is comparable to the vacuum width and not, e.g., an order of magnitude higher, then the broadening can be observable. Concerning baryon resonances, this explains why the total photoabsorption cross section on nuclei exhibits a clear broadening in the second-resonance region.\cite{Bianchi:1995vb}
A similar situation prevails in heavy-ion collisions where $\Gamma_{\rm tot}$ in the denominator of the branching ratio is canceled by $\Gamma_{\rm in}$, the width for producing the hadron in the entrance channel: While in an elementary
nuclear reaction $\Gamma_{\rm in}$ is selected by the specific experiment, the thermal
production of the considered hadron in a heavy-ion collision involves all possible
production channels, i.e.\ $\Gamma_{\rm in} \sim \Gamma_{\rm tot}$.

\subsection{Final-state interactions}
\label{sec:hadspec-obs}
In addition, the spectral functions observed in experiments by reconstructing them from hadronic decay products can be quite different from those of the original decaying meson. Final-state interactions can affect --- through rescattering --- the momenta and angles of the final-state hadrons, thus affecting also the spectral information reconstructed from the four-momenta of these hadrons. For example, the rescattering always leads to energy loss of the scattered outgoing hadron. This necessarily shifts the strength distribution towards smaller masses. In addition, hadrons can get absorbed or even reemitted (fsi); the former process will reduce the cross sections whereas the latter spoils any connection between the final observed momenta and the original spectral function.

While fsi do affect the actual observables in a significant way, if hadrons are among the decay products, their theoretical treatment is in most cases not up to the same degree of sophistication. Usually otherwise quite sophisticated in-medium calculations, which use state-of-the-art theoretical methods, do apply much more simplified methods to the treatment of fsi. Among the latter are the eikonal approximation or even a simple, absorption-only Glauber treatment. An obvious shortcoming of both of these methods is that it is ad hoc assumed that the particle that is initially hit in the very first interaction is the same as the one that ultimately leaves the nucleus on its way to the detector. However, detailed analyses have shown that there can be significant contributions from coupled-channel effects, in which a sidefeeding from an initially different channel into the final channel takes place. An example is charge transfer for pions where the more copiously produced charged pions can --- due to fsi --- be converted into uncharged ones. Such an effect plays a major role in particular in reactions with elementary probes in the incoming channel.

In addition, both the eikonal and the simple Glauber method take only flux out of a given channel; they do not yield any information on what happens with these particles. This is a major shortcoming for inclusive and semi-inclusive reactions.

An up-to-date method to treat fsi that is free of these shortcomings is provided by transport calculations. These transport calculations do take all the coupled channel effects into account, they allow for elastic and inelastic interactions and for sidefeeding and absorption. They are limited to inclusive, incoherent processes, so that exclusive particle production, for example, in coherent interactions cannot be described. However, for inclusive and semi-inclusive (or even semi-exclusive) reactions they are applicable and yield the desired results. They also provide a full dynamical simulation of the reaction and thus help to understand the reaction mechanism. State-of-the-art methods all rely on the Boltzmann-Uehling-Uhlenbeck (BUU) equation.\cite{Cassing:1990dr} A modern example of such an approach is provided by the GiBUU model.\cite{GiBUU}

\section{Experimental Approaches and Results}
\label{sec:exp}

The theoretically predicted medium modifications, discussed in the previous
sections, can be studied
experimentally by measuring the mass distribution of short-lived hadrons. The
life times of these hadrons have to be so short that, after being
produced in some nuclear reaction, they decay with
sizeable probability within the nuclear environment, i.e. within the atomic nucleus or in the collision
zone of a heavy-ion reaction.
Information on the in-medium mass $\mu$ of a hadron can be deduced from the four-momentum
vectors $p_1, p_2$ of its decay products for different three-momenta $\vec{p}$
of the hadron with respect to the nuclear medium. In general, the mass depends
on the baryon density $\rho$ and temperature $T$ of the medium:
\begin{equation}
\mu(\vec{p},\rho,T) = \sqrt{(p_1 + p_2)^2} \,. \label{VM:mass}
\end{equation}
The light vector mesons $\rho, \omega$, and $\phi$ are
particularly suited for these investigations. Produced in a nuclear
reaction at energies in the GeV regime, their decay lengths in the laboratory
$\beta \cdot \gamma \cdot c \cdot \tau \approx$  1.3 fm, 23 fm and 46 fm, respectively,
are of the order of magnitude of nuclear dimensions ($\beta \cdot \gamma $ is
of the order of 1 for the reactions of interest). For experimental investigations of
in-medium properties of the longer lived $\omega$ and $\phi$ mesons severe cuts on their
momentum will nevertheless have to be applied to ensure a sizeable fraction of
decays in the nuclear medium.

In principle, all decay modes of a meson can be used to reconstruct its
in-medium mass according to Eq.\ (\ref{VM:mass}).
As discussed in Sect. \ref{sec:hadspec-obs} a problem, however, arises when hadrons are involved as final states
since these hadrons may interact strongly with the surrounding
nuclear medium. As a consequence, their four-momentum vectors may be distorted
leading to an erroneous result for the invariant mass of the decaying meson.
In special cases, which have to be investigated individually, these effects
can, however, be reduced by specific cuts on kinematic variables like in the
$\omega \rightarrow \pi^0 \gamma$ decay (see Sect.\ \ref{omega_el}). Alternatively,
these final-state effects can be modeled with good accuracy as
outlined above.

Leptons are the preferred decay channel because they escape even a
compressed collision zone of a heavy-ion reaction without strong final-state
interactions. Vector mesons have such a decay branch into lepton pairs ($e^+ e^-$
or $\mu^+\mu^-$).
Unfortunately, the branching ratios are only of the order of $10^{-5}$-$ 10^{-4}$ which make
these measurements very difficult and sensitive to background subtraction.

Medium modifications of hadrons have been investigated experimentally in nuclear
reactions with elementary probes as well as in heavy-ion collisions. Both approaches have their advantages and
disadvantages. First observations of a change of the vector meson spectral function in dense matter were obtained in heavy-ion experiments. These experiments do reach high densities and temperatures but they suffer from an intrinsic limitation: every experimental signal is time-integrated over the whole collision history and thus also over very different states of strongly interacting matter. While the initial hard nucleon-nucleon (parton-parton) collisions in relativistic heavy-ion reactions are highly out of equilibrium, the system then develops to a thermalized quark-gluon plasma or hadronic phase with a temperature that decreases with time until freeze-out is reached. Any theoretical analysis of heavy-ion data with a focus on reconstructing the vector meson's spectral function thus also involves modeling the time-development of the reaction.
Indeed, the latest analysis of dilepton data shows a significant dependence of the momentum spectra and of the overall yield on the model parameters describing the time development of the reaction.\cite{vanHees:2009vk} Usually, the time development is not obtained from dynamical simulations, but instead is parametrized in terms of the times which the system spends in the various phases. This parametrization is then constrained by other observables, such as flow.

The advantage of heavy-ion collisions is that the regeneration of mesons in
the collision zone helps to enhance the in-medium effects: The dilepton yield
from meson decays in the hot and dense medium is determined by the partial
decay width $\Gamma_{l^+ l^-}$ and the life time of the fireball $\tau_F$,
i.e.\ $N_{l^+ l^-} \sim \Gamma_{l^+ l^-} \cdot \tau_F$, independent of the life time of
the meson.
In elementary reactions on nuclei the yield from meson decays within the nuclear
medium is again given by the partial decay width $\Gamma_{l^+ l^-}$ and --- in
contrast to heavy-ion collisions ---  proportional to the
in-medium life time of the meson $\tau = 1/\Gamma_{\rm tot}$ ,
i.e.\ $N_{l^+ l^-} \sim \Gamma_{l^+ l^-}/\Gamma_{\rm tot}$.
Since the total width $\Gamma_{\rm tot}$ is increased in a medium --- and as we will
discuss below this increase may be large --- the yield of the particles of interest
can sizeably drop as compared to the vacuum case, as discussed in Sect.\ \ref{subsec:obs-broad}.
Nevertheless, and even though the densities probed in the elementary nuclear reactions
are lower, the medium effects may still be comparable to those in
heavy-ion reactions.\cite{Mosel:1998rh}
In addition, there is no time dependence of the density
and temperature which makes the theoretical analysis of the results more
transparent.\cite{Mosel:1992rb}
Since the determination of in-medium hadron masses and life times
requires rather demanding experiments corresponding results have only recently
been published.

\subsection{Experimental challenges}
\label{sec:chall}
The detection and identification of electron or muon pairs in heavy-ion collisions but also
in photon and proton induced reactions represents a severe experimental
challenge. The main problem is the discrimination against $\pi^+  \pi^-$
pairs which are produced in these reactions with about 4-5 orders of
magnitude higher yields. All experimental setups working in this field use
Cherenkov detectors for electron/pion discrimination. In addition, information
from shower detectors or shower sizes in calorimeters is used. Pion-pair
suppression factors of the order of 10$^{-7}$  have been reported.\cite{Wood:2008ee}
\begin{figure}[th]
  \begin{center}
    \includegraphics[keepaspectratio,width=0.35\textwidth]{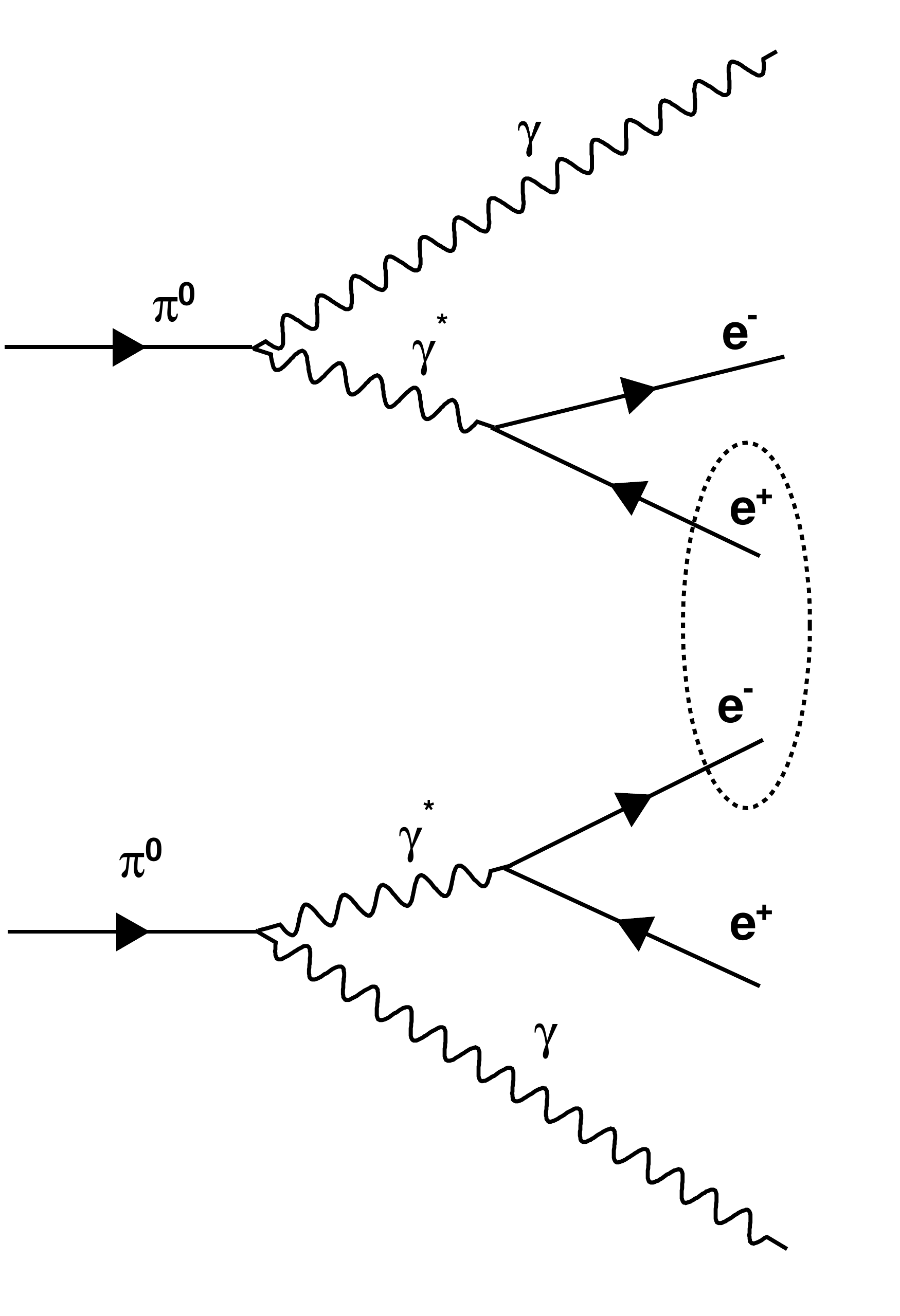}
  \end{center}
  \vspace*{8pt}
\caption{Feynman diagram illustrating the generation of unlike-sign
  combinatorial background through the Dalitz decay of two $\pi^0$ mesons
  produced in the same event.\protect\cite{Wood:2008ee}}
\label{fig:comb-background}
\end{figure}
Furthermore, conversion of photons into electron pairs represents a huge
background. To suppress this contribution the detectors have to be built with minimum
material budget. Thin multi-layer target slices rather than bulk targets have to be used
to suppress conversion already in the target. These processes are
characterized by small opening angles of the lepton pairs and can thus be
reduced by cuts on the angle between the two leptons.

A central problem in lepton spectroscopy are electrons and positrons of low
energy ($\le$ 50 MeV). They curl up in the magnetic field or may be deflected
into an angular range not covered by the detector, leading to limited track
reconstruction efficiencies, a reduced conversion-pair rejection, and acceptance problems.

There are additional dilepton ($e^+ e^-$, $\mu^+\mu^-$) sources like Dalitz decays of
hadrons, e.g., $\pi^0,\eta \rightarrow e^+e^- \gamma$; $\eta \rightarrow
\mu^+ \mu^- \gamma$, or decays of charged pions and kaons like $\pi^+,K^+
\rightarrow \mu^+ \nu$; $\pi^-,K^- \rightarrow \mu^-\bar{\nu}$
which contribute to the dilepton invariant-mass spectrum. They also give rise to a combinatorial
background when a positron ($\mu^+$) from one source is erroneously paired in the
analysis with an electron ($\mu^-)$ from another source within the same event, as
illustrated in Fig.~\ref{fig:comb-background} for combinatorial $e^+ e^-$ pairs.

\begin{figure}[th]
  \begin{minipage}[htb]{0.46\textwidth}
    \begin{center}
      \includegraphics[keepaspectratio,width=\textwidth]{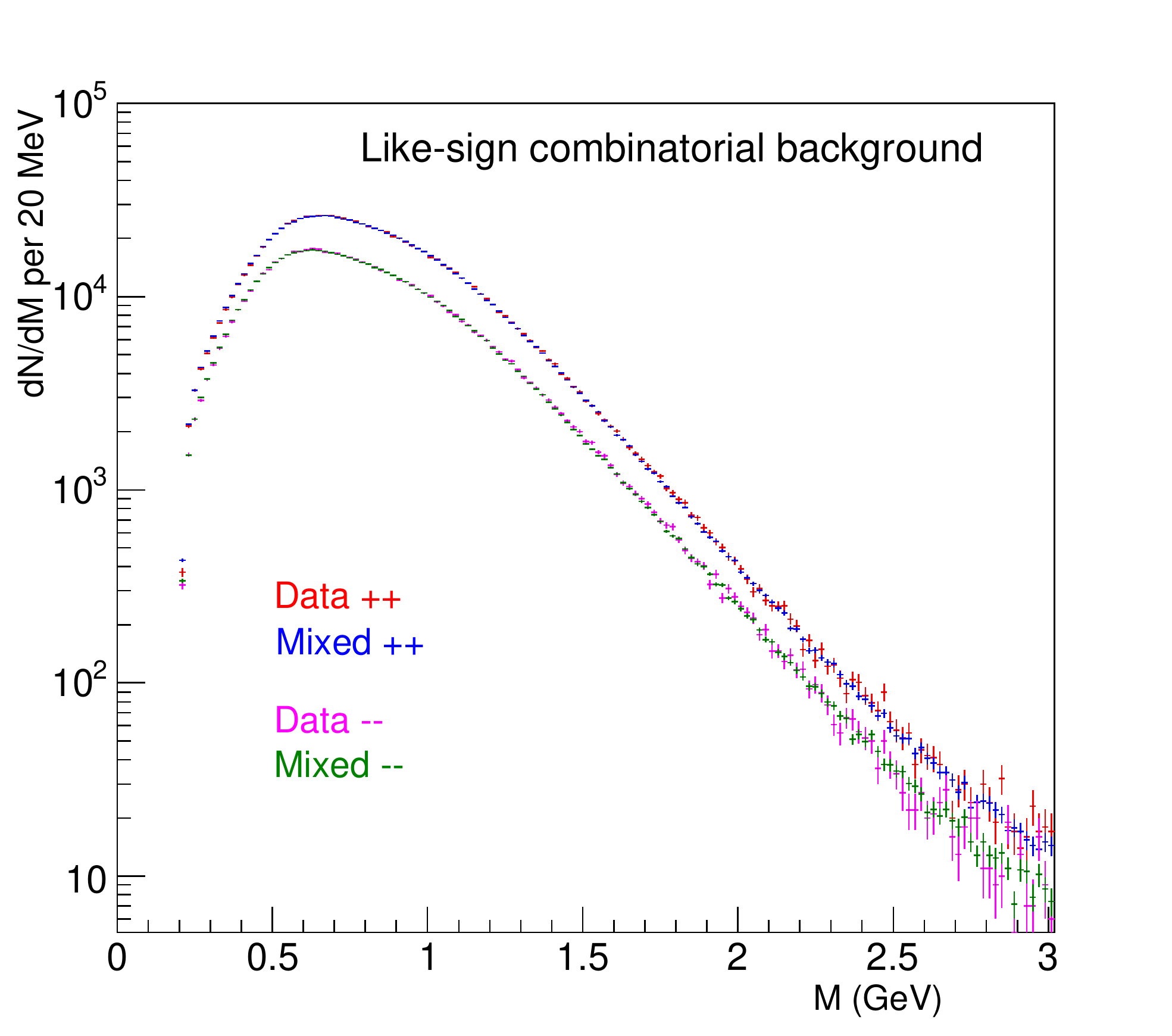}
    \end{center}
    \vspace*{8pt}
    \caption{Dilepton background generated by event mixing for In-In collisions at 158 AGeV in comparison to like-sign $\mu^+\mu^+$ and $\mu^-\mu^-$ invariant-mass spectra.\protect\cite{Arnaldi:2008er,Specht_Damjano}}
    \label{fig:NA60_mixed}
  \end{minipage}
  \hfill
  \begin{minipage}[h]{0.52\textwidth}
    \begin{center}
      \includegraphics[keepaspectratio,width=\textwidth]{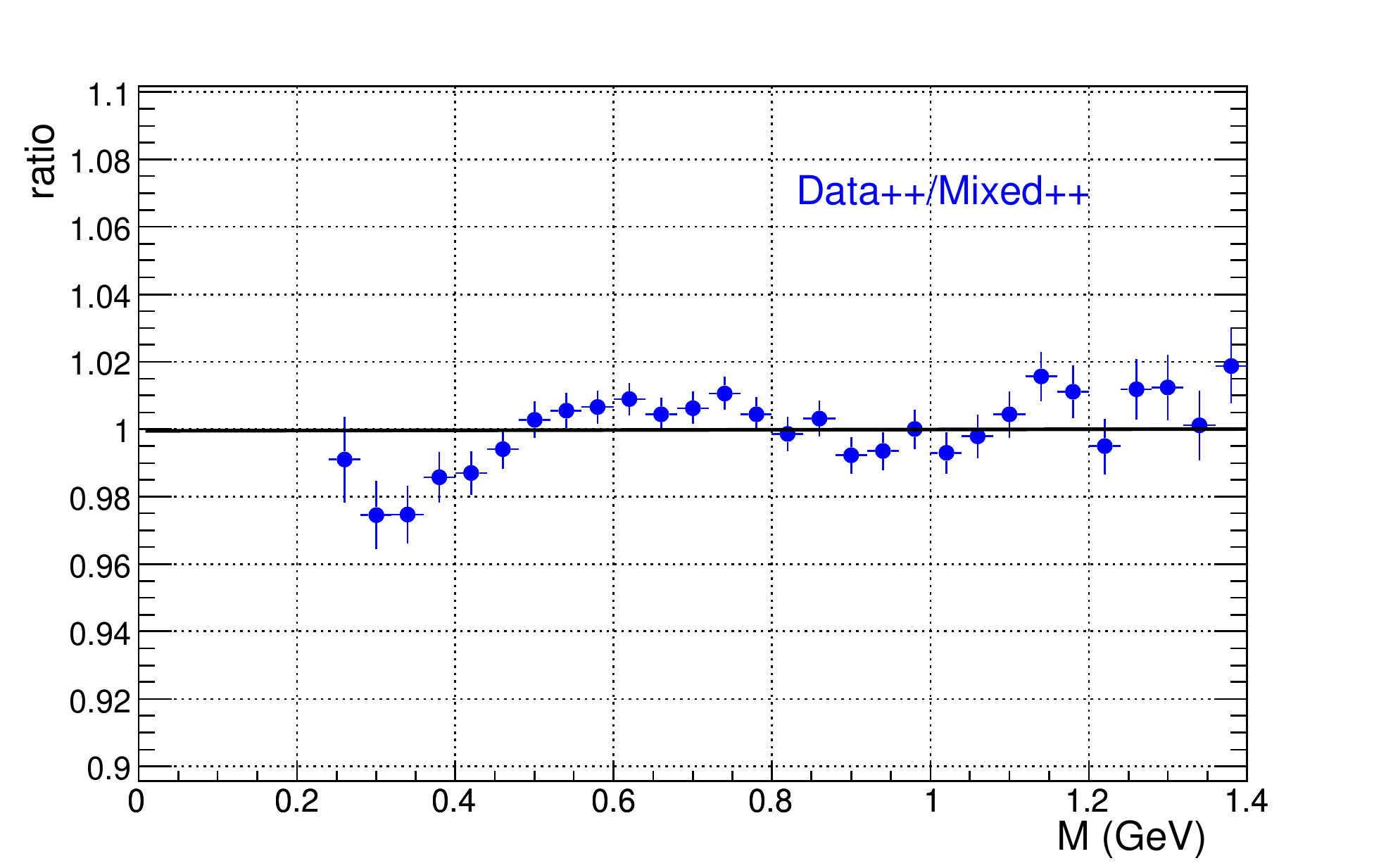}
      \includegraphics[keepaspectratio,width=\textwidth]{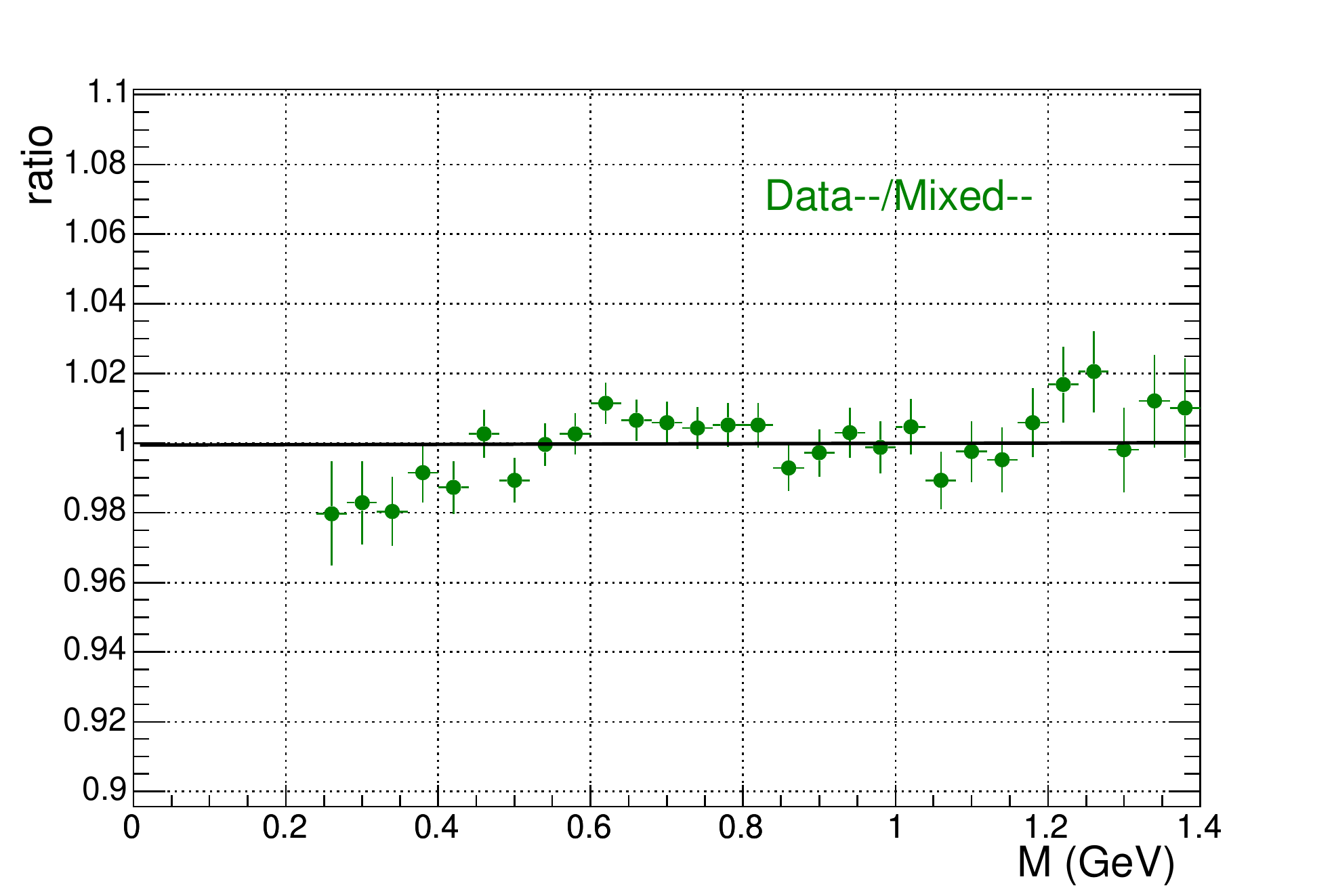}
    \end{center}
    \vspace*{8pt}
    \caption{Ratio of like-sign invariant-mass distribution over mixed-event
      distributions.\protect\cite{Specht_Damjano}}
    \label{fig:NA60_ratios}
  \end{minipage}
\end{figure}
In a procedure followed by most collaborations this combinatorial background
is determined by measuring like-sign dileptons. Assuming that the
like-sign pairs are uncorrelated and have the same acceptance as unlike-sign pairs
the combinatorial background $B$ and the signal $S$ are then given by
\begin{equation}
B = 2 \cdot \sqrt{N^{++} \cdot N^{--}}\quad, \hspace*{0.5cm}  S = N^{+-} - B \label{CB}
\end{equation}
\noindent
where $N^{++}, N^{--}$, and $N^{+-}$ are the number of measured $l^+ l^+$ ,
$l^-l^-$, and $l^+ l^-$ lepton pairs.
This approach has the disadvantage that for most detector systems the
acceptances for like-sign and unlike-sign pairs are not completely identical
and that the background is only determined with about the same statistical
significance as the signal. A high-statistics background can be obtained by the
event-mixing technique: like-sign tracks from different events are
paired whereby extreme care has to be applied in selecting events of similar
topology. The method is checked by comparing the mixed-event background to the
like-sign invariant-mass distribution. The procedure is illustrated in
Fig.~\ref{fig:NA60_mixed} with data from the NA60 collaboration who studied
dimuon production in ultra-relativistic heavy-ion collisions (see Sect.\ \ref{s:hirho}).
Like-sign and mixed-event distributions agree over nearly 3 1/2 orders of magnitude
on the level of less than 2$\%$ as shown by the ratios of the distributions
in Fig.~\ref{fig:NA60_ratios}.

In general, the mixed-event distribution is used for background subtraction
after normalization to the like-sign background. Mixed-event and like-sign pair
subtraction techniques have been developed to a
high degree of sophistication and have even been applied
in cases with signal-to-background ratios in the percent range.\cite{Toia:2008dj}

\subsection{Pioneering Experiments}
\label{pioneering_experiments}
Pioneering experiments on dilepton emission in heavy-ion and elementary nuclear
reactions started in the late 1980's at the Lawrence Berkeley Laboratory with the Dilepton
\begin{figure}[th]
  \begin{center}
    \includegraphics[keepaspectratio,width=0.55\textwidth]{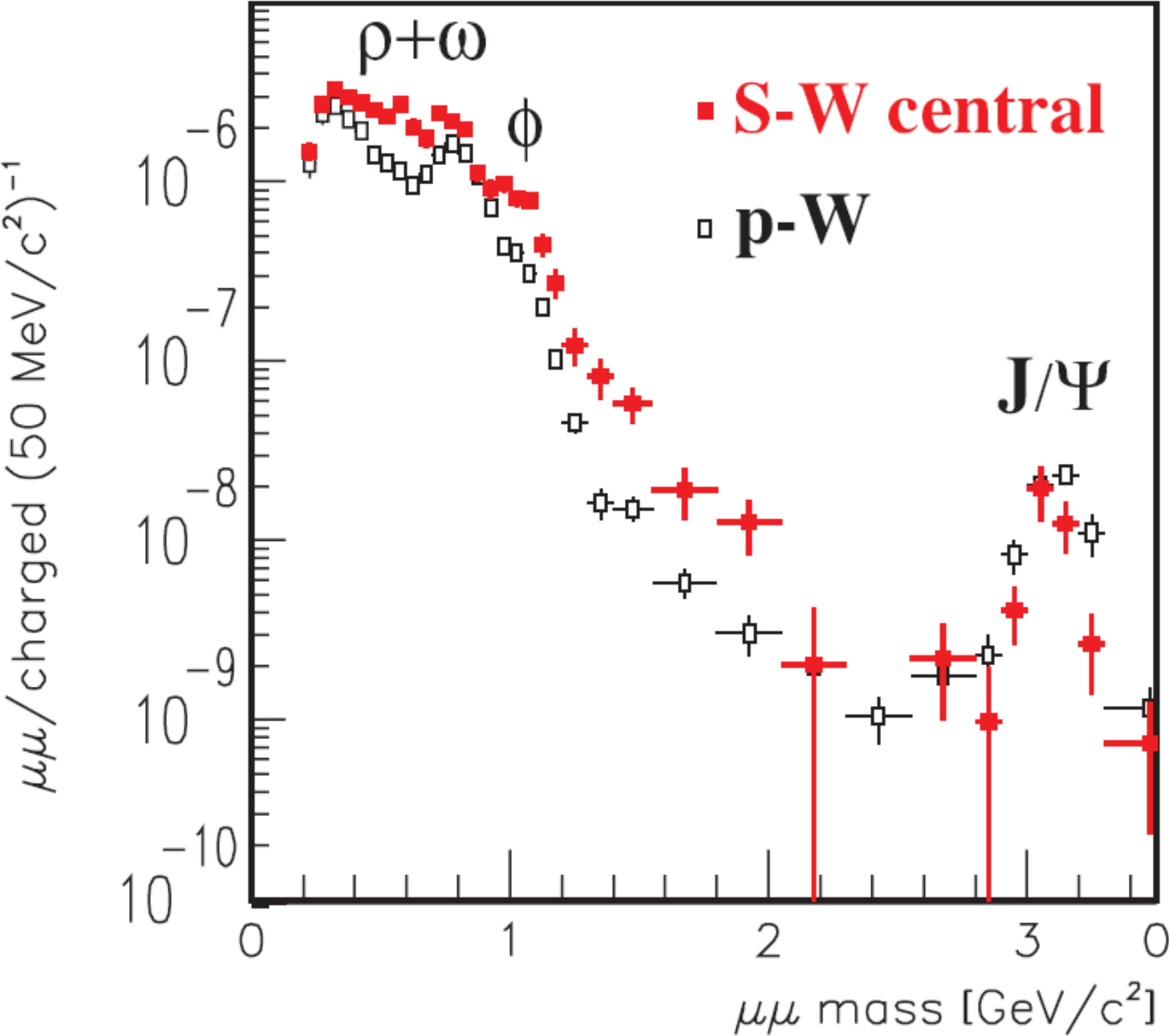}
  \end{center}
  \vspace*{8pt}
  \caption{$\mu^+\mu^-$ invariant-mass distribution for S+W (filled squares) and
    p+W (open squares) collisions at 200 GeV measured by the HELIOS-3
    collaboration.\protect\cite{Masera:1995ck}}
  \label{fig:HELIOS}
\end{figure}
Spectrometer (DLS)\cite{Roche:1988er,Naudet:1988kj,Porter:1997rc} and at the CERN SPS
with the CERES\cite{Agakishiev:1995xb} and HELIOS\cite{Masera:1995ck} detector
systems in the
energy ranges of $ \sqrt{s_{NN}}$ = 2-3 GeV and 17 GeV, respectively.

The DLS
data showed an unexpectedly high dielectron yield in C+C and Ca+Ca collisions
at 1 AGeV
which was much larger than could be explained in transport
calculations\cite{Bratkovskaya:1997mp} at that time. Also introducing medium
effects as discussed
in the previous sections did not remove this discrepancy which was then called
the {\it  DLS-puzzle}. Other calculations which also failed to reproduce the observed
dielectron
yields are summarized in Ref.\ \refcite{Cassing:1999es}. The DLS data will be discussed
together with recent results obtained with the HADES detector at
GSI in Sect.\ \ref{subsec:dil-1AGeV}.

 The HELIOS-3 collaboration\cite{Masera:1995ck}
reported an enhanced $\mu^+ \mu^-$ yield throughout the whole invariant-mass range
up to 3.5 GeV/c$^2$ in central S+W collisions compared to proton-induced
reactions on Tungsten at 200 GeV/c$^2$ (see Fig.\ \ref{fig:HELIOS}).

The CERES
collaboration\cite{Agakishiev:1995xb} focused on the low-mass part of the
dilepton spectrum below the $\phi$ mass, motivated by
discussions of possible medium modifications of the $\rho$ meson which
were initated by Pisarski\cite{Pisarski:1981mq}. In particular the CERES
results discussed below initiated widespread further experimental and
theoretical activities. New dedicated experiments were designed, optimized for
the detection of lepton pairs and theoretical tools were developed for
studying in-medium modifications of hadrons.

In the following sections the
wealth of data obtained in a series of experiments will be
confronted with the current status of our theoretical understanding developed
in the previous sections. The results on
the $\rho$, $\omega,$ and $\phi$ meson and on 2$\pi$ production will be
presented in separate subsections.

\subsection{In-medium properties of the $\rho$ meson}
\subsubsection{Heavy-ion experiments}
\label{s:hirho}
First information on medium modifications of vector mesons was derived from
ultra-relativistic heavy-ion reactions with the CERES detector.

\begin{figure}[th]
  \begin{center}
    \includegraphics[keepaspectratio,width=\textwidth]{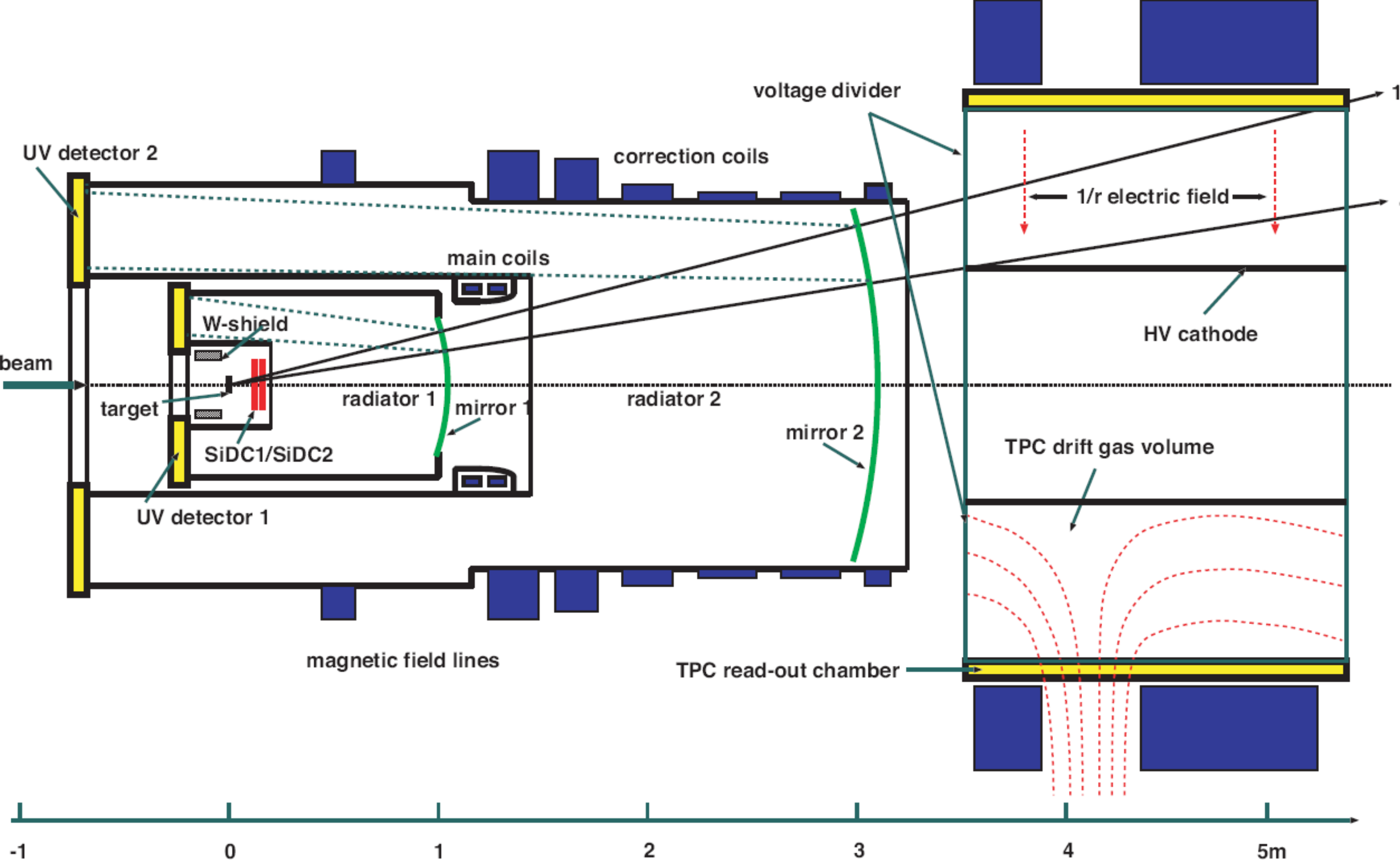} \\[1em]
    \includegraphics[keepaspectratio,width=0.55\textwidth]{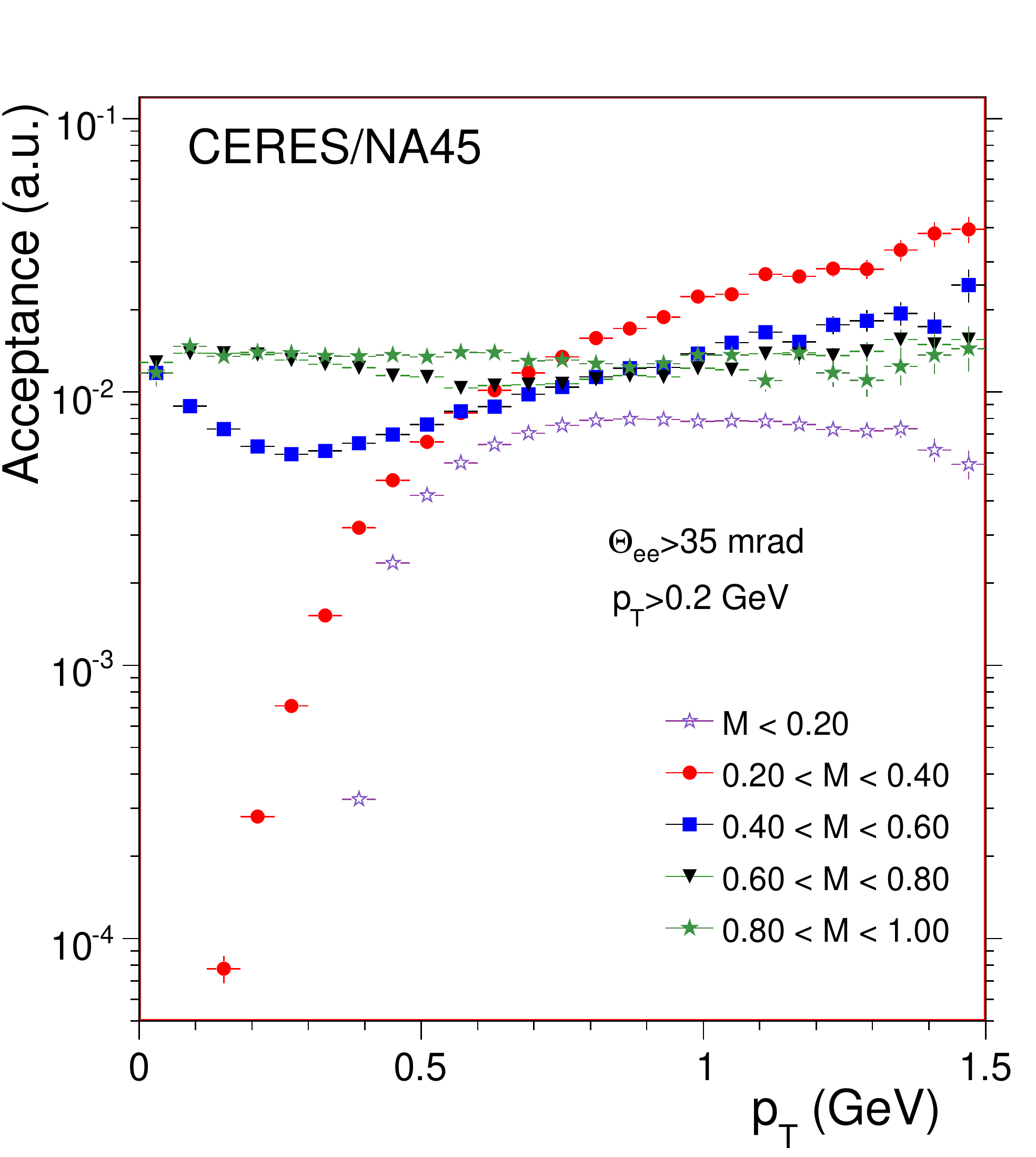}
  \end{center}
  \vspace*{8pt}
  \caption{Top: Schematic view of the CERES spectrometer, comprising two ring imaging
    Cherenkov detectors and a radial time projection
    chamber.\protect\cite{Marin:2004fx}
    Bottom: CERES acceptance (upper limit) for dileptons relative to the full phase space
    as a function of transverse momentum for different invariant-mass
    bins.\protect\cite{Agakichiev:2005ai,Specht_Damjano}}
  \label{fig:CERES_detector}
\end{figure}
The setup of the CERES experiment is shown in Fig.\ \ref{fig:CERES_detector}. It was
initially designed as a hadron blind detector consisting of two ring imaging
Cherenkov detectors. With a high Cherenkov threshold of $\gamma_{thr} \approx
32$ it was only sensitive to dileptons. The centrality of the heavy-ion
collisions was characterized by measuring the particle multiplicity in a Si
drift detector near the target. Later on a cylindrial time-projection chamber
was added to improve the invariant-mass resolution to $\frac{\delta m}{m} = 3.8\%$.
Fig.\ \ref{fig:CERES_detector} (bottom) shows the acceptance of CERES for
dileptons relative to the full phase space.\cite{Agakichiev:2005ai,Specht_Damjano}

\begin{figure}
  \begin{center}
    \includegraphics[keepaspectratio,width=0.465\textwidth]{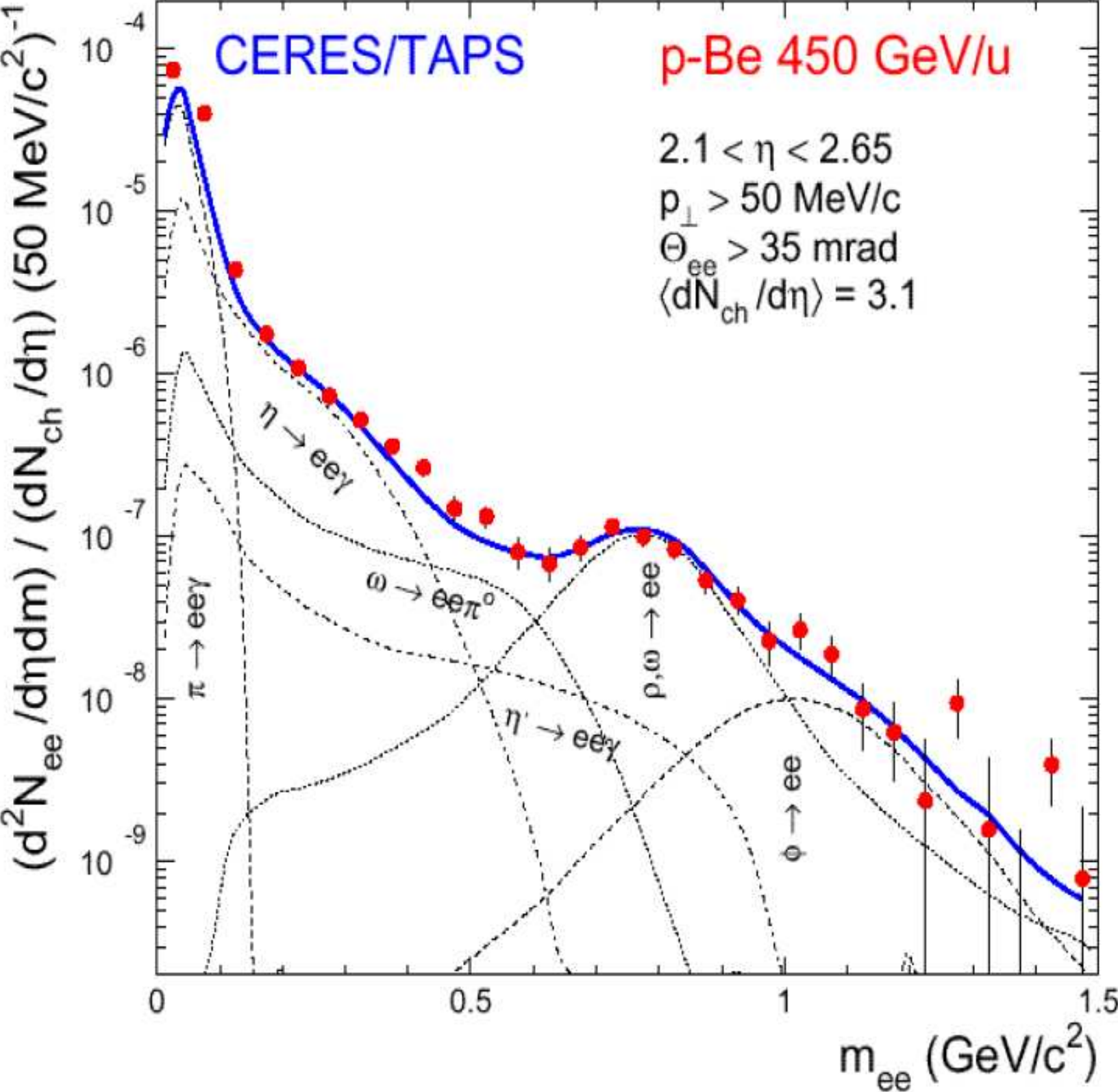}
    \hfill
    \includegraphics[keepaspectratio,width=0.51\textwidth]{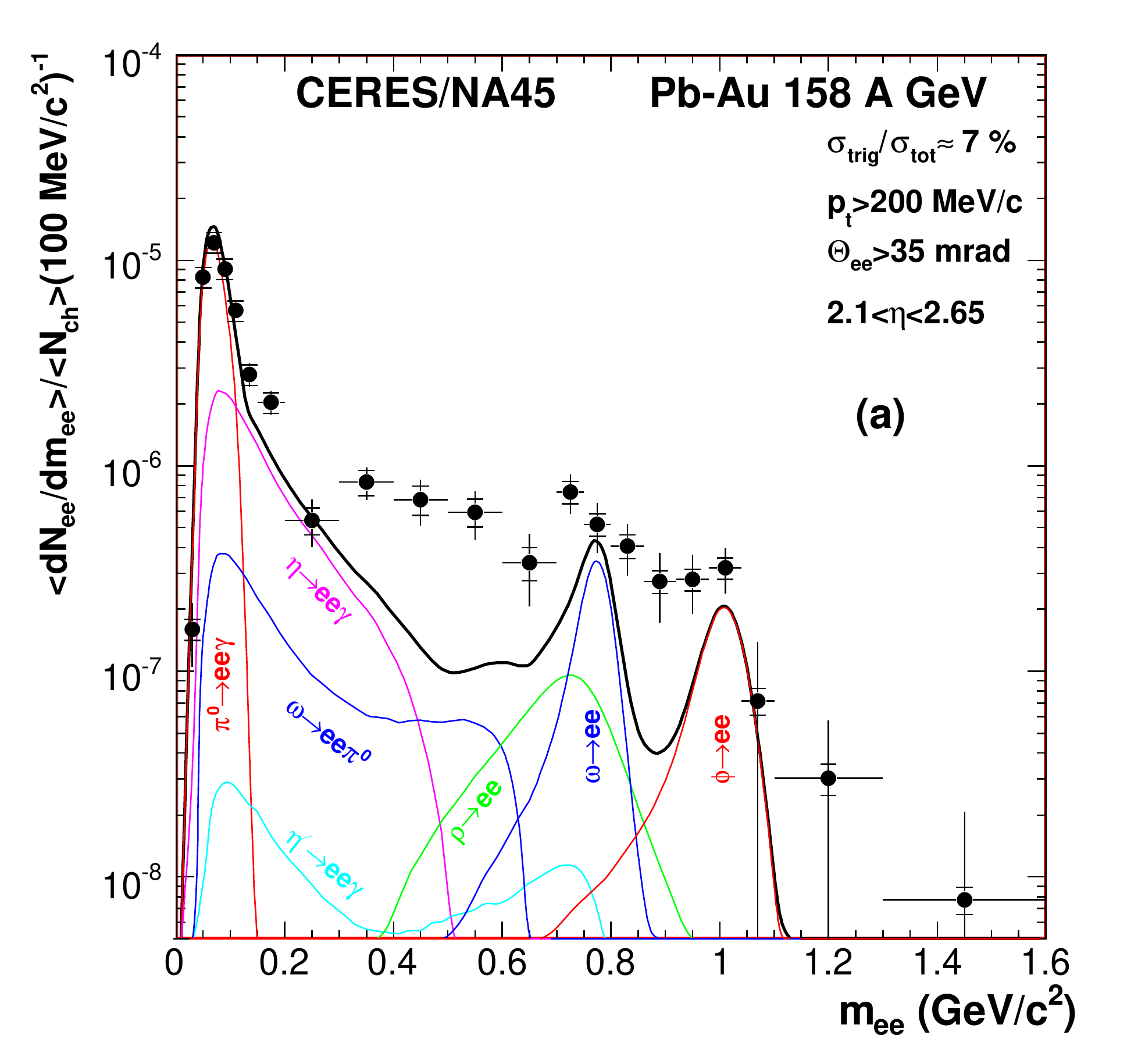}
  \end{center}
  \vspace*{8pt}
  \caption{Inclusive $e^+ e^-$ invariant-mass spectra for p+Be reactions at 450 GeV/c
    (left) and Pb+Au collisions at 158 AGeV (right) measured with the CERES
    detector. Contributions from different hadronic decays expected for p+p collisions
    are indicated. The sum of these contributions ({\it post-freeze-out cocktail})
    reproduces the observed dilepton yield in p+Be
    reactions\protect\cite{Agakishiev:1998mv} but
    leaves an excess unaccounted for in the Pb+Au collisions\protect\cite{Adamova:2006nu}.
    The figure on the right hand side represents the most
    recent CERES result making use of the additional radial time projection chamber as
    shown in Fig.~\ref{fig:CERES_detector}.}
  \label{fig:CERES_pBe_PbAu}
\end{figure}
\begin{figure}
  \begin{center}
    \includegraphics[keepaspectratio,width=0.49\textwidth]{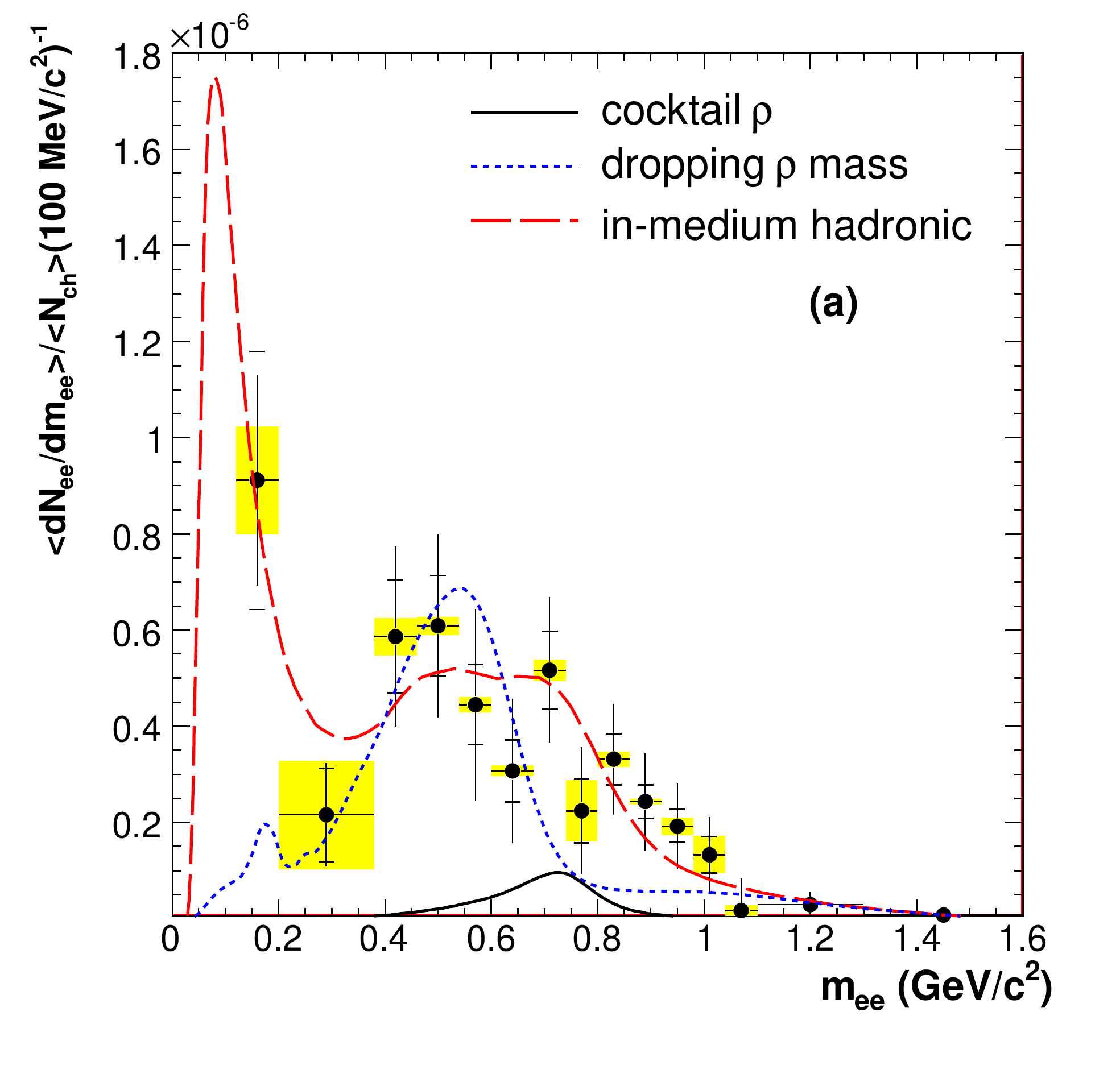}
    \hfill
    \includegraphics[keepaspectratio,width=0.49\textwidth]{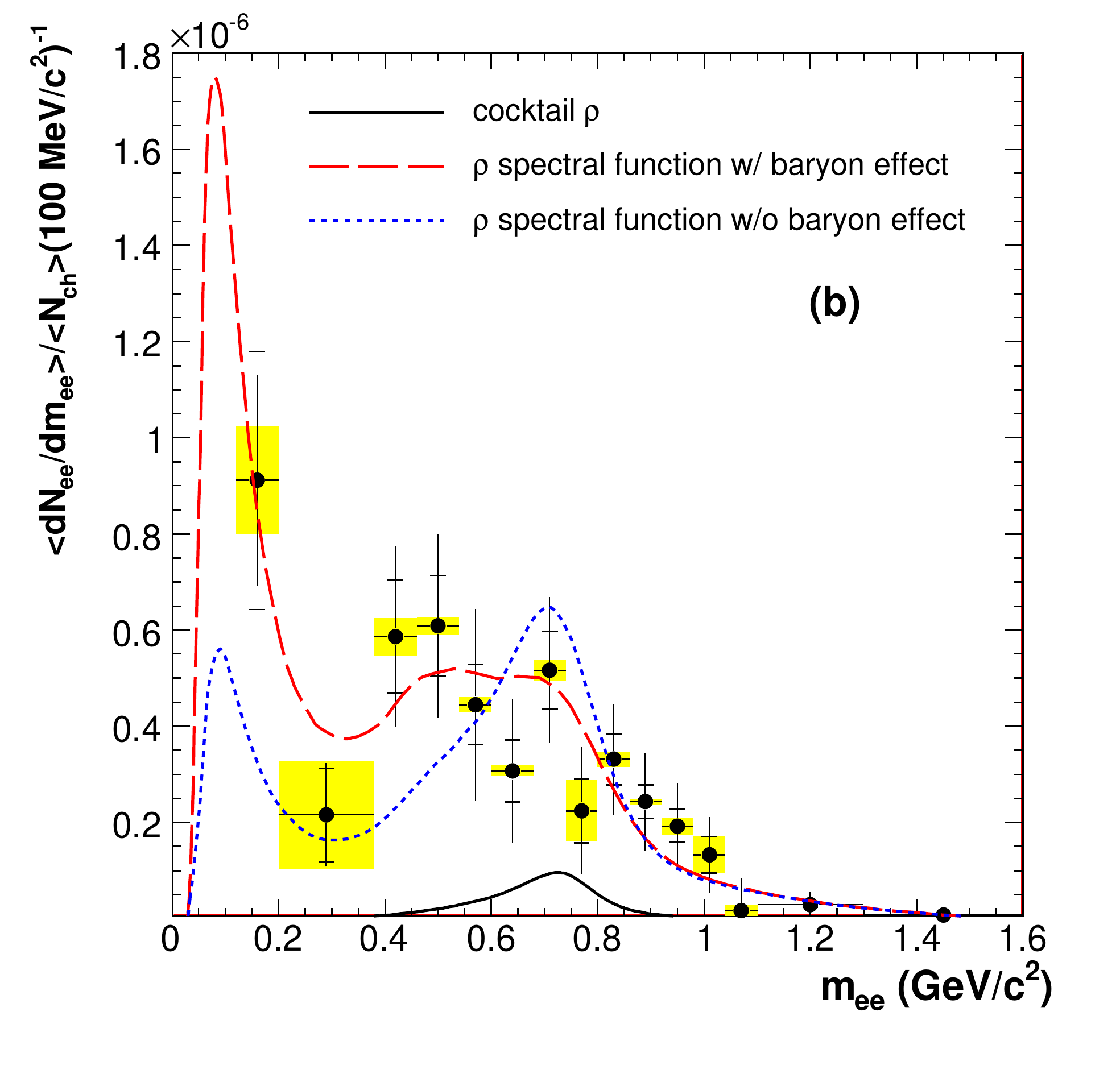}
  \end{center}
  \vspace*{8pt}
  \caption{$e^+ e^-$ invariant-mass distribution after subtraction of the
    post-freeze-out hadronic cocktail.
    Statistical errors, systematic errors (horizontal ticks) and the systematic
    uncertainty due to the cocktail subtraction (shaded boxes) are indicated.
    The broadening scenario\protect\cite{Rapp:1999us} (long-dashed curve) is
    compared to a calculation assuming a density dependent dropping $\rho$
    mass\protect\cite{Brown:1991kk} (dotted line in (a)) and to a broadening scenario
    excluding baryonic effects (dotted line in (b))\protect\cite{Adamova:2006nu}. }
  \label{fig:CERES_results}
\end{figure}
Fig.~\ref{fig:CERES_pBe_PbAu} shows a comparison of $e^+ e^-$ invariant-mass
spectra measured in p+Be at 450 GeV\cite{Agakishiev:1998mv} and Pb+Au collisions at 158
AGeV\cite{Adamova:2006nu}, respectively, after subtracting the combinatorial background.
For p+Be, the measured dilepton yield is reproduced by the sum of contributions
from Dalitz and direct decays of neutral mesons with known production
cross sections in p+p reactions. In contrast, the Pb+Au collisions reveal an
enhancement relative to this cocktail in the mass range below the $\phi$ mass. Here, the cocktail is
constructed from meson yields predicted in statistical model calculations (see
Ref.\ \protect\refcite{Agakichiev:2005ai}).
Similar di-electron excess yields had been observed in earlier CERES
experiments on S+Au and Pb+Au collisions, published in
    Refs.\ \protect\refcite{Agakishiev:1995xb,Agakishiev:1998mv,Agakichiev:2005ai}.
The enhancement was attributed to the annihiliation of charged
pions which are produced in the fireball of the heavy-ion collision with high
multiplicity.

\begin{figure}
  \begin{center}
    \includegraphics[keepaspectratio,width=0.49\textwidth]{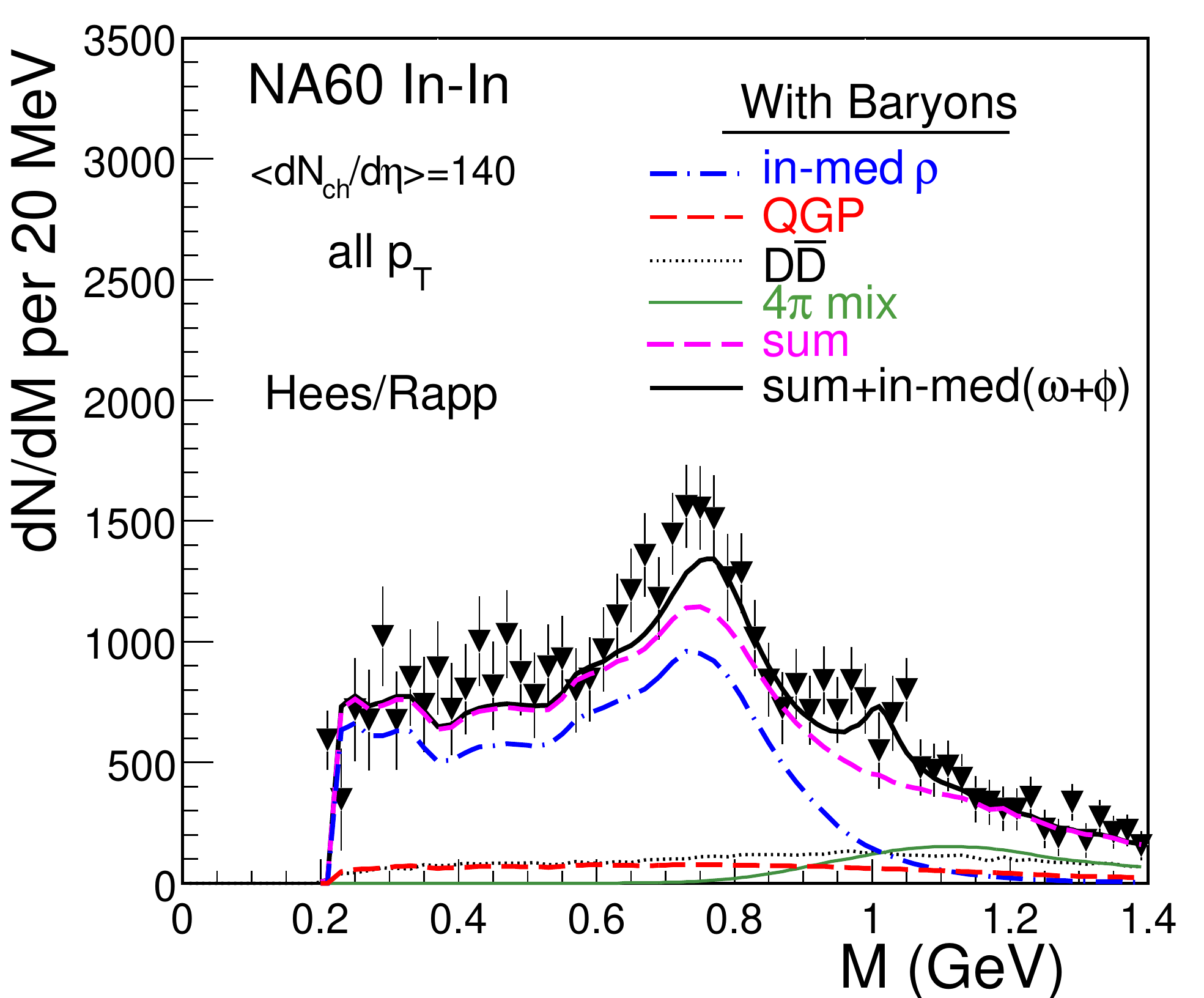}
    \hfill
    \includegraphics[keepaspectratio,width=0.49\textwidth]{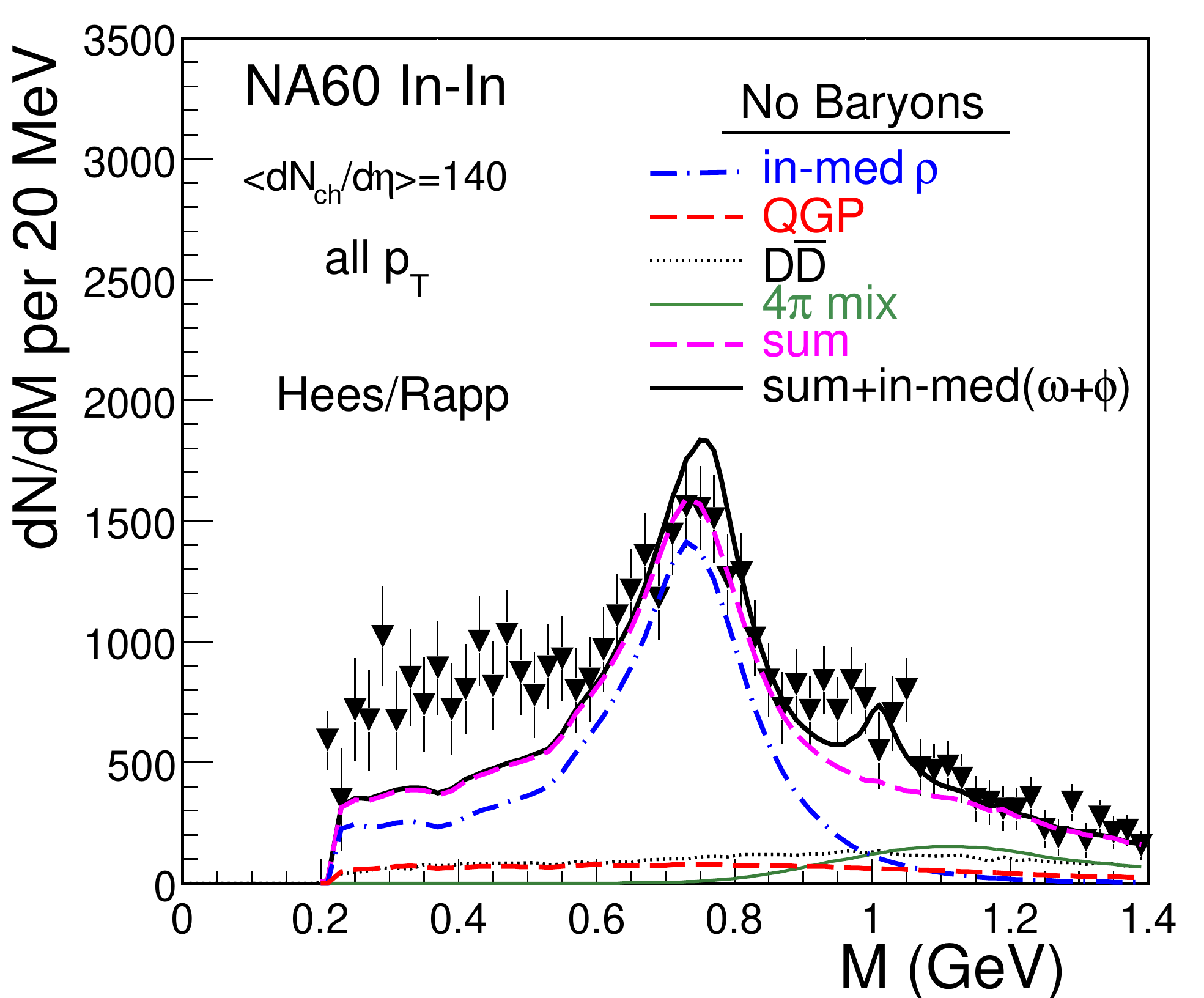}
  \end{center}
  \vspace*{8pt}
  \caption{NA60 data for semi-central In+In collisions in comparison to two model
    calculations\protect\cite{vanHees:2006ng} including and neglecting meson-baryon
    interactions, respectively.}
  \label{fig:NA60_Hees_Rapp}
\end{figure}

The annihilation proceeds via an intermediate $\rho$ meson,
$\pi^+\pi^- \rightarrow \rho \rightarrow e^+e^-$, and thereby provides
access to the properties of the $\rho$ meson at high densities and temperatures
reached in heavy-ion collisions.\cite{Cassing:1999es} As shown below (see
Figs.\ \ref{fig:CERES_results}, \ref{fig:NA60_Hees_Rapp}), however, not only pion
annihilation but also formation and decays of baryon resonances like $\pi N
\rightarrow N^*,\Delta,\Delta^*
\rightarrow N \rho \rightarrow N e^+e^-$  contribute to the observed dilepton
invariant-mass spectra, in particular at low masses.
For a more detailed study of the $\rho$
meson, the post-freeze-out dilepton cocktail has been subtracted in
Fig.\ \ref{fig:CERES_results} except for the $\rho$ contribution. The data are
compared to two theoretically proposed scenarios, a mass
shift\cite{Brown:1991kk} or an in-medium broadening\cite{Rapp:1999ej} of the
$\rho$ meson, respectively. With the limited statistics it is difficult to clearly
differentiate between the two cases.
The authors of Ref.\ \refcite{Adamova:2006nu}
conclude that the excess is smeared out over a wider mass range than expected
for a dropping mass scenario and therefore favor the interpretation of the
data in terms of an in-medium broadening of the $\rho$ meson over a density
dependendent $\rho$ mass shift. The relevance of high baryon densities for the
observed medium modifications indicated in Fig.~\ref{fig:CERES_results}
(right) has also been noted in a measurement of Pb+Au collisions at a lower
incident energy of 40 AGeV.\cite{Adamova:2002kf}

\begin{figure}
  \begin{center}
    \includegraphics[keepaspectratio,width=\textwidth]{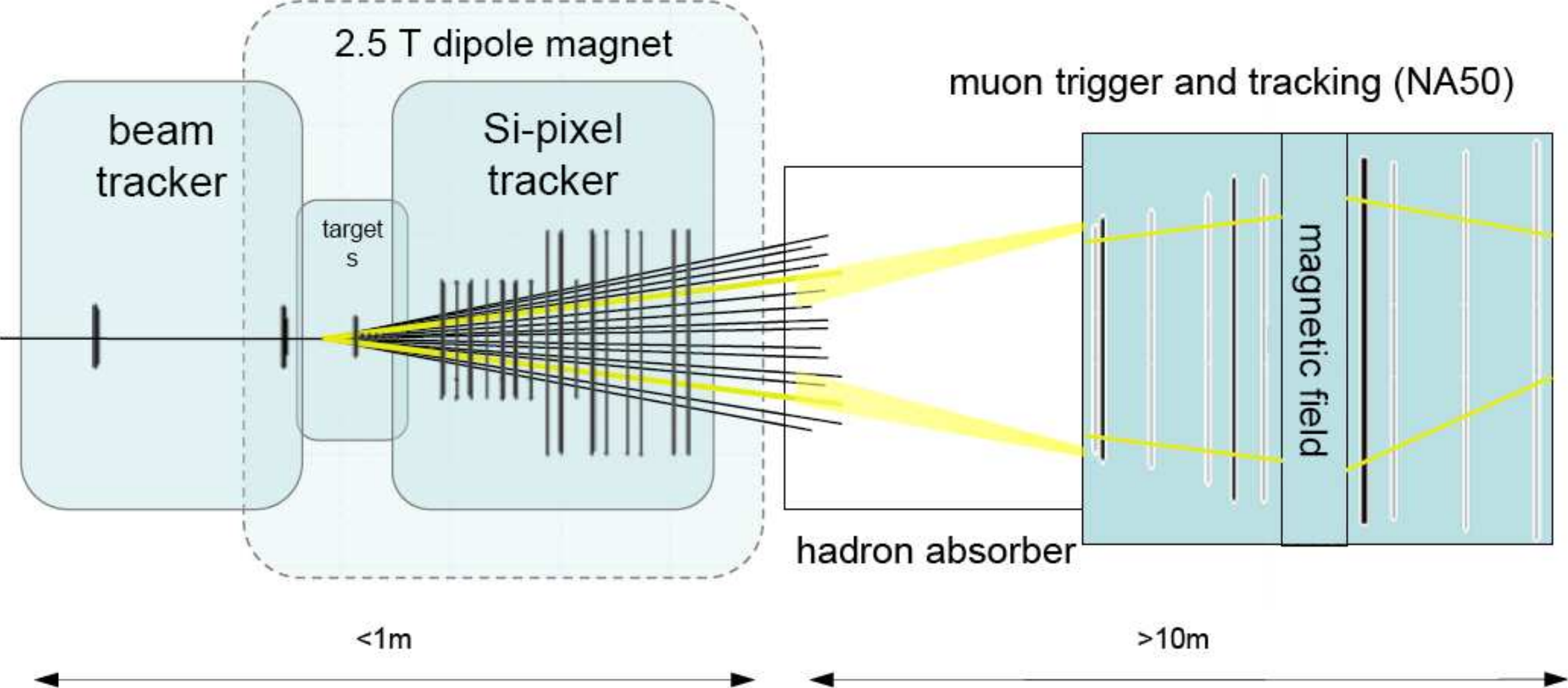}
    \\[1em]
    \includegraphics[keepaspectratio,width=0.5\textwidth]{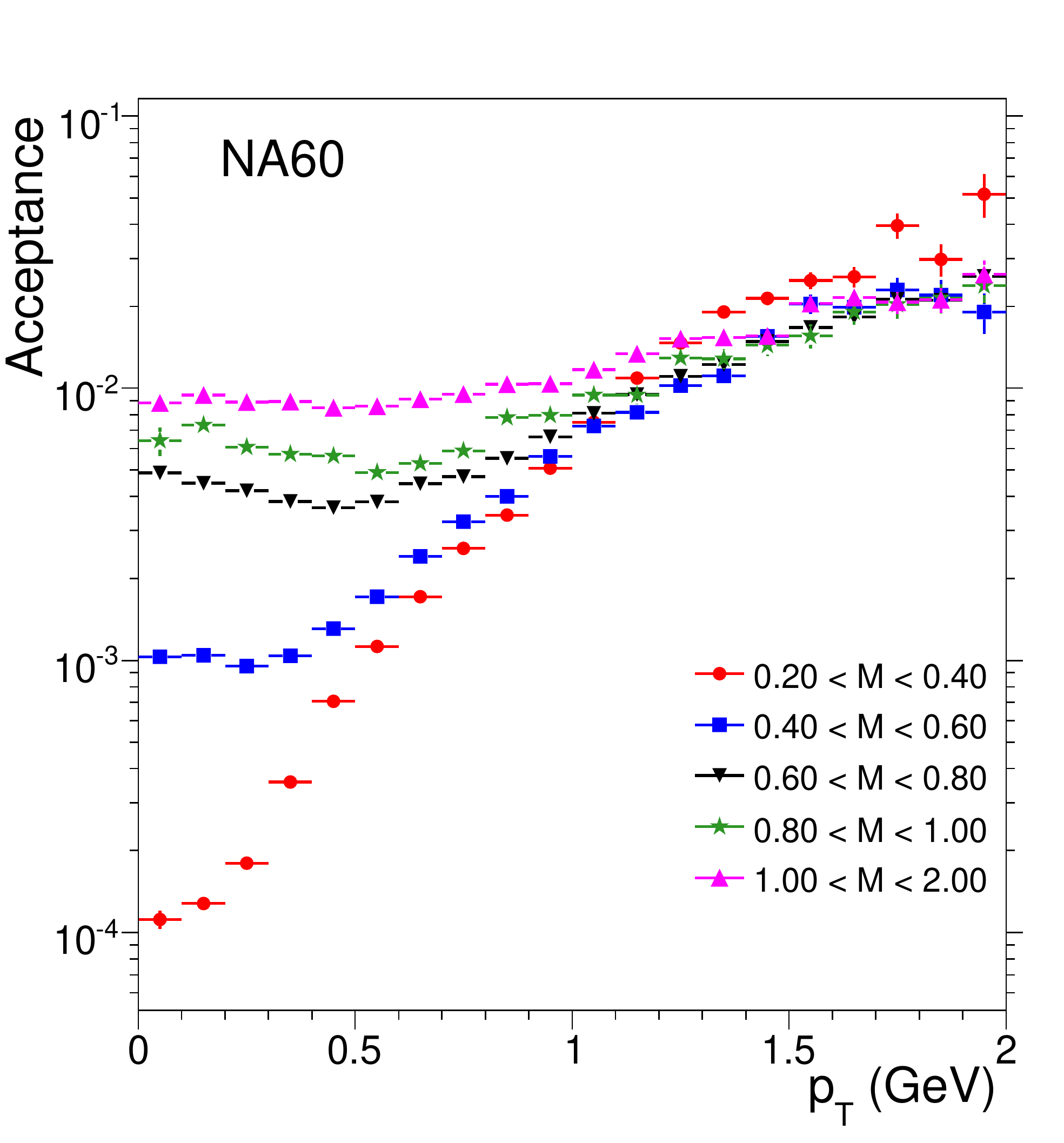}
  \end{center}
  \vspace*{8pt}
  \caption{Layout of the NA60 experiment: The radiation-hard Si-pixel detector before the hadron
    absorber was the essential addition to the former NA50 muon spectrometer to
    achieve a mass resolution of $\frac{\delta m}{m} = 2.5\%$. The bottom part
    of the figure shows the acceptance for dimuons relative to the full phase space
    as a function of transverse momentum for different invariant-mass
    bins.\protect\cite{Damjanovic:2007qm}}
  \label{fig:NA60_setup}
\end{figure}
A breakthrough in statistics and resolution in dilepton
spectroscopy of nucleus-nucleus collisions has been achieved by the NA60
collaboration\cite{Arnaldi:2006jq} who studied the $\mu^+\mu^-$ decay channel in the
In+In reaction at 158 AGeV. The experimental setup and the dilepton acceptance
of the detector relative to the full phase space are shown in Fig.\ \ref{fig:NA60_setup}.
\begin{figure}
  \begin{center}
    \includegraphics[keepaspectratio,width=0.47\textwidth]{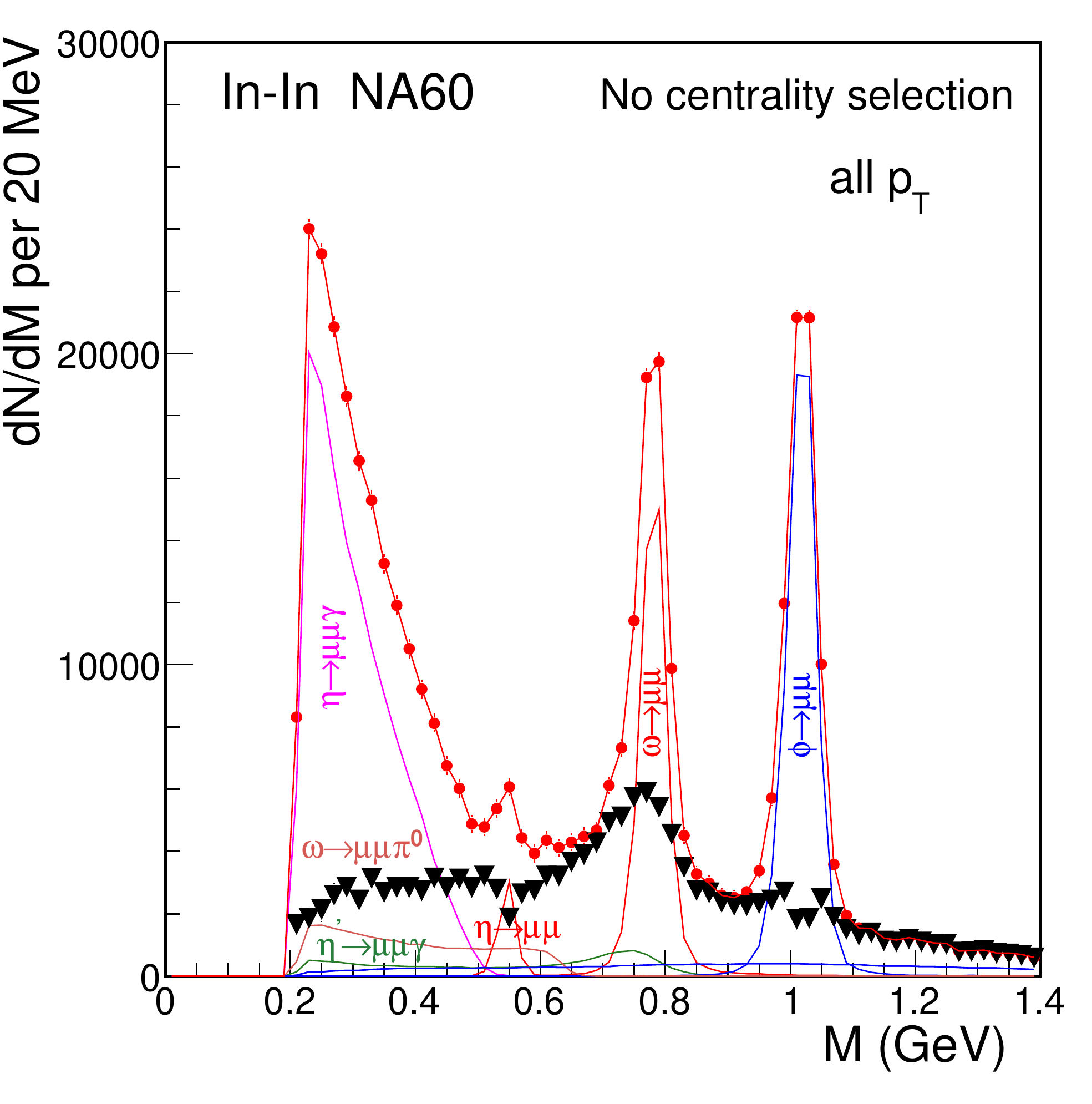}
    \hfill
    \includegraphics[keepaspectratio,width=0.52\textwidth]{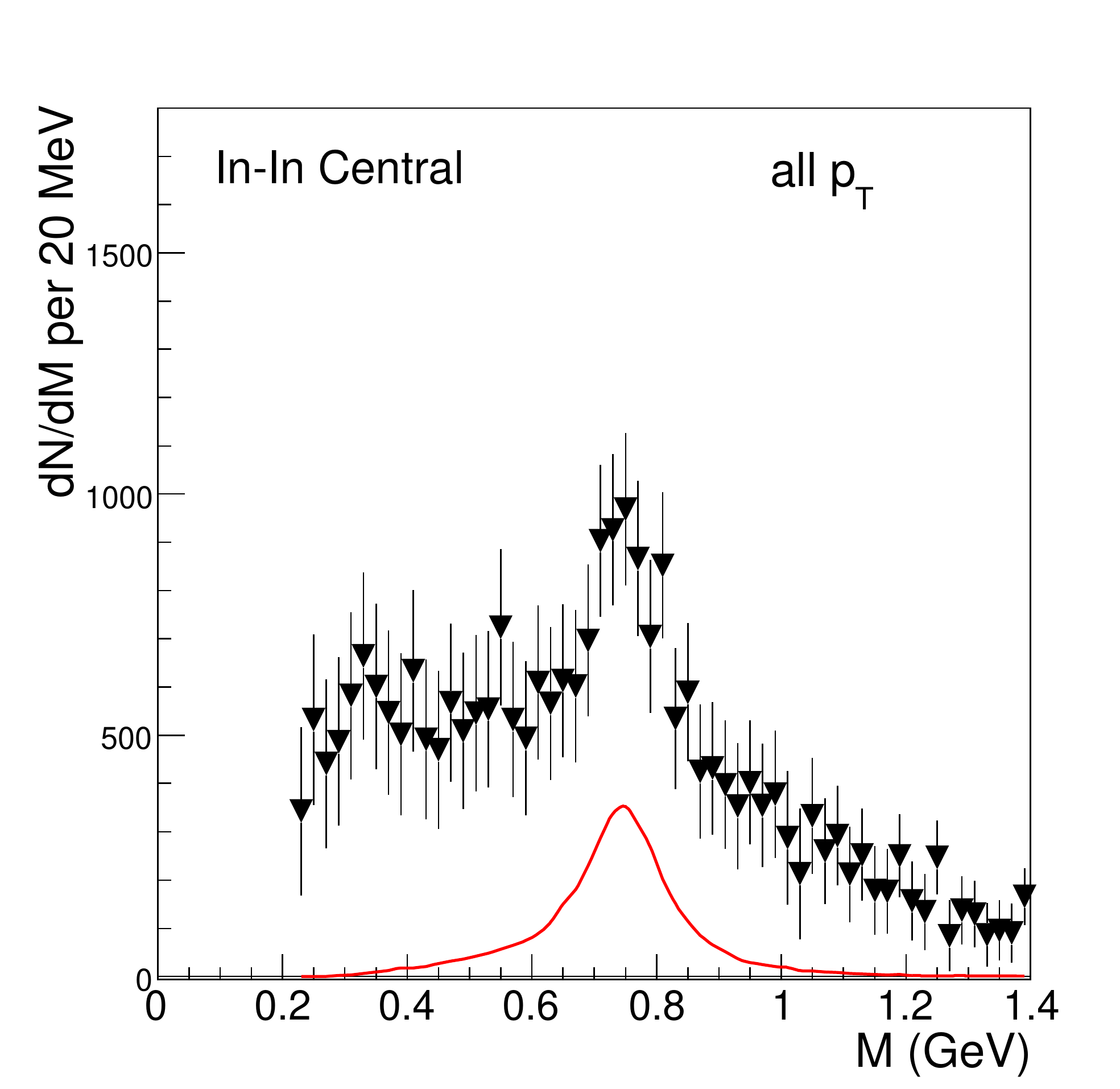}
  \end{center}
  \vspace*{8pt}
  \caption{Left: Background-subtracted $\mu^+ \mu^-$ invariant-mass spectrum before
    (dots) and after subtraction (triangles) of the known hadronic decay
    sources for In+In collisions at 158 AGeV. Right: $\mu^+ \mu^-$ invariant-mass
    spectrum for central collisons in comparison to the spectral function of a
    free (``cocktail'') $\rho$ meson.\protect\cite{Arnaldi:2006jq,Specht_Damjano}}
  \label{fig:NA60_excess}
\end{figure}
\begin{figure}
  \begin{center}
    \includegraphics[keepaspectratio,width=0.5\textwidth]{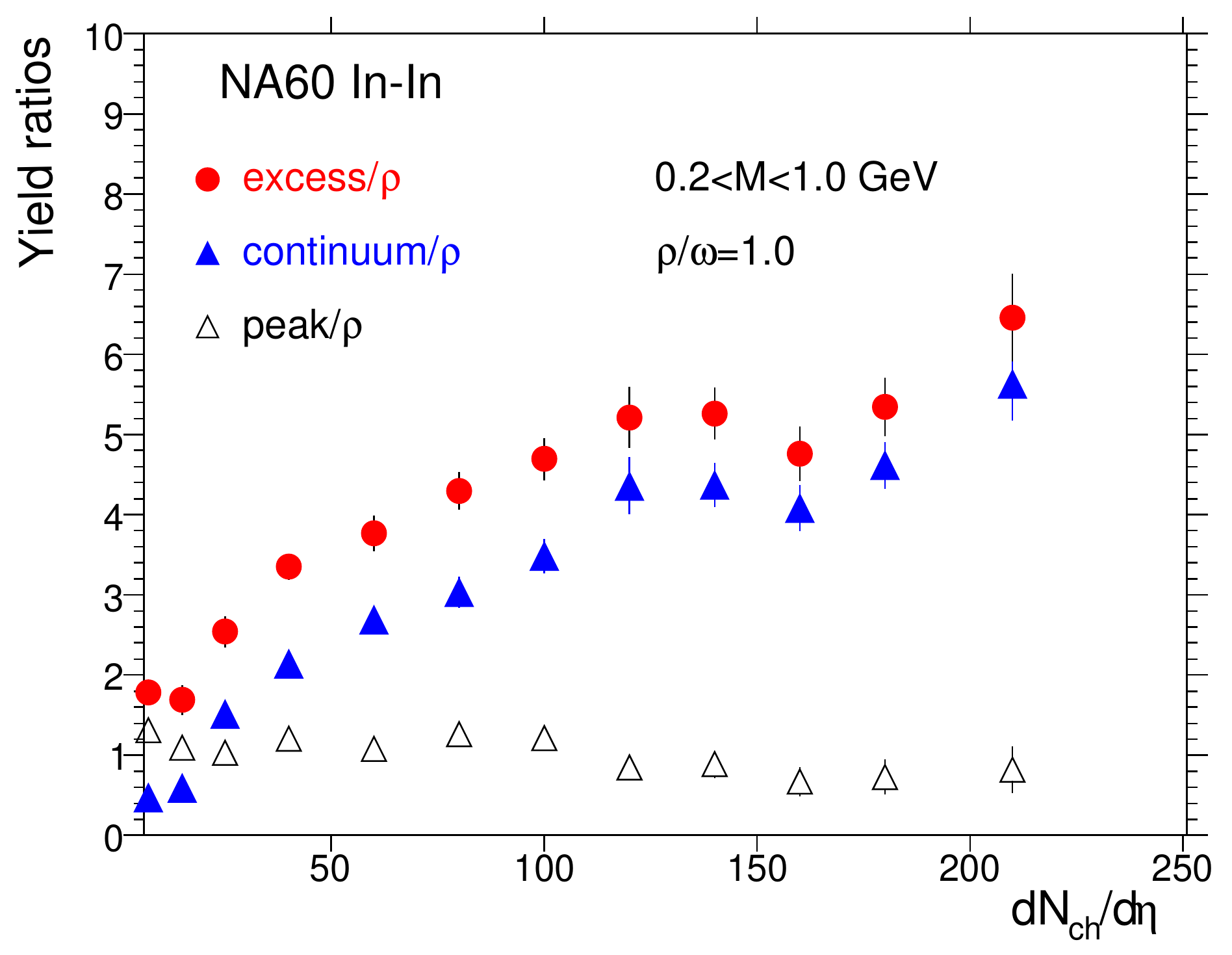}
  \end{center}
  \vspace*{8pt}
  \caption{Ratio of the dimuon-excess yield to the cocktail $\rho$ intensity
    as a function of centrality, assuming a production ratio $\rho/\omega$ =1. The
    centrality dependence is separately shown for the mass range 0.2-1.0
    GeV/c$^2$, the $\rho$ peak and the continuum.\protect\cite{Arnaldi:2008fw}}
  \label{fig:NA60_centrality_dep}
\end{figure}
A radiation-hard Si-pixel detector was installed between the target and the
hadron absorber of the former NA50 dimuon spectrometer. By matching the tracks
before and after the absorber in angle and momentum a mass resolution of
2.5$\%$ was achieved for vector mesons. As shown in
Fig.~\ref{fig:NA60_excess}, peaks from the $\omega$ and $\phi$ decays are cleanly
resolved in the $\mu^+\mu^-$ invariant-mass spectrum after subtraction of the
combinatorial background.
The quality of the data has allowed to subtract the
{\em measured} post-freeze-out dilepton cocktail separately for different
centrality bins and therefore does not rely on meson yields from statistical
model predictions. The remaining dimuon invariant-mass spectrum
is attributed mainly to the $\rho \rightarrow \mu^+\mu^-$ decay. A monotonic
broadening of the excess and a more than linear rise is observed with increasing
centrality of the heavy-ion reaction (see Fig.~\ref{fig:NA60_centrality_dep}).
The ratio of the dimuon-excess yield to the cocktail $\rho$ intensity is a
measure for the $\rho$ multiplicity which is found to be up to 6 times larger
in central In+In collisions than in elementary nucleon-nucleon reactions at the same energy.
As discussed in Refs.\ \refcite{vanHees:2006ng}, \refcite{Arnaldi:2008fw}
this ratio is related to the number of $\rho$
generations created by formation and decay during the fireball
evolution. Applying this ``$\rho$-clock'',\cite{Heinz:1990jw} information on the
life time of the fireball can be deduced and is estimated to be of the order of
$\approx$ 6 fm/c in central In+In collisions.\cite{vanHees:2007th}

For the most central collisions, Fig.~\ref{fig:NA60_excess} indicates a strong
in-medium broadening of the $\rho$ meson while the centroid of the distribution remains at the
nominal $\rho$ mass of 770 MeV/c$^2$. This observation has been interpreted by
the authors of Ref.\ \refcite{Arnaldi:2006jq}
as a melting of the $\rho$ meson indicative of a
restoration of chiral symmetry in the fireball of the collision zone which may
be associated with the phase transition into the quark-gluon plasma.

\begin{figure}
  \begin{center}
    \includegraphics[keepaspectratio,width=0.49\textwidth]{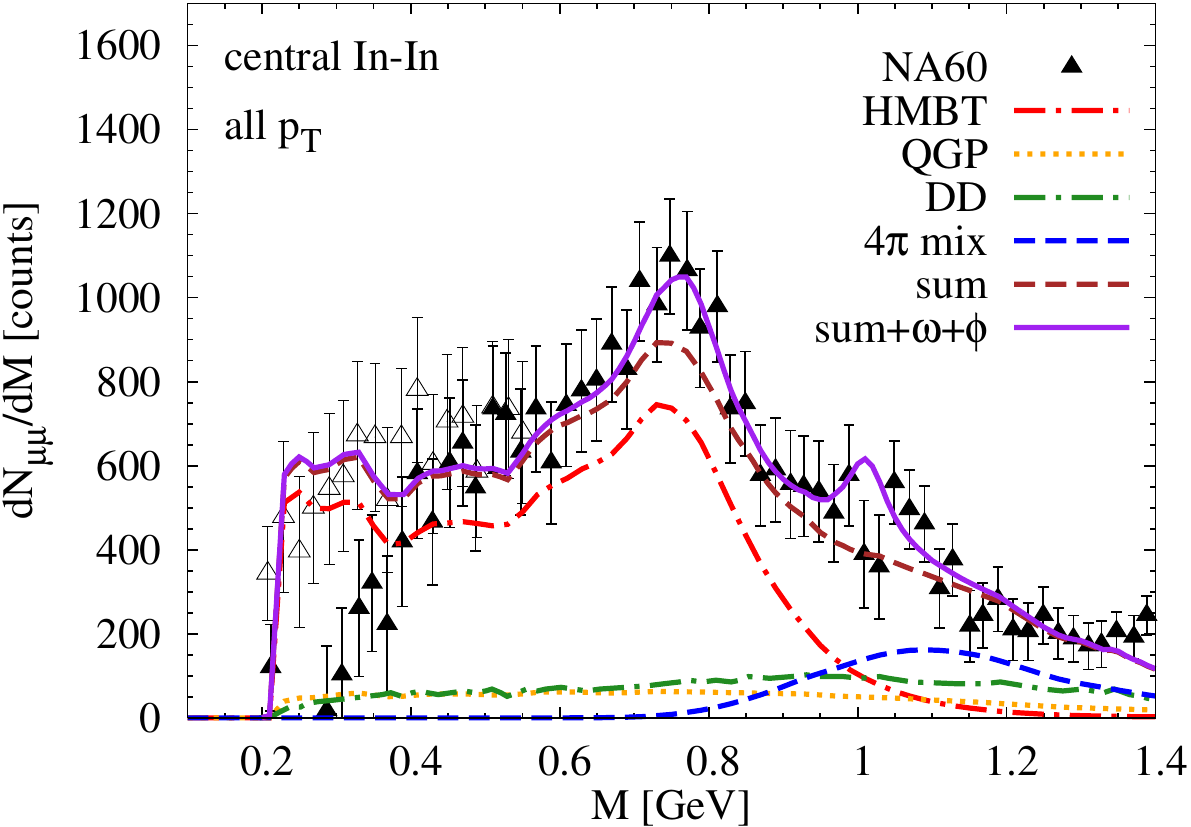}
    \includegraphics[keepaspectratio,width=0.49\textwidth]{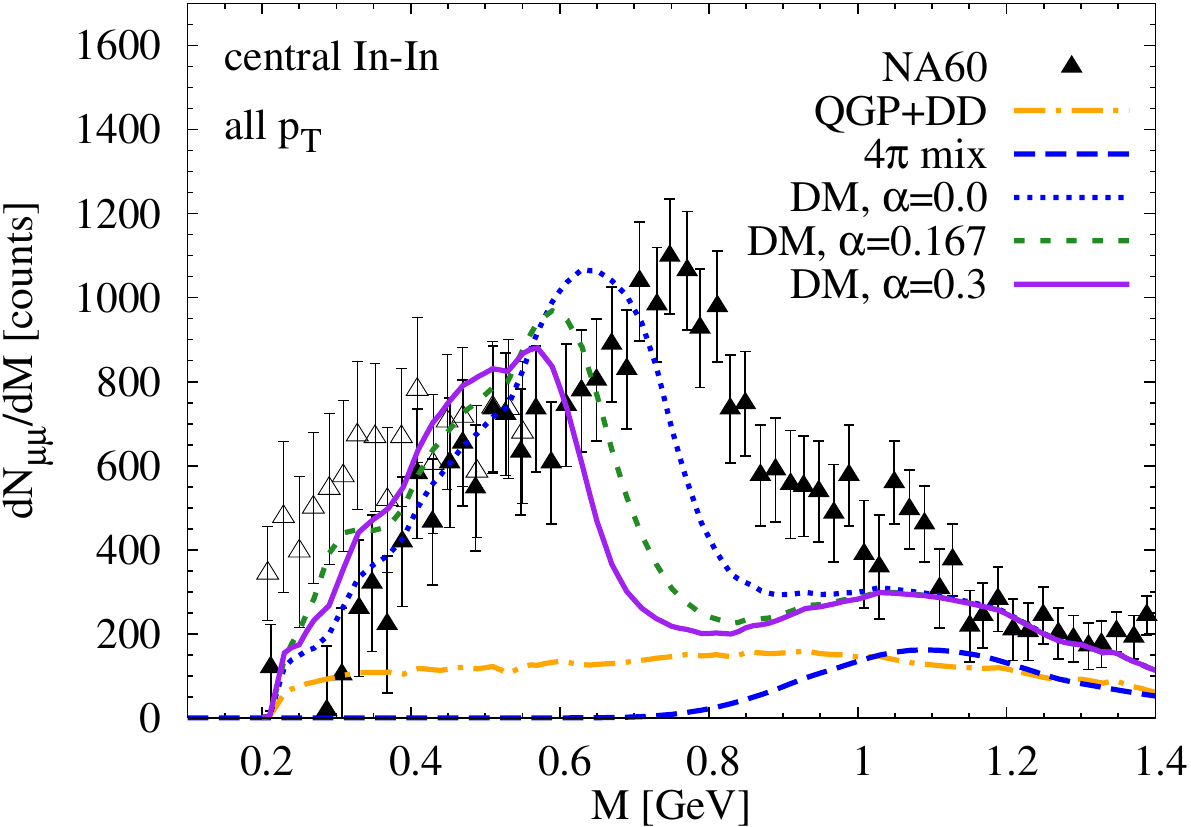}
  \end{center}
  \vspace*{8pt}
  \caption{Comparison of the $\mu^+ \mu^-$ excess for central In+In
    collisions to model predictions, (left) taking an in-medium broadening of the
    $\rho$ meson into account\protect\cite{Rapp:1999ej}, and (right) for
    a dropping mass scenario according to
    Ref.\ \protect\refcite{Brown:1991kk}. Both calcuations
    are performed for the same set of the fireball parameters assuming a transverse
    acceleration of $a_t=0.085$ c$^2$/fm. The parameter $\alpha$ controls the
    temperature dependence of the in-medium mass. The figure is taken from
    Ref. \protect\refcite{vanHees:2006iv}.}
  \label{fig:NA60_Broad_DM}
\end{figure}
The measured $\mu^+\mu^-$ invariant-mass spectrum is compared in Fig.~\ref{fig:NA60_Broad_DM}
with two theoretical predictions, a scenario taking the broadening of the
$\rho$ meson\cite{Rapp:1999ej} into account and another one assuming only a dropping of
the $\rho$ mass in the medium as proposed by Brown and
Rho\cite{Brown:1991kk}. Both scenarios are calculated for the same
parameter set of the fireball evolution\cite{vanHees:2006iv}.
We note, however, that several aspects of the decomposition of the dilepton spectrum
(primordial, thermal and freeze-out components) and of the fireball evolution
are still a matter of debate.\cite{vanHees:2007th,Specht_Damjano,Ruppert:2007cr}
The dropping mass scenario is found to be incompatible with the
NA60 data.\cite{Arnaldi:2006jq,Damjanovic:2008ta}

The decisive role of high baryon densities for the observed medium
modifications is convincingly demonstrated in Fig.~\ref{fig:NA60_Hees_Rapp} by
comparing the NA60 data to model
calculations by van Hees and Rapp\cite{vanHees:2006ng} including and neglecting the
contribution of (in-medium) baryon-resonance formation and decay processes,
$\pi N \rightarrow N^*,\Delta, \Delta^* \rightarrow N
\rho \rightarrow N\mu^+\mu^-$, to the dilepton spectrum, respectively.

\subsubsection{Nuclear reactions with elementary probes}

The properties of $\rho$ mesons in nuclei have also been investigated in
nuclear reactions using proton and photon beams. In early experiments at
DESY\cite{Alvensleben:1970uw} $\rho$ production from nuclei was studied in bremsstrahlung
induced reactions. The $\rho$ mesons were detected via their $\rho
\rightarrow \pi^+ \pi^-$ decays (the same group also did detect dileptons\cite{Alvensleben:1971rq}
from the light target $Be$) and had recoil momenta in the range of 3.5-7
GeV/c. For these high recoil momenta the in-medium width of the $\rho$ mesons is very
large.\cite{Bugg:1974cz} As discussed in Sect.\ \ref{subsec:obs-broad} such $\rho$ mesons
are outshined by the ones which decay in vacuum: Although a sizeable fraction
of $\rho$ mesons decays within the nuclear medium
even at these high recoil momenta, no strong deviation from the free $\rho$
line shape can be expected since the contributions to the $\rho$ signal from
higher nuclear densities are suppressed. It is therefore not surprising that
the experiment did not exhibit any change of the $\rho$ line shape with
increasing nuclear mass.

For $\rho$ mesons with lower recoil momenta this problem is reduced.
A nearly ideal experiment for the investigation of in-medium changes of vector mesons has been proposed in Ref.\ \refcite{Effenberger:1999ay}. In that reference all the cross sections as well as the expected sensitivities to in-medium changes had been worked out. Recently this proposed experiment has indeed been performed by the CLAS collaboration at JLAB in a $\gamma + A \to e^+e^- + X$ reaction\cite{:2007mga,Wood:2008ee}.
This is -- as the earlier exploratory studies of Ref.\ \refcite{Alvensleben:1971rq} -- an experiment
where electromagnetic interactions are exploited in the entrance as
well as in the exit channel, thereby avoiding initial as well as final state interactions. An additional advantage is that the partial decay width is known so that the spectral function itself can be extracted from the measured dilepton spectra.

\begin{figure}[ht]
  \begin{center}
    \includegraphics[keepaspectratio,width=0.49\textwidth]{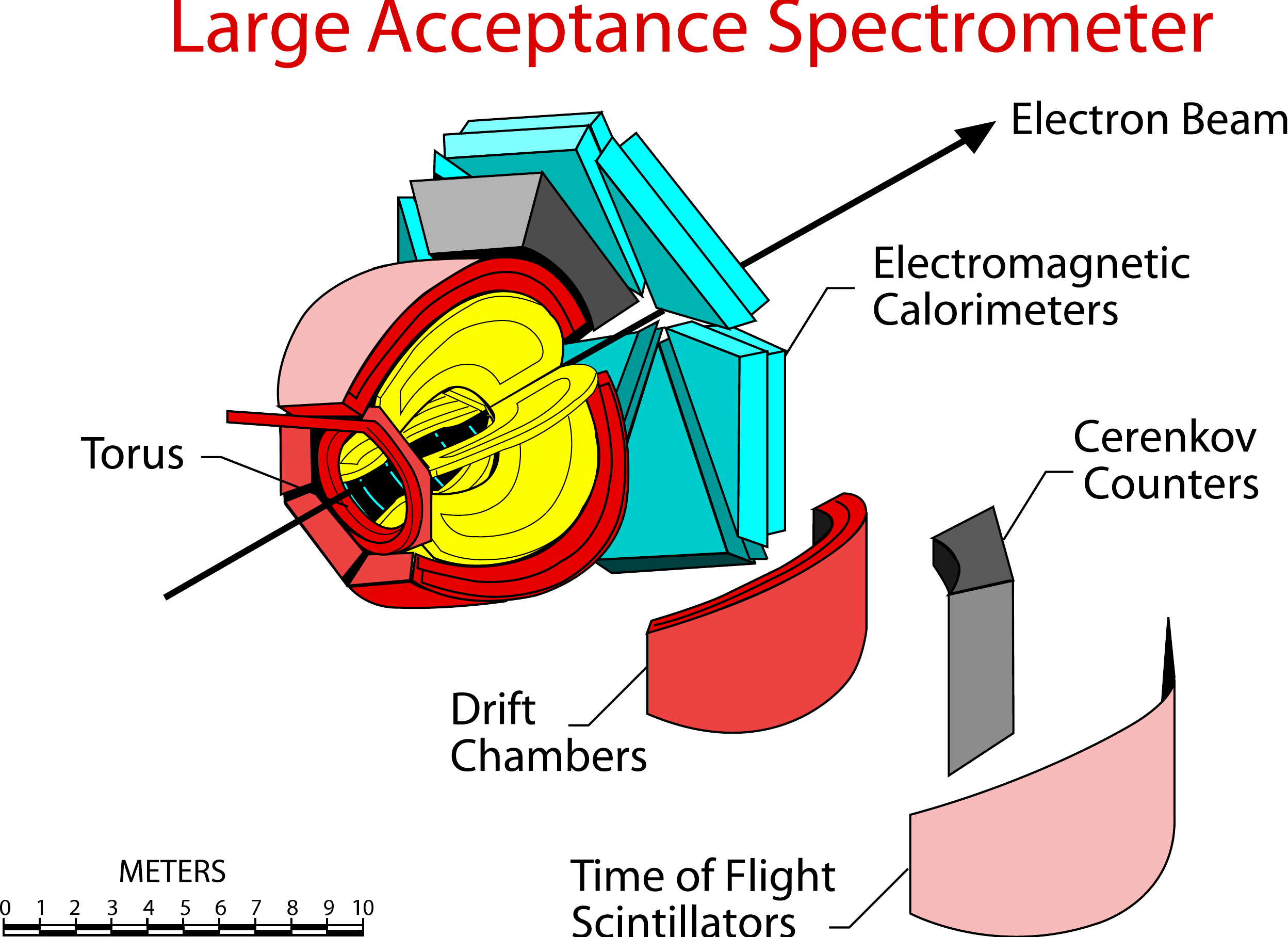}
    \hfill
    \includegraphics[keepaspectratio,width=0.49\textwidth]{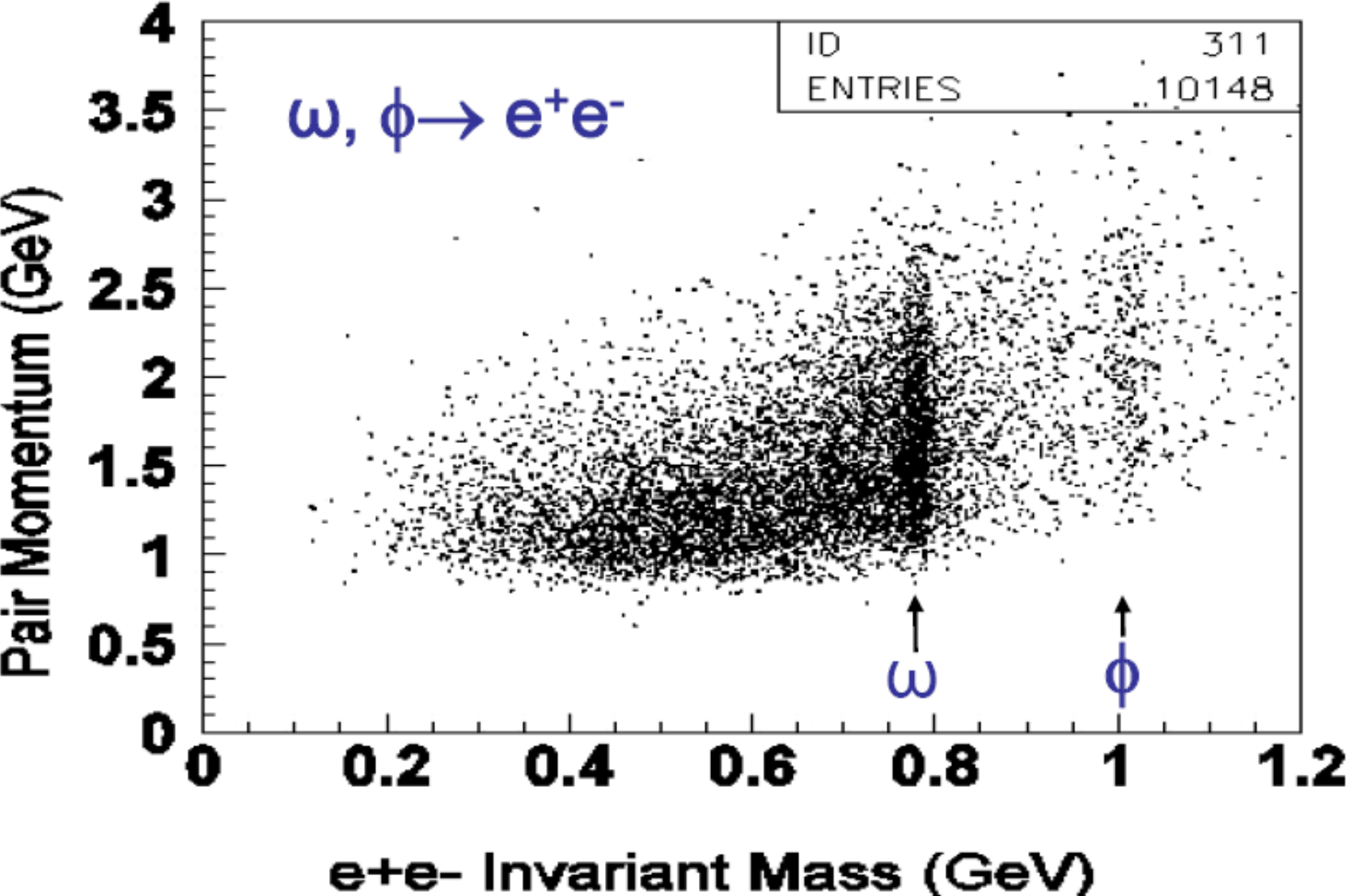}
  \end{center}
  \vspace*{8pt}
  \caption{Layout of the CLAS detector comprising a superconducting toriodal
    magnet, drift chambers, time-of-flight hodoscopes, an electromagnetic
    calorimeter and gas Cherenkov counters.\protect\cite{Mecking:2003zu} The plot on the
    right hand side illustrates that there is only acceptance for $e^+ e^-$
    pairs with momenta above about 800 MeV/c.}
  \label{fig:CLAS_setup}
\end{figure}

Compared to the DESY experiment the CLAS collaboration has studied the
photoproduction of vector mesons at lower incident energies
($E_{\gamma }= 0.6$-3.8 GeV) on a series of nuclear targets. Vector mesons
were identified via their $e^+ e^-$ decay. The detector system and the acceptance for
dileptons are shown in Fig.~\ref{fig:CLAS_setup}.
The detector covers
almost the full solid angle and comprises six identical sectors embedded in a
six-coil superconducting toroidal magnet. Each sector consists of drift
chambers, time-of-flight scintillators, Cherenkov counters and an
electromagnetic calorimeter. Low-energy $e^+ e^-$ background from pair
production in the target is suppressed by a ``mini-torus'' magnet. Suitable
trigger conditions were chosen to suppress coherent photoproduction of vector
mesons.

\begin{figure}
  \begin{minipage}[htb]{0.4\textwidth}
    \begin{center}
      \includegraphics[keepaspectratio,width=\textwidth]{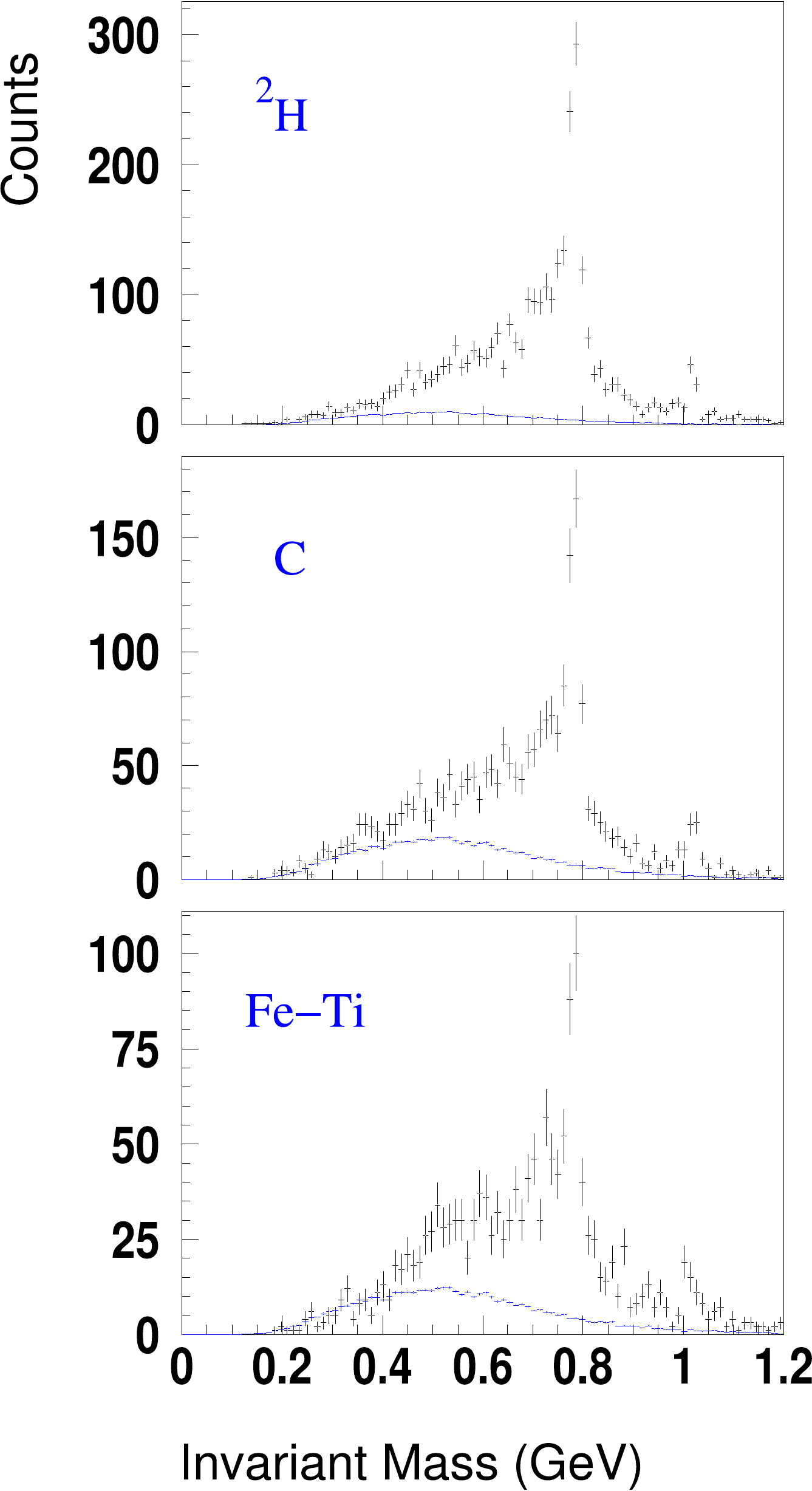}
    \end{center}
  \end{minipage}
  \hfill
  \begin{minipage}[h]{0.28\textwidth}
    \begin{center}
      \includegraphics[keepaspectratio,width=\textwidth]{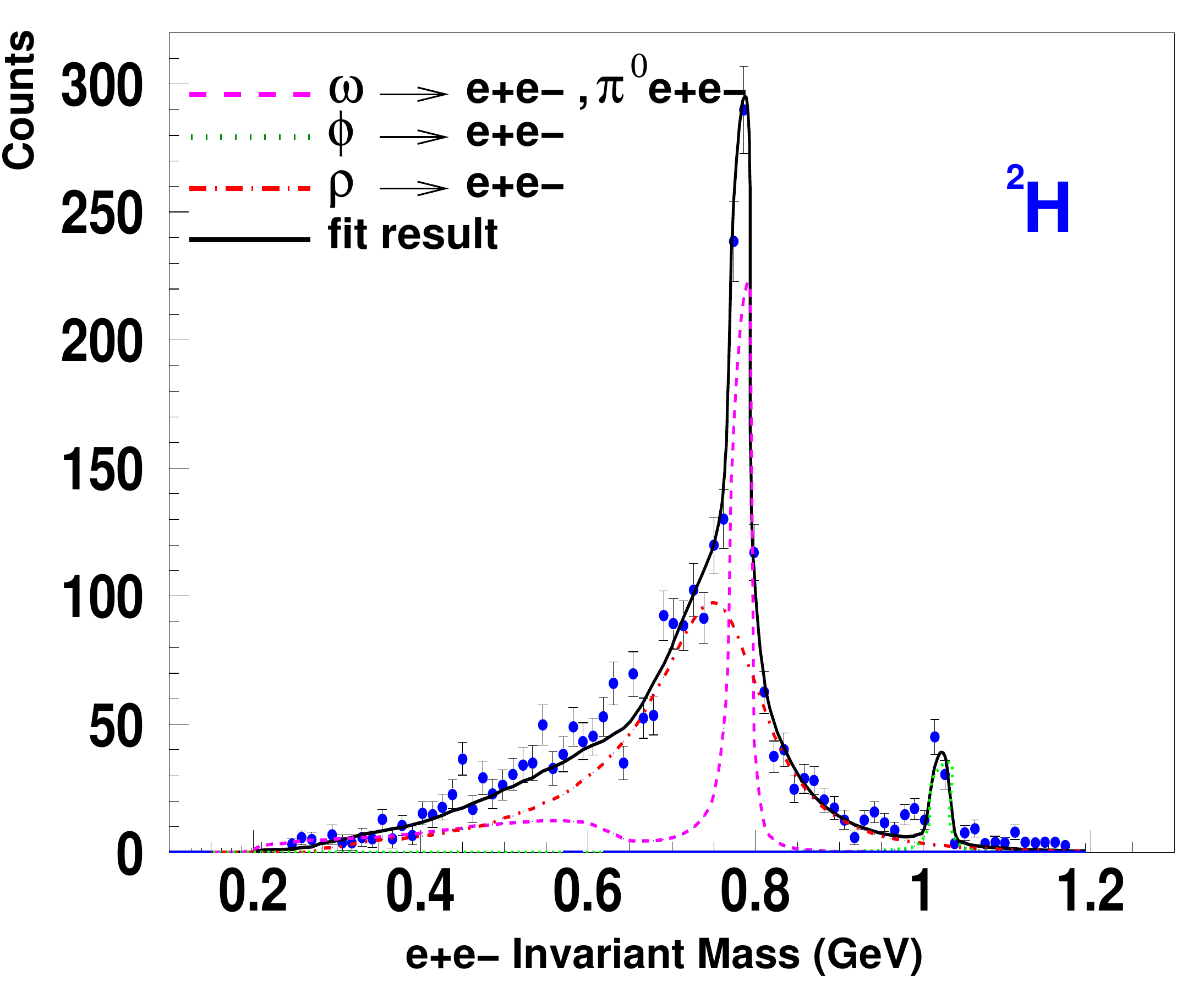}
      \includegraphics[keepaspectratio,width=\textwidth]{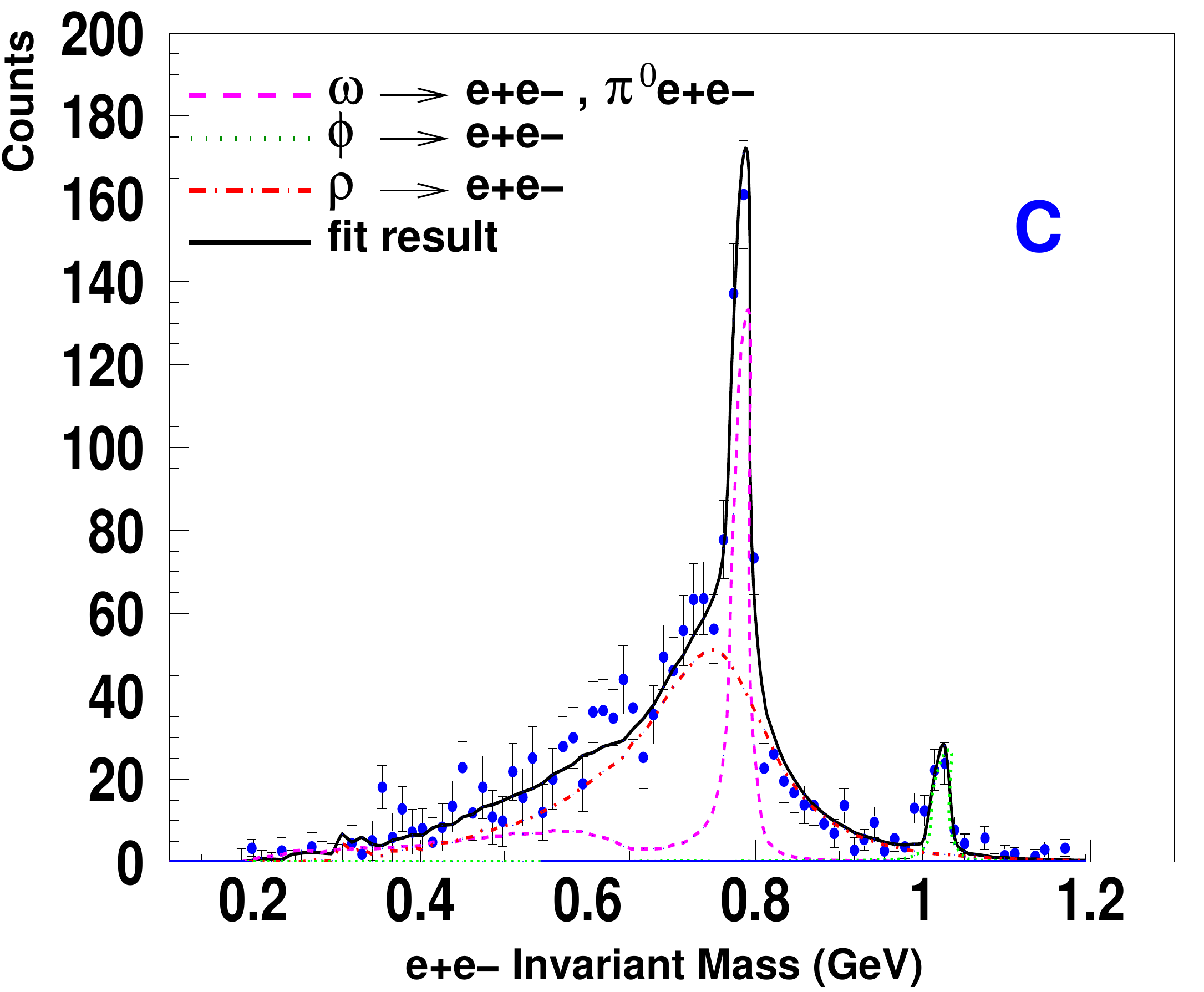}
      \includegraphics[keepaspectratio,width=\textwidth]{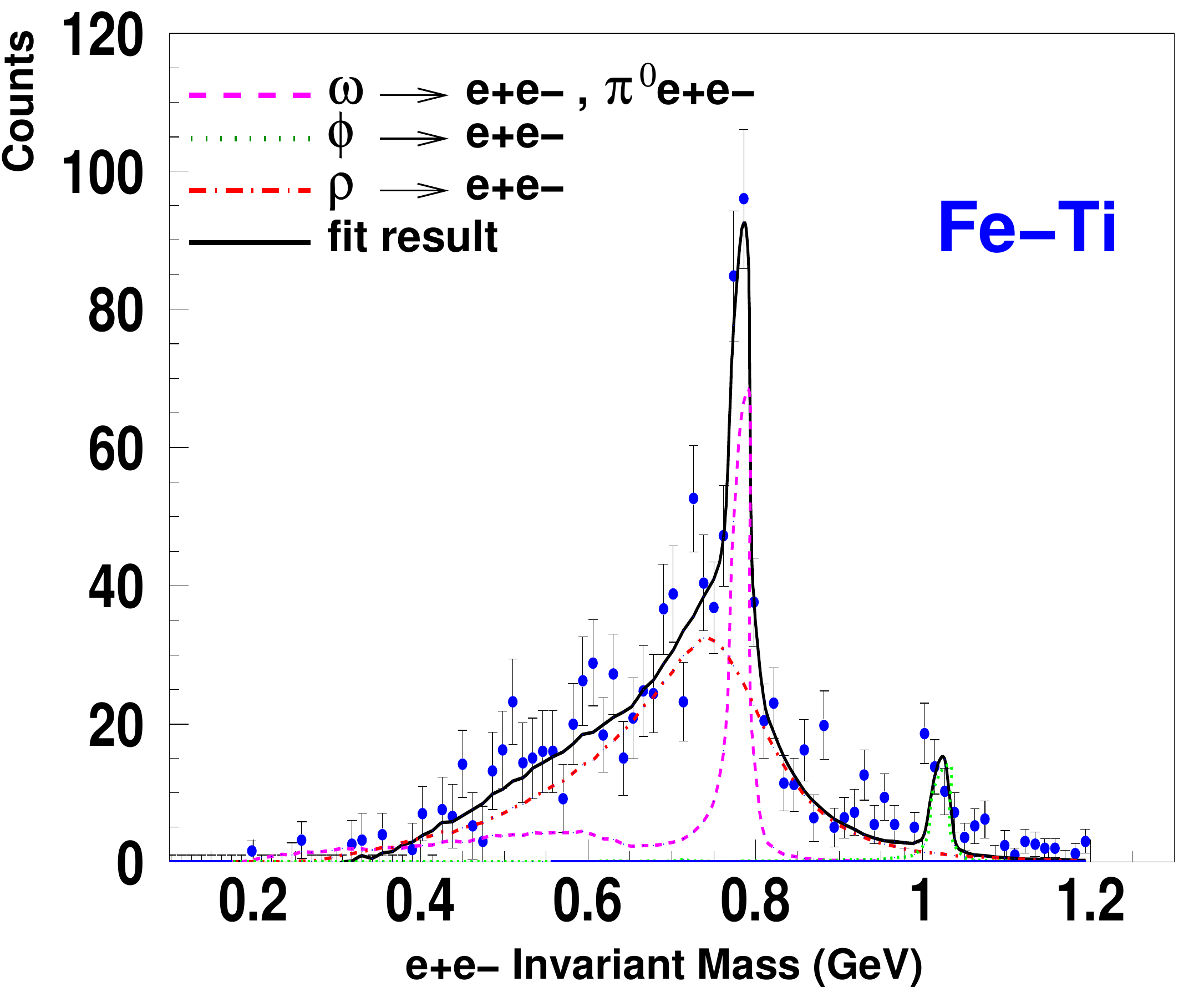}
    \end{center}
  \end{minipage}
  \hfill
  \begin{minipage}[h]{0.29\textwidth}
    \begin{center}
      \includegraphics[keepaspectratio,width=\textwidth]{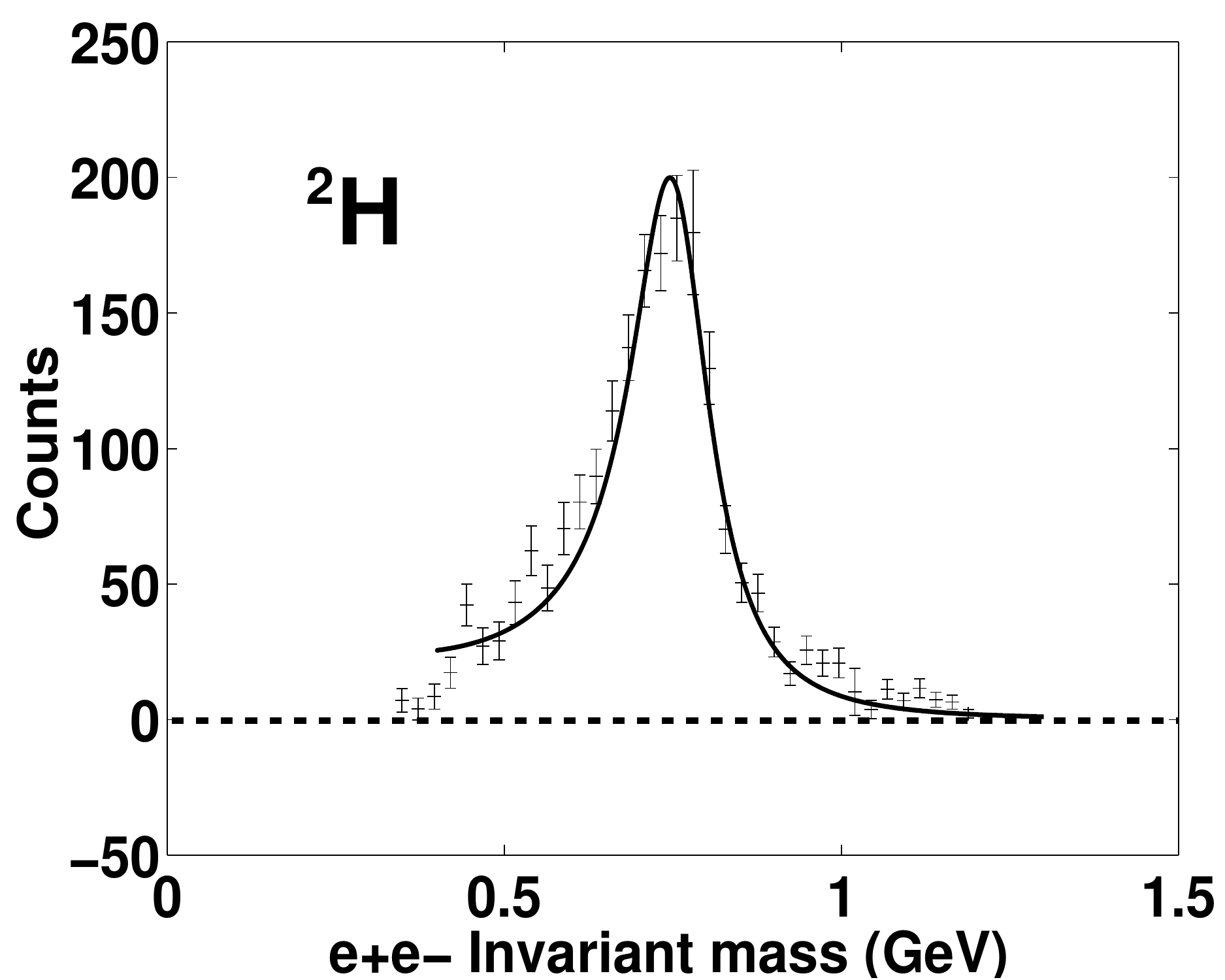}
      \includegraphics[keepaspectratio,width=\textwidth]{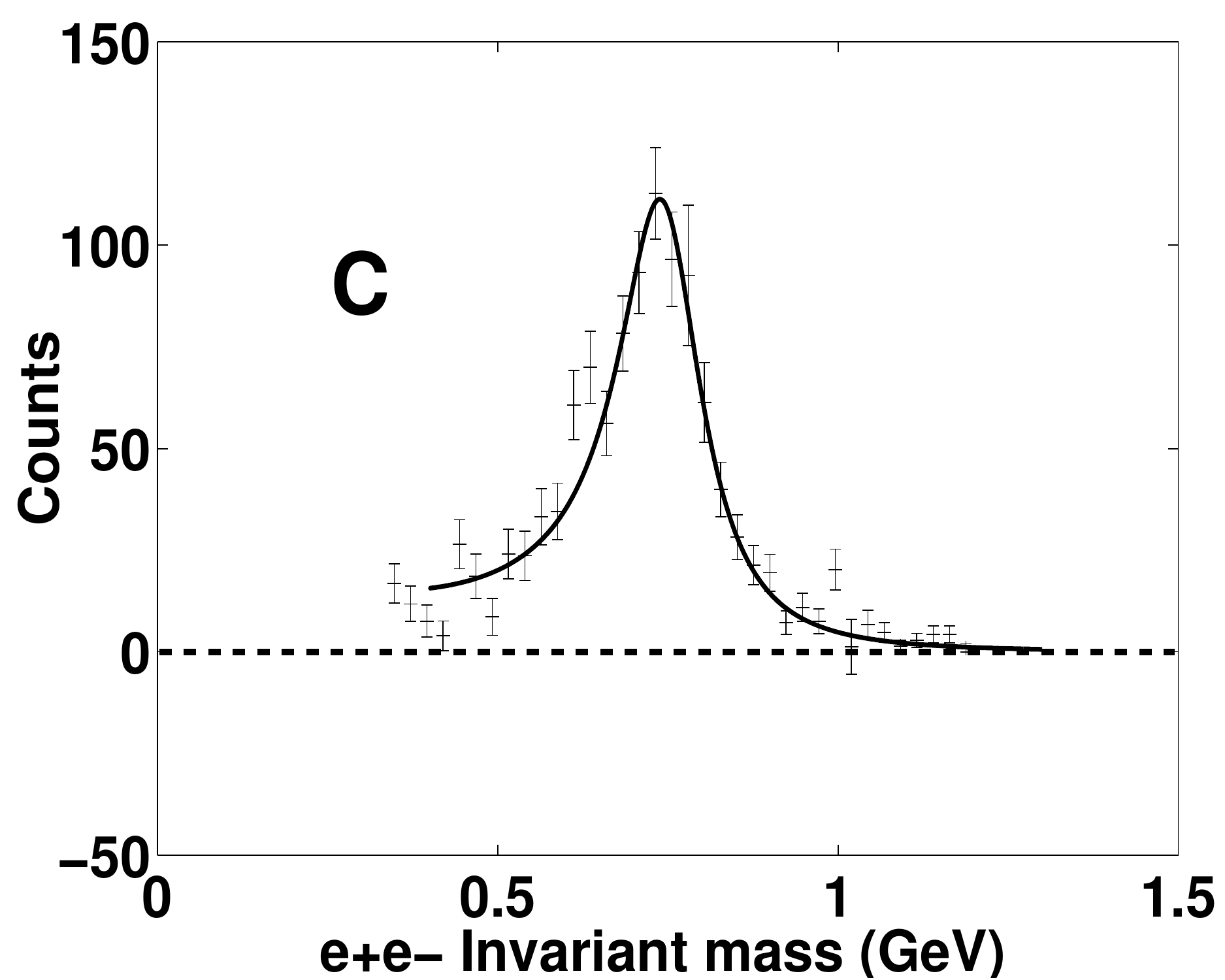}
      \includegraphics[keepaspectratio,width=\textwidth]{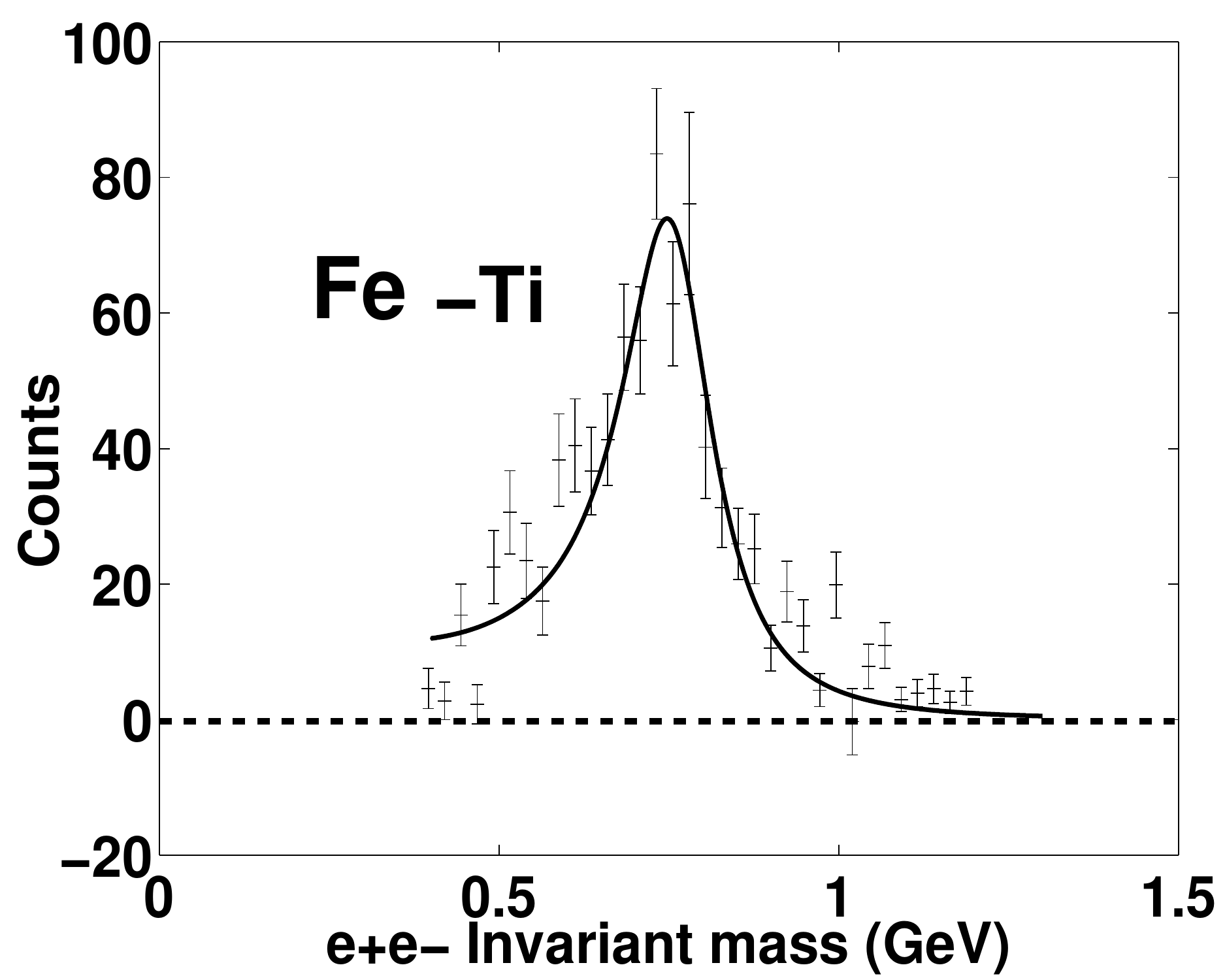}
    \end{center}
  \end{minipage}
  \vspace*{8pt}
  \caption{Left panels: $e^+ e^-$ spectra with combinatorial background
    normalized to like-sign pairs for different target nuclei.
    Middle panels: Same spectra after subtracting the combinatorial background,
    decomposed into contributions from different vector meson
    decays.\protect\cite{:2007mga,Wood:2008ee}
    The curves are transport model calculations with
    the GiBUU code.\protect\cite{GiBUU} Right panels: Same spectra with contributions
    from $\rho$ mesons only.}
  \label{fig:CLAS_fits}
\end{figure}

Resulting dilepton spectra are shown in Fig.~\ref{fig:CLAS_fits}.
The spectra on the left hand side include the combinatorial background which was
determined by the event-mixing technique and normalized to the number of
observed like-sign pairs, as described in Sect.\ \ref{sec:chall}.
After subtracting the
combinatorial background the spectra were decomposed into the $\omega$, $\phi$,
and $\rho$ contributions (see Fig.~\ref{fig:CLAS_fits}). The $e^+ e^-$
invariant-mass distributions attributed to the $\rho$ meson decay are shown
separately on the right hand side of the figure.

The experiment --- because of acceptance limitations ---  sees only vector mesons with rather high momenta beyond about 1 GeV (cf.\ Fig.\ \ref{fig:CLAS_setup}). The longer-lived vector mesons, i.e.\ the $\omega$ and the $\phi$ meson, decay mostly outside the nuclear target. However, the shorter-lived $\rho$ experiences a significant fraction of its decays inside the medium so that its in-medium properties can be extracted. For this region the results of Ref.\ \refcite{Post:2003hu} have predicted a slightly broadened $\rho$ spectral function, without any significant shift of the peak position (see Fig.\ \ref{rho_spectral}).

All measured spectra are indeed consistent with a collisional broadening of the $\rho$ meson without mass shift.
The width increases by about 70 MeV in medium size nuclei, as expected in BUU
simulations\cite{Muehlich:2006nn}
for densities of $\rho \approx \rho_0/2 $ and typical recoil momenta of 1-2 GeV/c.

\begin{figure}[th]
  \begin{center}
    \includegraphics[keepaspectratio,width=0.55\textwidth]{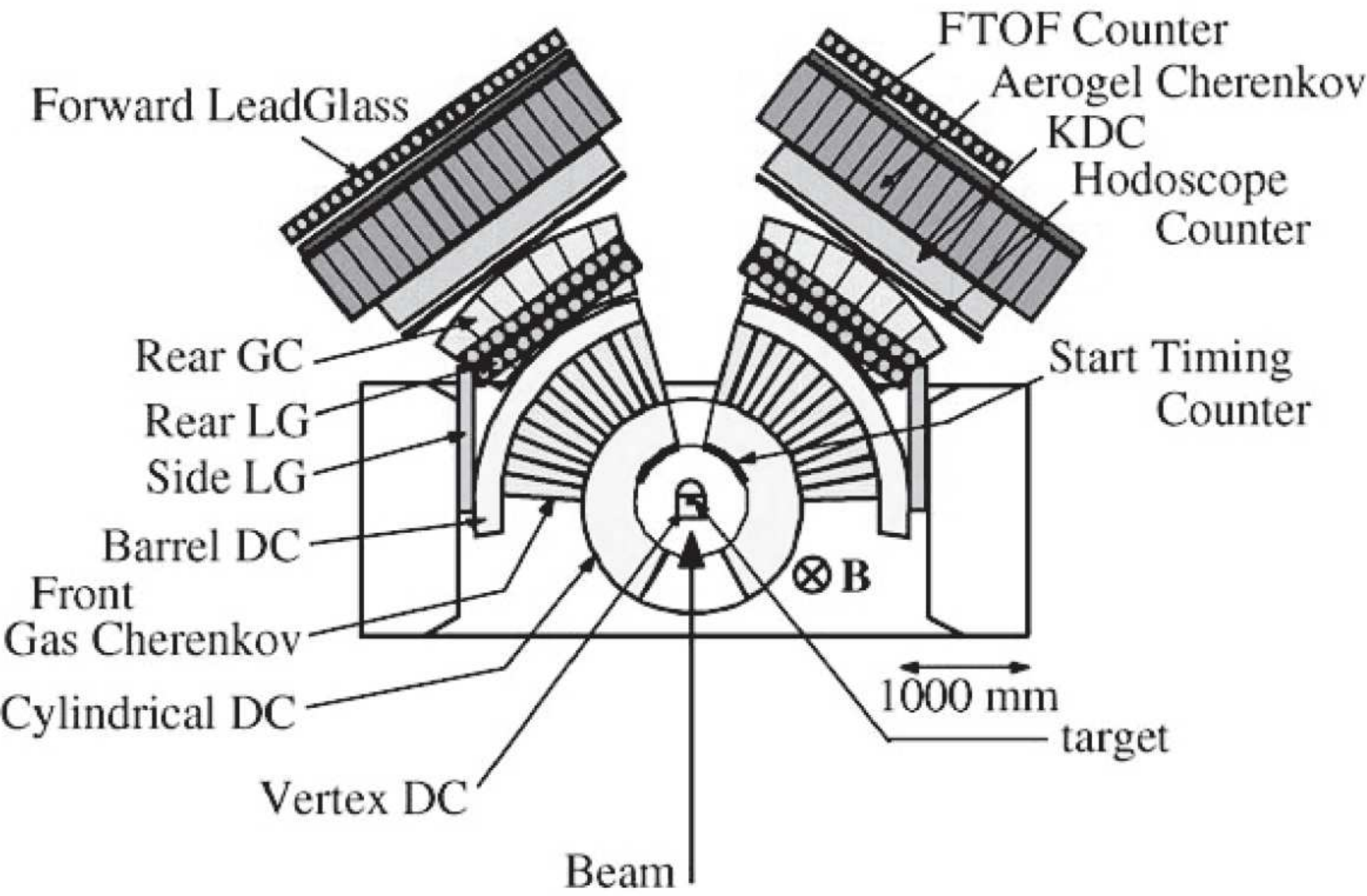}
    \hfill
    \includegraphics[keepaspectratio,width=0.43\textwidth]{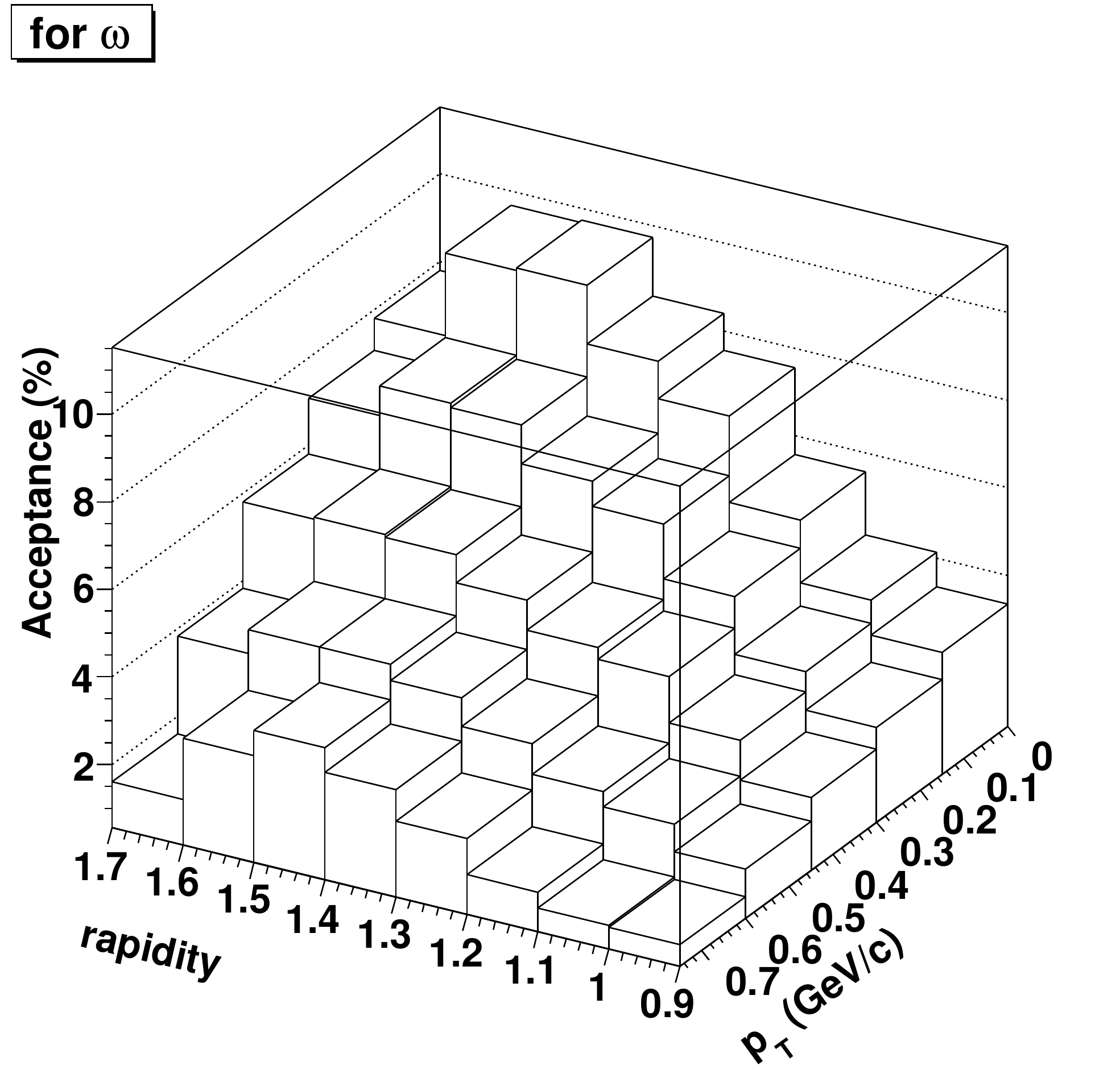}
  \end{center}
  \vspace*{8pt}
  \caption{Schematic view of the KEK-E325 experimental setup. The two-arm
    spectrometer comprises drift chambers, Cherenkov detectors, leadglass
    calorimeters, and time-of-flight hodoscopes.\protect\cite{Ozawa:2000iw} The right
    hand side shows the acceptance for lepton pairs from $\omega \rightarrow
    e^+ e^-$ decays as a function of the rapidity and transverse momentum of
    the $\omega$ meson.\protect\cite{Tabaru:2006az} }
  \label{fig:KEK_E325_setup}
\end{figure}
A quite different result has been obtained in the KEK-E325 experiment. Here, protons
of 12 GeV were used to produce vector mesons on different targets. This
experiment was the first to measure dileptons in nuclear reactions with elementary projectiles in the
search for medium modifications of vector mesons. The experimental arrangement
is shown in Fig.~\ref{fig:KEK_E325_setup}.
\begin{figure}[th]
  \begin{center}
    \includegraphics[keepaspectratio,width=0.33\textwidth]{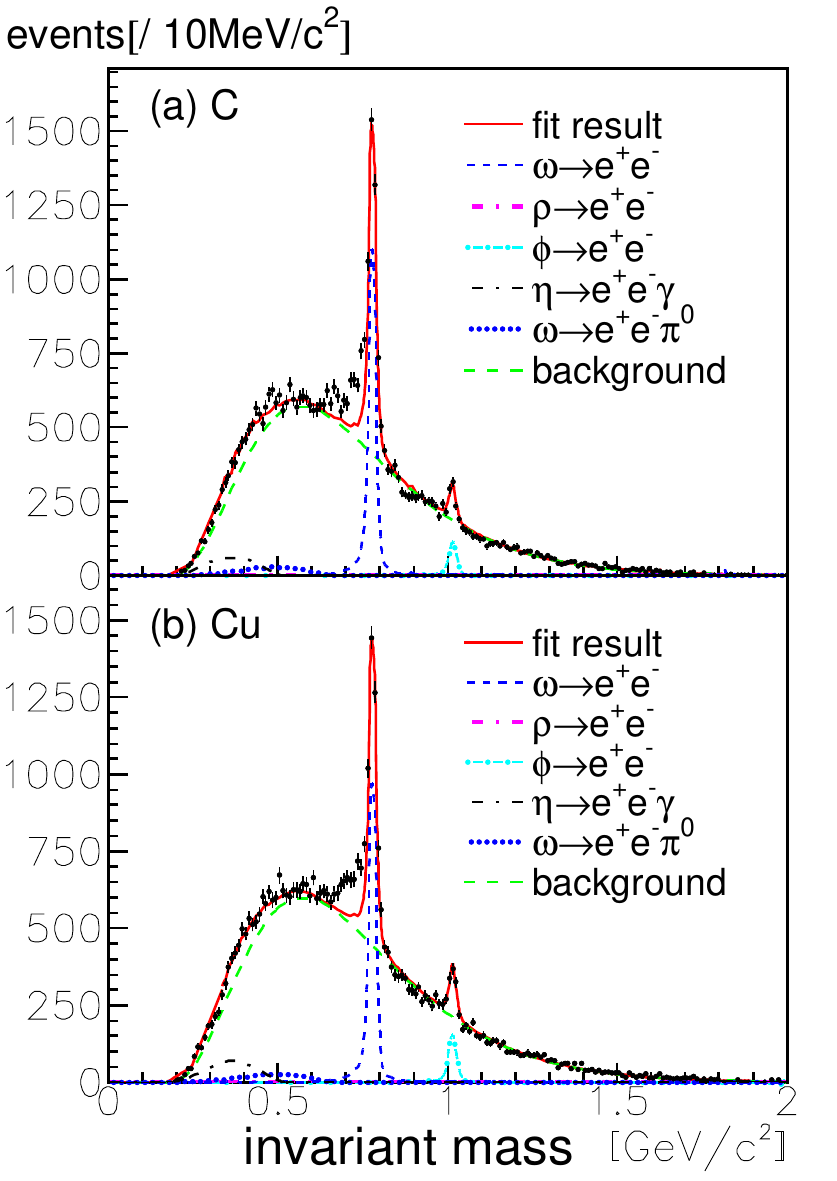}
    \hfill
    \includegraphics[keepaspectratio,width=0.65\textwidth]{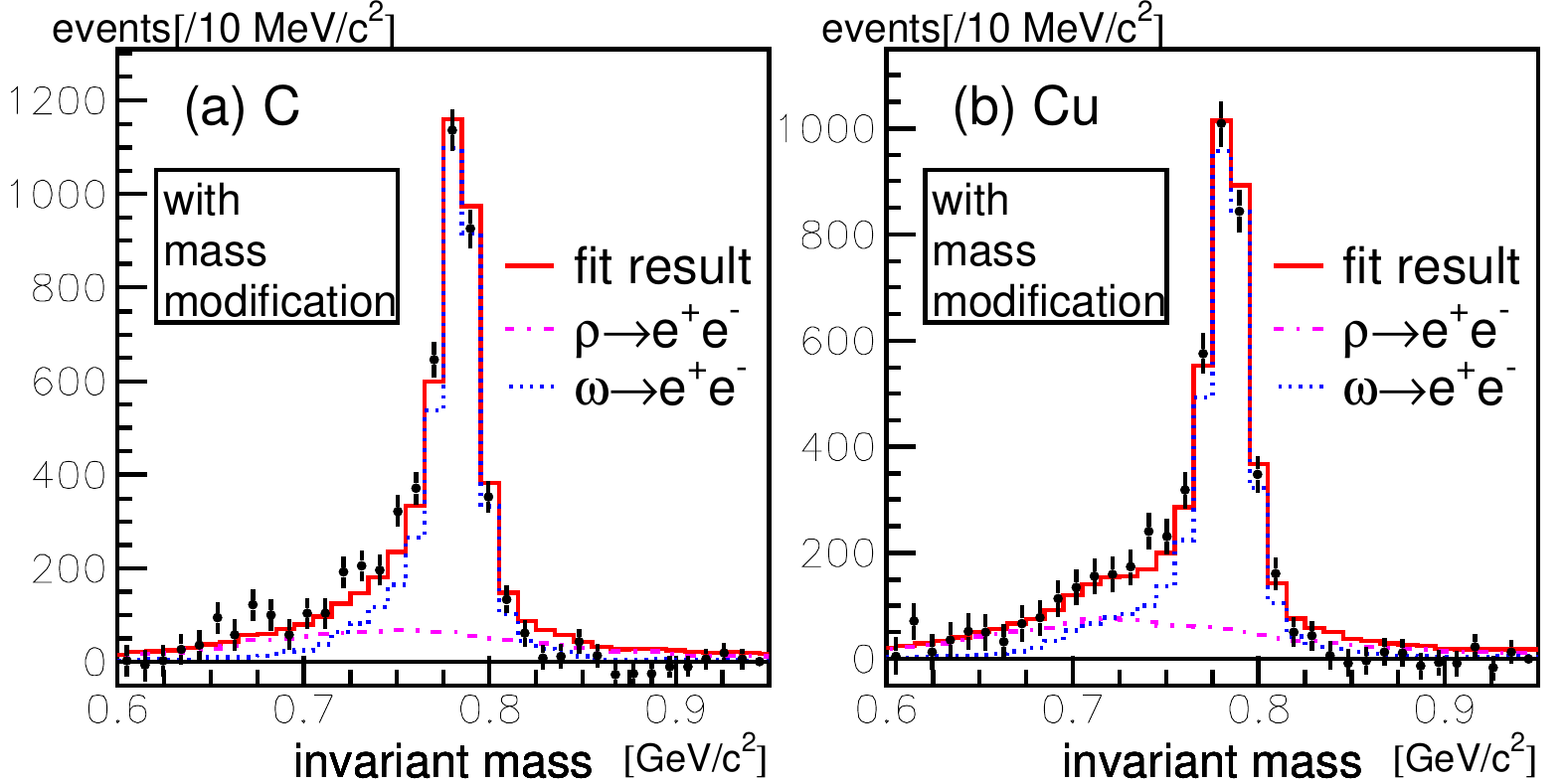}
  \end{center}
  \vspace*{8pt}
  \caption{Invariant-mass spectra of $e^+ e^-$ pairs for C and Cu nuclei. The
    fit shows the decomposition of the spectrum into the combinatorial
    background (long-dashed curve) and signals from the $\rho, \omega$, and
    $\phi$ decays into $e^+ e^-$ and the Dalitz decay modes $\eta \rightarrow
    e^+e^-\gamma$, and $\omega \rightarrow e^+e^- \pi^0$. The right hand side of the
    figure shows the same spectra for the $\rho, \omega$ mass range after
    subtraction of the combinatorial background. The shapes of the $\rho
    \rightarrow e^+e^-$ (dash-dotted) and $\omega\rightarrow e^+e^-$ (dotted curve)
    contributions are fitted by the formula
    $m_V(\rho)/m_V(0)= 1 - k \cdot (\rho/\rho_0)$ with k= 0.092.\protect\cite{Naruki:2005kd}}
  \label{fig:KEK_E325_rho_omega}
\end{figure}
It is a two arm configuration, each spectrometer arm being equipped with Cherenkov
detectors, drift chambers, time-of-flight counters, and an electromagnetic
calorimeter. The right hand side of Fig.\ \ref{fig:KEK_E325_setup} shows the
acceptance of the detector setup for lepton pairs from $\omega$ decay as a function of their
invariant mass.\cite{Muto_priv}

The dilepton spectra obtained for Carbon and Copper targets are shown in
Fig.~\ref{fig:KEK_E325_rho_omega}. The shape of the combinatorial background
was determined by event mixing and
its absolute magnitude by fitting the spectra together with contributions from
$\omega, \rho$ and $\phi \rightarrow e^+e^-$ and the Dalitz decays $\eta
\rightarrow e^+e^-\gamma$ and $\omega \rightarrow e^+e^- \pi^0$. A normalization of
the combinatorial background to the like-sign dilepton yield was not possible
because of the trigger condition, requiring unlike-sign particles in
the two arms of the spectrometer. After subtraction of the combinatorial background
a significant excess on the low-mass side of the $\omega$ peak is seen
(see Fig.~\ref{fig:KEK_E325_rho_omega}). For C and Cu, a best fit to the spectra is
obtained for a $\rho/\omega$ ratio of 0.7$\pm$ 0.1 and 0.9$\pm$ 0.2 and a drop
of the $\rho$ and $\omega$ mass by $9.2 \pm 0.2\%$ at normal nuclear matter
density without any in-medium broadening of the mesons. This result is
surprising as hadrons in the medium have additional ``decay'' options through
inelastic channels and as a consequence
their width is expected to increase in the nuclear environment. Inspection of
Fig.~\ref{fig:KEK_E325_rho_omega} shows that most of the data points above an
$e^+ e^-$ mass of 820 MeV are systematically below the fit curve which indicates
uncertainties in the background subtraction; a possible oversubtraction of the
background would remove $\rho$ strength in this mass range and could lead to an
apparent downward shift of the $\rho$ mass distribution.

It should be noted that the CLAS experiment is more sensitive to a
possible in-medium modification of the $\rho$ meson as the production of the
$\rho$ meson relative to the $\omega$ meson is about 3 times higher in the
photonuclear reaction.\cite{Wu:2005wf,Barth:2003kv}
Furthermore, it should be kept in mind that also the
KEK-E325 experiment --- just like the CLAS
experiment --- has acceptance only for mesons with
three-momenta of almost 1 GeV/c and higher and is thus not sensitive to these prominent
in-medium modifications theoretically expected at momenta less than 600 MeV/c,
as discussed in Sect.\ \ref{subsec:rho}.

Summarizing the experimental information on the $\rho$ meson, the majority of experiments, using
heavy-ion collisions as well as nuclear reactions with elementary probes,
observes a broadening of the $\rho$ meson which depends in magnitude on the density and
temperature of the nuclear environment. A mass shift of the $\rho$ meson is
reported by one experiment while all other measurements find that the
centroid of the mass distribution remains at the pole mass of the free $\rho$ meson.

\subsection{In-medium properties of the $\omega$ meson}
The in-medium properties of the $\omega$ meson have also been studied in
heavy-ion collisions as well as in elementary nuclear reactions.

\subsubsection{Heavy-ion reactions}
\begin{figure}[th]
  \begin{center}
    \includegraphics[keepaspectratio,width=0.48\textwidth]{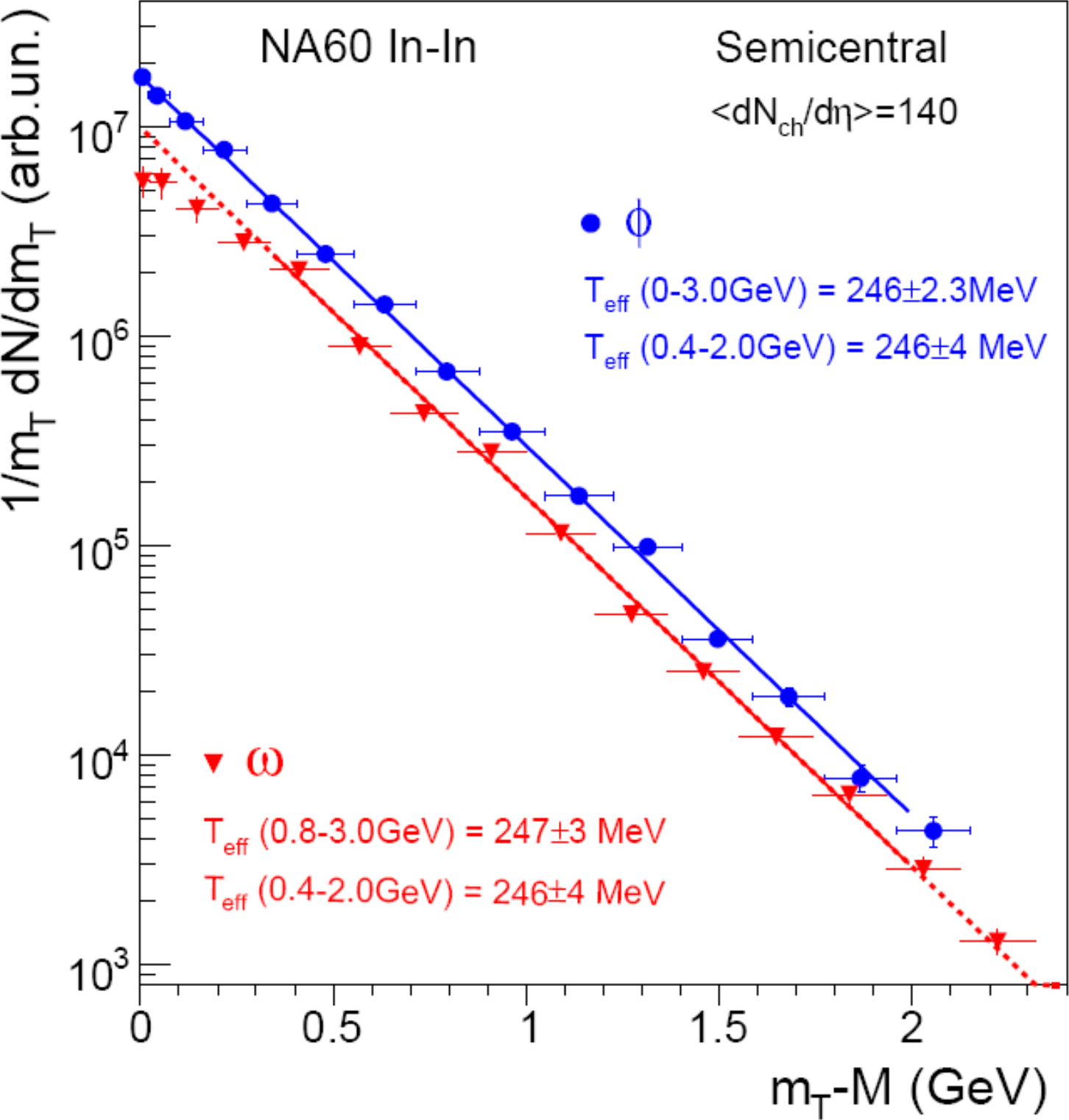}
    \hfill
    \includegraphics[keepaspectratio,width=0.48\textwidth]{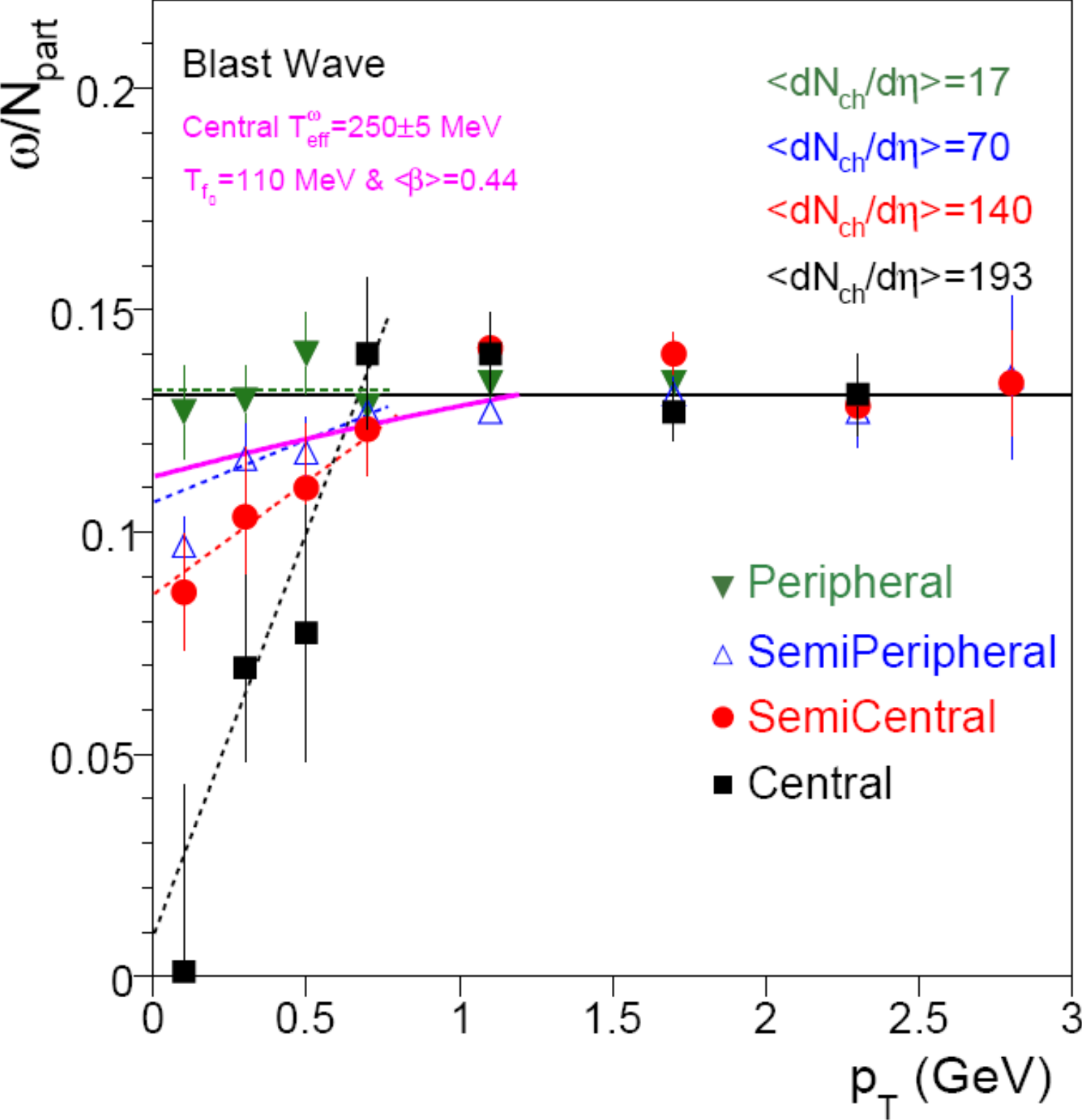}
  \end{center}
  \vspace*{8pt}
  \caption{Left: Acceptance-corrected transverse-mass spectra of the $\omega$ and
    $\phi$ meson for semi-central In+In collisions at 158 AGeV. A depletion of
    the §$\omega$ yield relative to the exponential fit is observed at low
    transverse masses. Right: $p_t$ dependence of the $\omega$ yield relative
    to the fit line for different centralities, absolutely normalized for the
    full phase space. The solid curve for p$_t \le $1GeV/c shows the result of a
    blast wave fit\protect\cite{Damjanovic:2008ta} to the $\omega$ data for central
    collisions (from Ref.\ \protect\refcite{Damjanovic:2006bd}). }
  \label{fig:NA60_omega}
\end{figure}

As shown in Fig.~\ref{fig:NA60_excess} a clean $\omega$ signal is observed by
the NA60 collaboration in In+In collisions at 158 AGeV. Many of the $\omega$
mesons live longer than the fireball and decay in vacuum after thermal
freeze-out. Medium effects can only be expected for low momentum $\omega$ mesons.
In a search for possible medium effects, the NA60 collaboration studied the
intensity of the $\omega$ and $\phi$ peaks in Fig.\ \ref{fig:NA60_excess} as a
function of the transverse momentum of the mesons. Both distributions are plotted
in Fig.~\ref{fig:NA60_omega} as a function of the transverse kinetic energy
$ m_t -m = \sqrt{p_t^2 + m^2}-m $ and fitted with an exponential
$1/m_t \cdot dN/dm_t = N_0 \cdot exp(-m_t/T_{{\rm eff}})$.

While the $\phi$ meson follows a straight exponential fall off a deviation
from the corresponding reference line is observed for the $\omega$ meson at
transverse momenta below 1 GeV/c. This suppression of the $\omega$ line intensity at low
momenta indicates
that slow $\omega$ mesons may either be broadened or shifted in mass in the
environment of the heavy-ion collision. This effect is more pronounced for
the more central collisions where the highest baryon densities and
temperatures and therefore also the largest medium modifications are expected.
This is illustrated on the right hand side of
Fig.~\ref{fig:NA60_omega} which shows the
$\omega$ yield divided by the respective exponential fit curve for different centrality bins.

While the experiment is sensitive to the disappearance of the $\omega$ yield
at low momenta the appearance of the intensity elsewhere in the $\mu^+\mu^-$ invariant-mass
spectrum is practically unmeasurable since it is masked by the
dominating  $\pi^+\pi^- \rightarrow \mu^+\mu^-$ and $ \pi N \to N^* \rightarrow
N\mu^+\mu^-$ contributions. The measurement
shows that slow $\omega$ mesons undergo medium modifications but it
cannot be concluded whether this is due to a mass shift, a broadening, or both.

\subsubsection{Nuclear reactions with elementary probes}
\label{omega_el}

In-medium properties of the $\omega$ meson have been deduced simultaneously
with information on the $\rho$ meson in the KEK-E325 experiment. Fitting
the spectrum in Fig.~\ref{fig:KEK_E325_rho_omega}, Naruki et
al.\cite{Naruki:2005kd} conclude that
not only the $\rho$ meson but also the $\omega$ meson drops in mass by 9$\%$
at normal nuclear density, although --- according to their model --- only 9$\%$ of
the $\omega$ mesons decay within a Cu nucleus, the heavier nucleus studied.

\begin{figure}[th]
  \begin{center}
    \includegraphics[keepaspectratio,width=0.48\textwidth]{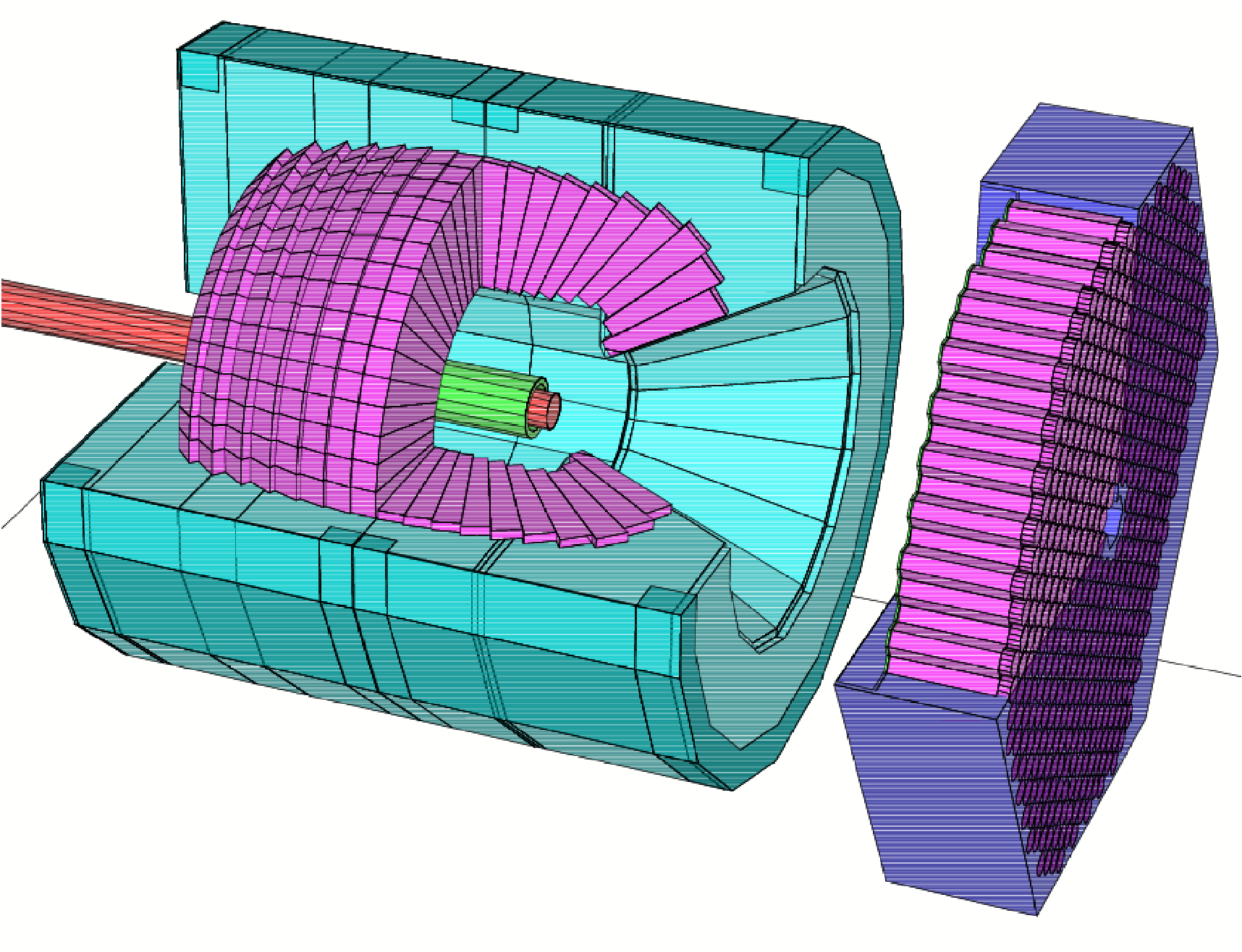}
    \hfill
    \includegraphics[keepaspectratio,width=0.48\textwidth]{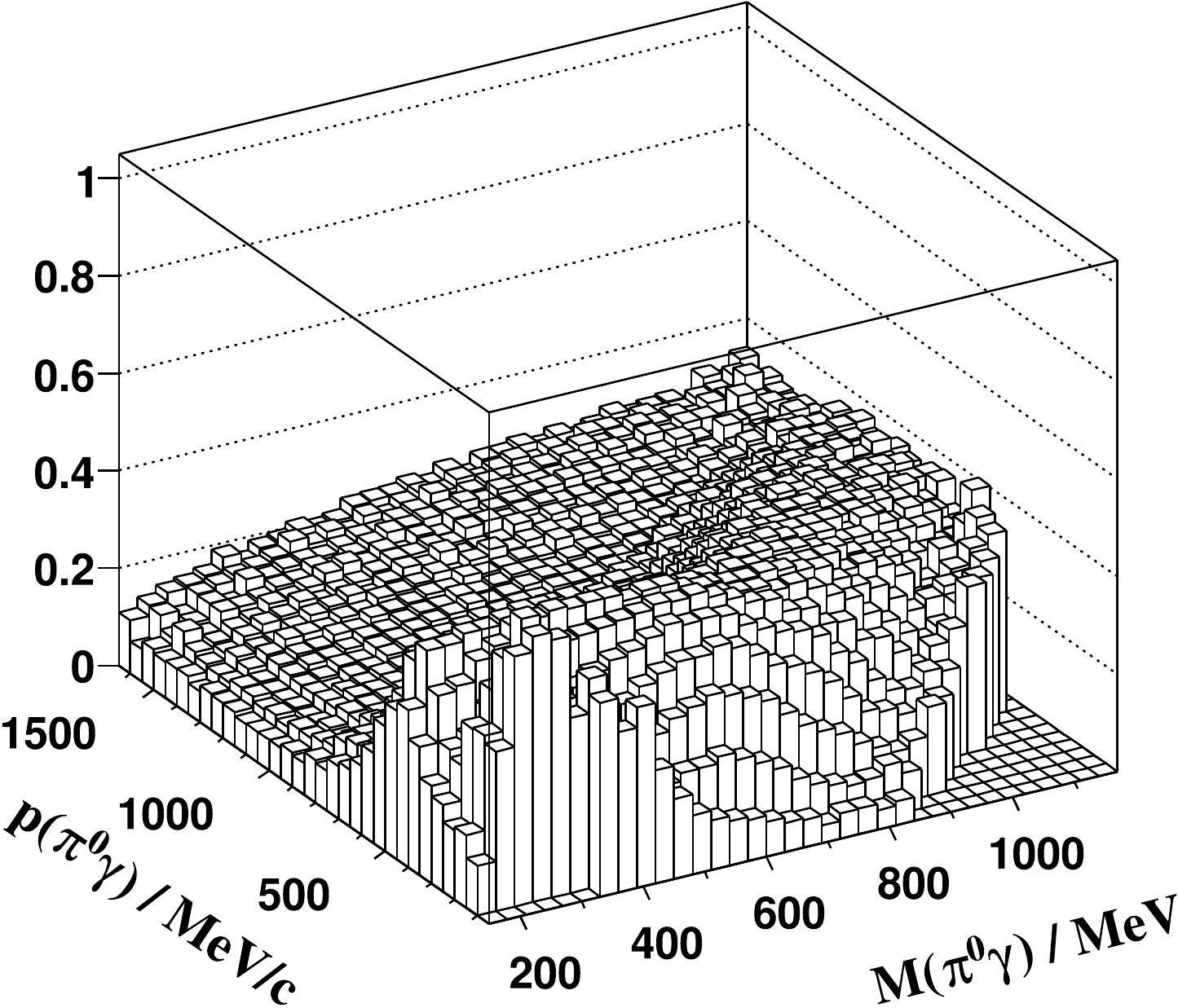}
  \end{center}
  \vspace*{8pt}
  \caption{Left: Experimental setup of the CBELSA/TAPS detector system, comprising the
    Crystal Barrel (1290 CsI crystals), the TAPS forward wall (528 BaF$_2$
    scintilators) and a scintillating fiber array for charged particle
    detection.\protect\cite{Metag:2007zz}
    Right: CBELSA/TAPS acceptance for $ p\pi^0 \gamma$ events.\protect\cite{Nanova}}
  \label{fig:CBELSA_TAPS_setup}
\end{figure}
In-medium modifications of the $\omega$ meson  have also been studied by the CBELSA/TAPS collaboration. In this experiment, photon beams
of 0.9-2.2 GeV were used to produce $\omega$ mesons on LH$_2$, C, Ca, Nb, and
Pb target nuclei. In contrast to all other experiments which
studied the dilepton decay of vector mesons this measurement exploited the $\omega
\rightarrow \pi^0 \gamma$ decay branch. This decay mode has the advantage of a
high branching ratio of 9$\%$ which gains about three orders of magnitude in
intensity compared
to the dilepton decay; furthermore this decay mode is insensitive to possible
in-medium effects of the $\rho$ meson as the $\rho \rightarrow \pi^0 \gamma$
branch is only 7$\cdot 10^{-4}$. A medium effect seen in the $\pi^0 \gamma$
channel can thus be attributed to the $\omega$ meson. A serious disadvantage
of this exit channel, however, is a possible strong final-state interaction of the
$\pi^0$ meson after the $\omega$ decay within the nuclear medium which may
distort the extracted invariant-mass distribution. Detailed Monte Carlo
simulations\cite{Messchendorp:2001pa} show that this effect is small in the mass
range of interest (600 MeV/c$^2 \le m_{\pi^0 \gamma} \le $ 800 MeV/c$^2$) and
that it can be further reduced by removing low-energy pions with kinetic energies less
than 150 MeV, typical of rescattered pions, as also found in Refs.\
\refcite{Muhlich:2003tj,Kaskulov:2006zc}.

Since the $\pi^0$ meson decays into two photons, a 3-photon final-state has
to be detected. The CBELSA/TAPS experiment is ideally suited for this
measurement as it is an almost hermetic photon detector (see
Fig.~\ref{fig:CBELSA_TAPS_setup}).
The acceptance for the $p\pi^0 \gamma$
channel as a function of mass and
momentum of the $\pi^0 \gamma $ pair varies in the range of 10-40$\%$ and is
particularly large for $\pi^0 \gamma$ momenta below 500 MeV/c.

A comparative study of photo production\cite{Trnka:2005ey} on the proton and on Nb showed a difference
in the $\omega$ line shape for the two targets which was considered by the
authors of Ref.\ \refcite{Trnka:2005ey} to be consistent with a lowering of the $\omega$ mass by 14$\%$ at
normal nuclear matter density. In order to enhance the
fraction of in-medium $\omega$ decays a cut on the momentum of the $ \pi^0
\gamma $ pair of less than 500 MeV/c had been applied.
Subsequently, the background subtraction used in this analysis was critized in the
literature.\cite{Kaskulov:2006zc} Meanwhile, a new method has been developed which allows a
model-independent background determination in shape and absolute magnitude
directly from the data. Applying this method, a re-analysis of the data is
being performed. Preliminary results\cite{Nanova} do not seem to confirm the earlier
claim by Trnka et al.\cite{Trnka:2005ey} It has been
questioned\cite{MuhlichDiss,Gallmeister:2007cm}
whether an experiment including incident photon energies up to 2.2 GeV is at
all sensitive to medium modifications of the $\omega$ meson because of the
high fraction of decays outside of the nuclear medium; a search for medium
effects would be much more promising for incident photon energies near the
production threshold of $E_{\gamma} \approx 1100$ MeV.
This is similar to the enhancement of subthreshold particle production in heavy-ion collisions.\cite{Cassing:1990dr,Mosel:1992rb}
At threshold a change in the mass and/or width leads to drastic changes of the cross section. On the other hand, if the incoming photon energies are too large the slow $\omega$ mesons that experience most of the in-medium interactions become relatively less abundant so that more and more stringent momentum cuts become necessary for any sensitivity to in-medium properties. A new measurement in this lower-energy regime with
much higher statistics has been performed which will hopefully clarify the
situation. At present, the experimental results are consistent with current
theoretical predictions of no mass shift.

Additional information will hopefully soon be provided by the HADES
collaboration who have recently studied the in-medium properties of the $\omega$ meson in the
$e^+ e^-$ channel in p+Nb reactions. At present, the analysis of the data is
still ongoing.

Because of the detector resolution and uncertainties in the decomposition of the
$\omega$ signal into in-medium and in-vacuum decay contributions it is
difficult to extract an in-medium $\omega$ width from the $\omega$ signal
reported in Ref.\ \refcite{Trnka:2005ey}.
An access to the in-medium width, i.e.\ the imaginary part of the self-energy, of the $\omega$ is provided by measuring
the transparency ratio\cite{Kaskulov:2006zc,Muhlich:2006ps}
\begin{equation}
T=\frac{\sigma_{\gamma A \rightarrow \omega X}}{A \cdot \sigma_{\gamma N \rightarrow \omega X}},
\end{equation}
i.e.\ the ratio of the $\omega$ production cross section on a nucleus divided
by the number of nucleons $A$ times the $\omega$ production cross section on a
free nucleon. As the $\omega$ photoproduction cross section on the neutron is
not yet known the transparency ratio is here normalized to
Carbon.
\begin{figure}[th]
  \begin{center}
    \includegraphics[keepaspectratio,width=0.8\textwidth]{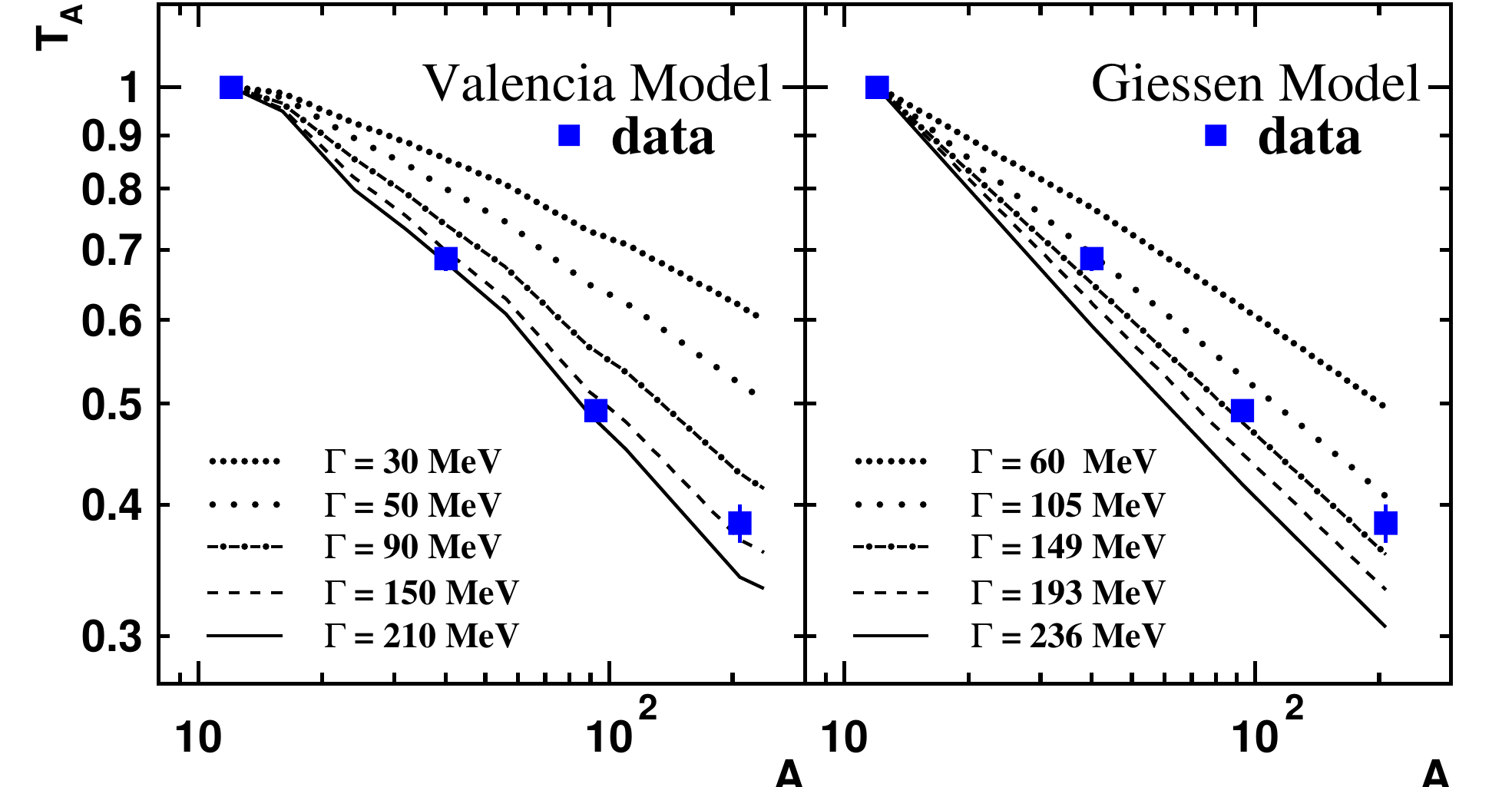}
  \end{center}
  \vspace*{8pt}
  \caption{Experimentally determined transparency ratio for $\omega$ mesons normalized to the Carbon
    data in comparison with a Monte-Carlo calculation\protect\cite{Kaskulov:2006zc} (left)
    and a BUU transport calculation\protect\cite{Muehlich:2006nn} (right). The width values
    are given in the nuclear rest frame.\protect\cite{:2008xy}}
  \label{fig:T_Gi_Val}
\end{figure}
\begin{figure}[th]
  \begin{center}
    \includegraphics[keepaspectratio,width=0.6\textwidth]{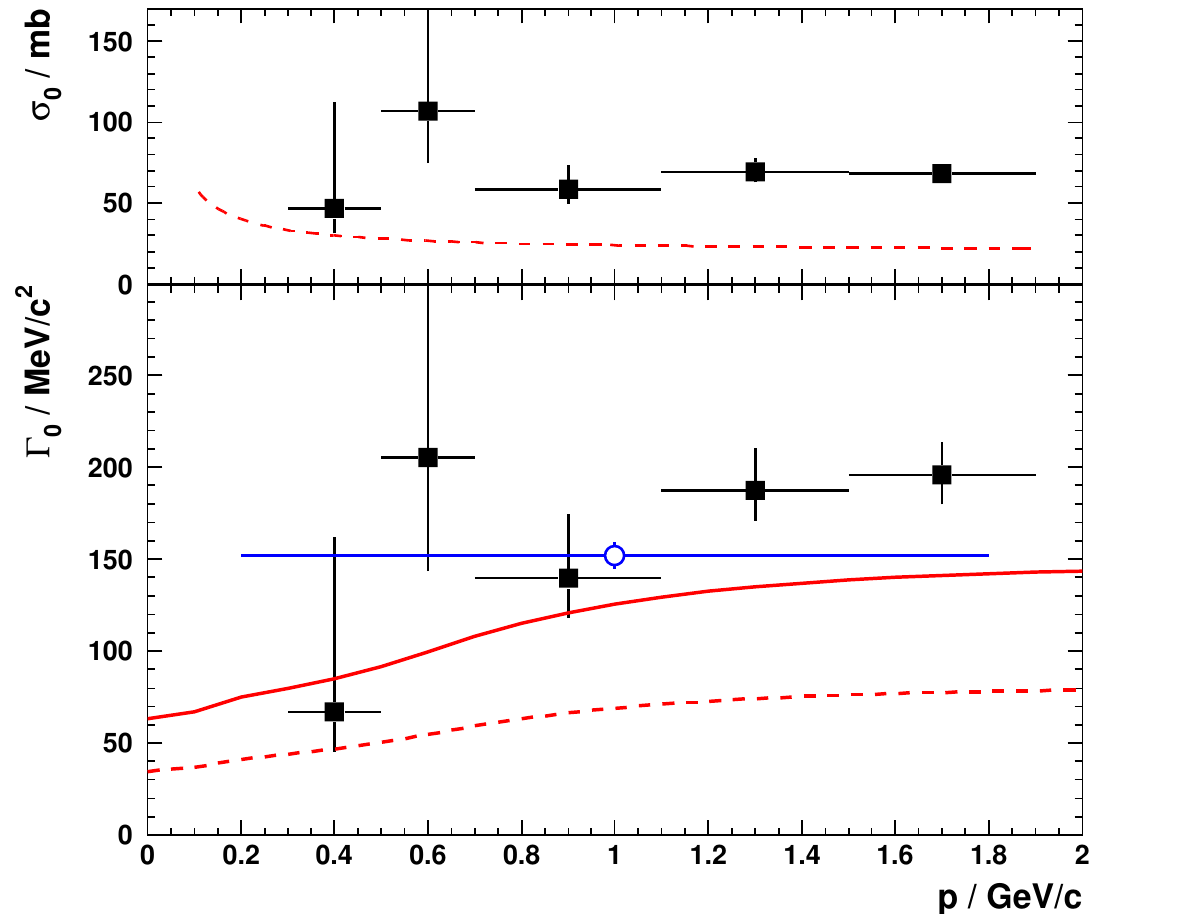}
  \end{center}
  \vspace*{8pt}
  \caption{Top: Inelastic $\omega$ N cross section deduced in a Glauber
    analysis of the data in comparison to the inelastic cross section used in
    BUU simulations. Bottom: Extracted width of the $ \omega$ meson (squares) in the nuclear rest
    frame as a function of the $\omega$ momentum in comparison to theoretical
    predictions by Refs.\ \protect\refcite{Kaskulov:2006zc} (blue circle)
    and \protect\refcite{Muehlich:2006nn} (dashed
    curve) as well as to a BUU calculation after a fit of the measured
    transparency ratio data (solid curve)\protect\cite{:2008xy}.}
  \label{fig:omega_absorption}
\end{figure}
If nuclei were completely transparent to $\omega$ mesons,
the transparency ratio would be $T=1$. Consequently, $T$ is a measure for the loss of
$\omega$ flux via inelastic processes in nuclei and can be determined in
attenuation experiments on nuclei of different mass number $A$. Within the low-density
approximation the $\omega$ absorption cross section is related to the
inelastic $\omega$ width by $\Gamma_{\omega} = v \rho \sigma$.
A comparison of  CBELSA/TAPS data\cite{:2008xy}
with calculations of the Valencia\cite{Kaskulov:2006zc} and Giessen\cite{Muhlich:2006ps}
theory groups (cf.\ Figs.\ \ref{fig:T_Gi_Val}, \ref{fig:omega_absorption})
yields an in-medium $\omega$ width in the nuclear reference frame of about
130-150 MeV at normal nuclear matter density and at an average $\omega$
momentum of 1100 MeV/c. This implies an in-medium broadening of the $\omega$
meson by a factor $\approx$ 16, i.e., the $\omega$ meson in the nuclear medium
is about as broad as the $\rho$ meson in free space.
Transport calculations, which involve only two-body collisions, can reproduce this result
only if the $\omega N$ cross section is significantly enhanced  up to 60 mb which is a
factor of 2.5-3 over previous estimates\cite{Muhlich:2006ps,MuhlichDiss} and
larger than measurements at $\omega$ momenta of 7-10 GeV/c\protect\cite{Behrend:1970}.
Assuming the momentum dependence of the
$\omega$ width given in Ref.\ \refcite{Muhlich:2006ps}
a width of 130-150 MeV at $\langle p_{\omega} \rangle \approx 1.1 $ GeV/c
would correspond to a
total width of the $\omega$ meson at rest in the medium of about 70 MeV, in line with theoretical expectatons (cf.\ Sect.\ \ref{subsec:omega}).
As discussed earlier the large width could, however, also imply that the in-medium collisional
width is generated by collisions of the $\omega$ meson with more than one nucleon. In any case, this large collisional width makes the determination of any in-medium change of the $\omega$ spectral function very difficult as we have discussed in Sect.\ \ref{subsec:obs-broad}.

Summarizing the experimental information on the $\omega$ meson, one has to
concede that a clear picture has not emerged as yet. Claims of an in-medium
mass shift are not confirmed. The large in-medium width deduced from a transparency ratio measurement is in line with the depletion of $\omega$ strength at low momenta observed in a heavy-ion experiment.

\subsection{In-medium properties of the $\phi$ meson}

Information on in-medium modifications of the $\phi$ meson has only been
reported from nuclear reactions with elementary probes. While $\gamma A$ reactions
may seem to be attractive because of the absence of any initial-state
interactions Muehlich et al.\ have shown\cite{Muhlich:2002tu} that in the dominant $K^+K^-$ decay channel 
final-state interactions hide any genuine in-medium changes of the $\phi$
meson. In particular no measurable effect caused by a shift of
the $\phi$-meson pole mass was found in these calculations which take
the kaon self-energies as well as the effects of the Coulomb force on
the outgoing kaons into account.

\begin{figure}[th]
  \begin{center}
    \includegraphics[keepaspectratio,width=0.9\textwidth]{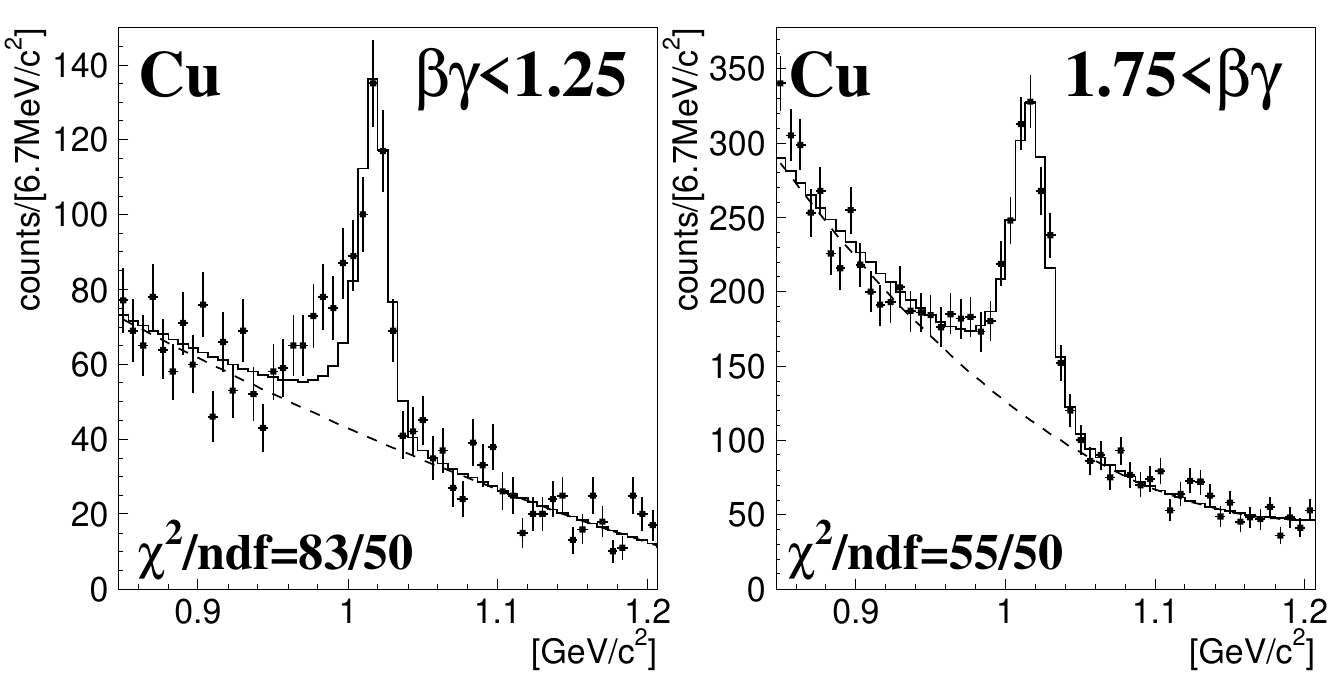}
  \end{center}
  \vspace*{8pt}
  \caption{$e^+ e^-$ invariant-mass distributions near the $\phi$ mass obtained
    in p+Cu for slow ($\beta \gamma \le 1.25$) and fast
    ($\beta \gamma \ge 1.25$) recoiling $\phi$ mesons. No difference in line
    shape is observed for the corresponding measurement on a
    C target.\protect\cite{Muto:2005za}}
  \label{fig:KEK_Phi}
\end{figure}
In $p A$ reactions the KEK-E325 experiment has also investigated possible in-medium modifications of the $\phi$ meson by looking both at the $K^+K^-$ channel as well as at the dilepton decays (see Fig.~\ref{fig:KEK_Phi}).\cite{Muto:2005za} The production
of $\phi$ mesons in proton-induced reactions at 12 GeV on a heavy (Cu) and a light
nucleus (C) have been measured and compared. In case of the light
C target no difference in line shape is observed for $\phi$ mesons recoiling
with different velocities. For the heavier Cu nucleus Fig.\ \ref{fig:KEK_Phi} shows a significant
excess on the low-mass side of the $\phi$ meson peak for slow
$\phi$ mesons ($\beta \cdot \gamma < 1.25$) which have a higher
probability to decay within the nucleus than fast ones.
From a careful analysis of the structure in the Cu spectrum
Muto et al.\cite{Muto:2005za} extracted a drop of the $\phi$ mass by 3.4$\%$ and an
increase of the $\phi$
width by a factor 3.6 at normal nuclear matter density $\rho_0$.

\begin{figure}[th]
  \begin{center}
    \includegraphics[keepaspectratio,width=0.4\textwidth]{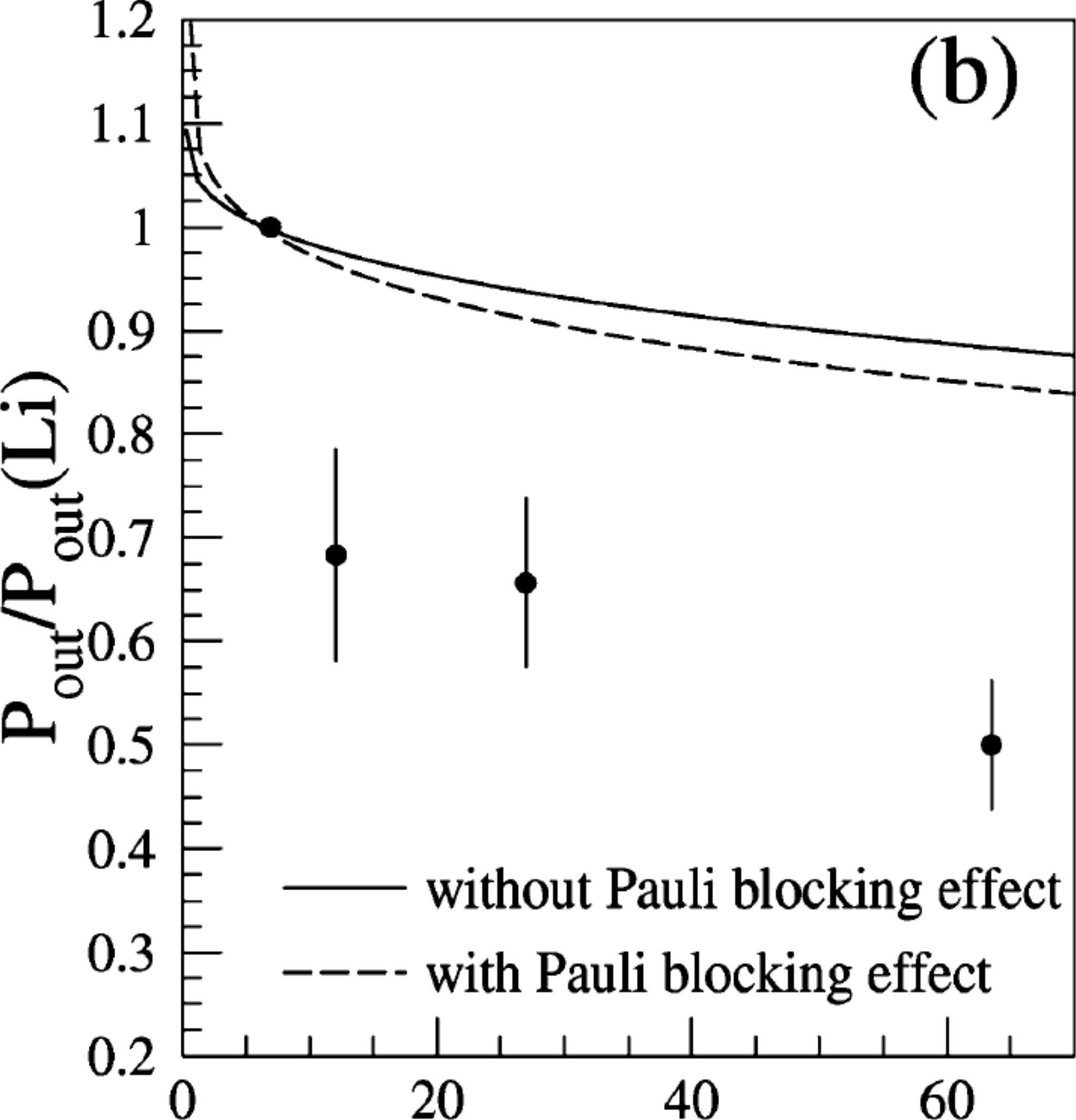}
    \hfill
    \includegraphics[keepaspectratio,width=0.58\textwidth]{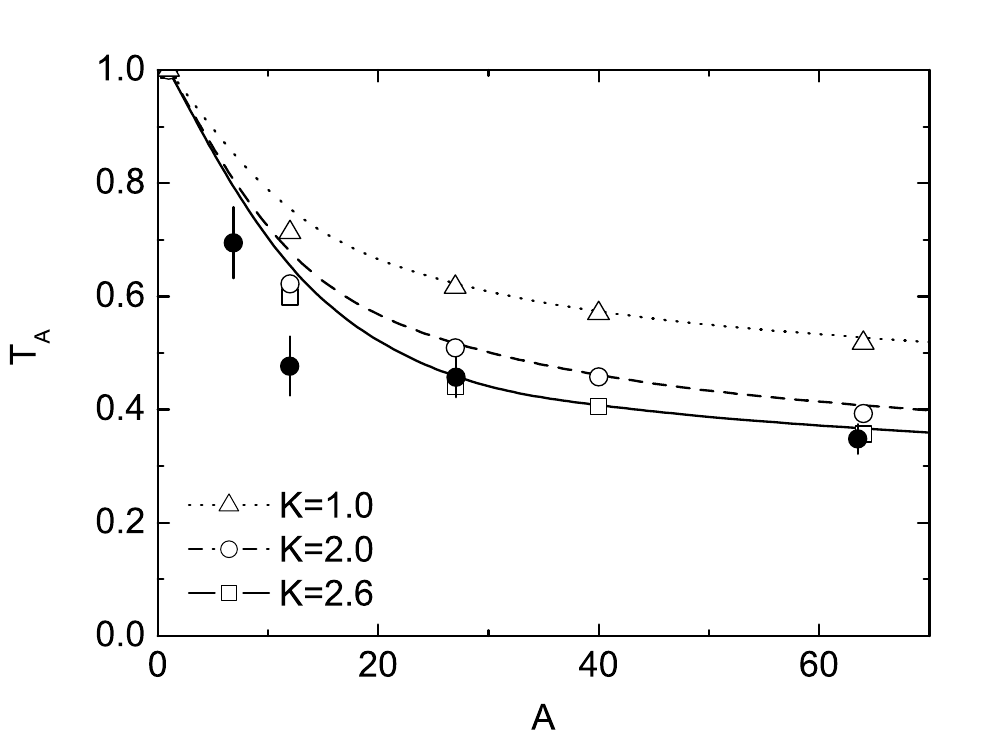}
  \end{center}
  \vspace*{8pt}
  \caption{Left: Transparency ratio for $\phi$ mesons as a function of mass,
    normalized to the Li data in comparison to a
    calculation\protect\cite{Cabrera:2003wb} with
    (dashed curve) and without (solid curve)
    Pauli-blocking\protect\cite{Ishikawa:2004id}.
    Right: The same data (normalized to hydrogen) in
    comparison to a BUU calculations.\protect\cite{Muhlich:2005kf} The K-factor indicates
    the increase in the $\phi N$ cross section needed to reproduce the
    experimentally determined transparency ratio.}
  \label{fig:Phi_Spring8}
\end{figure}
The in-medium width of the $\phi$ meson has also been determined in a
transparency ratio measurement.\cite{Ishikawa:2004id} At SPring8 the photoproduction of the
$\phi$ meson has been measured on a series of nuclei. The transparency ratio,
introduced in the previous section, was measured for Li, C, Ca, and Pb and is
shown in Fig.~\ref{fig:Phi_Spring8}.

As in the case of the $\omega$ meson, the transparency ratio --- here normalized to Li --- is much smaller than theoretically expected. If --- in a low-density approximation --- the collisional width is attributed to two-body collisions alone then a $\phi N$ cross section of $\approx 30$ mb is needed to explain the observed attenuation.\cite{Ishikawa:2004id,Magas:2004ui,Muhlich:2006ps} (see right hand side of Fig.\ \ref{fig:Phi_Spring8}). As discussed earlier this can either signal a significantly different in-medium cross section for the $\phi N$ interaction or the presence of strong $n$-body ($n>2$) interactions. A cross section of 30 mb corresponds to an in-medium width of the $\phi$ meson of $\approx$ 70 MeV which again is much larger than the width deduced by the KEK-E325 experiment from the $e^+ e^-$ invariant-mass spectrum in Fig.~\ref{fig:KEK_Phi}. One has to take into account, however, that the value reported in Ref.\ \refcite{Ishikawa:2004id} refers to a higher average recoil momentum than for the events selected in Fig.~\ref{fig:KEK_Phi}.

Summarizing the experimental information on the $\phi$ meson, a lowering of the $\phi$ mass in the medium and an increase in width has been reported. The amount of broadening, however, is different in the two analyses performed so far.

\subsection{Ongoing dielectron spectroscopy experiments}

There are ongoing dielectron-spectroscopy experiments at RHIC and GSI which also aim at extracting information on medium modifications of vector mesons. In both cases clean vector-meson signals have been observed in p+p reactions which serve as a reference. Corresponding measurements involving nuclei are currently being analyzed. No information on specific properties of $\rho$, $\omega$, and $\phi$ mesons is as yet available. The current status of these experiments is summarized in the following subsections.

\subsubsection{Vector-meson production at RHIC}

\begin{figure}[th]
  \begin{center}
    \includegraphics[keepaspectratio,width=0.52\textwidth]{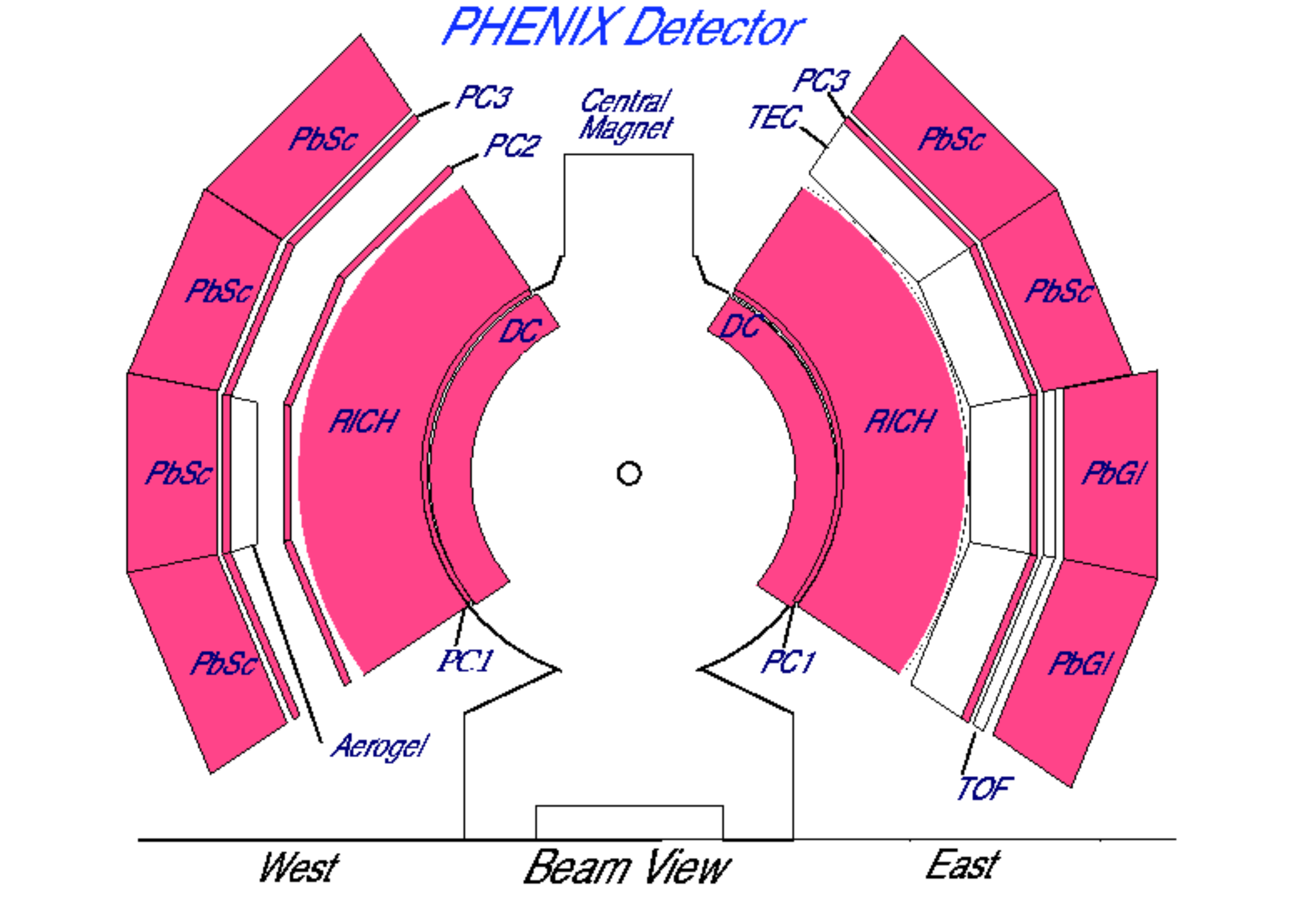}
    \hfill
    \includegraphics[keepaspectratio,width=0.46\textwidth]{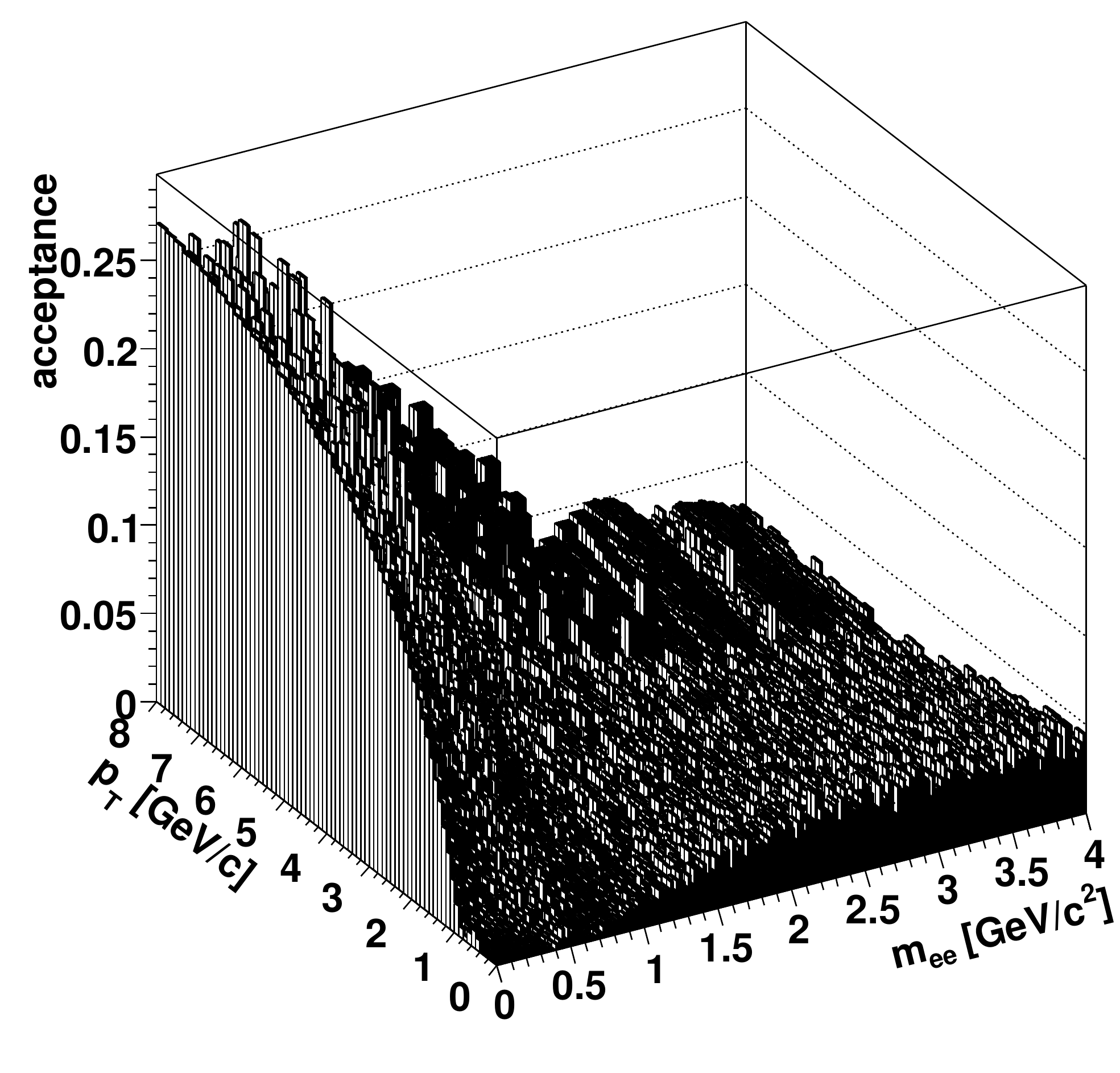}
  \end{center}
  \vspace*{8pt}
  \caption{Left: The PHENIX experiment at RHIC,\protect\cite{Adcox:2003zm}
    comprising drift chambers
    (DC), ring-imaging Cherenkov detectors (RICH), and electromagnetic
    calorimeters (PbSc,PbGl). Right: Acceptance
    for a thermal-like dilepton source.\protect\cite{Toia_priv}}
  \label{fig:PHENIX_setup}
\end{figure}

Vector-meson production and dielectron continua have been studied in p+p reactions
and ultra-relativistic heavy-ion collisions with the PHENIX detector at the
Relativistic Heavy-Ion Collider RHIC shown in Fig.~\ref{fig:PHENIX_setup}.
\begin{figure}[th]
  \begin{center}
    \includegraphics[keepaspectratio,width=0.7\textwidth]{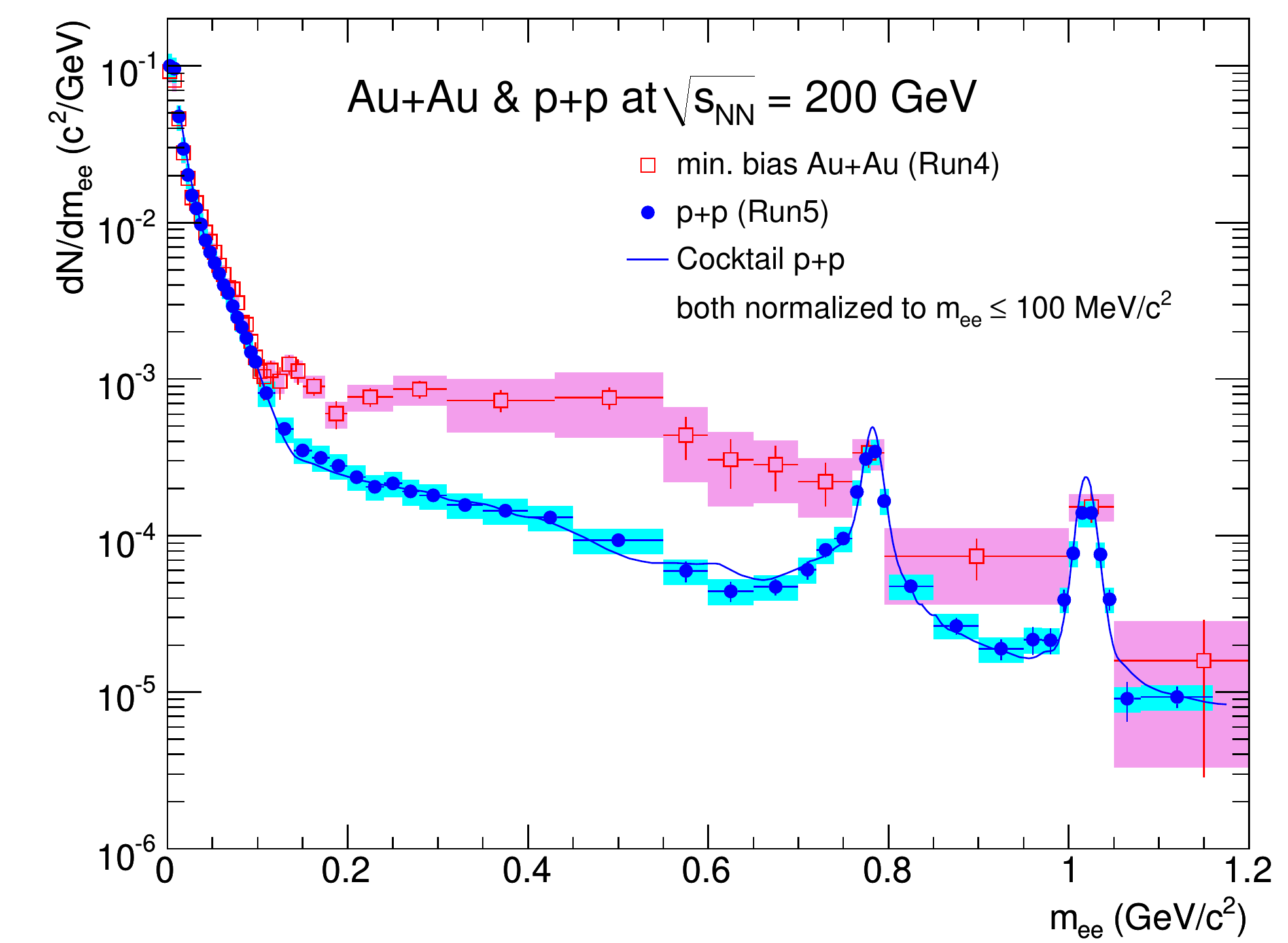}
  \end{center}
  \vspace*{8pt}
  \caption{Invariant $e^+ e^-$ mass distributions measured in p+p reactions and in Au+Au
    minimum-bias collisions at $\sqrt{s_{NN}}$ = 200 GeV. The combinatorial
    background has been subtracted using mixed event distributions after
    normalizing to the like-sign pair yields.\protect\cite{:2008asa,:2007xw,Toia:2008dj}}
  \label{fig:PHENIX_AuAu_pp}
\end{figure}
Electrons and positrons are detected in the two central arm spectrometers; the leptons are identified and  distinguished from other particles by ring imaging Cherenkov detectors and electromagnetic calorimeters; drift chambers measure their deflection angles in an axial magnetic field to determine their momenta. Fig.~\ref{fig:PHENIX_AuAu_pp} shows a comparison of the dielectron invariant-mass distributions measured in p+p reactions and Au+Au collisions at $\sqrt{s_{NN}}=200$ GeV. In the mass range between 150 and 750 MeV/$c^2$ a significant enhancement in the dilepton yield is observed for the heavy-ion reaction. However, all theoretical models which successfully explain the phenomena at SIS energies fail to reproduce this enhancement. Peaks from the decay of the vector mesons $\omega$ and $\phi$ are clearly resolved in the $p+p$ experiment. A detailed analysis of the dilepton continuum and possible in-medium modifications of the vector mesons is ongoing\cite{Milov:2008dd}.

\subsubsection{Dilepton emission in the 1 AGeV range}
\label{subsec:dil-1AGeV}

\begin{figure}[ht]
  \begin{center}
    \includegraphics[keepaspectratio,width=0.45\textwidth]{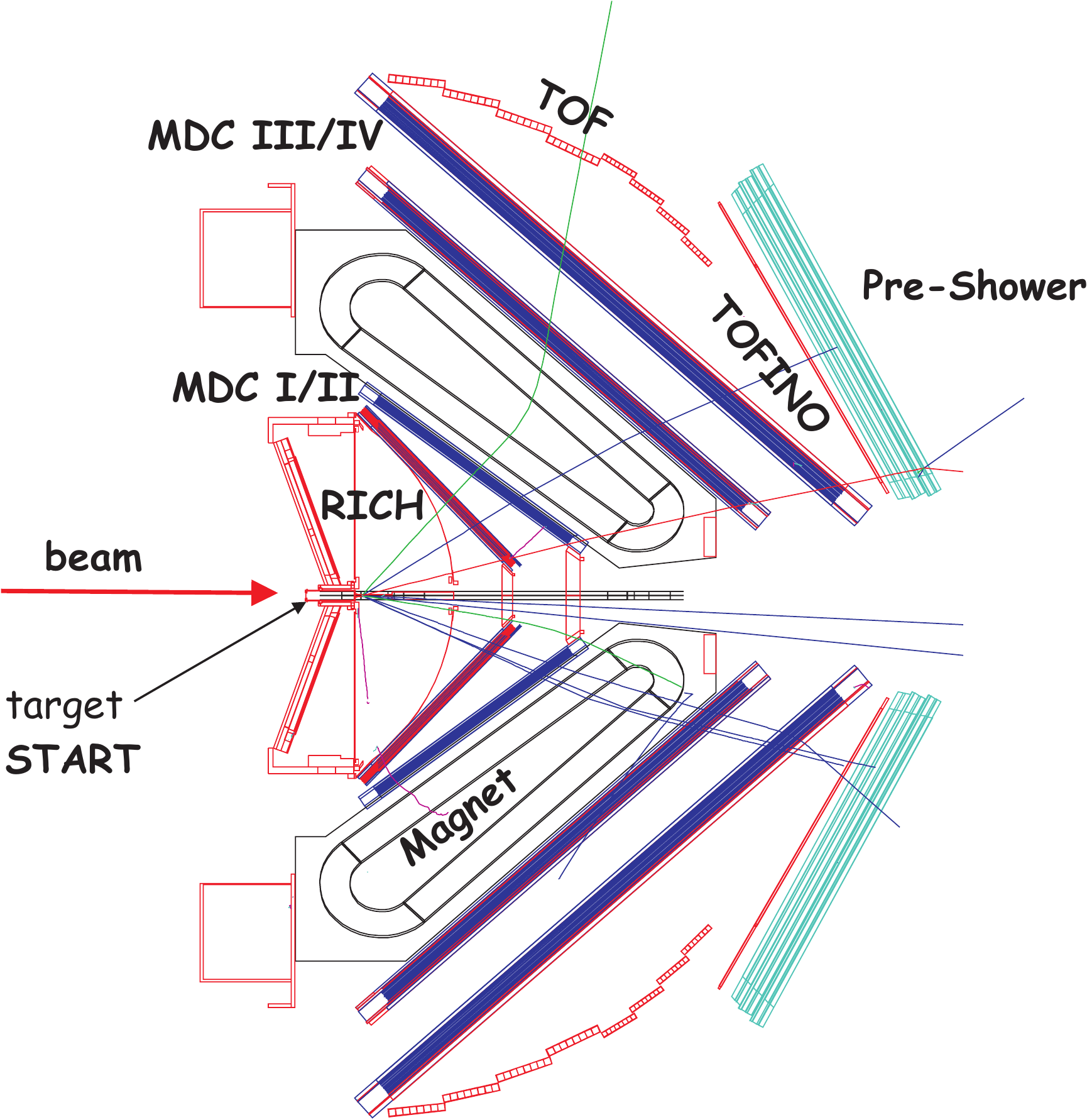}
    \hfill
    \includegraphics[keepaspectratio,width=0.54\textwidth]{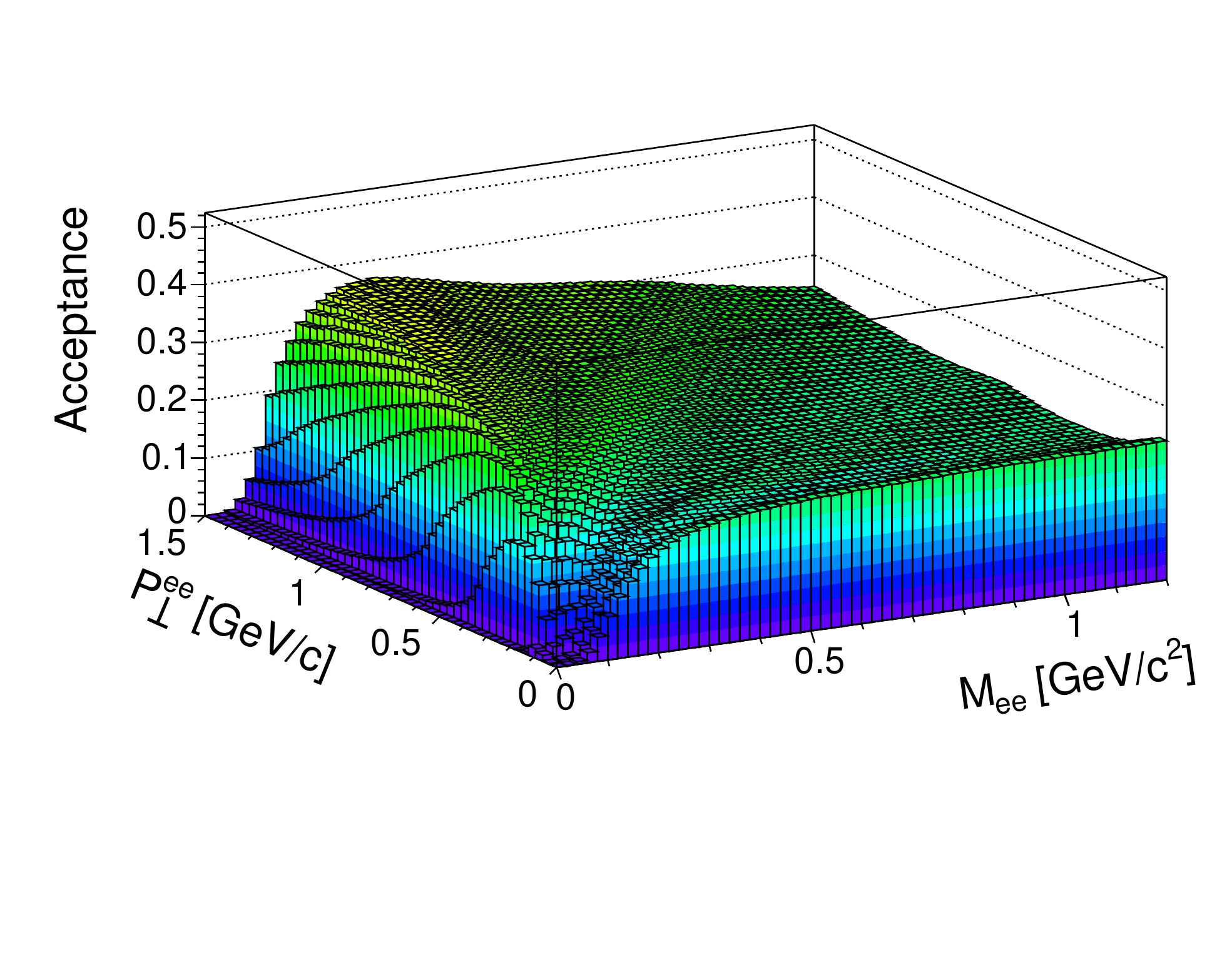}
  \end{center}
  \vspace*{8pt}
  \caption{Left: Cross section of the HADES experiment at
    GSI.\protect\cite{Agakishiev:2009am} It consists of 6 identical detection systems
    subtending polar angles between 18$^\circ$-85$^\circ$, Each sector comprises a ring-imaging
    Cherenkov detector (RICH), 4 planes of drift chambers (MDC I-IV), a
    time-of-flight detector (TOF/TOFINO) and a shower detector.
    Right: Acceptance
    for lepton pairs.\protect\cite{Agakishiev:2009am}}
  \label{fig:HADES_setup}
\end{figure}

The {\it DLS-puzzle} discussed in Sect.\ \ref{pioneering_experiments} was a
prime motivation for building the
{\bf H}igh {\bf A}cceptance {\bf D}i-{\bf E}lectron {\bf S}pectrometer (HADES)
which is shown in
Fig.\ \ref{fig:HADES_setup}.
As a second-generation experiment it was planned to cover a much larger solid
angle than the DLS spectrometer. The design of the detector is similiar to
that of the CLAS detector (cf.\ Fig.\ \ref{fig:CLAS_setup}).
It is azimuthally symmetric and consists of a
6-coil toroidal magnet centered on the beam axis and 6 identical
azimuthal detector sections mounted in between the coils.
Electron/pion discrimination is provided by the ring-imaging Cherenkov
detector and the shower detector combined with a matching of tracks in the
drift chambers. The resulting very flat acceptance for lepton pairs in the
relevant kinematic range is shown on the right hand side in
Fig.\ \ref{fig:HADES_setup}.

\begin{figure}[th]
  \begin{minipage}[c]{0.49\textwidth}
    \begin{center}
      \includegraphics[keepaspectratio,width=\textwidth]{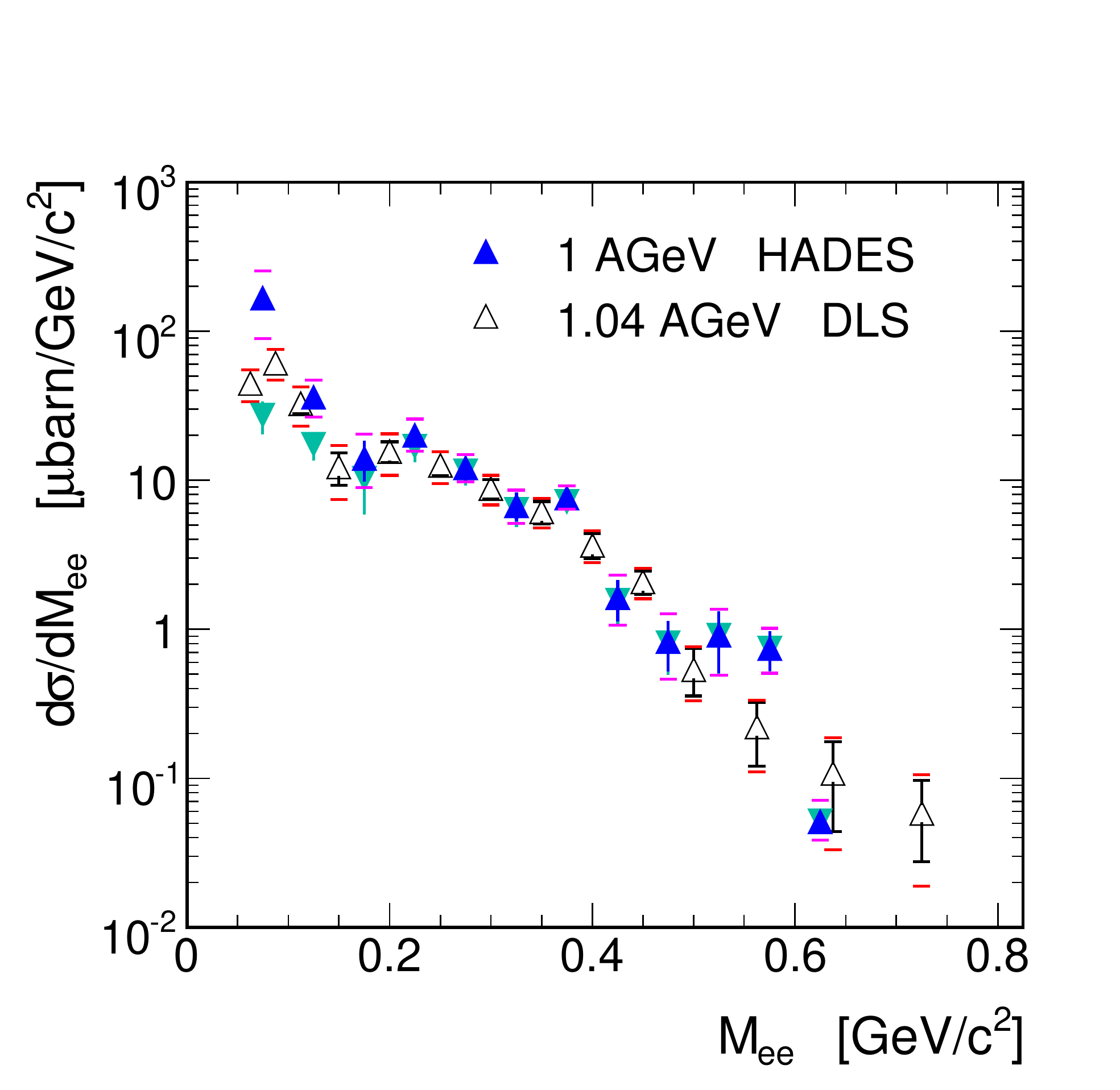}
    \end{center}
    \vspace*{8pt}
    \caption{Comparison within the DLS acceptance of the dilepton cross sections
      measured in C+C at 1 AGeV by HADES\protect\cite{Agakishiev:2007ts}
      and at 1.04 AGeV by DLS\protect\cite{Porter:1997rc}.}
    \label{fig:HADES_DLS_mass}
  \end{minipage}
  \hfill
  \begin{minipage}[c]{0.46\textwidth}
    \begin{center}
      \includegraphics[keepaspectratio,width=\textwidth,angle=-90]{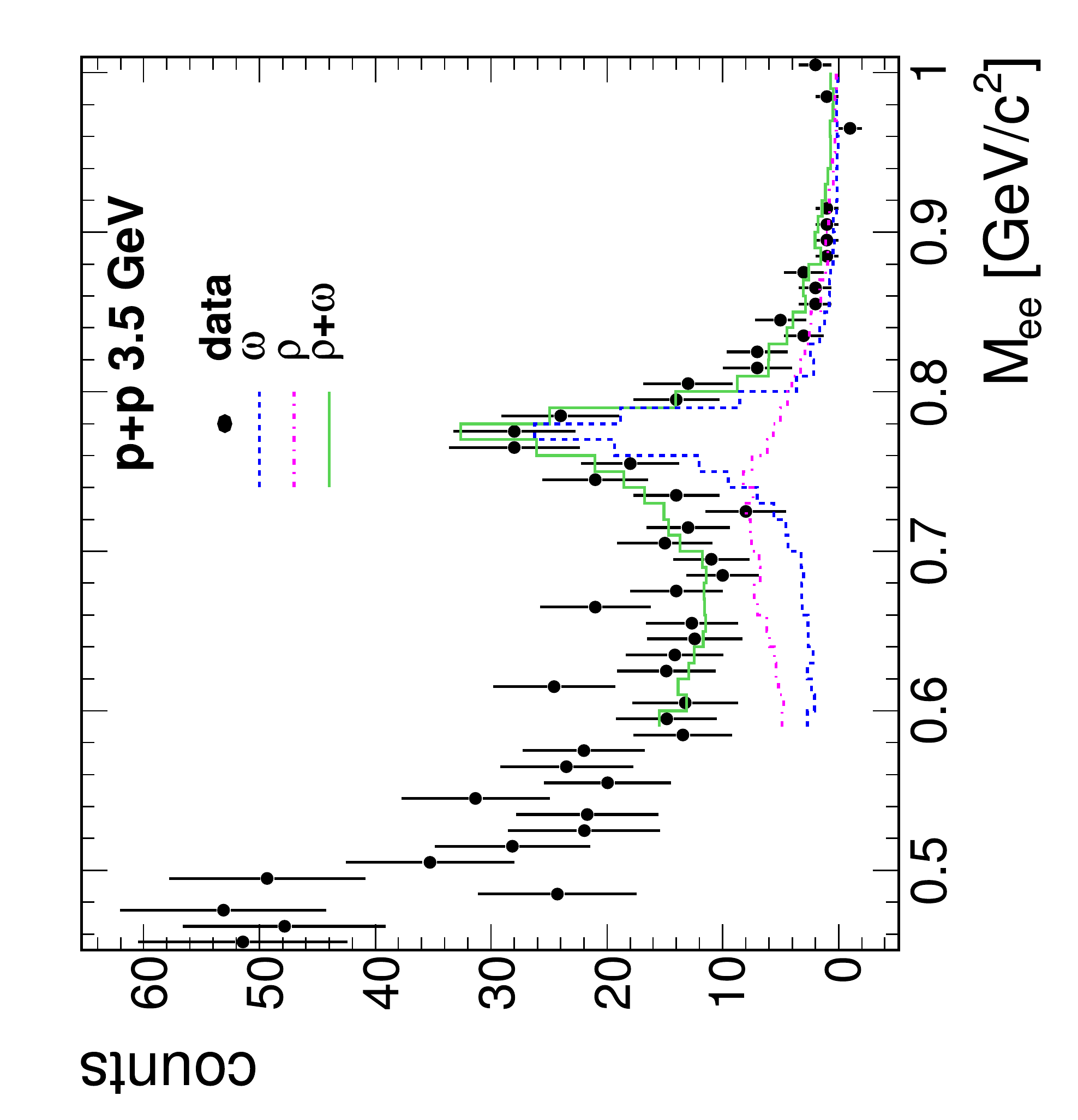}
    \end{center}
    \vspace*{8pt}
    \caption{Dielectron invariant-mass spectrum near the $\omega$ peak measured
      with the HADES detector in p+p reactions at 3.5 GeV. The
      spectrum is decomposed into an $\omega \rightarrow e^+ e^-$ signal and a
      $\rho \rightarrow e^+ e^-$ contribution.\protect\cite{{Agakishiev:2009am}}}
    \label{fig:HADES_omega_pp}
  \end{minipage}
\end{figure}

The most important result obtained so far with HADES is the confirmation of the
DLS data. This is demonstrated in Fig.\ \ref{fig:HADES_DLS_mass} which shows a
comparison of the pair-mass distributions in the DLS acceptance. Within
statistical and systematic uncertainties, the HADES and DLS data are in good
agreement.
Recent theoretical calculations\cite{Shyam:2003cn,Kaptari:2005qz}
indicate that dilepton production due to bremsstrahlung may have been
underestimated in previous studies. Indeed, it is found that implementing a larger bremsstrahlung contribution  may account for most of
the observed high dielectron yield.\cite{Bratkovskaya:2007jk}

Experiments at HADES will also contribute to the search for in-medium
modifications of vector mesons. Fig.\ \ref{fig:HADES_omega_pp} shows the
$\omega \rightarrow e^+ e^-$ signal on top of the
$\rho \rightarrow e^+ e^-$ contribution measured in proton-proton reactions at
3.5 GeV. This spectrum will be compared to the measurement of the p+Nb
reaction at 3.5 GeV which is currently being analyzed. Furthermore, in an
attempt to search for in-medium effects, $\omega$ production
has also been studied in Ar+KCl reaction at 1.756 AGeV. Corresponding results
are eagerly awaited.

\subsection{Summary of vector-meson experiments}

The experimental results on medium modifications of vector mesons reported in the previous sections
are compiled in Table \ref{tab:exp-vec}.
\begin{table}
\centering
\begin{footnotesize}
\begin{tabular}{|c|c|c|c|c|}
\hline
 experiment & momentum & $\rho$ & $\omega $ & $\phi$\\
 & acceptance & & &  \\
\hline
\hline
 KEK-E325 & & & & \\
 pA  & $p> 0.6$ GeV/c  & $\frac{\Delta m}{m} = - 9\%$ &
 $\frac{\Delta m}{m} = - 9\%$  & $\frac{\Delta m}{m} = - 3.4\%$\\[0.3em]
 12 GeV & & $\Delta \Gamma \approx 0$ & $\Delta \Gamma \approx 0$ & $\frac{\Gamma_{\phi}(\rho_0)}{\Gamma_{\phi}}= 3.6 $\\
\hline
 CLAS & & $\Delta m \approx 0$ & & \\
 $\gamma$A  & $p> 0.8$ GeV/c & $\Delta \Gamma \approx 70$ MeV & & \\
 0.6-3.8 GeV& & $(\rho \approx \rho_0/2)$ & & \\
\hline
 CBELSA & & & $ \Delta m \approx ? $ & \\
 /TAPS & & & $p_{\omega}< 0.5$ GeV/c & \\ \cline{4-4}
 $\gamma$A &$p>0$ MeV/c & & $ \Delta \Gamma(\rho_0) \approx 130$ MeV  &  \\
 0.9-2.2 GeV  & & & $\langle p_{\omega}\rangle = 1.1$ GeV/c & \\
\hline
 SPring8 & & & & \\
 $\gamma$A  & $p > 1.0$ GeV/c & & & $\Delta \Gamma(\rho_0) \approx 70$ MeV \\
 1.5-2.4 GeV& & & & $\langle p_{\phi}\rangle=1.8 $ GeV/c\\
\hline
CERES  & & broadening  &  &   \\
Pb+Au  & $p_t> 0$ GeV/c & favored over & & \\
158 AGeV &  & mass shift & & \\
\hline
NA60  & & $\Delta m \approx$ 0& & \\
In+In & $p_t>0$ GeV/c & strong & & \\
158 AGeV & &  broadening & & \\
\hline
\end{tabular}
\end{footnotesize}
\caption{Experimental results on in-medium modifications of $\rho$, $\omega$
  and $\phi$ mesons reported by different experiments. The reactions,
  incident energy ranges, and momentum acceptances of the detector systems are also given (adapted from Ref.\ \protect\refcite{Metag:2007zza}).}
\label{tab:exp-vec}
\end{table}
The table lists the reactions, momentum ranges and
results for $\rho$,$\omega$, and $\phi$ mesons obtained in the different
experiments discussed in this review. A fully consistent picture has so far not been achieved.
For the $\rho$ meson, the majority of experiments observes a broadening but no
mass shift in the nuclear medium. Conflicting results reported by the KEK
experiment will have to be confirmed. Earlier claims from one experiment\cite{Trnka:2005ey}
of a dropping $\omega$ mass could not be reproduced
in a re-analysis of the data. A broadening of the $\omega$ meson has been
observed in elementary nuclear reactions which is in line with a depletion of
the $\omega$ yield at low momenta observed in ultra-relativistic heavy-ion
reactions but which is again in conflict with the KEK-E325 result. For the
$\phi$ meson an in-medium mass shift and a broadening has been reported.

When comparing the experimental results it should be noted
that some detector systems have no or little acceptance for low meson momenta
for which the strongest medium modifications are expected.
In view of the existing inconsistencies further experiments are needed to clarify the
situation. Corresponding experiments are planned at CLAS, HADES, PHENIX, and
the JPARC facility.

\subsection{$2\pi$ correlations in nuclei: the $\sigma$ meson}
\label{subsec:exp-sigma}

In Sect.\ \ref{subsec:sigma-prod}
it was poined out that as a consequence of a conjectured partial restoration
of chiral symmetry in normal nuclear matter the mass of the $\sigma$ meson
would be expected to drop in the nuclear medium, combined with a reduction in
width for the $\sigma \rightarrow \pi^+\pi^-$ and $\sigma \rightarrow
\pi^0\pi^0$ decay channels as a result of the reduced phase space. Indeed, $\pi^+\pi^-$ production
experiments with pion beams on nuclear targets did show a striking shift of
the 2$\pi$ invariant-mass distributions towards lower
masses.\cite{Bonutti:1996ij,Bonutti:1999zz} Corresponding shifts were not observed in the
$\pi^+\pi^+$ channel, not accessible in the decay of the neutral $\sigma$
meson. These experiments, however, suffered from a very small acceptance and
correspondingly large corrections. A $\pi$ beam experiment using the Crystal
Ball as a large-acceptance detector system also showed a shift in the 2$\pi^0$
invariant-mass distributions with increasing nuclear mass
number.\cite{Starostin:2000cb}
It was pointed out, however, that in both experiments the pion initial-state interactions
led to a dominance of surface production thus complicating the interpretation of
the data.\cite{VicenteVacas:1999xx}

\begin{figure}[th]
  \begin{minipage}[c]{0.32\textwidth}
    \begin{center}
      \vspace*{-1cm}
      \includegraphics[keepaspectratio,width=\textwidth]{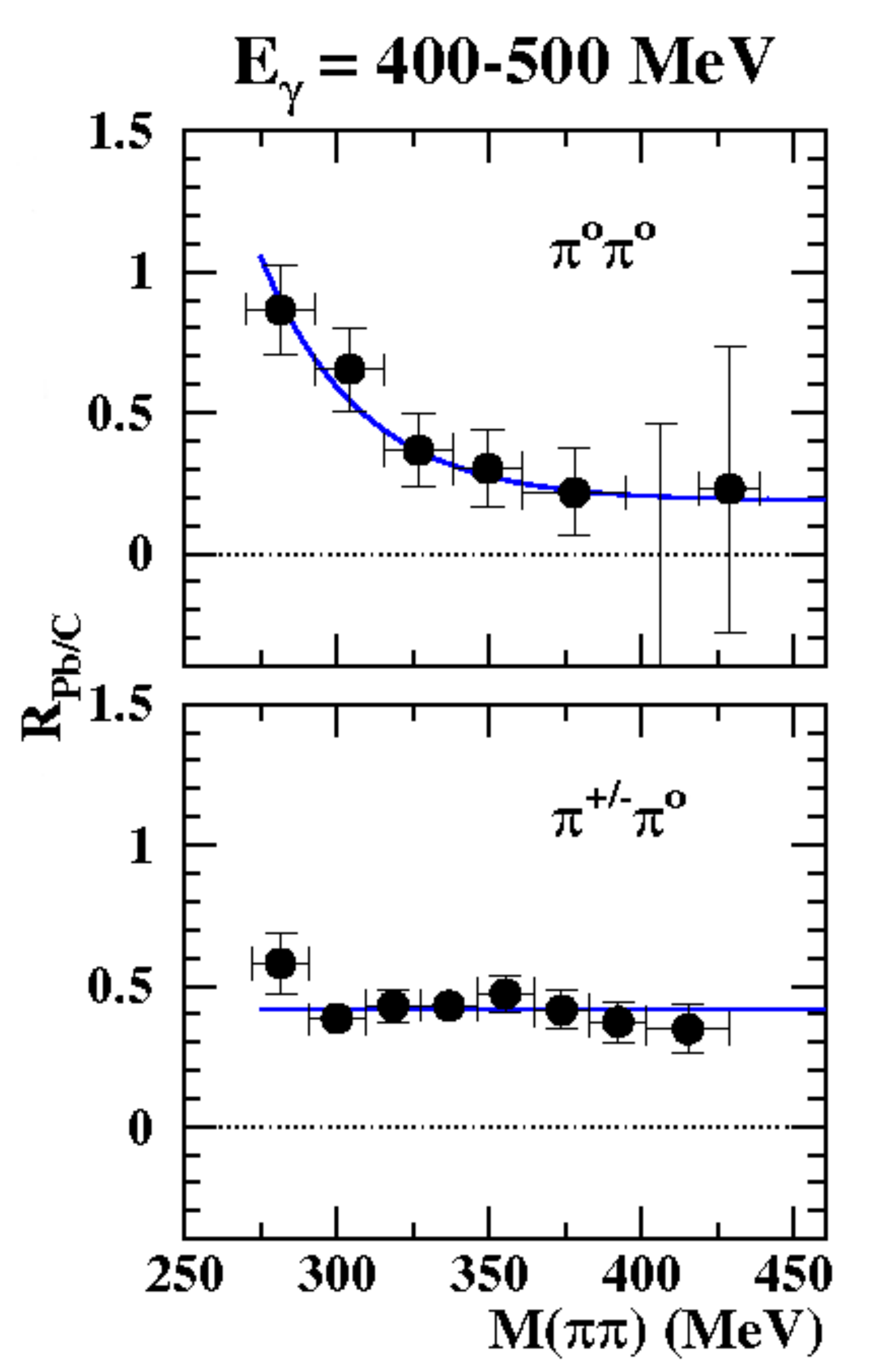}
    \end{center}
    \vspace*{8pt}
    \caption{Ratio of 2$\pi$ invariant-mass distributions for Pb and C targets in
      the 2$\pi^0$ and $\pi^{\pm}\pi^0$ channel.\protect\cite{Messchendorp:2001pa,Krusche}}
    \label{fig:pipi_Mesch}
  \end{minipage}
  \hfill
  \begin{minipage}[c]{0.64\textwidth}
    \begin{center}
      \includegraphics[keepaspectratio,width=\textwidth]{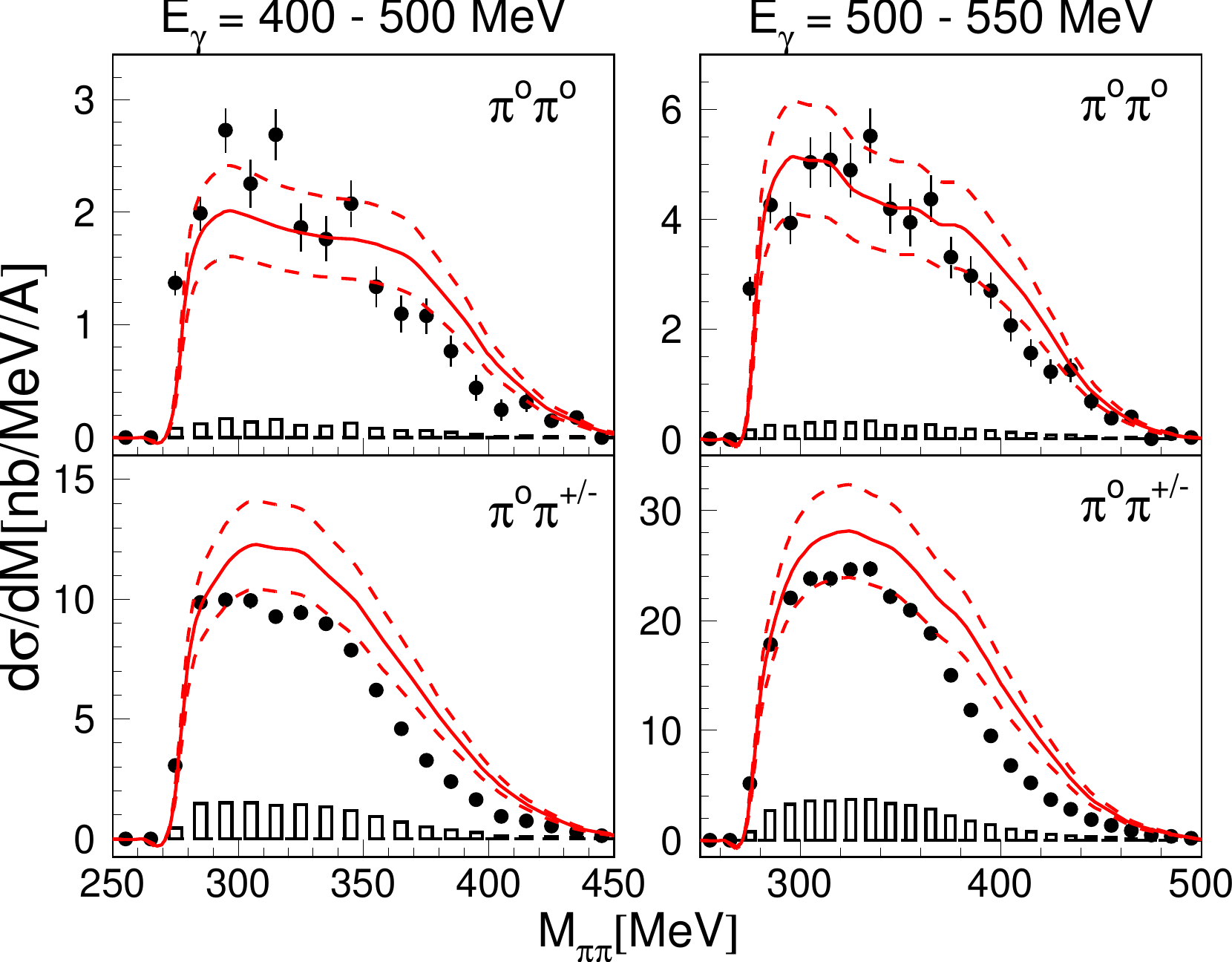}
    \end{center}
    \vspace*{8pt}
    \caption{Pion-pion invariant-mass distributions compared to results of the GiBUU
      calculation.\protect\cite{Muehlich:2006nn} The bars at the bottom represent systematic
      uncertainties of the data. The dashed lines represent the error band in
      the BUU calculations arising from uncertainties in the elementary 2$\pi$
      production cross sections (from Ref.\ \protect\refcite{Bloch:2007ka}).}
    \label{fig:Bloch}
  \end{minipage}
\end{figure}
A later photoproduction experiment,\cite{Messchendorp:2002au} where no initial-state
interactions are expected, did again show a striking shift of the 2$\pi^0$ mass
distributions with increasing nuclear target mass while the $\pi^{\pm}\pi^0$
channel, not possible in $\sigma$ decay, appeared to be less affected. This result is demonstrated in
Fig.~\ref{fig:pipi_Mesch} which shows the ratio of 2$\pi$ mass distributions
for Pb and C targets in the 2$\pi^0$ and $\pi^{\pm}\pi^0$ channels,
respectively. For the $\pi^{\pm}\pi^0$ channel this ratio is almost constant
as a function of the 2$\pi$ mass, indicating no change in the shape of the
invariant-mass spectra when going from C to Pb. In contrast, the ratio for the
2$\pi^0$ channel indicates an accumulation of strength near the 2$\pi$
threshold for the heavier Pb target.

While the theoretical models cited earlier\cite{Chiang:1997di,Rapp:1998fx,Chanfray:2004vb,Kaskulov:2005kr,Cabrera:2005wz} represent state-of-the-art treatments of pion-pion interactions and thus of the in-medium $\sigma$, they were much less sophisticated in their treatment of the actual observables, i.e.\ pions after final-state interactions, which were described in a simple eikonal approximation without allowance for charge transfer.
As discussed in Sect.\ \ref{sec:hadspec-obs} 
and investigated in transport
calculations,\cite{Muhlich:2004zj,Buss:2006vh} the different
behavior for both channels may, however, also arise from charge-exchange
reactions which shuffle strength from the dominant $\pi^{\pm}\pi^0$ to the
2$\pi^0$ channel. This reaction mechanism is associated with an energy loss
of the pions and consequently leads to a shift of strength towards lower
invariant masses which is more pronounced for the 2$\pi^0$ than for the $\pi^{\pm}\pi^0$
channel. Fig.~\ref{fig:Bloch} shows that these calculations provide a good
description of the data for a Ca target; remaining differences reflect
uncertainties in the elementary 2$\pi$ photoproduction cross sections on the
nucleon.

In order to establish differences in the 2$\pi^0$ and $\pi^{\pm}\pi^0$
channels beyond the charge-exchange effect data of much higher
statistics are needed. Corresponding data have been taken but are still under
analysis.\cite{Gregor} In addition, any interpretation of the data must involve state-of-the-art descriptions of the final-state interactions which tend to generate effects in the same direction as the predicted in-medium changes.

\section{Conclusions}
\label{sec:conclusion}

The study of in-medium modifications of hadrons is in the focus of many
ongoing theoretical and experimental investigations. This is a vibrant field which
has recently made a lot of progress as outlined in this review.
Eventually, these
investigations will provide a deeper and more quantitative understanding of
quantum chromodynamics in the non-perturbative regime.
In this review we have described the necessary steps one needs to go from
aspects of non-perturbative QCD via hadronic-model building towards actual observables.
Some of the links between these corner stones are much more involved than initially expected.

Starting from the symmetries of QCD we have first discussed the connection between non-perturbative properties of the vacuum and the spectral functions of hadrons; here QCD sum rules provide the link between the two worlds, the quark-gluon and the hadronic one.
We have discussed the changes of condensates in strongly interacting hadronic matter and have pointed out that they correspond to changes in ``standard'' hadronic models. There is thus a duality between these two descriptions of in-medium effects. We have also found, and we stress this again here, that QCD sum rules do not fix the properties of hadrons inside the nuclear medium, but they do provide important constraints for hadronic models.
Such models offer the only explicit access to hadronic in-medium properties and, in particular, their in-medium self-energies; the latter are often summarized in spectral functions. As for the well-known example of pion-nucleon-Delta interactions in nuclei, which have been studied since the 70's, also for vector mesons the coupling to nucleon resonances plays a major role. Foremost essential is the strength of the meson-nucleon-resonance coupling,
which determines the importance of resonance loops for the in-medium self-energies; this is where resonance physics on nucleons, preferentially studied at electron machines, meets heavy-ion physics with nuclei, which studies the dense and hot environment so interesting for in-medium effects.

We have pointed out that there is a clear consensus now that the $\rho$-meson properties should indeed change in a medium. This change is, however, subtle: the pole mass is not expected to shift significantly and the shape of the spectral function is strongly momentum dependent. Only for momenta (in the nuclear restframe) between 0 and about 800 MeV/c
the spectral function of the $\rho$ meson is predicted to become significantly skewed towards lower masses. Unfortunately this interesting regime has so far not yet been studied in nuclear reactions with elementary probes.
For momenta above 1 GeV/c, on the other hand, the spectral function resembles that of a free $\rho$ meson with some broadening and again no mass shift. For the $\omega$ meson, the situation is more complicated since there are no distinct resonances known at present that
couple significantly to the $\omega N$ decay channel. Here, the basic $\omega N$ scattering problem still has to be investigated and resonance couplings have to be uniquely identified. Calculations aiming at such a description again yield hardly any mass change, but predict a considerable broadening of the $\omega$ in cold nuclear matter. A similar situation prevails for the $\phi$ meson. Finally, we have discussed theoretical expectations for the in-medium changes of the $\sigma$ meson, which is the chiral partner of the pion and is thus particularly sensitive to  precursor phenomena of chiral symmetry restoration. For this meson, theories predict a significant shift and narrowing of the spectral function towards lower masses; this is a consequence of the closing of the $2\pi$ decay channel when $\sigma$ and $\pi$ spectral functions approach each other, caused by (partial) chiral symmetry restoration in nuclei.

Linking these theoretical in-medium properties with actual observables is a challenging task and involves some assumptions. Foremost one has to be aware of the fact that all hadronic in-medium properties are being calculated for mesons embedded in hadronic matter in equilibrium, i.e.\ infinitely long-lived without spatial boundaries and with equilibrated momentum distributions. This is clearly an idealization which, however, is probably more justified for the case of cold nuclear ground state targets than for heavy-ion collisions where the reaction proceeds through non-equilibrium and equilibrium phases, the latter possibly of different nature, and where all the experimental signals involve an integration in time over the collision history. In any case, profound transport calculations are needed that allow to link the fundamental hadronic properties with observables, in particular when the final decay products of the vector meson under study are hadrons. But even for dileptons in the final state a state-of-the-art coupled-channel description is necessary since the radiating mesons may be absorbed or recreated on their way through the nuclear medium.
As outlined in this review, in-medium properties of mesons are strongly
linked to the properties of baryon resonances due to the excitation of the latter in
collisions of the considered meson with the nucleons of the medium.
The future of the field is therefore closely linked to a better theoretical
understanding and experimental determination of baryon resonance parameters which
reflect their underlying quark-gluon structure.
Theoretical and experimental advancements in resonance spectroscopy are thus
essential for progress in the field of in-medium physics.

On the experimental side the following picture has emerged:
For the $\rho$, $\omega$, and $\phi$ meson a broadening in the nuclear medium
has been reported by almost all experiments in heavy-ion collisions as well as
in nuclear reactions with elementary probes.
This is in line with theoretical predictions. For the
$\rho$ meson even quantitative agreement between theory and experiment has
been achieved while for the $\omega$ and $\phi$ meson the broadening deduced from
transparency-ratio measurements is a factor 2-3 larger than theoretically
expected. With regard to possible in-medium mass shifts the experimental
situation is less clear. The majority of experiments does not find evidence for mass changes in the medium.
For the $\rho$ meson the centroid of the mass
distribution is reported by two experiments to remain at the pole mass of the
free $\rho$ meson, while in one experiment a drop in mass is claimed.
The search for in-medium mass shifts in elementary nuclear reactions turns
out to be more complicated than initially thought because of the observed
large in-medium broadening. While the $\rho$ meson is only broadened by a
factor of about 1.5-3, for the $\omega$ and $\phi$ meson this factor is
found to be in the range of 6 - 15. As a consequence, the branching ratio for
in-medium decays into the channels of interest is reduced
accordingly and makes these experiments less sensitive. Experiments with much
higher statistics will be needed to obtain reliable results. Furthermore,
the acceptance of some detector systems for low-momentum mesons will have to
be increased because the dominant in-medium modifications are expected at
recoil momenta of less than 1 GeV/c. Initiatives for corresponding new
experiments have been taken and experimental results are eagerly awaited.

\section{Acknowledgments}

We gratefully acknowledge many stimulating discussions and very helpful
comments on the manuscript from Sanja Damjanovic, Hideto En'yo, Bengt Friman,
Tetsuo Hatsuda, Hendrik van Hees, Burkhard K\"ampfer, Eulogio Oset, Ralf Rapp,
Piotr Salabura and Hans Specht.
We also thank Chaden Djalali, Megumi Naruki and Alberica Toia
for providing figures for this review.
This work has been supported by BMBF, DFG (SFB/TR16) and GSI.


\begin{thebibliography}{999}

\bibitem{Pisarski:1981mq}
  R.~D.~Pisarski,
  Phys.\ Lett.\  B {\bf 110} (1982) 155.

\bibitem{Brown:1991kk}
  G.~E.~Brown and M.~Rho,
  Phys.\ Rev.\ Lett.\  {\bf 66} (1991) 2720.

\bibitem{Hatsuda:1991ez}
  T.~Hatsuda and S.~H.~Lee,
  Phys.\ Rev.\  C {\bf 46} (1992) 34.

\bibitem{Bernard:1988db}
  V.~Bernard and U.~G.~Meissner,
  Nucl.\ Phys.\  A {\bf 489} (1988) 647.

\bibitem{Leutwyler:1990uq}
  H.~Leutwyler and A.~V.~Smilga,
  Nucl.\ Phys.\  B {\bf 342} (1990) 302.

\bibitem{Bugg:1974cz}
  D.~V.~Bugg,
  Nucl.\ Phys.\  B {\bf 88} (1975) 381.

\bibitem{Hayano:2008vn}
  R.~S.~Hayano and T.~Hatsuda,
  arXiv:0812.1702 [nucl-ex].

\bibitem{Cassing:1999es}
  W.~Cassing and E.~L.~Bratkovskaya,
  Phys.\ Rept.\  {\bf 308} (1999) 65.

\bibitem{Rapp:1999ej}
  R.~Rapp and J.~Wambach,
  Adv.\ Nucl.\ Phys.\  {\bf 25} (2000) 1
  [arXiv:hep-ph/9909229].

\bibitem{Rapp:2009yu}
  R.~Rapp, J.~Wambach and H.~van Hees,
  arXiv:0901.3289 [hep-ph].

\bibitem{Oset:1981ih}
  E.~Oset, H.~Toki and W.~Weise,
  Phys.\ Rept.\  {\bf 83} (1982) 281.

\bibitem{Peskin:1995ev}
  M.~E.~Peskin and D.~V.~Schroeder,
  ``An Introduction To Quantum Field Theory,''
{\it  Reading, USA: Addison-Wesley (1995).}

\bibitem{'tHooft:1977hy}
  G.~'t Hooft,
  Nucl.\ Phys.\  B {\bf 138} (1978) 1.

\bibitem{'tHooft:1979uj}
  G.~'t Hooft,
  Nucl.\ Phys.\  B {\bf 153} (1979) 141.

\bibitem{Polyakov:1978vu}
  A.~M.~Polyakov,
  Phys.\ Lett.\  B {\bf 72} (1978) 477.

\bibitem{Matsubara:1955ws}
  T.~Matsubara,
  Prog.\ Theor.\ Phys.\  {\bf 14} (1955) 351.

\bibitem{Karsch:2001cy}
  F.~Karsch,
  Lect.\ Notes Phys.\  {\bf 583} (2002) 209
  [arXiv:hep-lat/0106019].

\bibitem{Holland:2000uj}
  K.~Holland and U.~J.~Wiese,
  arXiv:hep-ph/0011193.

\bibitem{Scherer:2002tk}
  S.~Scherer,
  Adv.\ Nucl.\ Phys.\  {\bf 27} (2003) 277
  [arXiv:hep-ph/0210398].
  
\bibitem{MoselBuch}
  U.~Mosel,
  ``Fields, Symmetries, and Quarks'',
  {\it Berlin: Springer (1999)}

\bibitem{Amsler:2008zzb}
  C.~Amsler {\it et al.}  [Particle Data Group],
  Phys.\ Lett.\  B {\bf 667} (2008) 1.

\bibitem{GellMann:1968rz}
  M.~Gell-Mann, R.~J.~Oakes and B.~Renner,
  Phys.\ Rev.\  {\bf 175} (1968) 2195.


\bibitem{Aoki:2006we}
  Y.~Aoki, G.~Endrodi, Z.~Fodor, S.~D.~Katz and K.~K.~Szabo,
  Nature {\bf 443} (2006) 675
  [arXiv:hep-lat/0611014].

\bibitem{Leupold:2008ne}
  S.~Leupold, M.~F.~M.~Lutz and M.~Wagner,
  arXiv:0811.2398 [nucl-th].

\bibitem{Rapp:1997zu}
  R.~Rapp, T.~Schafer, E.~V.~Shuryak and M.~Velkovsky,
  Phys.\ Rev.\ Lett.\  {\bf 81} (1998) 53
  [arXiv:hep-ph/9711396].

\bibitem{Alford:1997zt}
  M.~G.~Alford, K.~Rajagopal and F.~Wilczek,
  Phys.\ Lett.\  B {\bf 422} (1998) 247
  [arXiv:hep-ph/9711395].

\bibitem{Alford:1998mk}
  M.~G.~Alford, K.~Rajagopal and F.~Wilczek,
  Nucl.\ Phys.\  B {\bf 537} (1999) 443
  [arXiv:hep-ph/9804403].

\bibitem{Schafer:1998ef}
  T.~Schafer and F.~Wilczek,
  Phys.\ Rev.\ Lett.\  {\bf 82} (1999) 3956
  [arXiv:hep-ph/9811473].

\bibitem{McLerran:2007qj}
  L.~McLerran and R.~D.~Pisarski,
  Nucl.\ Phys.\  A {\bf 796} (2007) 83
  [arXiv:0706.2191 [hep-ph]].

\bibitem{Gerber:1988tt}
  P.~Gerber and H.~Leutwyler,
  Nucl.\ Phys.\  B {\bf 321} (1989) 387.

\bibitem{Meissner:2001gz}
  U.~G.~Meissner, J.~A.~Oller and A.~Wirzba,
  Annals Phys.\  {\bf 297} (2002) 27
  [arXiv:nucl-th/0109026].

\bibitem{Gasser:1986vb}
  J.~Gasser and H.~Leutwyler,
  Phys.\ Lett.\  B {\bf 184} (1987) 83.

\bibitem{Gasser:1987zq}
  J.~Gasser and H.~Leutwyler,
  Nucl.\ Phys.\  B {\bf 307} (1988) 763.

\bibitem{Drukarev:1991fs}
  E.~G.~Drukarev and E.~M.~Levin,
  Prog.\ Part.\ Nucl.\ Phys.\  {\bf 27} (1991) 77.

\bibitem{Gasser:1990ce}
  J.~Gasser, H.~Leutwyler and M.~E.~Sainio,
  Phys.\ Lett.\  B {\bf 253} (1991) 252.

\bibitem{Kwon:2008vq}
  Y.~Kwon, M.~Procura and W.~Weise,
  Phys.\ Rev.\  C {\bf 78} (2008) 055203
  [arXiv:0803.3262 [nucl-th]].

\bibitem{Danielewicz:1982kk}
  P.~Danielewicz,
  Annals Phys.\  {\bf 152} (1984) 239.

\bibitem{Ecker:1989yg}
  G.~Ecker, J.~Gasser, H.~Leutwyler, A.~Pich and E.~de Rafael,
  Phys.\ Lett.\  B {\bf 223} (1989) 425.

\bibitem{Fearing:1999fw}
  H.~W.~Fearing and S.~Scherer,
  Phys.\ Rev.\  C {\bf 62} (2000) 034003
  [arXiv:nucl-th/9909076].

\bibitem{Friman:1997tc}
  B.~Friman and H.~J.~Pirner,
  Nucl.\ Phys.\  A {\bf 617} (1997) 496
  [arXiv:nucl-th/9701016].

\bibitem{Rapp:1997fs}
  R.~Rapp, G.~Chanfray and J.~Wambach,
  Nucl.\ Phys.\  A {\bf 617} (1997) 472
  [arXiv:hep-ph/9702210].

\bibitem{Lutz:2001mi}
  M.~F.~M.~Lutz, G.~Wolf and B.~Friman,
  Nucl.\ Phys.\  A {\bf 706} (2002) 431
  [Erratum-ibid.\  A {\bf 765} (2006) 431]
  [arXiv:nucl-th/0112052].

\bibitem{Post:2003hu}
  M.~Post, S.~Leupold and U.~Mosel,
  Nucl.\ Phys.\  A {\bf 741} (2004) 81
  [arXiv:nucl-th/0309085].

\bibitem{Riek:2004kx}
  F.~Riek and J.~Knoll,
  Nucl.\ Phys.\  A {\bf 740} (2004) 287
  [arXiv:nucl-th/0402090].

\bibitem{Muehlich:2006nn}
  P.~Muehlich, V.~Shklyar, S.~Leupold, U.~Mosel and M.~Post,
  Nucl.\ Phys.\  A {\bf 780} (2006) 187
  [arXiv:nucl-th/0607061].

\bibitem{vanHees:2007th}
  H.~van Hees and R.~Rapp,
  Nucl.\ Phys.\  A {\bf 806} (2008) 339
  [arXiv:0711.3444 [hep-ph]].

\bibitem{Shuryak:1993kg}
  E.~V.~Shuryak,
  Rev.\ Mod.\ Phys.\  {\bf 65} (1993) 1.

\bibitem{Shifman:1978bx}
  M.~A.~Shifman, A.~I.~Vainshtein and V.~I.~Zakharov,
  Nucl.\ Phys.\  B {\bf 147} (1979) 385.

\bibitem{Shifman:1978by}
  M.~A.~Shifman, A.~I.~Vainshtein and V.~I.~Zakharov,
  Nucl.\ Phys.\  B {\bf 147} (1979) 448.

\bibitem{Asakawa:2000tr}
  M.~Asakawa, T.~Hatsuda and Y.~Nakahara,
  Prog.\ Part.\ Nucl.\ Phys.\  {\bf 46} (2001) 459
  [arXiv:hep-lat/0011040].

\bibitem{Hatsuda:2005nw}
  T.~Hatsuda,
  Int.\ J.\ Mod.\ Phys.\  A {\bf 21} (2006) 688
  [arXiv:hep-ph/0509306].

\bibitem{Durr:2008zz}
  S.~Durr {\it et al.},
  Science {\bf 322} (2008) 1224.

\bibitem{Schael:2005am}
  S.~Schael {\it et al.}  [ALEPH Collaboration],
  Phys.\ Rept.\  {\bf 421} (2005) 191
  [arXiv:hep-ex/0506072].

\bibitem{Caldi:1975tx}
  D.~G.~Caldi and H.~Pagels,
  Phys.\ Rev.\  D {\bf 14} (1976) 809.

\bibitem{Dominguez:1992dw}
  C.~A.~Dominguez, M.~Loewe and J.~C.~Rojas,
  Z.\ Phys.\  C {\bf 59} (1993) 63.

\bibitem{Nambu:1961tp}
  Y.~Nambu and G.~Jona-Lasinio,
  Phys.\ Rev.\  {\bf 122} (1961) 345.

\bibitem{Bochkarev:1985ex}
  A.~I.~Bochkarev and M.~E.~Shaposhnikov,
  Nucl.\ Phys.\  B {\bf 268} (1986) 220.

\bibitem{Wilson:1969zs}
  K.~G.~Wilson,
  Phys.\ Rev.\  {\bf 179} (1969) 1499.

\bibitem{Asakawa:1993pq}
  M.~Asakawa and C.~M.~Ko,
  Phys.\ Rev.\  C {\bf 48} (1993) 526.

\bibitem{Leupold:1997dg}
  S.~Leupold, W.~Peters and U.~Mosel,
  Nucl.\ Phys.\  A {\bf 628} (1998) 311
  [arXiv:nucl-th/9708016].

\bibitem{Leupold:2003zb}
  S.~Leupold,
  Nucl.\ Phys.\  A {\bf 743} (2004) 283
  [arXiv:hep-ph/0303020].

\bibitem{Hatsuda:1992bv}
  T.~Hatsuda, Y.~Koike and S.~H.~Lee,
  Nucl.\ Phys.\  B {\bf 394} (1993) 221.

\bibitem{Leupold:2005eq}
  S.~Leupold,
  Phys.\ Lett.\  B {\bf 616} (2005) 203
  [arXiv:hep-ph/0502061].

\bibitem{Dey:1990ba}
  M.~Dey, V.~L.~Eletsky and B.~L.~Ioffe,
  Phys.\ Lett.\  B {\bf 252} (1990) 620.

\bibitem{Steele:1996su}
  J.~V.~Steele, H.~Yamagishi and I.~Zahed,
  Phys.\ Lett.\  B {\bf 384} (1996) 255
  [arXiv:hep-ph/9603290].

\bibitem{Urban:2001uv}
  M.~Urban, M.~Buballa and J.~Wambach,
  Phys.\ Rev.\ Lett.\  {\bf 88} (2002) 042002
  [arXiv:nucl-th/0110005].

\bibitem{Hatsuda:1995dy}
  T.~Hatsuda, S.~H.~Lee and H.~Shiomi,
  Phys.\ Rev.\  C {\bf 52} (1995) 3364
  [arXiv:nucl-th/9505005].

\bibitem{Leupold:1998bt}
  S.~Leupold and U.~Mosel,
  Phys.\ Rev.\  C {\bf 58} (1998) 2939
  [arXiv:nucl-th/9805024].

\bibitem{Eletsky:1996jg}
  V.~L.~Eletsky and B.~L.~Ioffe,
  Phys.\ Rev.\ Lett.\  {\bf 78} (1997) 1010
  [arXiv:hep-ph/9609229].

\bibitem{Hatsuda:1997gj}
  T.~Hatsuda and S.~H.~Lee,
  arXiv:nucl-th/9703022.

\bibitem{Eletsky:1997rz}
  V.~L.~Eletsky and B.~L.~Ioffe,
  arXiv:hep-ph/9704236.

\bibitem{Leupold:2001hj}
  S.~Leupold,
  Phys.\ Rev.\  C {\bf 64} (2001) 015202
  [arXiv:nucl-th/0101013].

\bibitem{Leupold:2004gh}
  S.~Leupold and M.~Post,
  Nucl.\ Phys.\  A {\bf 747} (2005) 425
  [arXiv:nucl-th/0402048].

\bibitem{Steinmueller:2006id}
  B.~Steinmueller and S.~Leupold,
  Nucl.\ Phys.\  A {\bf 778} (2006) 195
  [arXiv:hep-ph/0604054].

\bibitem{Klingl:1997kf}
  F.~Klingl, N.~Kaiser and W.~Weise,
  Nucl.\ Phys.\  A {\bf 624} (1997) 527
  [arXiv:hep-ph/9704398].

\bibitem{DuttMazumder:2000ys}
  A.~K.~Dutt-Mazumder, R.~Hofmann and M.~Pospelov,
  Phys.\ Rev.\  C {\bf 63} (2001) 015204
  [arXiv:hep-ph/0005100].

\bibitem{Thomas:2005dc}
  R.~Thomas, S.~Zschocke and B.~Kampfer,
  Phys.\ Rev.\ Lett.\  {\bf 95} (2005) 232301
  [arXiv:hep-ph/0510156].

\bibitem{Hilger:2008jg}
  T.~Hilger, R.~Thomas and B.~Kampfer,
  Phys.\ Rev.\  C {\bf 79} (2009) 025202
  [arXiv:0809.4996 [nucl-th]].

\bibitem{Birse:1996qp}
  M.~C.~Birse and B.~Krippa,
  Phys.\ Lett.\  B {\bf 381} (1996) 397
  [arXiv:hep-ph/9603387].

\bibitem{Thomas:2007gx}
  R.~Thomas, T.~Hilger and B.~Kampfer,
  Nucl.\ Phys.\  A {\bf 795} (2007) 19
  [arXiv:0704.3004 [hep-ph]].

\bibitem{Asakawa:1992ht}
  M.~Asakawa, C.~M.~Ko, P.~Levai and X.~J.~Qiu,
  Phys.\ Rev.\  C {\bf 46}, 1159 (1992).

\bibitem{Herrmann:1992kn}
  M.~Herrmann, B.~L.~Friman and W.~Noerenberg,
  Nucl.\ Phys.\  A {\bf 545} (1992) 267C.

\bibitem{Rapp:1995zy}
  R.~Rapp, G.~Chanfray and J.~Wambach,
  Phys.\ Rev.\ Lett.\  {\bf 76} (1996) 368
  [arXiv:hep-ph/9508353].

\bibitem{Rapp:1997ei}
  R.~Rapp, M.~Urban, M.~Buballa and J.~Wambach,
  Phys.\ Lett.\  B {\bf 417} (1998) 1
  [arXiv:nucl-th/9709008].

\bibitem{Klingl:1997tm}
  F.~Klingl, T.~Waas and W.~Weise,
  Phys.\ Lett.\  B {\bf 431} (1998) 254
  [arXiv:hep-ph/9709210].

\bibitem{Klingl:1996ps}
  F.~Klingl and W.~Weise,
  Nucl.\ Phys.\  A {\bf 606} (1996) 329.

\bibitem{Eichstaedt:2007zp}
  F.~Eichstaedt, S.~Leupold, U.~Mosel and P.~Muehlich,
  Prog.\ Theor.\ Phys.\ Suppl.\  {\bf 168} (2007) 495
  [arXiv:0704.0154 [nucl-th]].

\bibitem{Peters:1997va}
  W.~Peters, M.~Post, H.~Lenske, S.~Leupold and U.~Mosel,
  Nucl.\ Phys.\  A {\bf 632} (1998) 109
  [arXiv:nucl-th/9708004].

\bibitem{Post:2000qi}
  M.~Post, S.~Leupold and U.~Mosel,
  Nucl.\ Phys.\  A {\bf 689} (2001) 753
  [arXiv:nucl-th/0008027].

\bibitem{Post:2001am}
  M.~Post and U.~Mosel,
  Nucl.\ Phys.\  A {\bf 699} (2002) 169
  [arXiv:nucl-th/0108017].

\bibitem{Cabrera:2000dx}
  D.~Cabrera, E.~Oset and M.~J.~Vicente Vacas,
  Nucl.\ Phys.\  A {\bf 705} (2002) 90
  [arXiv:nucl-th/0011037].

\bibitem{Manley:1992yb}
  D.~M.~Manley and E.~M.~Saleski,
  Phys.\ Rev.\  D {\bf 45} (1992) 4002.

\bibitem{Zabrodin:1999sq}
  A.~Zabrodin {\it et al.},
  Phys.\ Rev.\  C {\bf 60} (1999) 055201.

\bibitem{Langgartner:2001sg}
  W.~Langgartner {\it et al.},
  Phys.\ Rev.\ Lett.\  {\bf 87} (2001) 052001.

\bibitem{Lehr:2001ju}
  J.~Lehr and U.~Mosel,
  Phys.\ Rev.\  C {\bf 64} (2001) 042202
  [arXiv:nucl-th/0105054].

\bibitem{Bianchi:1995vb}
  N.~Bianchi {\it et al.},
  Phys.\ Rev.\  C {\bf 54} (1996) 1688.

\bibitem{Rapp:1999us}
  R.~Rapp and J.~Wambach,
  Eur.\ Phys.\ J.\  A {\bf 6} (1999) 415
  [arXiv:hep-ph/9907502].

\bibitem{Shklyar:2004ba}
  V.~Shklyar, H.~Lenske, U.~Mosel and G.~Penner,
  Phys.\ Rev.\  C {\bf 71} (2005) 055206
  [Erratum-ibid.\  C {\bf 72} (2005) 019903]
  [arXiv:nucl-th/0412029].

\bibitem{Barth:2003kv}
  J.~Barth {\it et al.},
  Eur.\ Phys.\ J.\  A {\bf 18} (2003) 117.

\bibitem{:2008gs}
  F.~Klein {\it et al.}  [The CBELSA/TAPS Collaboration],
  Phys.\ Rev.\  D {\bf 78} (2008) 117101
  [arXiv:0807.0594 [hep-ex]].

\bibitem{Lutz:1999jn}
  M.~Lutz, B.~Friman and G.~Wolf,
  Nucl.\ Phys.\  A {\bf 661} (1999) 526.

\bibitem{Wolf:2004hr}
  G.~Wolf, M.~F.~M.~Lutz and B.~Friman,
  Acta Phys.\ Hung.\  A {\bf 19} (2004) 301.

\bibitem{Penner:2002ma}
  G.~Penner and U.~Mosel,
  Phys.\ Rev.\  C {\bf 66} (2002) 055211
  [arXiv:nucl-th/0207066].

\bibitem{Penner:2002md}
  G.~Penner and U.~Mosel,
  Phys.\ Rev.\  C {\bf 66} (2002) 055212
  [arXiv:nucl-th/0207069].

\bibitem{MuhlichDiss}
P.~M\"uhlich, Dissertation, Giessen University, 2007; \\
http://www.uni-giessen.de/cms/fbz/fb07/fachgebiete/physik/einrichtungen/\\
theorie/theorie1/publications/dissertation/muehlich\_diss/at\_download/file

\bibitem{Zschocke:2002mp}
  S.~Zschocke, O.~P.~Pavlenko and B.~Kampfer,
  Phys.\ Lett.\  B {\bf 562} (2003) 57
  [arXiv:hep-ph/0212201].

\bibitem{Saito:1998wd}
  K.~Saito, K.~Tsushima, A.~W.~Thomas and A.~G.~Williams,
  Phys.\ Lett.\  B {\bf 433} (1998) 243
  [arXiv:nucl-th/9804015].

\bibitem{Oset:2000eg}
  E.~Oset and A.~Ramos,
  Nucl.\ Phys.\  A {\bf 679} (2001) 616
  [arXiv:nucl-th/0005046].

\bibitem{Cabrera:2002hc}
  D.~Cabrera and M.~J.~Vicente Vacas,
  Phys.\ Rev.\  C {\bf 67}, 045203 (2003)
  [arXiv:nucl-th/0205075].

\bibitem{Hatsuda:1985eb}
  T.~Hatsuda and T.~Kunihiro,
  Phys.\ Rev.\ Lett.\  {\bf 55}, 158 (1985).

\bibitem{Bernard:1987im}
  V.~Bernard, U.~G.~Meissner and I.~Zahed,
  Phys.\ Rev.\ Lett.\  {\bf 59}, 966 (1987).

\bibitem{Bernard:1987sx}
  V.~Bernard and U.~G.~Meissner,
  Phys.\ Rev.\  D {\bf 38} (1988) 1551.

\bibitem{Hatsuda:1999kd}
  T.~Hatsuda, T.~Kunihiro and H.~Shimizu,
  Phys.\ Rev.\ Lett.\  {\bf 82} (1999) 2840.

\bibitem{Chiang:1997di}
  H.~C.~Chiang, E.~Oset and M.~J.~Vicente-Vacas,
  Nucl.\ Phys.\  A {\bf 644}, 77 (1998)
  [arXiv:nucl-th/9712047].

\bibitem{Rapp:1998fx}
  R.~Rapp {\it et al.},
  Phys.\ Rev.\  C {\bf 59}, 1237 (1999)
  [arXiv:nucl-th/9810007].

\bibitem{Bonutti:1999zz}
  F.~Bonutti {\it et al.},
  Phys.\ Rev.\  C {\bf 60} (1999) 018201.

\bibitem{Starostin:2000cb}
  A.~Starostin {\it et al.}  [Crystal Ball Collaboration],
  Phys.\ Rev.\ Lett.\  {\bf 85} (2000) 5539.

\bibitem{Messchendorp:2002au}
  J.~G.~Messchendorp {\it et al.},
  Phys.\ Rev.\ Lett.\  {\bf 89} (2002) 222302
  [arXiv:nucl-ex/0205009].

\bibitem{Chanfray:2004vb}
  G.~Chanfray, D.~Davesne, M.~Ericson and M.~Martini,
  Eur.\ Phys.\ J.\  A {\bf 27} (2006) 191
  [arXiv:nucl-th/0406003].

\bibitem{Kaskulov:2005kr}
  M.~M.~Kaskulov, E.~Oset and M.~J.~Vicente Vacas,
  Phys.\ Rev.\  C {\bf 73} (2006) 014004
  [arXiv:nucl-th/0506031].

\bibitem{Cabrera:2005wz}
  D.~Cabrera, E.~Oset and M.~J.~Vicente Vacas,
  Phys.\ Rev.\  C {\bf 72} (2005) 025207
  [arXiv:nucl-th/0503014].

\bibitem{Schenke:2005ry}
  B.~Schenke and C.~Greiner,
  Phys.\ Rev.\  C {\bf 73} (2006) 034909
  [arXiv:hep-ph/0509026].

\bibitem{Muhlich:2005kf}
  P.~Muhlich and U.~Mosel,
  Nucl.\ Phys.\  A {\bf 765} (2006) 188
  [arXiv:nucl-th/0510078].

\bibitem{Oset:1986yi}
  E.~Oset, Y.~Futami and H.~Toki,
  Nucl.\ Phys.\  A {\bf 448} (1986) 597.

\bibitem{Sibirtsev:2006yk}
  A.~Sibirtsev, H.~W.~Hammer, U.~G.~Meissner and A.~W.~Thomas,
  Eur.\ Phys.\ J.\  A {\bf 29} (2006) 209
  [arXiv:nucl-th/0606044].

\bibitem{Effenberger:1999jc}
  M.~Effenberger and U.~Mosel,
  Phys.\ Rev.\  C {\bf 62} (2000) 014605
  [arXiv:nucl-th/9908078].

\bibitem{Gallmeister:2007cm}
  K.~Gallmeister, M.~Kaskulov, U.~Mosel and P.~Muhlich,
  Prog.\ Part.\ Nucl.\ Phys.\  {\bf 61} (2008) 283
  [arXiv:0712.2200 [nucl-th]].

\bibitem{Cassing:1990dr}
  W.~Cassing, V.~Metag, U.~Mosel and K.~Niita,
  Phys.\ Rept.\  {\bf 188} (1990) 363.

\bibitem{GiBUU}
For details on GiBUU see: http://gibuu.physik.uni-giessen.de/GiBUU


\bibitem{vanHees:2009vk}
  H.~van Hees and R.~Rapp,
  arXiv:0901.2316 [nucl-th].


\bibitem{Mosel:1998rh}
  U.~Mosel,
  Prog.\ Part.\ Nucl.\ Phys.\  {\bf 42} (1999) 163
  [arXiv:nucl-th/9812067].

\bibitem{Mosel:1992rb}
  U.~Mosel,
  Ann.\ Rev.\ Nucl.\ Part.\ Sci.\  {\bf 41} (1991) 29.

\bibitem{Wood:2008ee}
  M.~H.~Wood {\it et al.}  [CLAS Collaboration],
  Phys.\ Rev.\  C {\bf 78} (2008) 015201
  [arXiv:0803.0492 [nucl-ex]].

\bibitem{Arnaldi:2008er}
  R.~Arnaldi {\it et al.}  [NA60 Collaboration],
  Eur.\ Phys.\ J.\  C {\bf 59} (2009) 607
  [arXiv:0810.3204 [nucl-ex]].

\bibitem{Specht_Damjano} S. Damjanovic and H.J. Specht, priv. communication 2009.

\bibitem{Toia:2008dj}
  A.~Toia,
  J.\ Phys.\ G {\bf 35} (2008) 104037
  [arXiv:0805.0153 [nucl-ex]].

\bibitem{Roche:1988er}
  G.~Roche {\it et al.}  [DLS Collaboration],
  Phys.\ Rev.\ Lett.\  {\bf 61} (1988) 1069.

\bibitem{Naudet:1988kj}
  C.~Naudet {\it et al.},
  Phys.\ Rev.\ Lett.\  {\bf 62} (1989) 2652.

\bibitem{Porter:1997rc}
  R.~J.~Porter {\it et al.}  [DLS Collaboration],
  Phys.\ Rev.\ Lett.\  {\bf 79} (1997) 1229
  [arXiv:nucl-ex/9703001].

\bibitem{Agakishiev:1995xb}
  G.~Agakishiev {\it et al.}  [CERES Collaboration],
  Phys.\ Rev.\ Lett.\  {\bf 75} (1995) 1272.

\bibitem{Masera:1995ck}
  M.~Masera  [HELIOS Collaboration],
  Nucl.\ Phys.\  A {\bf 590} (1995) 93C.


\bibitem{Bratkovskaya:1997mp}
  E.~L.~Bratkovskaya, W.~Cassing, R.~Rapp and J.~Wambach,
  Nucl.\ Phys.\  A {\bf 634} (1998) 168
  [arXiv:nucl-th/9710043].

\bibitem{Agakichiev:2005ai}
  G.~Agakichiev {\it et al.}  [CERES Collaboration],
  Eur.\ Phys.\ J.\  C {\bf 41} (2005) 475
  [arXiv:nucl-ex/0506002].

\bibitem{Agakishiev:1998mv}
  G.~Agakishiev {\it et al.},
  Eur.\ Phys.\ J.\  C {\bf 4} (1998) 231.

\bibitem{Adamova:2006nu}
  D.~Adamova {\it et al.},
  Phys.\ Lett.\  B {\bf 666} (2008) 425
  [arXiv:nucl-ex/0611022].

\bibitem{Marin:2004fx}
  A.~Marin  [CERES Collaboration],
  J.\ Phys.\ G {\bf 30} (2004) S709
  [arXiv:nucl-ex/0406007].


\bibitem{Adamova:2002kf}
  D.~Adamova {\it et al.}  [CERES/NA45 Collaboration],
  Phys.\ Rev.\ Lett.\  {\bf 91} (2003) 042301
  [arXiv:nucl-ex/0209024].

\bibitem{Arnaldi:2006jq}
  R.~Arnaldi {\it et al.}  [NA60 Collaboration],
  Phys.\ Rev.\ Lett.\  {\bf 96} (2006) 162302
  [arXiv:nucl-ex/0605007].


\bibitem{Arnaldi:2008fw}
  R.~Arnaldi {\it et al.}  [NA60 Collaboration],
  arXiv:0812.3053 [nucl-ex].

\bibitem{vanHees:2006ng}
  H.~van Hees and R.~Rapp,
  Phys.\ Rev.\ Lett.\  {\bf 97} (2006) 102301
  [arXiv:hep-ph/0603084].

\bibitem{Heinz:1990jw}
  U.~W.~Heinz and K.~S.~Lee,
  Phys.\ Lett.\  B {\bf 259} (1991) 162.


\bibitem{vanHees:2006iv}
  H.~van Hees and R.~Rapp,
  arXiv:hep-ph/0604269.

\bibitem{Damjanovic:2008ta}
  S.~Damjanovic  [NA60 Collaboration],
  J.\ Phys.\ G {\bf 35} (2008) 104036
  [arXiv:0805.4153 [nucl-ex]].


\bibitem{Ruppert:2007cr}
  J.~Ruppert, C.~Gale, T.~Renk, P.~Lichard and J.~I.~Kapusta,
  Phys.\ Rev.\ Lett.\  {\bf 100} (2008) 162301
  [arXiv:0706.1934 [hep-ph]].


\bibitem{Damjanovic:2007qm}
  S.~Damjanovic {\it et al.}  [NA60 Collaboration],
  Nucl.\ Phys.\  A {\bf 783} (2007) 327
  [arXiv:nucl-ex/0701015].


\bibitem{Alvensleben:1970uw}
  H.~Alvensleben {\it et al.},
  Phys.\ Rev.\ Lett.\  {\bf 24} (1970) 786.
  
\bibitem{Alvensleben:1971rq}
  H.~Alvensleben {\it et al.},
  Nucl.\ Phys.\  B {\bf 25} (1971) 342.


\bibitem{Effenberger:1999ay}
  M.~Effenberger, E.~L.~Bratkovskaya and U.~Mosel,
  Phys.\ Rev.\  C {\bf 60} (1999) 044614
  [arXiv:nucl-th/9903026].

\bibitem{:2007mga}
  R.~Nasseripour {\it et al.}  [CLAS Collaboration],
  Phys.\ Rev.\ Lett.\  {\bf 99} (2007) 262302
  [arXiv:0707.2324 [nucl-ex]].


\bibitem{Mecking:2003zu}
  B.~A.~Mecking {\it et al.}  [CLAS Collaboration],
  Nucl.\ Instrum.\ Meth.\  A {\bf 503} (2003) 513.

\bibitem{Ozawa:2000iw}
  K.~Ozawa {\it et al.}  [E325 Collaboration],
  Phys.\ Rev.\ Lett.\  {\bf 86} (2001) 5019
  [arXiv:nucl-ex/0011013].

\bibitem{Tabaru:2006az}
  T.~Tabaru {\it et al.},
  Phys.\ Rev.\  C {\bf 74} (2006) 025201
  [arXiv:nucl-ex/0603013].

\bibitem{Muto_priv} R. Muto, priv. communication 2009.

\bibitem{Naruki:2005kd}
  M.~Naruki {\it et al.},
  Phys.\ Rev.\ Lett.\  {\bf 96} (2006) 092301
  [arXiv:nucl-ex/0504016].

\bibitem{Wu:2005wf}
  C.~Wu {\it et al.},
  Eur.\ Phys.\ J.\  A {\bf 23} (2005) 317.


\bibitem{Damjanovic:2006bd}
  S.~Damjanovic,
  Eur.\ Phys.\ J.\  C {\bf 49} (2007) 235
  [arXiv:nucl-ex/0609026].


\bibitem{Messchendorp:2001pa}
  J.~G.~Messchendorp, A.~Sibirtsev, W.~Cassing, V.~Metag and S.~Schadmand,
  Eur.\ Phys.\ J.\  A {\bf 11} (2001) 95
  [arXiv:hep-ex/0106041].

\bibitem{Muhlich:2003tj}
  P.~Muhlich, T.~Falter and U.~Mosel,
  Eur.\ Phys.\ J.\  A {\bf 20} (2004) 499
  [arXiv:nucl-th/0310067].

\bibitem{Kaskulov:2006zc}
  M.~Kaskulov, E.~Hernandez and E.~Oset,
  Eur.\ Phys.\ J.\  A {\bf 31} (2007) 245
  [arXiv:nucl-th/0610067].

\bibitem{Trnka:2005ey}
  D.~Trnka {\it et al.}  [CBELSA/TAPS Collaboration],
  Phys.\ Rev.\ Lett.\  {\bf 94} (2005) 192303
  [arXiv:nucl-ex/0504010].

\bibitem{Metag:2007zz}
  V.~Metag,
  Prog.\ Theor.\ Phys.\ Suppl.\  {\bf 168} (2007) 503.

\bibitem{Nanova} M. Nanova, proceedings Eur. Phys. Soc. Nucl. Phys. conference, Bochum, 2009.


\bibitem{Muhlich:2006ps}
  P.~Muhlich and U.~Mosel,
  Nucl.\ Phys.\  A {\bf 773} (2006) 156
  [arXiv:nucl-th/0602054].

\bibitem{:2008xy}
  M.~Kotulla {\it et al.}  [CBELSA/TAPS Collaboration],
  Phys.\ Rev.\ Lett.\  {\bf 100} (2008) 192302
  [arXiv:0802.0989 [nucl-ex]].


\bibitem{Muhlich:2002tu}
  P.~Muhlich, T.~Falter, C.~Greiner, J.~Lehr, M.~Post and U.~Mosel,
  Phys.\ Rev.\  C {\bf 67} (2003) 024605
  [arXiv:nucl-th/0210079].

\bibitem{Behrend:1970} H.-J.\ Behrend {\it et al.}, Phys.\ Rev.\ Lett.\  {\bf 24}
  (1970) 1246.

\bibitem{Muto:2005za}
  R.~Muto {\it et al.}  [KEK-PS-E325 Collaboration],
  Phys.\ Rev.\ Lett.\  {\bf 98} (2007) 042501
  [arXiv:nucl-ex/0511019].


\bibitem{Ishikawa:2004id}
  T.~Ishikawa {\it et al.},
  Phys.\ Lett.\  B {\bf 608} (2005) 215
  [arXiv:nucl-ex/0411016].


\bibitem{Magas:2004ui}
  V.~K.~Magas, L.~Roca and E.~Oset,
  Nucl.\ Phys.\  A {\bf 755} (2005) 495
  [arXiv:nucl-th/0412066].

\bibitem{Cabrera:2003wb}
  D.~Cabrera, L.~Roca, E.~Oset, H.~Toki and M.~J.~Vicente Vacas,
  Nucl.\ Phys.\  A {\bf 733} (2004) 130
  [arXiv:nucl-th/0310054].


\bibitem{Adcox:2003zm}
  K.~Adcox {\it et al.}  [PHENIX Collaboration],
  Nucl.\ Instrum.\ Meth.\  A {\bf 499} (2003) 469.

\bibitem{Toia_priv} A. Toia, priv. communication 2009.

\bibitem{:2008asa}
  A.~Adare {\it et al.}  [PHENIX Collaboration],
  Phys.\ Lett.\  B {\bf 670}, 313 (2009)
  [arXiv:0802.0050 [hep-ex]].

\bibitem{:2007xw}
  S.~Afanasiev {\it et al.}  [PHENIX Collaboration],
  arXiv:0706.3034 [nucl-ex].

\bibitem{Milov:2008dd}
  A.~Milov,
  arXiv:0809.3880 [nucl-ex].


\bibitem{Agakishiev:2007ts}
  G.~Agakishiev {\it et al.}  [HADES Collaboration],
  Phys.\ Lett.\  B {\bf 663} (2008) 43
  [arXiv:0711.4281 [nucl-ex]].


\bibitem{Agakishiev:2009am}
  G.~Agakishiev {\it et al.}  [HADES Collaboration],
  Eur.\ Phys.\ J.\  A {\bf 41} (2009) 243
  [arXiv:0902.3478 [nucl-ex]].

\bibitem{Shyam:2003cn}
  R.~Shyam and U.~Mosel,
  Phys.\ Rev.\  C {\bf 67} (2003) 065202
  [arXiv:hep-ph/0303035].

\bibitem{Kaptari:2005qz}
  L.~P.~Kaptari and B.~Kampfer,
  Nucl.\ Phys.\  A {\bf 764} (2006) 338
  [arXiv:nucl-th/0504072].

\bibitem{Bratkovskaya:2007jk}
  E.~L.~Bratkovskaya and W.~Cassing,
  Nucl.\ Phys.\  A {\bf 807} (2008) 214
  [arXiv:0712.0635 [nucl-th]].

\bibitem{Metag:2007zza}
  V.~Metag,
  J.\ Phys.\ G {\bf 34} (2007) S397.


\bibitem{Bonutti:1996ij}
  F.~Bonutti {\it et al.}  [CHAOS Collaboration],
  Phys.\ Rev.\ Lett.\  {\bf 77} (1996) 603.


\bibitem{VicenteVacas:1999xx}
  M.~J.~Vicente Vacas and E.~Oset,
  Phys.\ Rev.\  C {\bf 60} (1999) 064621
  [arXiv:nucl-th/9907008].

\bibitem{Muhlich:2004zj}
  P.~Muhlich, L.~Alvarez-Ruso, O.~Buss and U.~Mosel,
  Phys.\ Lett.\  B {\bf 595} (2004) 216
  [arXiv:nucl-th/0401042].

\bibitem{Buss:2006vh}
  O.~Buss, L.~Alvarez-Ruso, P.~Muhlich and U.~Mosel,
  Eur.\ Phys.\ J.\  A {\bf 29} (2006) 189
  [arXiv:nucl-th/0603003].


\bibitem{Krusche}
 B. Krusche, 
  Int.\ J.\ Mod.\ Phys.\ A {\bf 22} (2007) 406
  [arXiv:nucl-ex/0608044].

\bibitem{Bloch:2007ka}
  F.~Bloch {\it et al.},
  Eur.\ Phys.\ J.\  A {\bf 32} (2007) 219
  [arXiv:nucl-ex/0703037].

\bibitem{Gregor} R. Gregor, Dissertation, Giessen University, 2007,
and priv. communication 2009.







\end{thebibliography}
\end{document}